\documentclass[a4paper,notitlepage,eqsecnum,nofootinbib]{revtex4-1}
\usepackage{slashed,bbold,bm,amsmath,amssymb,mathtools,wrapfig,epsfig} 
\usepackage[usenames]{color}
\usepackage{hyperref}
\usepackage{tcolorbox}
\hypersetup{
    colorlinks=true,       % false: boxed links; true: colored links
    linkcolor=red,          % color of internal links (change box color with linkbordercolor)
    citecolor=magenta,        % color of links to bibliography
    filecolor=green,      % color of file links
    urlcolor=blue           % color of external links
}

\newcommand{\crt}{\\[2mm]}
\newcommand{\nn}{\nonumber}
\newcommand{\beq} {\begin{equation}}
\newcommand{\eeq} {\end{equation}}
\newcommand{\beqa} {\begin{eqnarray}}
\newcommand{\eeqa} {\end{eqnarray}}

\newcommand{\mrm}[1] {{\mathrm{#1}}}

\newcommand{\bs}[1]{\boldsymbol{#1}}

\newcommand{\cf}{{\it cf.}}
\newcommand{\ie}{{\it i.e.}}

\newcommand{\eg}{{\it e.g.}}
\newcommand{\etal}{{\it et al.}}

\newcommand{\gev}{{\mrm{GeV}}}
\newcommand{\as}{{\alpha_s}}
\newcommand{\lqcd}{\Lambda_{QCD}}
\newcommand{\la}{\Lambda}
\newcommand{\lm}{\lambda}

\newcommand{\ieps}{i\varepsilon}
\newcommand{\vphi}{\varphi}
\newcommand{\veps}{\varepsilon}
\newcommand{\order}[1]{${\cal O}\left(#1 \right)$}
\newcommand{\morder}[1]{{\cal O}\left(#1 \right)}
\newcommand{\eq}[1]{(\ref{#1})}
\newcommand{\fig}[1]{Fig.~\ref{#1}}
\newcommand{\lsim}{\lesssim}   % \alt
\newcommand{\gsim}{\gtrsim}   % \agt
\newcommand{\ovl}{\overline}

\newcommand{\inv}[1]{\frac{1}{#1}}
\newcommand{\halft}{{\textstyle \frac{1}{2}}}
\newcommand{\quart}{{\textstyle \frac{1}{4}}}
\newcommand{\sfrac}[2]{{\textstyle\frac{#1}{#2}}}
\newcommand{\intt}{{\textstyle \int}}
\newcommand{\ket}[1]{\left\vert{#1}\right\rangle}
\newcommand{\bra}[1]{\langle{#1}\vert}
\newcommand{\ave}[1]{\langle{#1}\rangle}
\newcommand{\com}[2]{\left[{#1},{#2}\right]}
\newcommand{\comb}[2]{\big[{#1},{#2}\big]}
\newcommand{\acom}[2]{\left\{{#1},{#2}\right\}}
\newcommand{\acomb}[2]{\big\{{#1},{#2}\big\}}
\newcommand{\tr}{\mathrm{Tr}\,}

\newcommand{\im}{{\rm Im}}
\newcommand{\re}{{\rm Re}}

\newcommand{\psl}{{\slashed{p}}}

\newcommand{\Asl}{{\slashed{A}}}

\newcommand{\lsl}{{\slashed{\ell}}}

\newcommand{\mB}{\mathcal{B}}
\newcommand{\mC}{\mathcal{C}}
\newcommand{\bC}{\mathbb{C}}
\newcommand{\mD}{\mathcal{D}}
\newcommand{\mE}{\mathcal{E}}

\newcommand{\mH}{\mathcal{H}}
\newcommand{\mHV}{\mathcal{H}_V}
\newcommand{\mJ}{\mathcal{J}}
\newcommand{\mK}{\mathcal{K}}
\newcommand{\mL}{\mathcal{L}}
\newcommand{\mM}{\mathcal{M}}
\newcommand{\mO}{\mathcal{O}}
\newcommand{\mP}{\mathcal{P}}
\newcommand{\bP}{\mathbb{P}}
\newcommand{\mS}{\mathcal{S}}

\newcommand{\xv}{{\bs{x}}}
\newcommand{\xvh}{{\bs{\hat{\xv}}}}
\newcommand{\yv}{{\bs{y}}}
\newcommand{\zv}{{\bs{z}}}
\newcommand{\rv}{{\bs{r}}}
\newcommand{\pv}{{\bs{p}}}
\newcommand{\kv}{{\bs{k}}}
\newcommand{\qv}{{\bs{q}}}

\newcommand{\dv}{{\bs{d}}}
\newcommand{\ev}{{\bs{e}}}
\newcommand{\Av}{{\bs{A}}}

\newcommand{\Ev}{{\bs{E}}}
\newcommand{\Jv}{{\bs{J}}}

\newcommand{\Lv}{{\bs{L}}}
\newcommand{\Pv}{{\bs{P}}}
\newcommand{\Pva}{{\bs{P}_A}}
\newcommand{\Pvb}{{\bs{P}_B}}
\newcommand{\Rv}{{\bs{R}}}
\newcommand{\Sv}{{\bs{S}}}
\newcommand{\ellv}{{\bs{\ell}}}
\newcommand{\sv}{\bs{\sigma}}
\newcommand{\gv}{\bs{\gamma}}
\newcommand{\delv}{\bs{\delta}}
\newcommand{\vepsv}{{\bs{\varepsilon}}}
\newcommand{\smu}{s}
\newcommand{\gz}{\gamma^0}
\newcommand{\go}{\gamma^1}

\newcommand{\gf}{\gamma_5}

\newcommand{\kum}{{\,{_1}F_1}}
\newcommand{\xbj}{{x_{Bj}}}
\newcommand{\wt}{\widetilde}

\newcommand{\rar}{\rightarrow}
\newcommand{\lar}{\leftarrow}
\newcommand{\lra}{\leftrightarrow}
\newcommand{\rder}{{\buildrel\rar\over{\partial}}}
\newcommand{\lder}{{\buildrel\lar\over{\partial}}}

\newcommand{\nv}{\bs{\nabla}}
\newcommand{\ntr}{\bs{\nabla}_\perp}
\newcommand{\rnab}{{\overset{\rar}{\nv}}\strut}
\newcommand{\lnab}{{\overset{\lar}{\nv}}\strut}
\newcommand{\rorb}{{\overset{\rar}{\bs{L}}}\strut}
\newcommand{\lorb}{{\overset{\lar}{\bs{L}}}\strut}
\newcommand{\rla}{{\overset{\rar}{\la}}\strut}  % \roarrow \la
\newcommand{\lla}{{\overset{\lar}{\la}}\strut}  % \loarrow \la

\newcommand{\lh}{{\overset{\lar}{\mathfrak{h}}}\strut}
\newcommand{\rh}{{\overset{\rar}{\mathfrak{h}}}\strut}

\newcommand{\xtr}{\bs{x}_\perp}

\newcommand{\atr}{{\bs{\alpha}_\perp}}
\newcommand{\alv}{{\bs{\alpha}}}

\newcommand{\aly}{{\alpha_2}}
\newcommand{\alz}{{\alpha_3}}

\newcommand{\wfr}{\Phi^{(0)}}

\newcommand{\phip}{\phi^{\scriptscriptstyle{(P)}}}
\newcommand{\Phip}{\Phi^{(P)}}
\newcommand{\Phipb}{\Phi_{\mathcal{B}}^{(\Pv)}}
\newcommand{\Npb}{N_{\mathcal{B}}^{(P)}}
\newcommand{\Phirb}{\Phi_{\mathcal{B}}^{(0)}}
\newcommand{\Fp}{F^{(\Pv)}}
\newcommand{\phir}{\phi^{\scriptscriptstyle(0)}}

\newcommand{\sph}{{Y_{j\lambda}}}

\begin{document}
%{\par\raggedleft \texttt{\vspace{-.3cm} 16 January 2021}\par} \bigskip{}

\title{Journey to the Bound States\footnote{
\copyright\ The Author, under exclusive license to Springer Nature Switzerland AG 2021.\\
P. Hoyer, Journey to the Bound States, SpringerBriefs in Physics. \\
Expanded version of lectures presented at the University of Pavia in January 2020. \\ Slides are available at https://www.mv.helsinki.fi/home/hoyer/Talks.html}}

\author{Paul Hoyer}
\affiliation{ \vspace{1mm} Department of Physics, POB 64, FIN-00014 University of Helsinki, Finland}
\email{paul.hoyer@helsinki.fi}

\begin{abstract} 

Guided by the observed properties of hadrons I formulate a perturbative bound state method for QED and QCD. The expansion starts with valence Fock states ($e^+e^-,\ q\bar q,\ qqq,\ gg$) bound by the instantaneous interaction of temporal gauge ($A^0=0$). The method is tested on Positronium atoms at rest and in motion, including hyperfine splitting at \order{\alpha^4}, electromagnetic form factors and deep inelastic scattering. Relativistic binding is studied for QED in $D=1+1$ dimensions, demonstrating the frame independence of the DIS electron distribution and its sea for $\xbj \to 0$. In QCD a homogeneous solution of Gauss' constraint in $D=3+1$ implies \order{\alpha_s^0} confining potentials for $q\bar q,\ q\bar qg,\ qqq$ and $gg$ states, whereas $q\bar q\,q\bar q$ is unconfined. Meson states lie on linear Regge trajectories and have the required frame dependence. A scalar bound state with vanishing four-momentum causes spontaneous chiral symmetry breaking when mixed with the vacuum.

 These lecture notes assume knowledge of field theory methods, but not of bound states. Brief reviews of existing bound state methods and Dirac electron states are included. Solutions to the exercises are given in the Appendix.

\end{abstract}

\maketitle

\vspace{-.5cm}
\tableofcontents

\parindent 0cm
\vspace{-.2cm}

\section{Motivations and Outline} \label{secI}
%%%%%%%%%%%%%%%%%%%%%%%%%%

%%%%%%%%%%%%%%%%%%%%%%%%%%%
\subsection{Motivations} \label{secI.A}
%%%%%%%%%%%%%%%%%%%%%%%%%%%

Hadrons differ qualitatively from atoms due to their relativistic binding, confinement and spontaneous chiral symmetry breaking. Deep inelastic scattering shows that hadrons have significant sea quark and gluon constituents. Nevertheless, the hadron spectrum has atomic features, with quantum numbers determined by the valence quarks only. Nuclei are multiquark states analogous to molecules, being comparatively loosely bound states of nucleons. The more recently discovered heavy multi-quark ($X,Y,Z$) states tend to be associated with hadron thresholds, as would be expected for weakly bound states of hadrons.

These properties should emerge in a correct description of hadrons as QCD bound states. Many aspects have indeed been confirmed by numerical studies, see the reviews of  lattice QCD by the Particle Data Group \cite{Zyla:2020zbs} and FLAG \cite{Aoki:2019cca}. Valuable insights have been obtained also through studies of models, especially the quark model \cite{Zyla:2020zbs}. The QCD2019 Workshop Summary \cite{Brodsky:2020vco} gives an overview of the experimental and theoretical status of hadron physics. We still lack even a qualitative understanding of main aspects, \eg, why the degrees of freedom of the non-valence constituents are not manifest in the hadron spectrum. The contrast to the dense excitation spectrum of nuclei due to rotational and vibrational modes is striking.

The observed, puzzling similarities between hadrons and atoms allows us to benefit from the understanding of QED bound states, which gradually emerged since the beginnings of quantum field theory. Unfortunately, the fields of QED and QCD bound states have grown apart \cite{Blum:2017diy}. Modern textbooks on the applications of QFT to particle physics hardly mention bound states. There are seemingly solid reasons to believe that QED methods are irrelevant for hadrons. My lectures are motivated by a concern that this conclusion might be premature. Let me briefly indicate why I do not find some of the common arguments completely conclusive.

\paragraph{Hadrons are non-perturbative, whereas QED is perturbative.}\ \\ 
In their review of QED bound state calculations Bodwin \etal\ \cite{RevModPhys.57.723} remark that \textit{``precision bound-state calculations are essentially nonperturbative''}. Perturbation theory for atoms needs to expand around an approximate bound state, whose wave function is necessarily non-polynomial in $\alpha$. This means that there is no unique perturbative expansion for bound states, since a polynomial in $\alpha$ may be shifted between the initial wave function and the higher order terms. Measurable quantities, such as the binding energy, have nevertheless unique expansions, being independent of the initial wave function. For example, the hyperfine splitting between Orthopositronium ($J^{PC}=1^{--}$) and Parapositronium ($0^{-+}$) is impressively known up to \order{\alpha^7} corrections, and is in agreement with accurate data \cite{Murota:1988hr,Penin:2014bea,Adkins:2015wya,Adkins:2018lvj}
\begin{align} \label{mo1}
\frac{\Delta E_b}{m_e} =& \frac{7}{12}\alpha^4 - \Big(\frac{8}{9}+\frac{\ln2}{2}\Big)\frac{\alpha^5}{\pi}-\frac{5}{24}\alpha^6\ln\alpha +\bigg[\frac{1367}{648}-\frac{5197}{3456}\pi^2 +\Big(\frac{221}{144}\pi^2+\inv{2}\Big)\ln2-\frac{53}{32}\zeta(3)\bigg]\frac{\alpha^6}{\pi^2} \nn\crt
&-\frac{7\alpha^7}{8\pi}\ln^2\alpha + \Big(\frac{17}{3}\ln2-\frac{217}{90}\Big)\frac{\alpha^7}{\pi}\ln\alpha +\morder{\alpha^7}
\end{align}
The freedom of choice of the initial wave function was used in the evaluation of the higher order terms, by expanding around states given by the Schr\"odinger equation. Analogously, hadrons may have a perturbative expansion based on an initial state that incorporates the relevant features, including confinement.

\paragraph{Confinement requires a scale $\lqcd$, which can arise only from renormalization.}\ \\ 
The form of the classical atomic potential, $V(r) = -\alpha/r$, follows from $\alpha$ being dimensionless ($\hbar=c=1$). A confining potential requires a parameter with dimension (the confinement scale), but the QCD action has no such parameter.
The properties of heavy quarkonia are well described by
the Schr\"odinger equation with the ``Cornell potential'' \cite{Eichten:1979ms,Eichten:2007qx},
\begin{align} \label{mo3}
V(r) = V'r-\frac{4}{3}\frac{\as}{r} \ \ \ \text{with}\ \ V' \simeq 0.18\ \text{GeV}^2, \ \ \as \simeq 0.39
\end{align}
This suggests that $V'r$, similarly to the $1/r$ potential, should be determined by Gauss' law.
In section \ref{secIV.C2} I consider a homogeneous solution of Gauss' law that has a spatially constant color field energy density. It gives the classical $q\bar q$ potential \eq{mo3}, with $V'$ determined by the energy density. This and the corresponding instantaneous potentials for $qqq,\ q\bar qg$ and other Fock states are derived in section \ref{secVII.A}.

\paragraph{The QCD coupling $\as(Q)$ is large at low scales Q, excluding a perturbative expansion.}\ \\ 
Standard perturbative determinations of $\as(Q)$ are restricted to $Q \gsim m_\tau = 1.78$ GeV, with $\alpha_s^{\ovl{MS}}(m_\tau) \simeq 0.33$ \cite{Zyla:2020zbs}. Since $\as$ is not directly measurable its value at low $Q$ depends on the theoretical framework. 
A dispersive approach indicates that $\as(0) \simeq 0.5$ (\cite{Gehrmann:2012sc}, section \ref{secIIa4}). Due to confinement no low momentum (IR) singularities are expected in loop integrals. 
Thus $\as$ may freeze at low scales, allowing a perturbative expansion. Strong binding is then due to the confining potential $V'r$ in \eq{mo3}, not to the Coulomb potential $\propto\as$.

The above observations, together with other theoretical and phenomenological arguments \cite{tHooft:2002pmx,Dokshitzer:2003bt,Dokshitzer:2004ie,Dokshitzer:2010zza} prompt me to consider the possibility of a perturbative expansion for QCD bound states. Perturbation theory is our main analytic tool in the Standard Model, and merits careful consideration. Bound states are interesting in their own right, providing insights into the structure of QFT which are complementary to those of scattering. QED methods for atoms have been developed over a long time, and may now be close to optimal. A perturbative approach to hadrons raises issues which so far have received little attention.

\subsection{Outline} \label{secI.B}
%%%%%%%%%%%%%%%%%%%%%%%%%%

To help the reader navigate through these fairly extensive lectures I provide here brief characterizations of the various chapters and sections. Those marked with a star * may be skipped in a first reading. Students are welcome to try the exercises, whose solutions are given in Appendix A.

\textbf{Chapter} \ref{secIIa} summarizes features of hadron dynamics which motivate the bound state approach of these lectures. Heavy quarkonia have atomic characteristics, are nearly non-relativistic yet display confinement (\ref{secIIa1}). Regge behavior and duality reveal a close connection between high energy scattering and bound states (\ref{secIIa2}). The $Q^2$-dependence of Deep Inelastic Scattering distinguishes partons that are are intrinsic to the hadron from those that are created by the hard scattering (\ref{secIIa3}). Data is consistent with a QCD coupling $\as(Q^2)$ which ``freezes'' at low scales (\ref{secIIa4}).

\textbf{Chapter} \ref{secII} is a brief survey of established QED bound state methods. The reduction of a non-relativistic two-particle bound state to one particle in a central potential is recalled (\ref{secII.A}). The Schr\"odinger equation is derived by summing Feynman ladder diagrams (\ref{secII.B}). The derivation of the Bethe-Salpeter bound state equation is sketched, and its non-uniqueness noted (\ref{secII.C}). The non-relativistic effective field theory method NRQED is introduced (\ref{secII.D}). The corresponding heavy quark effective theories HQET, NRQCD and pNRQCD are noted (\ref{secII.E}).

\textbf{Chapter} \ref{secIII} covers aspects of the Dirac bound states (\ref{secIII.A}).
Time-ordered Feynman $Z$-diagrams give rise to virtual pairs (\ref{secIII.B}). A Bogoliubov transformation of the free creation and annihilation operators allows to define the Dirac state as a single fermion state (\ref{secIII.C}). The Dirac wave functions for central $A^0(r)$ potentials are given in terms of radial and angular functions (\ref{secIII.D}), and the explicit example of the Coulomb potential worked out (\ref{secIII.E}). The case of a linear potential, for which the spectrum is continuous, is considered in section \ref{secIII.F}.

\textbf{Chapter} \ref{secIV} motivates and defines the approach to bound states used here: Quantization at equal time in temporal $(A^0=0)$ gauge. A perturbative expansion around valence Fock states, bound by the instantaneous gauge potential (\ref{secIV.A}). Comparison of quantization procedures using covariant, Coulomb and temporal gauges in QED (\ref{secIV.B}). The vanishing of the color octet electric field $\Ev_L^a$ for color singlet states allows to include a homogeneous solution of Gauss' constraint for each color component of the state. The boundary condition introduces a universal constant $\la$ (\ref{secIV.C}).

\textbf{Chapter} \ref{secV} applies the bound state method to Positronium atoms. The states and wave functions are defined (\ref{secV.A}). The Schr\"odinger equation is derived for atoms at rest (\ref{secV.B}) and in a general frame (\ref{secV.C}). The \order{\alpha^4} Hyperfine splitting between Ortho- and Parapositronia is evaluated in section \ref{secV.D}. The Poincare covariance of the Positronium form factor is demonstrated (\ref{secV.E}), as well as that of deep inelastic scattering on Parapositronium (\ref{secV.F}).

\textbf{Chapter} \ref{secVI} considers $e^-e^+$ bound states in $D=1+1$ dimensions (QED$_2$). The bound state equation is solved analytically in a general frame (\ref{secVI.A}). The gauge and Lorentz invariance of electromagnetic form factors is verified (\ref{secVI.B}). The electron distribution given by deep inelastic scattering is numerically evaluated in the rest frame of the target and shown to agree with an earlier result in the Breit frame. DIS has a sea contribution for $\xbj \to 0$ (\ref{secVI.C}).

\textbf{Chapter} \ref{secVII} applies the bound state method to QCD hadrons. 
The instantaneous \order{\alpha_s^0} potential due to the homogeneous solution of Gauss' constraint is evaluated for several Fock states ($q\bar q,\ qqq,\ gg,\ q\bar qg$ and $q\bar q\,q\bar q$) (\ref{secVII.A}). The wave functions of $q\bar q$ states in the rest frame are determined for all $J^{PC}$ quantum numbers (\ref{secVII.B}). The bound state equation for states with general momentum $\Pv \neq 0$ is formulated (\ref{secVII.C}). The states lie on nearly linear Regge trajectories with parallel daughter trajectories. Highly excited states have a non-vanishing overlap with multi-hadron states, with features that are consistent with the parton model, string breaking and duality (\ref{secVII.D}). The glueball ($gg$) spectrum has features similar to that of $q\bar q$ mesons (\ref{secVII.E}). There is a massless $0^{++}$ $q\bar q$ state which has vanishing four-momentum in all frames. It may mix with the vacuum without violating Poincar\'e invariance, giving rise to a spontaneous breaking of chiral symmetry (\ref{secVII.F}).

\textbf{Chapter} \ref{secVIII} is a recapitulation and discussion of the principles followed in these lectues. Experienced readers may profit from reading this chapter before the more technical parts.

\section{Features of hadrons} \label{secIIa}
%%%%%%%%%%%%%%%%%%%%%%%%%%

The approach to QCD bound states presented here is guided by experimental information and its interpretation in models. Hadrons have properties that could not have been anticipated by our experience with QED. The quark and gluon constituents are strongly bound into color singlets. Colored states apparently have infinite excitation energies. An abundance of sea quarks and gluons in the nucleon has been revealed by deep inelastic scattering.

An approximation scheme for QCD bound states should, even at lowest order, be compatible with the general features of hadron dynamics, including confinement, linear Regge trajectories and duality. Crucially, Nature has provided us with heavy (charm and bottom) quarks whose bound states (quarkonia) are approximately non-relativistic. Quarkonia reveal features of confinement without the added complication of relativistic binding.

In this chapter I briefly review some central features of hadrons and the descriptive models they have inspired.

\subsection{Quarkonia} \label{secIIa1}
%%%%%%%%%%%%%%%%%%%%%%%%%% 

\begin{figure}[h] \centering
\includegraphics[width=1.\columnwidth]{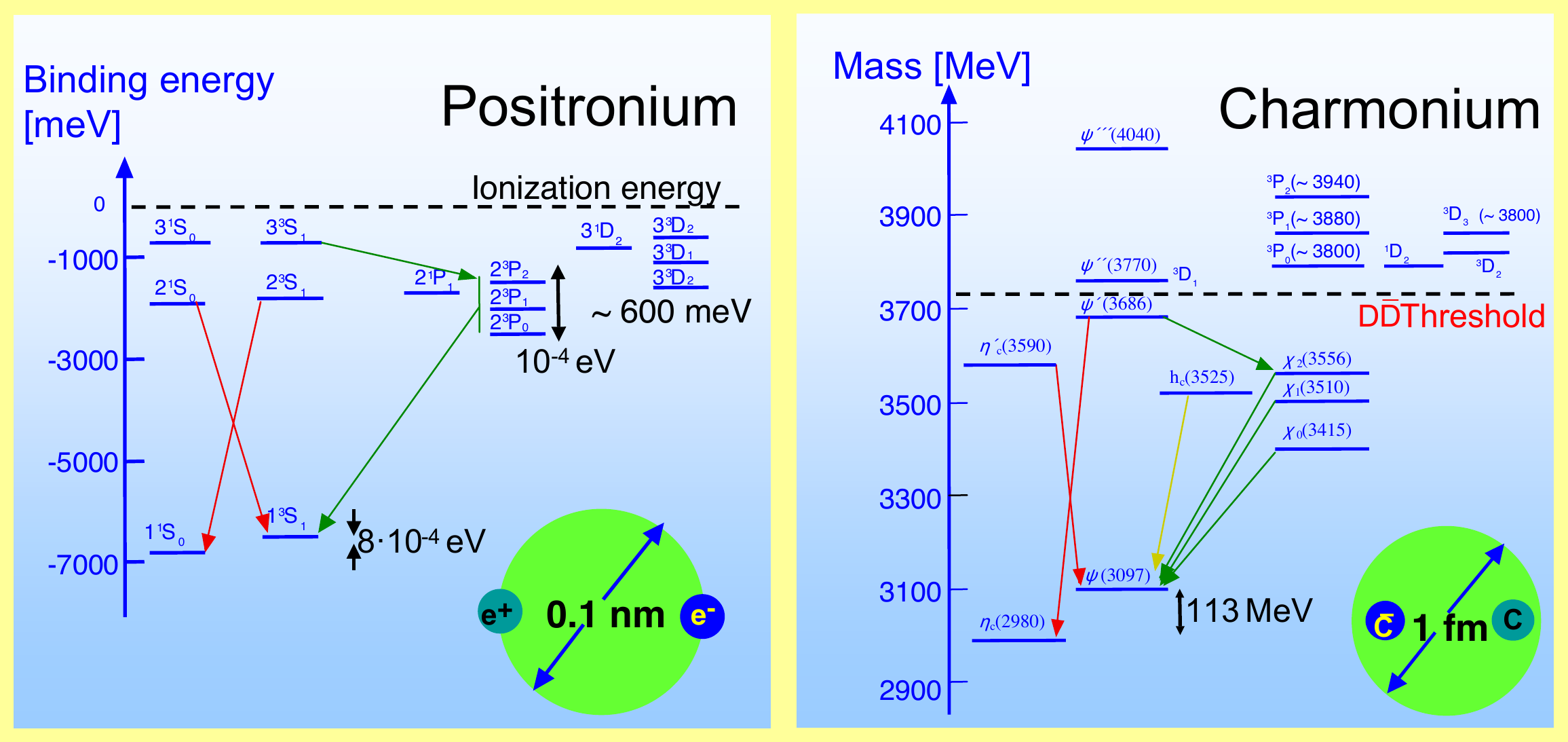}
\caption{Comparison of Positronium and Charmonium $n^{\,2S+1}L_J$ states and transitions. Notice the \order{10^{11}} difference in the hyperfine mass splitting $M(1^{\,3}S_1) - M(1^{\,1}S_0)$. \label{fpc}}
\end{figure}

I refer to \cite{Eichten:2007qx} for a review of quarkonium phenomenology. The charm quark mass $m_c \sim 1.5$ GeV is larger than the confinement scale indicated by the nucleon radius, $1\ \rm{fm}^{-1} \simeq 200$ MeV. Charmonium ($c\bar c$) bound states are nearly non-relativistic, with average constituent velocities $\ave{v^2} \simeq 0.24$. As seen in \fig{fpc} the spectrum is qualitatively similar to that of Positronium ($e^-e^+$) atoms, although with mass splittings differing in scale by up to $10^{11}$. This motivated studies of charmonia based on the Schr\"odinger equation. The short distance potential was expected to be given by single gluon exchange, $V_1(r)=-\sfrac{4}{3}\as/r$. The data constrained the confining part of the potential (in the relevant range of $r$) to be close to linear, $V_0(r) \simeq V'r$. This led to the Cornell potential $V(r) = V_0(r)+V_1(r)$ \eq{mo3}.

The phenomenology based on the Cornell potential turned out to be successful. Not only the mass splittings but also the many transitions (electromagnetic via photons as well as strong via gluons) are fairly described. The early hope that the ``The $J/\psi$ is the Hydrogen atom of QCD'' has to a large extent been fulfilled.

%%%%%%%%%%%%%%%%%%%%%%%%%%%%%%%%%%%%%%%%%%%%%%%%%%%%%%%%
%%%%%%%%%%%%%%%%%%%%%%%%%%%%%%%%%%%%%%%%%%%%%%%%%%%%%%%%%

\begin{figure}[h] \centering
\includegraphics[width=1.\columnwidth]{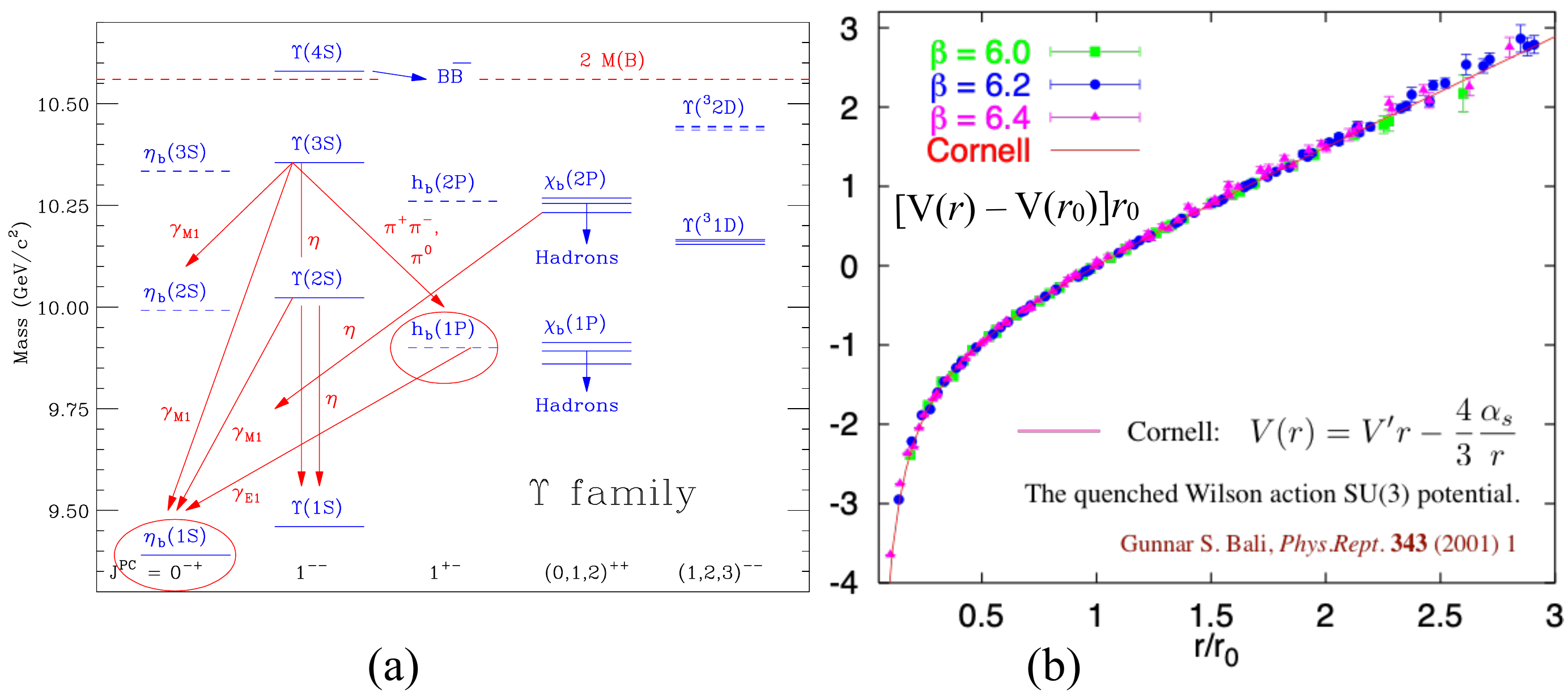}
\caption{(a) The bottomonium spectrum with some transitions indicated. From \cite{Rosner:2011eg}. (b) The static potential between heavy quarks calculated using quenched lattice QCD ($r_0 \simeq 0.5$ fm) compared to the phenomenological potential \eq{mo3}. From \cite{Bali:2000gf}. \label{fups}}
\end{figure}

The charmonium phenomenology faced a non-trivial test when applied to bottomonium ($b\bar b$) states (\fig{fups}(a)). Due to the larger bottom quark mass $m_b \simeq 4.9$ GeV the non-relativistic approximation is better justified, with velocities $\ave{v^2} \simeq 0.08$. Since the QCD interactions are flavor blind the same potential \eq{mo3} (probed at lower $r$) should describe the bottomonium spectrum and transitions. This was indeed found to be the case. The linear part of the potential is essential, contributing 50 \% for charmonia and 35\% for bottomonia \cite{Godfrey:1985xj}. Moreover, the phenomenological potential \eq{mo3} closely agrees (\fig{fups}(b)) with that calculated between static (infinitely heavy) quarks using lattice QCD in the quenched approximation \cite{Bali:2000gf}. In a calculation with dynamical quarks the creation of a light quark pair (``string breaking'') is expected to terminate the linear rise of the potential at large $r$. See \cite{Bali:2005fu,Bali:2005bg} for a lattice QCD study of string breaking.

%%%%%%%%%%%%%%%%%%%%%%%%%%%%%%%%%%%%%%%%%%%%%%%%%%%%%%%%
%%%%%%%%%%%%%%%%%%%%%%%%%%%%%%%%%%%%%%%%%%%%%%%%%%%%%%%%%

It is reasonable to assume that the description of Positronia in QED and Quarkonia in QCD, based on the Schr\"odinger equation, should follow from analogous approximations of the underlying gauge theory. Yet this raises the question of how the confining potential can appear in QCD. The Coulomb $-\alpha/r$ potential of QED is a solution of Gauss' law for $A^0$ without loop corrections. The same potential is given by the classical Maxwell equations, and its dependence $\propto 1/r$ is mandated by dimensional analysis. The linear potential in the quarkonium potential \eq{mo3} has a parameter $V'$ with dimension GeV$^2$, which does not appear in the QCD action. The QCD scale $\lqcd$ is thought to arise via ``dimensional transmutation'' \cite{Coleman:1973jx}, related to the renormalization of loop integrals. The scale is not expected at the classical (no loop) level.

It should be possible to settle this issue by scrutinizing the derivation of the Schr\"odinger equation from the QED action, and considering its applicability for QCD. This is a main motivation of the present study. It is not quite as simple as it sounds -- bound state perturbation theory is viewed as something of an ``art'' even in QED \cite{Itzykson:1980rh,RevModPhys.57.723}. A confinement scale in the solution of Gauss' law can (at the classical level) arise only due to a boundary condition. I shall argue that this possibility may exist for color singlet states in QCD, and study its consequences. Including a homogeneous solution of Gauss' law implies a departure from standard methods. Feynman diagrams are based on free propagators and vanishing gauge fields at spatial infinity. Dyson-Schwinger equations are derived without boundary contributions to the functional integral of a total derivative \cite{Itzykson:1980rh}.

\begin{figure}[h] \centering
\includegraphics[width=.5\columnwidth]{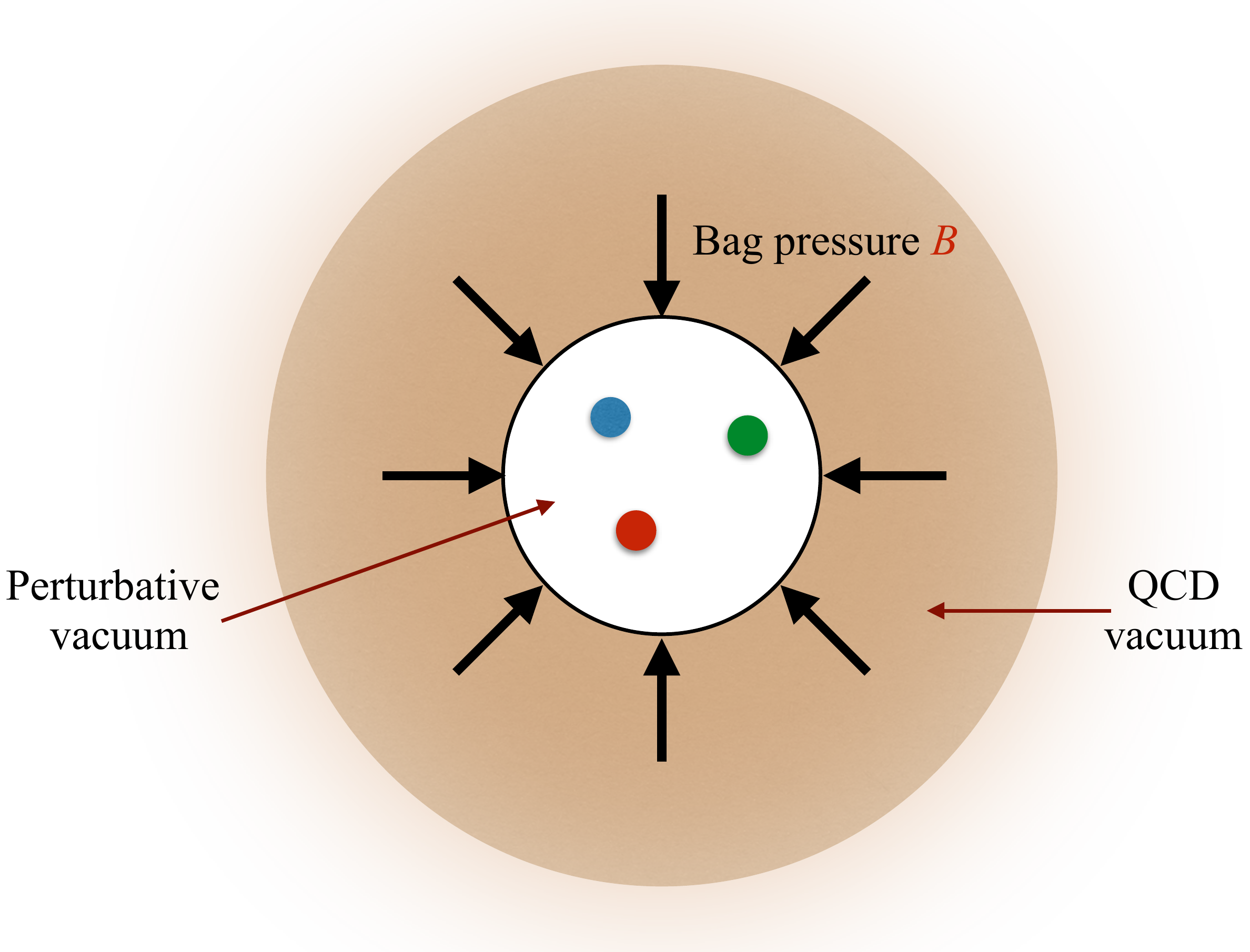}
\caption{Sketch of the MIT Bag Model \cite{Chodos:1974je}. The kinetic pressure of the quarks balances the pressure $B$ of the color field in the QCD vacuum. \label{fbag}}
\end{figure}

The notion of a non-vanishing vacuum gluon field has been around since the beginnings of QCD. The MIT Bag Model \cite{Chodos:1974je} describes hadrons as free quarks in a of perturbative vacuum bubble, confined by a QCD vacuum pressure $B^{1/4} \simeq 200$ MeV (as illustrated \fig{fbag}). The present approach, described below in section \ref{secIV.C}, agrees in spirit with the Bag Model but differs in its realization. There is no perturbative vacuum bubble, instead the quarks interact with the vacuum gluon field in the whole volume of the bound state. The universal energy density arises from a boundary condition on Gauss' law which in temporal gauge concerns only the longitudinal gluon field.

%%%%%%%%%%%%%%%%%%%%%%%%%%%%%%%%%%%%%%%%%%%%%%%%%%%%%%%%
%%%%%%%%%%%%%%%%%%%%%%%%%%%%%%%%%%%%%%%%%%%%%%%%%%%%%%%%%

\subsection{Regge behavior and duality} \label{secIIa2}
%%%%%%%%%%%%%%%%%%%%%%%%%%

\begin{figure}[h] \centering
\includegraphics[width=1.\columnwidth]{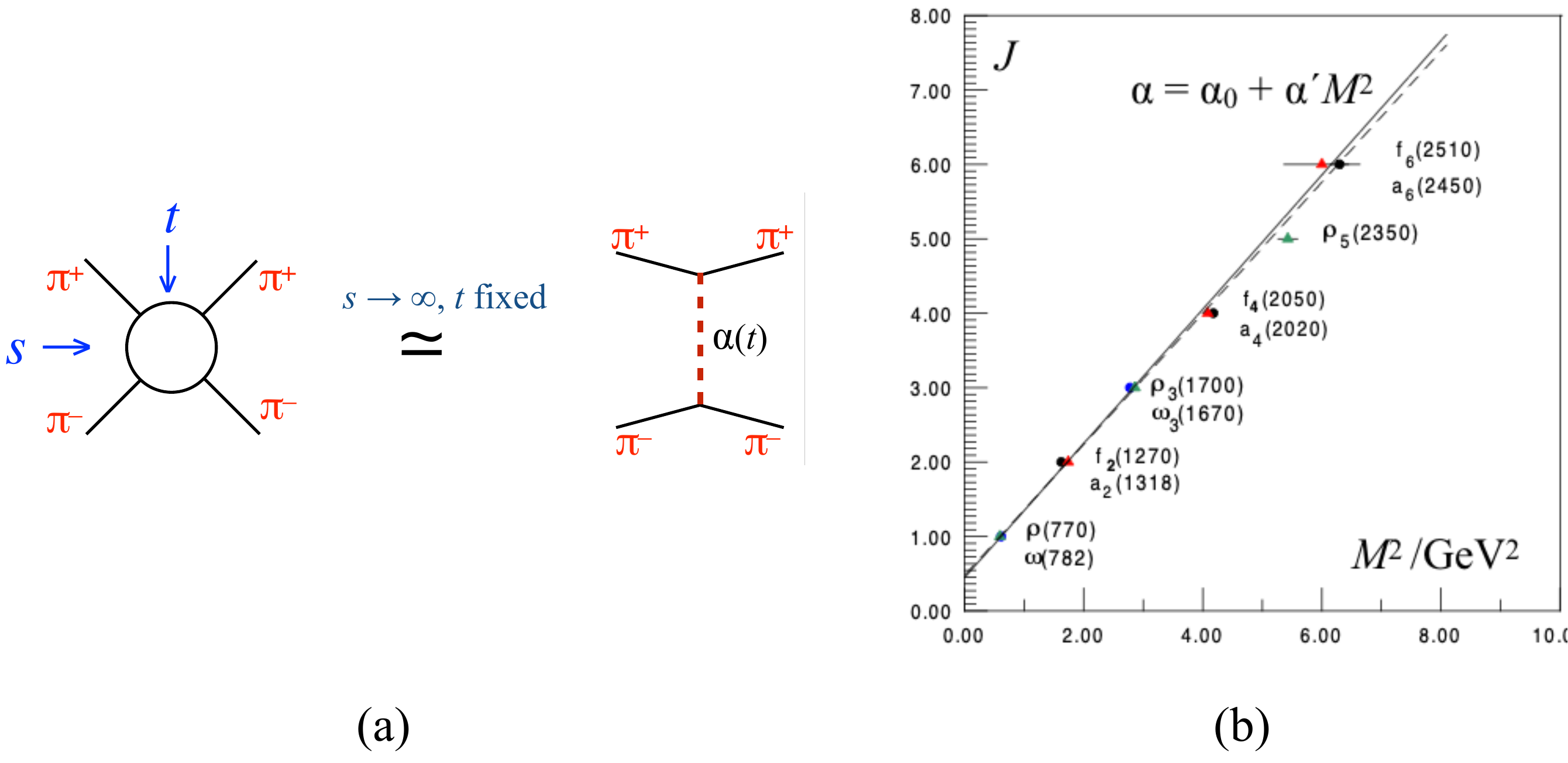}
\caption{(a) Scattering amplitude for $\pi^+\pi^- \to \pi^+\pi^-$ with $\rho$ Regge exchange at high energies. (b) Chew-Frautschi plot of hadron spins $J$ vs. their $M^2$, and the Regge trajectory  $\alpha_\rho(t)$. Plot from \cite{Desgrolard:2000sf}. \label{freg}}
\end{figure}

The main features of hadron scattering amplitudes were uncovered already in the 1960-70's, see \cite{Eden:1971jm,Phillips:1974mr,Melnitchouk:2005zr,Kopeliovich:2008fy} for reviews of Regge behavior and duality. Hadron-hadron scattering $a+b \to c+d$ is described by two variables, often taken to be the Lorentz invariants $s= E_{CM}^2 = (p_a+p_b)^2 \geq (m_a+m_b)^2$ and $t = (p_a-p_c)^2 \lsim 0$. With increasing $s$ the scattering amplitude $A(s,t)$ tends to peak in the forward direction, $t \simeq 0$. This is described by ``Regge exchange'',
\begin{align} \label{had1}
A(s\to\infty,\ t\lsim 0\ {\rm fixed}) \simeq \beta(t)s^{\alpha(t)} \hspace{2cm} \alpha(t) = \alpha_0+\alpha't
\end{align}
The exchanged "Reggeon" may be viewed as an off-shell ($t\leq 0$) hadron. Data shows that Regge trajectories are approximately linear, with a universal slope $\alpha' \simeq 0.9\ \gev^2$. Regge exchange is illustrated in \fig{freg}(a) for $\pi^+\pi^- \to \pi^+\pi^-$, to which the $\rho$ trajectory $\alpha_\rho(t) \simeq .5+.9\,t/\gev^2$ contributes.

In a Chew-Frautschi plot the spin $J$ of hadrons is plotted versus their squared masses $M^2$. Remarkably, the hadrons lie on the linear Regge trajectories determined by scattering data for $t \leq 0$, \ie, $\alpha(M^2) = J$. This is shown for the $\rho$ trajectory states in \fig{freg}(b)). Other hadrons with light ($u,d,s$) valence quarks such as nucleons and hyperons similarly lie on linear Regge trajectories. The reason for this is not understood, but it has inspired string-like models of hadrons, with the valence (di)quarks connected by a color flux tube \cite{Greensite:2011zz,Selem:2006nd}.

%%%%%%%%%%%%%%%%%%%%%%%%%%%%%%%%%%%%%%%%%%%%%%%%%%%%%%%%
%%%%%%%%%%%%%%%%%%%%%%%%%%%%%%%%%%%%%%%%%%%%%%%%%%%%%%%%%

\begin{figure}[h] \centering
\includegraphics[width=1.\columnwidth]{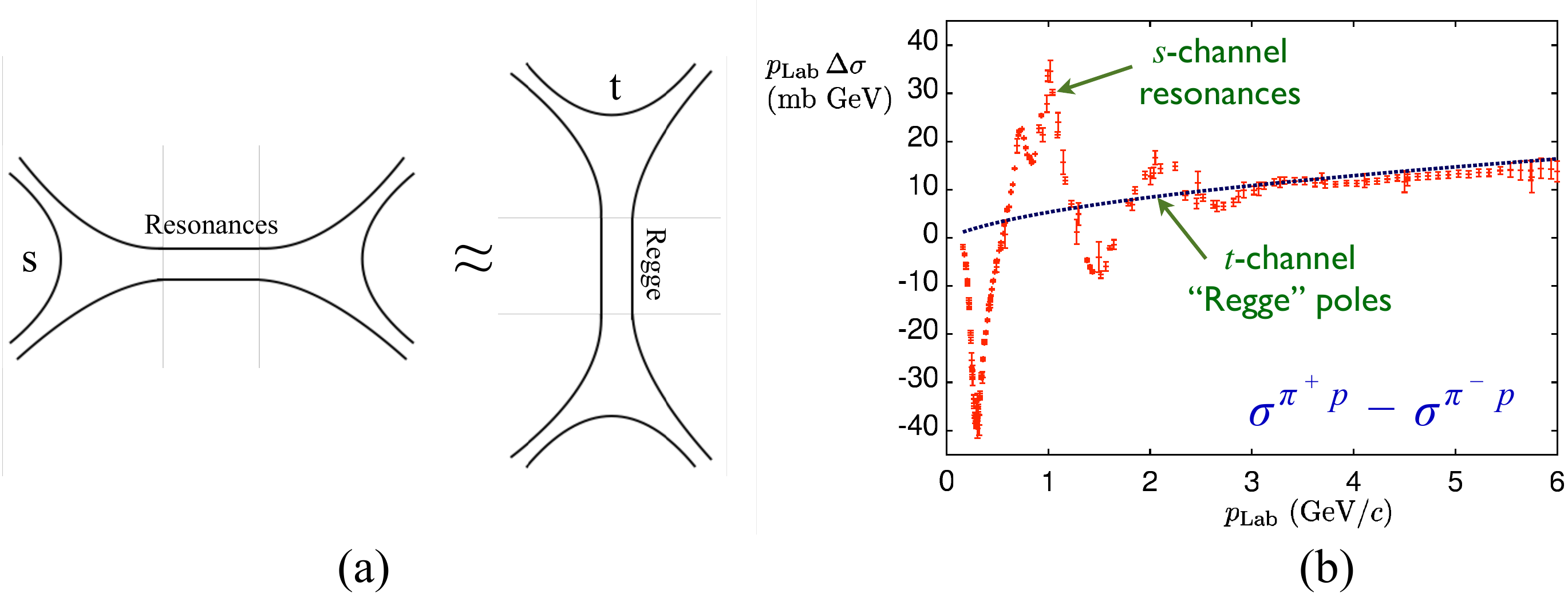}
\caption{(a) Diagrams illustrating the duality between $s$-channel resonances and $t$-channel Regge exchange. (b) Data on the forward $\pi N \to \pi N$ amplitude compared to $\rho$ Regge exchange extrapolated to low energy. Plot from \cite{Melnitchouk:2005zr} and W. Melnitchouk, private communication. \label{fdua}}
\end{figure}

Duality is a pervasive feature of hadron dynamics. In hadron scattering duality implies that $s$-channel resonances build (the imaginary part of) $t$-channel Regge exchange. This is illustrated by the flow of valence quarks in the dual diagrams \cite{Harari:1981nn,Rosner:1981np,Zweig:2015gpa} of \fig{fdua}(a). These diagrams may be ``stretched'' to emphasize either the $s$-channel resonances or the equivalent $t$-channel exchanges. Duality requires that the high energy Regge exchange amplitude \eq{had1} averages the resonance contributions when extrapolated to low energy, as shown for the $t=0$ $\pi N$ amplitude in \fig{fdua}(b). 

%%%%%%%%%%%%%%%%%%%%%%%%%%%%%%%%%%%%%%%%%%%%%%%%%%%%%%%%
%%%%%%%%%%%%%%%%%%%%%%%%%%%%%%%%%%%%%%%%%%%%%%%%%%%%%%%%%

\begin{figure}[h] \centering
\includegraphics[width=1.\columnwidth]{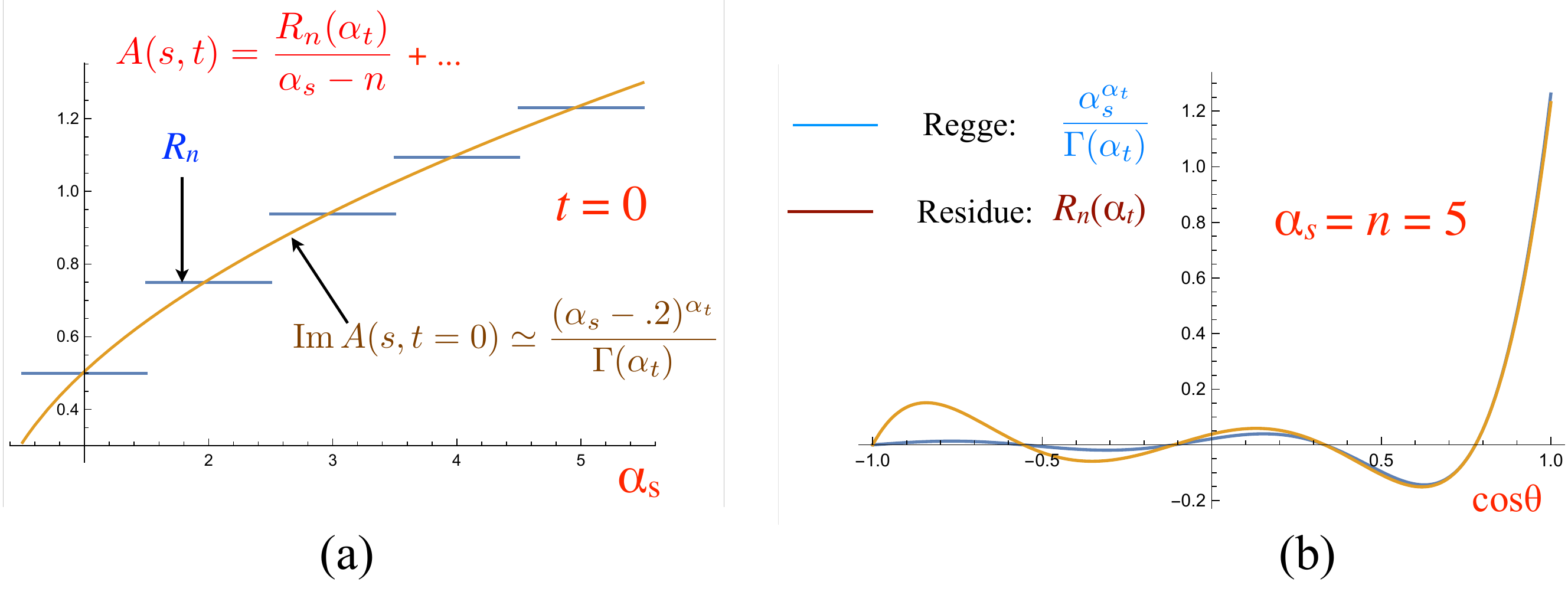}
\caption{(a) Comparison of the smeared resonance contributions \eq{had4} with the imaginary part of the Regge behavior \eq{had3} at $t=0$ (a common factor of $-\pi$ is omitted). (b) Comparison of the $t$-dependence of the residue $R_5(t)$ in \eq{had4} with that of the Regge behavior. Here $t=-\halft s\,(1-\cos\theta)$ (the pion mass is neglected, $M_\pi=0$). Both figures are from my lectures at the \textit{2015 International Summer Schools on Reaction Theory}, http://cgl.soic.indiana.edu/jpac/schools.html . \label{fven}}
\end{figure}

Dual models \cite{Schwarz:1973yz,Veneziano:1974dr} provide a mathematical illustration of duality. The amplitude for $\pi^+\pi^- \to \pi^+\pi^-$ is \cite{Lovelace:1969se,Shapiro:1969km},
\begin{align} \label{had2}
A(s,t) = \frac{\Gamma[1-\alpha(s)]\Gamma[1-\alpha(t)]}{\Gamma[1-\alpha(s)-\alpha(t)]}
\end{align}
Here $\Gamma(x)$ is the Euler Gamma function and the $\rho$ trajectory $\alpha(s) = \sfrac{1}{2}+ s$ (the scale is set by $\alpha'=1$). The asymptotic behavior of the $\Gamma$-function for large argument ensures the Regge behavior \eq{had1}. Taking first $s\to -\infty$ and then $s\to s\,e^{-i\pi}$ to reach the positive real $s$-axis from above gives
\begin{align} \label{had3}
\lim_{s\to \infty+\ieps}A(s,t) = \frac{\pi}{\Gamma[\alpha(t)]}\, \frac{e^{-i\pi\alpha(t)}}{\sin[\pi\alpha(t)]}\,s^{\alpha(t)}
\end{align}
The poles of $\Gamma[1-\alpha(s)]$ at $\alpha(s) = n,\ n=1,2,\ldots$ represent (zero-width) $s$-channel resonances, contributing $\delta$-functions to the imaginary part of the amplitude,
\begin{align} \label{had4}
\lim_{\alpha(s)\to n+\ieps}{\rm Im}\,A(s,t) =\lim_{\alpha(s)\to n+\ieps}{\rm Im}\Big[\inv{\alpha(s)-n}\Big]\, \frac{\Gamma[\alpha(t)+n]}{\Gamma(n)\Gamma[\alpha(t)]} \equiv -\pi\delta\big[\alpha(s)-n\big]R_n(t)
\end{align}
In \fig{fven}(a) the imaginary part of the Regge amplitude \eq{had3} is seen to agree with the resonance contributions \eq{had4} at $t=0$. The $\delta$-functions are smeared over $n-\halft < \alpha(s) < n+\halft$. This demonstrates semilocal duality.

The residue $R_n(t)$ of the pole at $\alpha(s) = n$ is an $n$th order polynomial in $t$. Expanding the residue into a sum of Legendre polynomials $P_J(\cos\theta)$, where $\theta$ is the CM scattering angle, shows that each pole is a superposition of resonances with spins $J= 0,\ldots n$. Their coherent sum builds the $t$-dependence of the Regge exchange. This is demonstrated in \fig{fven}(b) for the residue at $\alpha(s)=5$. The Regge and resonance contributions are practically indistinguishable in the forward peak, $\cos\theta \gsim 0.8$. 

%%%%%%%%%%%%%%%%%%%%%%%%%%%%%%%%%%%%%%%%%%%%%%%%%%%%%%%%
%%%%%%%%%%%%%%%%%%%%%%%%%%%%%%%%%%%%%%%%%%%%%%%%%%%%%%%%%

All resonance contributions to the elastic $\pi^+\pi^- \to \pi^+\pi^-$ amplitude must be positive at $\cos\theta = 1$, since they are proportional to the square of their coupling to $\pi^+\pi^-$. It so happens that (with $M_\pi=0$) at least the first 230 coefficients of the Legendre polynomials for $J\leq 20$ are positive. I do not know of a general proof, but see \cite{Shapiro:1969km} for a discussion.

Soon after the first dual model with four external hadrons was discovered \cite{Veneziano:1968yb}, corresponding $N$-point amplitudes were found. It was realized that these amplitudes describe string-like states \cite{Nielsen:2009ci}, and that they could be relevant in a totally different context, including gravity \cite{Scherk:1974ca}. The further developments of string theory were not connected to hadron physics.

\subsection{DIS: Deep Inelastic Scattering} \label{secIIa3}
%%%%%%%%%%%%%%%%%%%%%%%%%%

Deep Inelastic Scattering of leptons $\ell$ on nucleons $N$ (DIS, $\ell N \to \ell'X$) \cite{Abramowicz:1998ii,CooperSarkar:2012tx} probes the quark and gluon structure of the target. At large momentum transfers $-(\ell-\ell')^2 = Q^2 \gg M_N^2$ the exchanged virtual photon resolution is $\sim 1/Q$ in the direction transverse to the beam momentum $\ell$. This ensures that the lepton scatters (coherently) on a single target constituent, up to ``higher twist'' corrections of \order{1/Q^2}. At \order{\alpha_s^0} the ``inclusive'' cross section, summed over all states $X$, determines the fraction $x$ of the nucleon momentum carried by the struck quark (in a frame where the nuclon momentum is large). This should be independent of the probe and hence of $Q^2$, which is referred to as ``Bjorken scaling'' and is approximately satisfied by the data. 

%%%%%%%%%%%%%%%%%%%%%%%%%%%%%%%%%%%%%%%%%%%%%%%%%%%%%%%%
\begin{figure}[h] \centering
\includegraphics[width=1.\columnwidth]{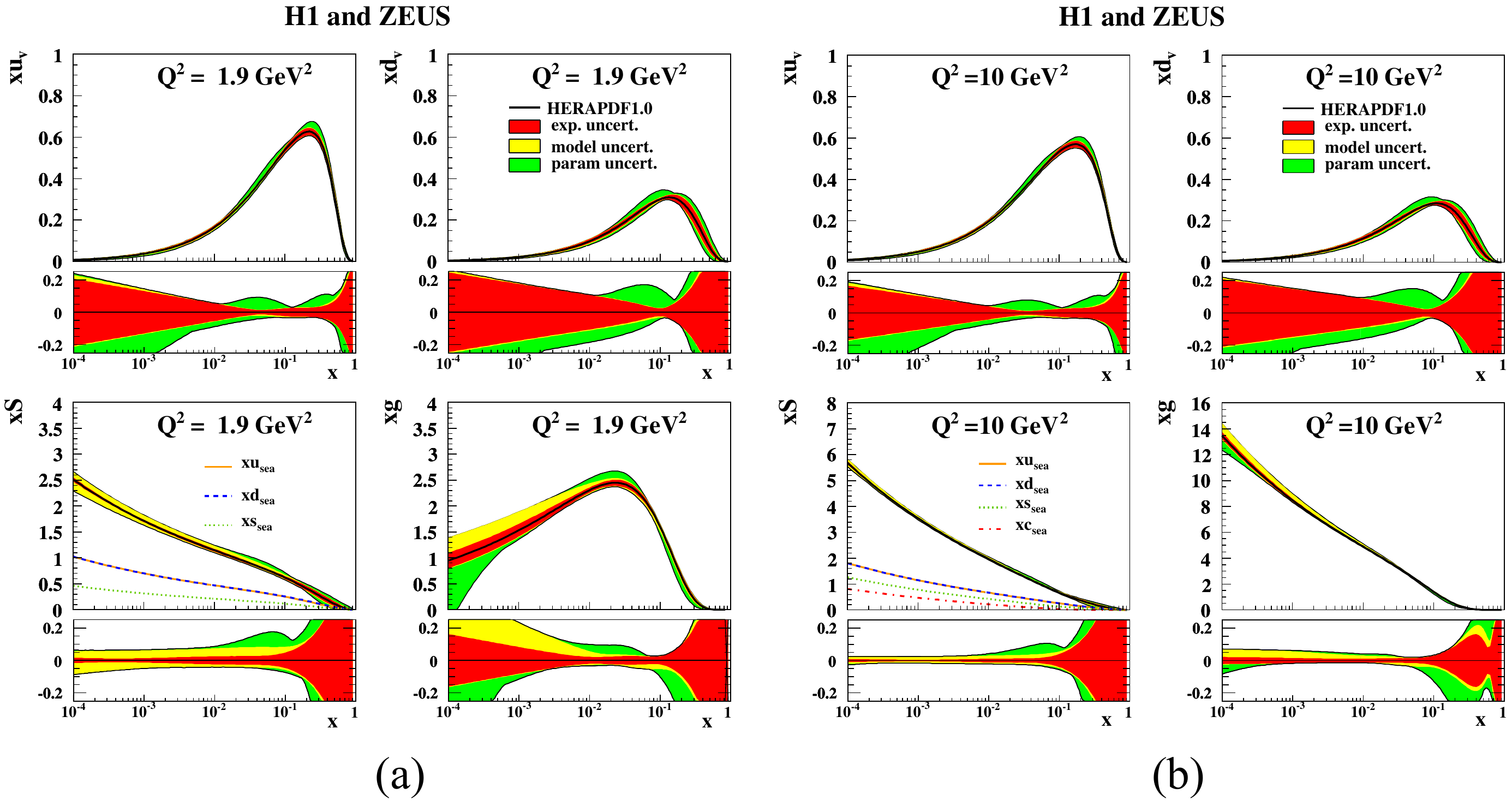}
\caption{Quark and gluon distributions determined from HERA data \cite{CooperSarkar:2012tx} at (a) $Q^2=1.9\ \gev^2$ and (b) at $Q^2=10\ \gev^2$. \label{fevo}}
\end{figure}
%%%%%%%%%%%%%%%%%%%%%%%%%%%%%%%%%%%%%%%%%%%%%%%%%%%%%%%%%

DIS has \order{\as} contributions due to gluons of momentum $k$ emitted by the quarks. Ever harder emissions with $|k^2| \lsim Q^2$ are resolved with increasing $Q^2$. This gives rise to calculable ``scaling violations'', which have a logarithmic dependence on $Q^2$ and were found to agree with data. This (together with numerous other predictions for hard scattering processes) has established QCD as the theory of the strong interactions. It also implies that DIS provides a reliable measurement of quark and gluon parton distributions in the nucleon, $q(x,Q^2)$ and $g(x,Q^2)$. The data agrees with the leading twist $Q^2$-dependence down to remarkable low values of $Q^2 \simeq 2$ GeV$^2$.

The large gluon contribution, and its steep increase for $x \to 0$, is a striking feature of DIS at high $Q^2$, as shown by \fig{fevo}(b) for $Q^2= 10\ \gev^2$. Even when multiplied by $x$ the gluon distribution $xg(x)$ dominates at low $x$, and is many times larger than the valence quark distributions $xu_V(x)$ and $xd_V(x)$. Sea quarks can arise from gluon splitting, $g \to q\bar q$. Hence $xS(x)$ is expected to follow the trend of $xg(x)$, as is confirmed in \fig{fevo}(b).

The valence quark distributions hardly change as $Q^2$ decreases to $1.9\ \gev^2$ (\fig{fevo}(a)). Photons of lower virtuality resolve fewer gluons, so the gluon distribution decreases quickly with $Q^2$. The sea quark distribution on the other hand evolves more slowly and maintains its rise at low $x$ down to $Q^2=1.9\ \gev^2$. 

The trend of the parton distributions with decreasing $Q^2$ indicates which hadron constituents are intrinsic to the bound state, and which may be associated with the hard scattering vertex. DIS suggests that most (low-$x$) gluons are created by the interaction with the virtual photon. Hadrons may thus have no valence gluons, which is consistent with their observed quantum numbers. However, sea quarks seem to be present even at the hadronic scale \cite{CooperSarkar:2009xz}.
 
%%%%%%%%%%%%%%%%%%%%%%%%%%%%%%%%%%%%%%%%%%%%%%%%%%%%%%%%
\vspace{2mm}
\begin{figure}[h] \centering
\includegraphics[width=.6\columnwidth]{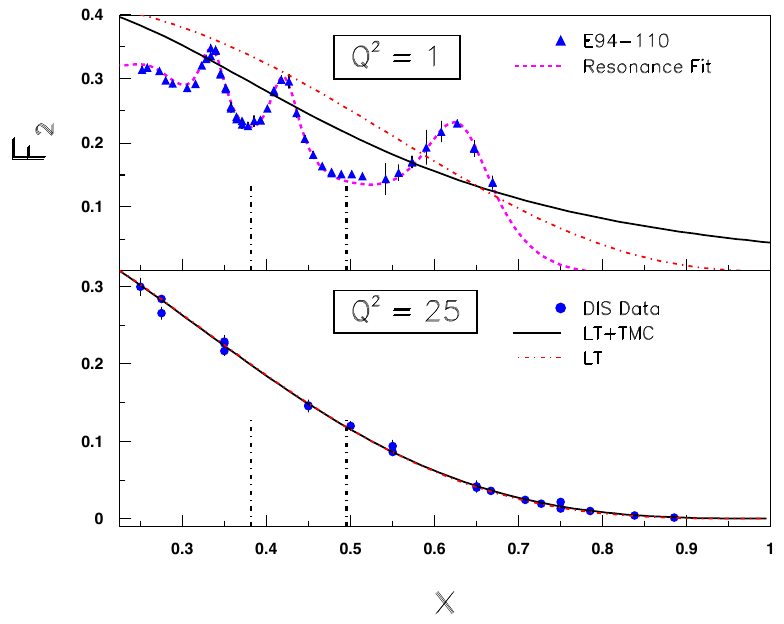}
\caption{Lower panel: A global fit of parton distributions at leading twist (LT) is shown to agree with DIS data for $F_2^p(x,Q^2=25\ \gev^2)$. Upper panel: The same fit evaluated at $Q^2= 1\ \gev^2$ (dot-dashed curve) is compared to $ep\to e X$ data. The solid line includes kinematic target mass corrections (TMC). Figure from \cite{Melnitchouk:2005zr,Melnitchouk:2011aa}. \label{fblo}}
\end{figure}
%%%%%%%%%%%%%%%%%%%%%%%%%%%%%%%%%%%%%%%%%%%%%%%%%%%%%%%%%

DIS experiments uncovered a surprising new form of duality, first noted by Bloom and Gilman in 1970 \cite{Bloom:1970xb}. The relation between $Q^2$, the momentum fraction $x$ and the mass squared of the inclusive system, $W^2 = M_X^2$, is
\begin{align} \label{had5}
x = 1-\frac{W^2-M_N^2}{Q^2+W^2-M_N^2}
\end{align}
$W$ decreases with decreasing $Q$ when $x$ is fixed, reaching $W \sim M_{N^*}$ at low $Q$. The lower panel of \fig{fblo} demonstrates that a global fit to DIS data agrees with measurements of the $F_2^p$ structure function at $Q^2= 25\ \gev^2$. In the upper panel the same fit, evolved to $Q^2 = 1\ \gev^2$, is compared to data of $F_2^p$ at this lower value of $Q^2$. The inclusive system $X$ is now in the resonance region, as seen from the contributions of the $\Delta(1232)$ at $x\simeq 0.62$ and the $S_{11}(1535)$ (between the vertical dashed lines). The fit determined from data at high $Q^2$ averages the resonance contributions at low $Q^2$. This ``Bloom-Gilman duality'' implies an unexpected relation between the parton distributions and the transition form factors $\gamma^*p \to N^*$. 

Analogous features of duality have been observed in other aspects of lepton scattering \cite{Melnitchouk:2005zr}, in $e^+e^-$ annihilation to hadrons and in hard hadron-hadron collisions \cite{Fantoni:2006im,Dokshitzer:2010zza}. Duality reflects a basic principle of hadron dynamics, which relates bound states to high energy scattering.

\subsection{The QCD coupling at low scales} \label{secIIa4}
%%%%%%%%%%%%%%%%%%%%%%%%%%

The coupling $g$ in the quark and gluon interaction terms of the QCD action is not a well-defined parameter. Its higher order corrections involve divergent loop integrals which need to be regularized. In renormalizable theories (such as QCD) the divergences arise from infinitely large loop momenta and are universal, \ie, the same for all physical processes. Removing the common divergence in $g$ leaves, however, a renormalization scale $\mu$ dependence in the coupling $\as(\mu^2) = g^2(\mu)/4\pi$. This scale may intuitively be thought of as the momentum at which the loop integrals are cut off. Results summed to all orders in $\as(\mu)$ are independent of the choice of $\mu$.

\begin{figure}[h] \centering
\includegraphics[width=.8\columnwidth]{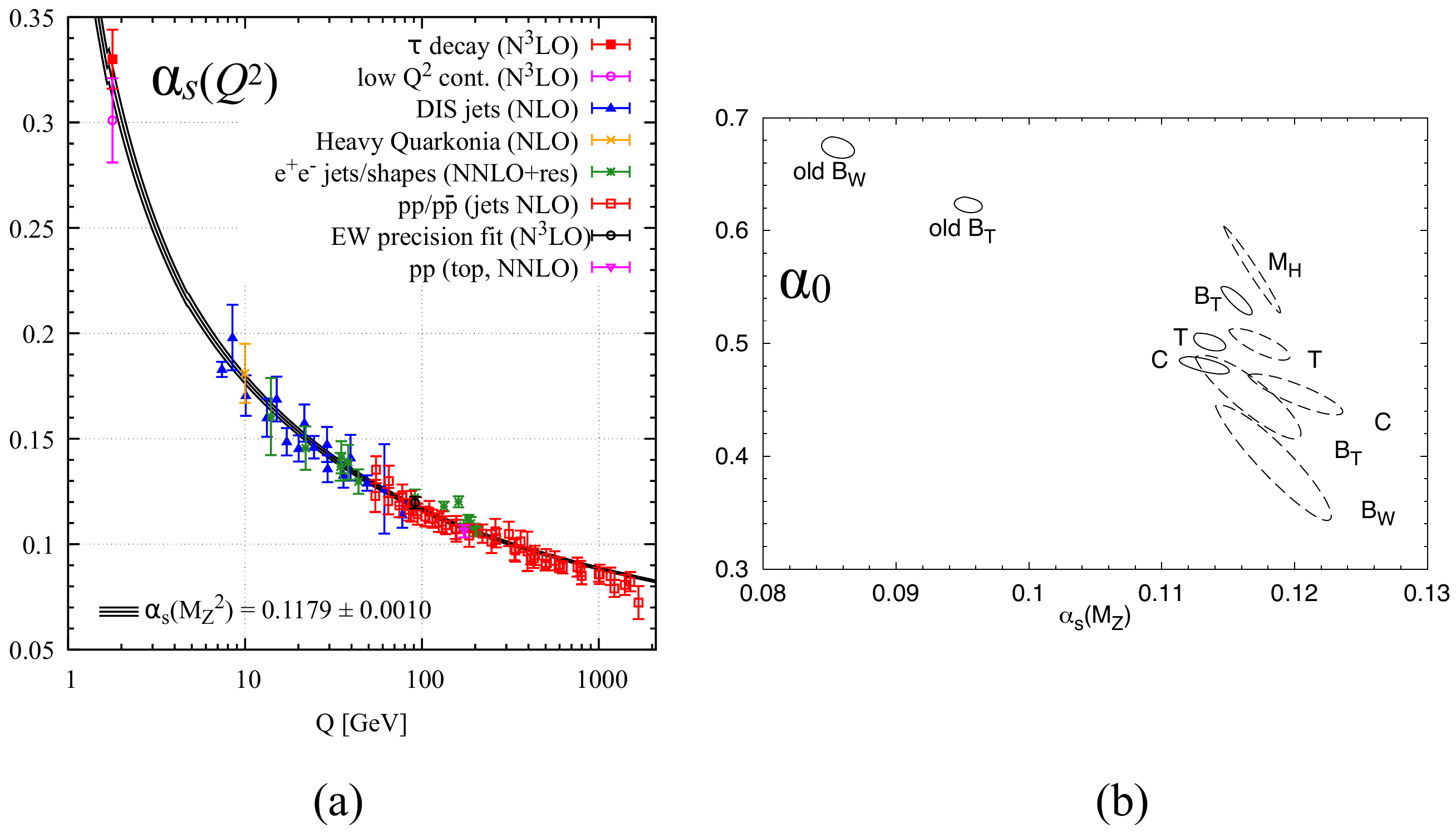}
\caption{(a) Measurements of the QCD coupling $\as(Q^2)$ confirm its expected $Q^2$-dependence. From \cite{Zyla:2020zbs}. (b) Results for the effective low energy coupling $\alpha_0(2\ \gev)$ \eq{had7} and for $\as(M_Z^2)$ obtained from a fit to event shapes in $e^+e^-$ annihilations. The solutions labeled ``old'' were obtained with an incorrect analysis, see \cite{Dokshitzer:1998nz}. From \cite{Dokshitzer:1998qp}. \label{falp}}
\end{figure}

Processes with momentum transfers $Q >> \mu$ probe the part of the loop integrals that were not included in the definition of $\as(\mu)$. This gives rise to factors of $\log(Q^2/\mu^2)$ which enhance higher order contributions. These factors may be absorbed in the coupling by making it $Q^2$-dependent \cite{Zyla:2020zbs,Dokshitzer:1998nz,Deur:2016tte}. Since QCD is ``asymptotically free'' its running coupling (for $n_f$ flavors) decreases logarithmically with $Q^2$,
\begin{align} \label{had6}
\as(Q^2) = \frac{12\pi}{(33-2n_f)\log(Q^2/\Lambda_{QCD}^2)} + \morder{\frac{\log\log(Q^2)}{\log(Q^2)}}
\end{align}
As shown in \fig{falp}(a) data on a variety of processes involving large scales $Q^2$ agree on the value of the QCD coupling and verify its predicted $Q^2$-dependence.

%%%%%%%%%%%%%%%%%%%%%%%%%%%%%%%%%%%%%%%%%%%%%%%%%%%%%%%%
\vspace{2mm}
%%%%%%%%%%%%%%%%%%%%%%%%%%%%%%%%%%%%%%%%%%%%%%%%%%%%%%%%%

The perturbative analysis of $\as(Q^2)$ in \fig{falp}(a) is restricted to $Q \geq m_\tau = 1.78$ GeV, with $\as(m_\tau^2) \simeq 0.33$. It is remarkable that the expression \eq{had6} for $\as(Q^2)$ (with higher order perturbative corrections) works down to $Q \simeq 2~\gev$. Perturbative results for DIS and other hard processes are found to be valid down to similar values of $Q$, and to join smoothly with the distributions at lower $Q$ \cite{Abt:2017nkc,Dokshitzer:2010zza}. There is no abrupt ``phase transition'' to non-perturbative physics.

There are many studies (reviewed in \cite{Deur:2016tte}) of the value of $\as$ in soft processes, where confinement effects dominate. Since $\as$ is not a physically measurable quantity the answer depends on the theoretical framework. A fairly model-independent result has been obtained using a dispersive approach \cite{Dokshitzer:1995qm,Dokshitzer:1997ew,Dokshitzer:1998qp}. The observed $1/Q$ power corrections to event shapes in $e^+e^-$ annihilations determine an average low energy coupling,
\begin{align} \label{had7}
\alpha_0(\mu_I) \equiv \inv{\mu_I}\int_0^{\mu_I} dk\,\as(k^2)
\end{align}
This coupling should be universal, \ie, independent of the shape parameter considered. Data on several shape measures give consistent values, see \fig{falp}(b). An analysis of the Thrust distribution at higher order gave \cite{Gehrmann:2012sc},
\begin{align} \label{had8}
\alpha_0(2\ \gev) = 0.538\;{+0.102\atop -0.047}
\end{align}

Hadron data is compatible with a framework where the coupling stays perturbative down to $Q=0$ \cite{Dokshitzer:1998nz}. Imposing a boundary condition on Gauss' law gives an \order{\alpha_s^0} confining potential which for $q\bar q$ Fock states agrees with \eq{mo3}, determined by quarkonium phenomenology and lattice QCD (section \ref{secVII.A1}). The states bound by this potential may serve as the basis for a perturbative expansion in $\as$, progressively including higher Fock states.

In this scenario the QCD coupling is renormalized at an initial scale $\mu \simeq 2\ \gev$. Loop corrections (higher Fock state fluctuations) with momenta $|k| < \mu$ are, due to the confining potential, free of infrared singularities. Thus the coupling is fixed for $Q < \mu$, at a value compatible with \eq{had8}. Its running for $Q > \mu$ is essentially perturbative, being insensitive to the confining potential.

\section{Brief survey of present QED approaches to atoms} \label{secII}
%%%%%%%%%%%%%%%%%%%%%%%%%%

\subsection{Recall: The Hydrogen atom in introductory Quantum Mechanics} \label{secII.A}
%%%%%%%%%%%%%%%%%%%%%%%%%%

In Introductory Quantum Mechanics we write the Hamiltonian for the Hydrogen atom as a sum of the electron and proton kinetic energies, plus the potential energy. Since the atom is stationary in time the wave function may be expressed as $\exp(-iEt)\Phi(\xv_e,\xv_p)$. The Schr\"odinger equation (including the mass contributions) is then
\begin{align} \label{pro1}
\Big[m_e+m_p-\frac{\nv_e^2}{2m_e} -\frac{\nv_p^2}{2m_p} + V(\xv_e-\xv_p)\Big]\Phi(\xv_e,\xv_p) = E\Phi(\xv_e,\xv_p)
\end{align}
We then transform to CM and relative coordinates,
\begin{align} \label{pro2}
\Rv = \frac{m_e\rv_e+m_p\rv_p}{m_e+m_p} \hspace{2cm} \rv= \rv_e-\rv_p
\end{align}
and get,
\begin{align} \label{pro3}
\Big[m_e+m_p-\frac{\nv_\Rv^2}{2(m_e+m_p)} -\frac{\nv_\rv^2}{2\mu} + V(\rv)\Big]\Phi(\Rv,\rv) = E\Phi(\Rv,\rv)  \hspace{2cm} \mu= \frac{m_em_p}{m_e+m_p}
\end{align}
The dependence of the wave function on $\Rv$ and $\rv$ may now be separated. Denoting the CM momentum of the bound state by $\Pv$ we have $\Phi(\Rv,\rv)  = \exp(i\Pv\cdot\Rv)\Phi(\rv)$. The total energy is given by the electron and proton masses, the kinetic energy of the CM motion and an \order{\alpha^2} binding energy, $E=m_e+m_p+\Pv^2/2(m_e+m_p)+E_b$ with,
\begin{align} \label{pro4} 
\Big[-\frac{\nv_\rv^2}{2\mu} + V(\rv)\Big]\Phi(\rv) = E_b\Phi(\rv) 
\end{align}
The transformation has reduced the dynamics to that of one particle in an external potential, and the bound states are determined by the Schr\"odinger equation \eq{pro4}. The two-to-one particle reduction is possible only for non-relativistic kinematics. For relativistic motion one needs to transform the times together with the positions of the electron and proton in \eq{pro2}. The wave function of a Hydrogen atom with large CM momentum $|\Pv| \gsim m_e+m_p$ can nevertheless be determined. The energy is $E=\sqrt{\Pv^2+(m_e+m_p+E_b)^2}$ and the wave function $\Phi(\rv)$ depends on $\Pv$ (Lorentz contraction), as we shall see in section \ref{secV.C}.

\subsection{The Schr\"odinger equation from Feynman diagrams (rest frame)} \label{secII.B}
%%%%%%%%%%%%%%%%%%%%%%%%%%

\subsubsection{Bound states vs. Feynman diagrams} \label{secII.B1}
%%%%%%%%%%%%%%%%%%%%%%%%%%

Bound states are (by definition) stationary in time and thus eigenstates of the Hamiltonian. The eigenstate condition gives the Schr\"odinger equation \eq{pro4}. On general grounds, bound states also appear as poles in scattering amplitudes at $E_{CM} = M-i\Gamma$, where $M$ is the bound state mass and $\Gamma$ its width. \textit{E.g.,} the $e^-p \to e^-p$ scattering amplitude has poles at the masses of the ground and all excited states of the Hydrogen atom. Since the binding energy $E_b<0$ the poles are below the threshold for scattering, $M<m_e+m_p$. QED scattering amplitudes can be calculated perturbatively, in terms of Feynman diagrams. The expansion is defined by the perturbative $S$-matrix,
\begin{align} \label{mo2}
S_{fi}={}_{out}\bra{f,\,t\to \infty}&\left\{ {\rm T}\exp\Big[-i\int_{-\infty}^\infty dt\,H_I(t)\Big]\right\}\ket{i,\,t\to -\infty}_{in} \nn\crt
& H_I = e\int d\xv\, \bar\psi\,e\Asl\,\psi \ \ \ \ \mbox{(in\ QED)}
\end{align}
where the $in$ and $out$ states at  $t=\pm\infty$ are free. Feynman diagrams of any finite order in the coupling $e$ for the process $i \to f$ are generated by expanding the time ordered exponential of the interaction Hamiltonian $H_I$. The interaction vertices are connected by free propagators.

Unfortunately the $S$-matrix boundary condition of free states excludes bound states, which are bound due to interactions. There is no overlap between, say, an $e^+e^-$ Positronium atom, which has finite size, and a free $e^+e^-$ state, which has infinite size. As a consequence, there are no Positronium poles in any Feynman diagram for the $e^+e^- \to e^+e^-$ scattering amplitude. This is why (as I mentioned above) atoms may be called ``non-perturbative'', and their perturbative expansion differs from that of the $S$-matrix.

\begin{figure}[h] \centering
\includegraphics[width=.8\columnwidth]{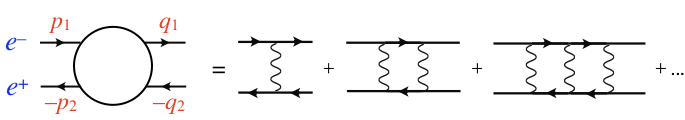}
\caption{Ladder diagram expansion of the $e^-e^+ \to e^-e^+$ scattering amplitude. Momenta are shown in the fermion direction, \eg, the positron energy $p_2^0 > 0$. \label{f1}}
\end{figure}

Nevertheless, it turns out that we can generate Positronium poles by (implicitly or explicitly) summing an infinite set of Feynman diagrams. The poles then arise through the divergence of the sum. The simplest set of diagrams to sum are the so-called ``ladder diagrams'' shown in \fig{f1}.
 
%%%%%%%%%%%%%%%%%%%%%%%%%%%%%%%%%%%%%%%%%%%%%%%%%%%%%%%%
%%%%%%%%%%%%%%%%%%%%%%%%%%%%%%%%%%%%%%%%%%%%%%%%%%%%%%%%%

At first sight it seems curious that all the ladder diagrams of \fig{f1} can be of the same order in $\alpha$, allowing the series to diverge at any value of the coupling. This is indeed true only for the special kinematics of bound states. In the rest frame all 3-momenta are of the order of the Bohr momentum, \eg, $|\pv_1|$ is of \order{\alpha m}, and its kinetic energy $E_{p_1} = \sqrt{\pv_1^2+m^2} \simeq m+\pv_1^2/2m$ differs from $m$ by \order{\alpha^2 m}. The exchanged momentum is similar, $(q_1-p_1)^0 \sim \alpha^2 m$, $|\qv_1-\pv_1| \sim \alpha m$. Each propagator contributes a factor of \order{1/\alpha^2}, making the diagram with a single ladder of \order{1/\alpha}. In fact all ladder diagrams are of \order{1/\alpha} for bound state kinematics, whereas all non-ladder diagrams are of higher order in $\alpha$.

%%%%%%%%%%%%%%%%%%%%%%%%%%
\begin{tcolorbox}
\textit{Exercise \ref{e1}:} Convince yourself that the diagram with two ladders in \fig{f1} is of \order{1/\alpha}, like the single ladder diagram. \textit{Hint:} The relevant loop momenta are commensurate with bound state kinematics.
\end{tcolorbox}
%%%%%%%%%%%%%%%%%%%%%%%%%%

In processes where the momenta are even lower than in bound states the propagators are further enhanced and the ladder series in \fig{f1} diverges more strongly. This is the kinematic region where classical fields dominate, and Feynman diagrams give non-leading contributions. Bound states are at the borderline between quantum and classical physics.

\subsubsection{Forming an integral equation} \label{secII.B2}
%%%%%%%%%%%%%%%%%%%%%%%%%%

The expression for the sum of all ladder diagrams at leading \order{1/\alpha} may be formulated as an integral equation. Bound state poles are just below threshold, $2m-M \sim \alpha^2 m$, so also the initial and final $e^\pm$ must be off-shell by \order{\alpha^2}. Their propagators may be expressed using
\begin{align} \label{pro5}
\frac{\slashed{p}+m}{p^2-m^2+\ieps} = \frac{1}{2E_p}\sum_\lm\left[\frac{u(\pv,\lm)\bar u(\pv,\lm)}{p^0-E_p+\ieps}+ \frac{v(-\pv,\lm)\bar v(-\pv,\lm)}{p^0+E_p-\ieps}\right]
\hspace{2cm} E_p = \sqrt{\pv^2+m^2}
\end{align}
At leading order in $\alpha$ we need only retain the $e^-$ pole in the electron propagator and the $e^+$ pole for the positron, \eg, $1/(p_1^0-E_{p_1}+\ieps) \propto \alpha^{-2}$ and $1/(-p_2^0+E_{p_2}-\ieps) \propto \alpha^{-2}$. In the following I show for conciseness only the spinors of the external propagators, \eg, $u(\pv_1,\lm_1)$ for the incoming electron. The analysis is done in the rest frame, $\pv_1+\pv_2=0$.

For bound state kinematics the spinors are trivial at leading order in $\alpha$,
\begin{align} \label{pro6}
u(\pv,\lm) &\equiv \frac{\psl+m}{\sqrt{E_p+m}}\left(\begin{array}{c}\chi_\lm \crt 0 \end{array} \right)
= \sqrt{2m} \left(\begin{array}{c}\chi_\lm \crt 0 \end{array} \right) + \morder{\alpha} \nn\crt
v(\pv,\lm) &\equiv \frac{-\psl+m}{\sqrt{E_p+m}}\left(\begin{array}{c}0\crt \bar\chi_\lm  \end{array} \right)
= \sqrt{2m} \left(\begin{array}{c}0\crt \bar\chi_\lm  \end{array} \right) + \morder{\alpha} \hspace{1cm}
\bar\chi_\lm=i\sigma_2\chi_\lm
\end{align}
The relation between $\bar\chi_\lm$ and $\chi_\lm$ follows from charge conjugation, see \eq{A13} below. In the single ladder diagram of \fig{f2}(a) the Dirac structure of the electron line is $\bar u(\qv_1,\lm_1')\gamma^\mu u(\pv_1,\lm_1) = 2m \delta_{\lm_1,\lm_1'}\delta^{\mu,0} + \morder{\alpha}$. The positron line gives a similar result. In the photon propagator $(q_1-p_1)^2 = (q_1^0-p_1^0)^2-(\qv_1-\pv_1)^2 = -(\qv_1-\pv_1)^2 + \morder{\alpha^4 m^2}$. The amplitude for this diagram is then, at lowest order in $\alpha$, denoting $\pv \equiv \pv_1,\ \qv \equiv \qv_1$ and suppressing the conserved helicities,
\begin{align} \label{pro7}
A_1(\pv,\qv) = (2m)^2 \frac{-e^2}{(\qv-\pv)^2} \equiv (2m)^2\, V(\qv-\pv)
\end{align}
where the notation indicates that $V(\pv-\qv)$ is the single photon exchange potential in momentum space. The factor $(2m)^2$ is due to my normalization of the spinors in \eq{pro6}. 
%%%%%%%%%%%%%%%%%%%%%%%%%%%%%%%%%%%%%%%%%%%%%%%%%%%%%%%%
\begin{figure}[h] \centering
\includegraphics[width=.6\columnwidth]{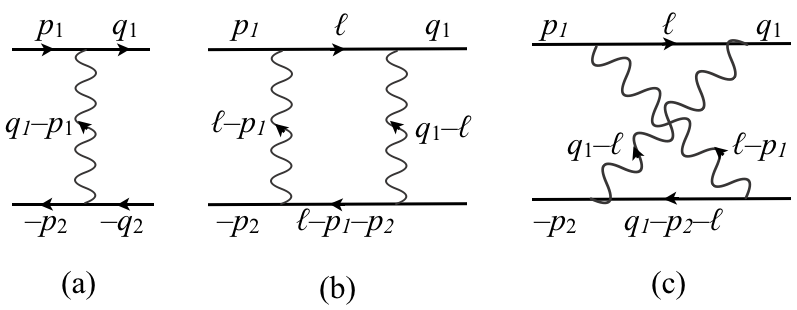}
\caption{(a) Single ladder diagram for the $e^-(p_1)e^+(p_2) \to e^-(q_1)e^+(q_2)$ scattering amplitude. (b) Double ladder diagram. (c) A diagram with crossed exchanges. Momenta are shown in the fermion (arrow) direction. \label{f2}}
\end{figure}
%%%%%%%%%%%%%%%%%%%%%%%%%%%%%%%%%%%%%%%%%%%%%%%%%%%%%%%%%

A similar calculation of the double ladder amplitude in \fig{f2}(b) gives, with $P^0 \equiv p_1^0+p_2^0$ the CM energy,
\begin{align} \label{pro8}
A_2(\pv,\qv) = \int\frac{d^3\ellv}{(2\pi)^3}\, A_1(\pv,\ellv) \inv{P^0-2E_\ellv+\ieps}\, V(\qv-\ellv)
\end{align}

%%%%%%%%%%%%%%%%%%%%%%%%%%
\begin{tcolorbox}
\textit{Exercise \ref{e2}:} Derive the expression \eq{pro8} for $A_2(\pv,\qv)$. Why does diagram \ref{f2}(c) contribute only at a higher order in $\alpha$?
\end{tcolorbox}
%%%%%%%%%%%%%%%%%%%%%%%%%%

It is now straightforward to see that the amplitude with $n$ ladders may be expressed as
\begin{align} \label{pro9}
A_n(\pv,\qv) = \int\frac{d^3\ellv}{(2\pi)^3}\, A_{n-1}(\pv,\ellv) \inv{P^0-2E_\ell+\ieps}\, V(\qv-\ellv) 
\equiv A_{n-1}(\pv,\ellv)\,S(\ellv)\,V(\qv-\ellv)
\end{align}
where a convolution over $\ellv$ is understood in the last expression. Summing over all ladder diagrams we get
\begin{align} \label{pro10}
A(\pv,\qv) = \sum_{n=1}^\infty A_n(\pv,\qv) = A_1(\pv,\qv)+A(\pv,\ellv)\,S(\ellv)\,V(\qv-\ellv)
\end{align}
which has the form of a Dyson Schwinger equation \cite{Itzykson:1980rh}.
A bound state pole in the $e^-e^+ \to e^-e^+$ amplitude has in the rest frame the structure
\begin{align} \label{pro10b}
A(\pv,\qv) = \frac{\Phi^\dag(\pv)\Phi(\qv)}{P^0-M} + \ldots
\end{align}
where $M$ is the bound state mass. $\Phi^\dag(\pv)$ and $\Phi(\qv)$ are the bound state wave functions, expressing the coupling to the initial and final states with relative momenta $\pv$ and $\qv$. Eq. \eq{pro10} gives a bound state equation for the wave function since $A_1(\pv,\qv)$ has no pole. Cancelling the factor $\Phi^\dag(\pv)/(P^0-M)$ on both sides and extracting a factor $P^0-2E_{\qv}$ from the wave function (which gives the ``truncated'' wave function) we have (at $P^0=M$)
\begin{align} \label{pro11}
\Phi(\qv)(M-2E_q)= \int\frac{d^3\ellv}{(2\pi)^3}\, \Phi(\ellv)(M-2E_\ell)\, \inv{M-2E_\ell+\ieps}\, \frac{-e^2}{(\qv-\ellv)^2}
= \int\frac{d^3\ellv}{(2\pi)^3}\, \Phi(\ellv)\, \frac{-e^2}{(\qv-\ellv)^2}
\end{align}
where I used the explicit expression of the potential from \eq{pro7}. This is the Schr\"odinger equation in momentum space. We can go to coordinate space using
\begin{align} \label{pro12}
\inv{(\ellv-\qv)^2}&= \int d^3\xv\, \frac{e^{i(\qv-\ellv)\cdot\xv}}{4\pi|\xv|} \nn\crt
\Phi(\qv) &\equiv \int d^3\xv\, \Phi(\xv)\,e^{-i\qv\cdot\xv}
\end{align}
Defining the binding energy $E_b$ by $M=2m+E_b$ and expanding $E_q \simeq m+\qv^2/2m$ on the lhs. of \eq{pro11} we have as in \eq{pro4},
\begin{align} \label{pro13}
\Big(E_b+\frac{\nv^2}{m}\Big)\Phi(\xv) = V(\xv)\Phi(\xv) \hspace{2cm} V(\xv) = -\frac{\alpha}{|\xv|}
\end{align}

\subsection{The Bethe-Salpeter equation} \label{secII.C}
%%%%%%%%%%%%%%%%%%%%%%%%%%

The Bethe-Salpeter equation \cite{Salpeter:1951sz,Itzykson:1980rh,Silagadze:1998ri} is a generalization of the integral equation \eq{pro11}, obtained by considering all Feynman diagrams (not just the ladder ones), and without assuming non-relativistic kinematics. It is thus a formally exact framework for bound states with explicit Poincar\'e covariance, and so far is the only bound state equation which applies in any frame. Boost covariance requires that the relative time of the constituents in the wave function is frame dependent. It is possible to project on constituents at equal time in any one frame. I give a brief summary here, following \cite{Lepage:1978hz}. A comprehensive review may be found in \cite{Nakanishi:1969ph,Nakanishi:1988hp}.

\begin{figure}[h] \centering
\includegraphics[width=.8\columnwidth]{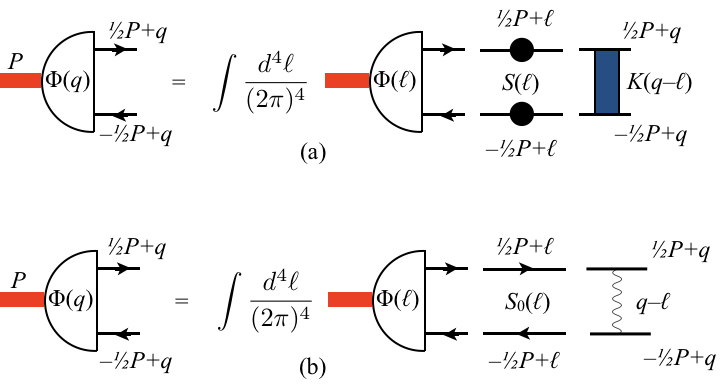}
\caption{(a) The exact Bethe-Salpeter equation \eq{pro15} for a bound state of momentum $P$. The black dots on the fermion propagators in $S$ represent full self-energy corrections, and the irreducible kernel $K$ has contributions of all orders in $\alpha$. (b) The Bethe-Salpeter equation with free propagators and a single photon exchange kernel. \label{f3}}
\end{figure}

Let $G_T$ be a truncated Green function (\ie, without external propagators) for a $2 \to 2$ process. Denote by $K$ a $2 \to 2$ truncated ``kernel'' and by $S$ a 2-particle propagator. Then if $K$ is 2-particle irreducible, \ie, does not have two parts that are only connected by $S$, we have the Dyson-Schwinger identity
\begin{align} \label{pro14}
G_T = K + G_T\,S\,K
\end{align}
By construction, any Feynman diagram on the lhs. is either contained in $K$ or has the structure of $G_T\,S\,K$. Eq. \eq{pro14} may be regarded as an exact equation for $G_T$ since it holds for the complete sum of Feynman diagrams.
The product $G_T\,S\,K$ implies a convolution over the momenta and helicities of the two particles in the propagator $S$. If $G_T$ has a bound state pole it must have the form \eq{pro10b}. Identifying the residues on both sides of  \eq{pro14} gives the Bethe-Salpeter equation for the truncated wave function shown in \fig{f3}(a),
\begin{align} \label{pro15}
\Phi(P,q) = \int \frac{d^4\ell}{(2\pi)^4}\Phi(P,\ell)\,S(\ell)\,K(q-\ell)
\end{align}
%%%%%%%%%%%%%%%%%%%%%%%%%%%%%%%%%%%%%%%%%%%%%%%%%%%%%%%%
%%%%%%%%%%%%%%%%%%%%%%%%%%%%%%%%%%%%%%%%%%%%%%%%%%%%%%%%%
The bound state momentum $P=(P^0,\Pv)$ satisfies $P^0 = \sqrt{\Pv^2+M^2}$ by Poincar\'e invariance. The wave function $\Phi(P,q)$ depends on the relative energy $q^0$ of the constituents or, quivalently, on the difference in time $x^0$ of the constituents in coordinate space,
\begin{align} \label{pro16}
\Phi(P,x) = \int \frac{d^4q}{(2\pi)^4}\Phi(P,q)\,e^{-iq\cdot x}
\end{align}
The B-S wave function for the $e^-e^+$ component can be expressed as a matrix element of electron field operators between the bound state $\ket{P}$ and the vacuum,
\begin{align} \label{pro17}
\bra{0}\mathrm{T}\{\bar\psi_\beta(x_2)\psi_\alpha(x_1)\}\ket{P} \equiv e^{-iP\cdot(x_1+x_2)/2}\Phi_{\alpha\beta}(P,x_1-x_2)
\end{align}
It thus describes a component of the bound state which has an electron at time $x_1^0$ at position $\xv_1$, and a positron at time $x_2^0$ at position $\xv_2$. For $x_1^0 = x_2^0$ the B-S wave function describes an $e^-e^+$ Fock state, belonging to the Hilbert space of states defined at equal time and expressed in the free field basis.

A Lorentz transformation $\la P = (P'^{\,0},\Pv')$ transforms the electron field as $\psi'(x') = S^{-1}(\la)\psi(\la x)$, where the $4\times 4$ matrix $S(\la)$ is the Dirac spinor representation of the transformation, familiar from the Dirac equation \cite{Itzykson:1980rh}. Hence the B-S wave function transforms as
\begin{align} \label{pro18}
\Phi'(P',x'_1-x'_2) = S(\la) \Phi(P,x_1-x_2) S^{-1}(\la)
\end{align}
Poincar\'e covariance thus allows the constituents be taken at equal time ($x_1^0 = x_2^0$) in at most one frame.
The B-S equation has ``abnormal'' solutions \cite{Nakanishi:1969ph,Karmanov:2020epm,Carbonell:2021gxp}, which vanish in the non-relativistic limit and seem related to the dependence on relative time. Their physical significance is not fully understood.

Expanding the propagator $S$ and the kernel $K$ in powers of $\alpha$ allows to solve the B-S equation perturbatively. There are many formally equivalent expansions. The Dyson-Schwinger equation \eq{pro14} determines $K$ in terms of the truncated Green function $G_T$ and $S$,
\begin{align} \label{pro19}
K = \inv{1+G_T\,S}\,G_T = G_T-G_T\,S\,G_T +\ldots
\end{align}
The choice of $S$ together with the standard perturbative expansion of $G_T$ fixes the expansion of the kernel. My remark in section \ref{secI.A} that the perturbative expansion for bound states is not unique refers to this more precise statement.

The B-S equation is difficult to solve when the kernel $K(\ell,q)$ depends on $(\ell-q)^0$, which implies retarded interactions. Already the single photon exchange kernel has denominator $(\ell^0-q^0)^2-(\ellv-\qv)^2$. The $(\ell-q)^0$ dependence arises from the propagation of transversely polarized photons, which create intermediate $e^-e^+\gamma$ states that affect the $e^-e^+$ B-S wave function. No analytic solution of the B-S equation is known even for single photon exchange with free propagators, illustrated in \fig{f3}(b).

Caswell and Lepage \cite{Caswell:1978mt} noted that $S$ may be chosen so that the kernel $K(\ell,q)$ is static (independent of $\ell^{\,0}$ and $q^0$) at lowest order. This reduced the B-S equation to a Schr\"odinger equation which has an analytic solution, simplifying the calculation of higher order corrections in the rest frame.

Bound states with arbitrary CM momenta are needed for scattering processes, \eg, form factors. It is then non-trivial to take into account the frame dependence of the wave function \cite{Brodsky:1968ea}. Model dependent assumptions are often made, but there are few studies based on field theory. The B-S framework was used in \cite{Jarvinen:2004pi} to determine the frame dependence of equal-time Positronium wave functions. It showed the importance of the $\ket{e^+e^-\gamma}$ intermediate state, and demonstrateed (apparently for the first time) that the standard Lorentz contraction of the $\ket{e^+e^-}$ Fock component holds at lowest order. I verify this result using a Fock state expansion in section \ref{secV.C}.

\subsection{Non-relativistic QED} \label{secII.D}
%%%%%%%%%%%%%%%%%%%%%%%%%%

The realization that there are many formally equivalent versions of the Bethe-Salpeter equation underlined the need for physical judgement in the choice of perturbative expansion. The most accurate data for atoms relates to their binding energies. These may be calculated in the rest frame, where the constituents have mean velocities of \order{\alpha}. It has been estimated that the probability for Positronium electrons to have 3-momenta $|\pv| \gsim m$ is of order $\alpha^5 \sim 10^{-11}$ \cite{Kinoshita:1990ai}. It is thus well motivated to expand the QED action in powers of $|\pv|/m$. This defines the effective theory of non-relativistic QED (NRQED) \cite{Caswell:1985ui,Kinoshita:1998jfa}. The constraints of gauge and rotational invariance allow only a limited number of terms at each order of $|\pv|/m$ in the Lagrangian. The expansion begins as,
\begin{align} \label{pro32}
\mL_{NRQED} = &- \quart F_{\mu\nu}F^{\mu\nu} +
  \chi^\dag \Big\{ i\partial_t-eA^0 + \frac{\bs{D}^2}{2m} + \frac{\bs{D}^4}{8m^3}
 + c_1\frac{e}{2m}\,\sv\cdot\bs{B} + c_2\frac{e}{8m^2}\,\nv \cdot \bs{E} \nn\crt
 & + c_3\,\frac{ie}{8m^2} \,\sv\cdot(\bs{D}\times\bs{E} - \bs{E}\times\bs{D})\Big\} \chi 
 + \frac{d_1}{m^2}\,(\chi^\dag\chi)^2 + \frac{d_2}{m^2}\,(\chi^\dag\sv\chi)^2 + \ldots  \nn\crt
 &+ \mbox{~positron and positron-electron terms}.
\end{align}
The photon action $-F_{\mu\nu}F^{\mu\nu}/4$ is as in QED, since photons are relativistic. The field $\chi$ is a two-component Pauli spinor, representing the electron part (upper components) of the QED Dirac field. There are further terms involving the lower (positron) components, as well as terms mixing the positron and electron fields. $\bs{D} = \nv-ie\Av$ is the covariant derivative, $\Ev$ and $\bs{B}$ are the electric and magnetic field operators.

The NRQED action implies a finite momentum cutoff $\la \sim m$. Contributions of momenta $|\pv|\gsim\la$ to low energy dynamics are included in the UV-divergent terms in $\mL_{NRQED}$. Their coefficients $c_i$ and $d_i$ are process-independent and may thus be determined (as expansions in powers of $\alpha$) by comparing the results of QED and NRQED for selected processes, such as a scattering amplitude close to threshold. Since both theories are gauge invariant, one may use different gauges in their calculations. Coulomb gauge $\nv\cdot\Av=0$ has been found to be convenient for bound state calculations in NRQED, while covariant gauges (\eg, Feynman gauge) is efficient for scattering amplitudes.

The expansion in powers of $|\pv|/m$ shows that the Coulomb field $A^0$ is the dominant interaction. In \eq{pro32} the vector potential $\Av$, although contributing at the same order in $\alpha$ as $A^0$, is suppressed by a power of $m$. The choice of initial bound state approximation is then evident: The lowest order terms in \eq{pro32} give the familiar non-relativistic Hamiltonian of the Hydrogen atom in Quantum Mechanics. The Schr\"odinger equation with the $A^0$ potential is solved exactly, and the terms of higher orders in $|\pv|/m$ are included using Rayleigh-Schr\"odinger perturbation theory.

NRQED has turned out to be an efficient calculational method for the binding energies of atoms. It has, in particular, allowed the impressive expression \eq{mo1} for the hyperfine splitting of Positronium. The evaluation of the higher order corrections are discussed in \cite{Caswell:1985ui,Kinoshita:1990ai,Kinoshita:1995mt,Pachucki:1997zza,Czarnecki:1999mw,Adkins:2018lvj,Haidar:2019kcp}. The NRQED approach is limited to the rest (or non-relativistic) frames of weakly bound states.

\subsection{Effective theories for heavy quarks} \label{secII.E}
%%%%%%%%%%%%%%%%%%%%%%%%%%

The large masses $m_Q \gg \lqcd$ of the charm and bottom quarks allow the formulation of effective theories for QCD that are analogous to NRQED. Heavy Quark Effective Theory (HQET, reviewed in \cite{Neubert:1993mb}) expands the heavy quark contribution to the QCD action in powers of $1/m_Q$. In a heavy-light bound state the heavy quark velocity is (in the $m_Q \to \infty$ limit) unaffected by soft, \order{\lqcd} hadronic interactions. The light quark and gluon dynamics is in turn independent of the heavy quark flavor and spin. This implies mass degeneracies in the spectrum, such as between the pseudoscalar and vector mesons ($D$ and $D^*$). In leptonic decays $B \to D\ell\nu$ the light system does not feel the sudden change of heavy quark flavor, constraining the decay form factor in the recoilless limit. HQET provides many tests and constraints on the dynamics of heavy hadrons.

Charmonia and bottomonia ($c\bar c$ and $b\bar b$) resemble Positronia, being nearly non-relativistic, compact bound states. This indicates that the coupling $\as$ is perturbative and the Bohr momentum is small, $\as m_Q \sim v\,m_Q \ll m_Q$. The QCD action can then be expanded in powers of $1/m_Q$ similarly as in NRQED. This defines the effective theory of Non-Relativistic QCD (NRQCD, reviewed in \cite{Brambilla:2004jw,Pineda:2011dg}). The interactions of NRQCD are determined by matching with QCD at the cut-off scale $m_Q$. 

NRQCD has light quarks and gluons with momenta of \order{\as\,m_Q}, but also ``ultrasoft'' fields at the binding energy scale \order{\alpha_s^2\,m_Q}. In order to further reduce the number of scales  the \order{v\,m_Q} interactions of NRQCD may be integrated out, defining ``potential NRQCD'' (pNRQCD) \cite{Brambilla:2004jw,Pineda:2011dg} at the \order{v^2\,m_Q} scale. Confinement effects of \order{\lqcd} do not appear in the perturbative framework and their relative importance is unclear. If one assumes that $\as\,m_Q \gg \lqcd$ the matching between NRQCD and pNRQCD can be made perturbatively at \order{\as\,m_Q}. The pNRQCD action has thus been determined, including non-leading orders in $\as$ and $1/\as\,m_Q$. The resulting heavy quark potential is found to agree with the one calculated using lattice methods at short distances ($\lsim 0.25$ fm). Quantitative applications to quarkonia suffer from uncertainties concerning the influence of confinement.

\section{Dirac bound states} \label{secIII}
%%%%%%%%%%%%%%%%%%%%%%%%%%

\subsection{Weak \textit{vs.} strong binding} \label{secIII.A}
%%%%%%%%%%%%%%%%%%%%%%%%%%

The QED atoms discussed above were weakly coupled ($\alpha \ll 1$). We have only a limited understanding of the dynamics of strong binding in QFT. Some features are known in $D=1+1$ dimensions (QED$_2$), where the dimensionless parameter is $e/m$ \cite{Schwinger:1962tp,Coleman:1975pw,Coleman:1976uz}. For $e/m \ll 1$ the $e^+e^-$ states are weakly bound and approximately described by the Schr\"odinger equation. For $e/m \gg 1$ on the other hand the spectrum is that of weakly interacting bosons. This may be qualitatively understood since the large coupling locks the fermion degrees of freedom into compact neutral bound states. In the limit of $e/m \to\infty$ (the massless Schwinger model) QED$_2$ has only a pointlike, non-interacting massive ($M=e/\sqrt{\pi}$) boson. The physical hadron spectrum does not resemble the strong binding limit of QED$_2$.

Solving the relativistic Bethe-Salpeter equation is complicated by the dependence of the kernel on the relative time of the constituents (section \ref{secII.C}). The time dependence is due to the exchange of transversely polarized photons. In chapter \ref{secIV} I take this into account through a Fock expansion of the bound state, keeping the instantaneous (Coulomb) part of the interaction within each Fock state.

The Dirac equation has no retardation effects since the potential $A^\mu$ is external, \ie, fixed. A space-dependent potential $A^\mu(\xv)$ breaks translation invariance, so there are no eigenstates of 3-momentum. Nevertheless, Dirac solutions with large potentials give insights into relativistic binding. For a linear potential $eA^0(\xv) = V'|\xv|$ it has long been known \cite{Plesset:1930zz} (but is rarely mentioned) that the Dirac spectrum is continuous. I discuss this case in section \ref{secIII.F}.

Klein's paradox \cite{Itzykson:1980rh,Klein:1929zz,Hansen:1980nc} signals an essential difference between the Schr\"odinger and Dirac equations. For potentials of the order of the electron mass (\ie, relativistic binding) the Dirac wave function does not describe a single electron. The state has $e^+e^-$ pairs which are not constituents in the usual (non-relativistic) sense. As noted in \cite{Weinberg:1995mt} the Dirac wave function should (when possible) be normalized to unity, regardless of the number of pairs. The Dirac pairs do not add degrees of freedom to the Dirac spectrum, which corresponds to that of a single electron. This motivates the study of the states described by the Dirac wave functions in section \ref{secIII.C}.

\subsection{The Dirac equation} \label{secIII.B}
%%%%%%%%%%%%%%%%%%%%%%%%%%

The Dirac equation  
\begin{align} \label{pro33}
(i\slashed{\partial}-m-e\Asl)\psi(x) =0
\end{align}
should be distinguished from the operator equation of motion for the electron field, given by $\delta\mS_{QED}/\delta\bar\psi(x)=0$. The $c$-numbered equation \eq{pro33} studied by Dirac in 1928 \cite{Dirac:1928hu,Dirac:1928ej} is a relativistic version of the Schr\"odinger equation, where $A^\mu(x)$ is an external, classical field. The condition \eq{pro33} implies that propagation in the field $A^\mu(x)$ is singular for electrons with wave function $\psi(x)$. Scattering in the field is explicit in a perturbative expansion,
\begin{align} \label{pro33b}
\frac{i}{i\slashed{\partial}-m-e\Asl} = \frac{i}{i\slashed{\partial}-m} - \frac{i}{i\slashed{\partial}-m}ie\Asl \frac{i}{i\slashed{\partial}-m} + \ldots
\end{align}

For time-independent potentials $A^\mu(\xv)$ the static solutions $\psi(t,\xv)= \exp(-itM)\Psi(\xv)$ have both positive and negative energy eigenvalues $M$. The corresponding wave functions $\Psi$ and $\ovl\Psi$ satisfy
\begin{align}
\big[-i\nv\cdot\gv+m+e\Asl(\xv)\big]\Psi_n(\xv) &=M_n\gz\Psi_n(\xv)  \label{dir1} \crt
\big[-i\nv\cdot\gv+m+e\Asl(\xv)\big]\ovl\Psi_n(\xv) &=-\ovl M_n\gz\ovl\Psi_n(\xv) \label{dir2}
\end{align}
where $M_n,\ \ovl M_n \geq 0$. The free ($A^\mu = 0$) solutions are given by the spinors \eq{pro6} as $\psi(x)=e^{-itp^0}\Psi_{\pv\lm}(\xv)=u(\pv,\lm)e^{-ip\cdot x}$ and $\psi(x)=e^{itp^0}\ovl\Psi_{\pv\lm}(\xv)=v(\pv,\lm)e^{ip\cdot x}$ with $M_p=\ovl M_p =p^0=\sqrt{\pv^2+m^2}$. The solutions with negative kinetic energy $-\ovl M_p$ are related to positrons. For potentials $A^\mu \gsim m$ the wave function $\psi(x)$ has both positive and negative energy components, due to contributions of $e^-e^+$ pairs (section \ref{secIII.C}).

The Dirac equation with a Coulomb potential can be obtained from a sum of Feynman ladder diagrams, analogously to the Schr\"odinger equation \cite{Brodsky:1971we,Gross:1982nz,Neghabian:1983vm}. There are some instructive differences, however. Relativistic two-particle dynamics cannot be reduced to that of a single particle in an external field, as in \eq{pro2}. We must therefore consider a limit where the mass of one particle goes to infinity. The recoil of the heavy particle may then be neglected. The heavy particle gives rise to a static potential in its rest frame.

Consider again the diagrams in \fig{f2}. Let the mass of the lower (antifermion) line be $m_T$ and its charge be $eZ$. We take $m_T\to\infty$ keeping the electron (fermion) momenta $\pv_1,\qv_1$ fixed. The initial momentum of the antifermion is $p_2=(m_T,\bs{0})$. Since $\qv_2 = \pv_1-\qv_1$ is fixed as $m_T\to\infty$ the energy $q_2^0 = \sqrt{m_T^2+\qv_2^2} = m_T + \morder{1/m_T}$. Thus kinematics ensures that no energy is transferred from the heavy target to the electron, \ie, $p_1^0 = q_1^0$ up to \order{1/m_T}.

In the diagrams of \fig{f2}(b,c) the loop integral converges even without the antifermion propagator. Hence the limit $m_T\to\infty$ can be taken in the integrand. The antifermion spinors are non-relativistic so $\bar v(\pv_2,\lm_2)(-ieZ)\gamma^\mu v(\qv_2,\lm_2') \simeq -ieZ\,2m_T \delta^{\mu,0}\delta_{\lm_2,\lm_2'}$. The Born diagram of \fig{f2}(a) is then, for large $m_T$ and relativistic electron momenta,
\begin{align} \label{pro35}
A_1(\pv_1,\qv_1) = -ieZ\, 2m_T\bar u(\qv_1,\lm'_1)\frac{-ie\gz}{(\qv_1-\pv_1)^2} u(\pv_1,\lm_1)
\end{align}
This corresponds to single scattering in the field of of the heavy particle with charge $-eZ$. 

When the electron is non-relativistic the positive energy pole of its propagator \eq{pro5} dominates. Then the diagram of \fig{f2}(c) with crossed photons is suppressed compared to the uncrossed diagram of \fig{f2}(b). Now the crossed diagram does contribute and is required to get the result
\begin{align} \label{pro36}
A_2(\pv_1,\qv_1) = i(eZ)^2 2m_T \int\frac{d^3\ellv}{(2\pi)^3}\,\bar u(\qv_1,\lm'_1)(-ie\gz)\inv{(\ellv-\pv_1)^2} \frac{i(\lsl+m)}{\ell^2-m^2+\ieps}\inv{(\qv_1-\ellv)^2}(-ie\gz)u(\pv_1,\lm_1)
\end{align}
corresponding to double scattering in the external potential.

%%%%%%%%%%%%%%%%%%%%%%%%%%
\begin{tcolorbox}
\textit{Exercise \ref{e6}:} Derive \eq{pro36}, and convince yourself that also the exchange of three photons reduces to scattering in an external potential. \textit{Hint:} You need only consider the antifermion line, since the upper part is the same for the uncrossed and crossed diagrams.
\end{tcolorbox}
%%%%%%%%%%%%%%%%%%%%%%%%%%

For ladders with $n$ exchanges all $n!$ diagrams with arbitrary crossings of the photons contribute. This means that the Bethe-Salpeter equation \eq{pro15} reduces to the Dirac equation as $m_T\to\infty$ only for kernals of infinite degree in $\alpha$ (containing arbitrarily many crossed photons). The B-S equation can, however, be modified so that it does reduce to the Dirac equation even for finite kernels \cite{Gross:1982nz}.

\begin{figure}[h] \centering
\includegraphics[width=0.5\columnwidth]{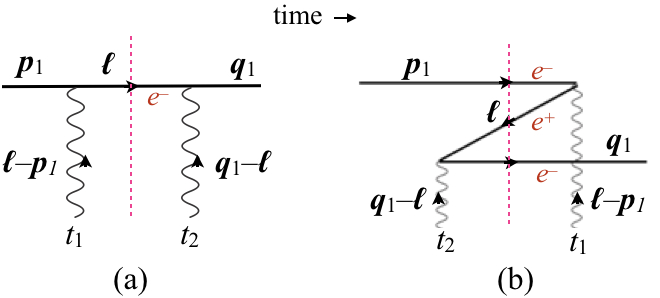}
\caption{Time ordered diagrams for the double scattering \eq{pro36} in an external, static potential. (a) The intermediate electron has positive energy. (b) The intermediate electron has negative energy, corresponding to the creation and subsequent annihilation of an $e^+e^-$ pair. This is often referred to as a ``$Z$-diagram''. \label{f4}}
\end{figure}

In full QED a large charge $eZ$ is screened by the creation of $e^+e^-$ pairs. The $2 \to 2$ ladder diagrams that give the Dirac equation do not describe true pair production. The $e^+e^-$ pairs in the Dirac state which are implied by Klein's paradox \cite{Itzykson:1980rh,Klein:1929zz,Hansen:1980nc} must therefore be virtual. The pairs only arise when the diagrams are time ordered, which is required to determine a state at an instant of time. Time ordering the electron propagator \eq{pro5} gives a positive and negative energy part,
\begin{align} \label{pro37}
S(t,\pv) \equiv \int\frac{dp^0}{2\pi}\,i\frac{(\psl+m)e^{-ip^0t}}{p^2-m^2+\ieps} = \inv{2E_p}\sum_\lm\Big[\theta(t)\,u(\pv,\lm)\bar u(\pv,\lm)\,e^{-itE_p}-\theta(-t)\,v(-\pv,\lm)\bar v(-\pv,\lm)\,e^{itE_p}\Big]
\end{align}
In strong potentials the electron can scatter into a negative energy state which evolves backward in time. This corresponds to an intermediate $e^-e^+e^-$ state, as illustrated in \fig{f4}(b). In weakly coupled bound states, described by the Schr\"odinger equation, such higher Fock components are suppressed.
%%%%%%%%%%%%%%%%%%%%%%%%%%%%%%%%%%%%%%%%%%%%%%%%%%%%%%%%
%%%%%%%%%%%%%%%%%%%%%%%%%%%%%%%%%%%%%%%%%%%%%%%%%%%%%%%%%

Multiple scattering gives rise to Fock states with any number of intermediate $e^+e^-$ pairs. Despite its apparent one-particle nature the Dirac wave function describes many pairs in the free Fock state basis. In order to see this more explicitly we need to define the Dirac states in terms field operators. The following study is based on \cite{Blaizot:2016} and previously published in \cite{Hoyer:2016aew}.

\subsection{Dirac states} \label{secIII.C}
%%%%%%%%%%%%%%%%%%%%%%%%%%

The Dirac wave functions define eigenstates of the Dirac Hamiltonian,
\beq\label{ham1}
H_D(t)=\int d\xv\,\bar\psi(t,\xv)\big[-i\rnab\cdot\gv+m+e\slashed{A}(\xv)\big]\psi(t,\xv)
\eeq
where $\psi(t,\xv)$ now is the electron field with the canonical anticommutation relation
\beq\label{canon}
\acom{\psi^\dag_\alpha(t,\xv)}{\psi_\beta(t,\yv)}=\delta_{\alpha,\beta}\,\delta^3(\xv-\yv)
\eeq
This field may be expanded in the standard operator basis, which creates/annihilates free $e^\pm$ states,
\beqa\label{eop}
\psi_\alpha(t=0,\xv)&=&\int \frac{d\kv}{(2\pi)^32E_k}\sum_\lambda\Big[u_\alpha(\kv,\lambda)e^{i\kv\cdot\xv}b_{\kv,\lambda}+v_\alpha(\kv,\lambda)e^{-i\kv\cdot\xv}d^\dag_{\kv,\lambda}\Big]\\[2mm]
\acom{b_{\pv,\lambda}}{b^\dag_{\qv,\lambda'}} &=& \acom{d_{\pv,\lambda}}{d^\dag_{\qv,\lambda'}} = 2E_p\,(2\pi)^3 \delta^3(\pv-\qv) \delta_{\lambda,\lambda'} \label{acom2}
\eeqa
I take the classical, $c$-numbered potential $A^\mu(\xv)$ to be time independent. There are no physical (propagating) photons. Since the Hamiltonian is quadratic in the fermion fields it can be diagonalized \cite{Blaizot:WjVlxh3f}.

The positive \eq{dir1} and negative \eq{dir2} energy Dirac wave functions determine $e^-$ and $e^+$ states defined at $t=0$ by
\beqa
\ket{M_n} &=& \int d\xv\sum_\alpha\psi^\dag_\alpha(\xv)\Psi_{n,\alpha}(\xv)\ket{\Omega}
\equiv c_n^\dag\ket{\Omega} \label{state1} \\[2mm]
\ket{\ovl M_n} &=& \int d\xv\sum_\alpha{\ovl\Psi}_{n,\alpha}^{\,\dag}(\xv)\psi_\alpha(\xv)\ket{\Omega} \equiv \bar c_n^\dag\ket{\Omega} \label{state2}
\eeqa

Charge conjugation transforms the electron field as
\begin{align} \label{dir3}
\mC\psi^\dag(t,\xv)\mC^\dag = i\psi^T(t,\xv)\gamma^2
\end{align}
Hence
\begin{align} \label{dir4}
\mC\ket{M} = \int d\xv\,\psi^T(\xv)i\gamma^2\Psi(\xv)\ket{\Omega} = \int d\xv\,\Psi^T(\xv)i\gamma^2\psi(\xv)\ket{\Omega}
\end{align}
has the form of $\ket{\ovl M}$ in \eq{state2}, with wave function $\ovl\Psi(\xv)=i\gamma^2\Psi^*(\xv)$. This wave function satisfies the Dirac equation \eq{dir1} with $M \to -M$ and $eA^\mu \to -eA^\mu$, as expected for a positron.

The vacuum state $\ket{\Omega}$ is an eigenstate of the Hamiltonian with eigenvalue taken to be zero,
\beq\label{vac1}
H_D\ket{\Omega} =0
\eeq
Two equivalent expressions for $\ket{\Omega}$ are given in \eq{vac2} below. Using
\beq\label{hdcom}
\com{H_D}{\psi^\dag(\xv)} = \psi^\dag(\xv)\gamma^0(i\lnab\cdot\gv+m+e\Asl) \hspace{2cm}
\com{H_D}{\psi(\xv)} = -\gamma^0(-i\rnab\cdot\gv+m+e\Asl)\psi(\xv)
\eeq
we see that both states \eq{state1} and \eq{state2} are eigenstates of the Dirac Hamiltonian with {\it positive} eigenvalues,
\beqa\label{hdeigen}
H_D\ket{M_n} &=& M_n\ket{M_n} \hspace{1cm} M_n>0 \nn \\[2mm]
H_D\ket{\ovl M_n} &=& \ovl M_n\ket{\ovl M_n} \hspace{1cm} \ovl M_n>0
\eeqa

In terms of the wave functions in momentum space,
\beq\label{fourierdef}
\Psi_n(\xv) = \int \frac{d\pv}{(2\pi)^3}\, \Psi_n(\pv) e^{i\pv\cdot\xv}
 \hspace{2cm} \ovl\Psi_n(\xv) = \int \frac{d\pv}{(2\pi)^3}\, \ovl\Psi_n(\pv) e^{i\pv\cdot\xv}
\eeq
the eigenstate operators defined in \eq{state1} and \eq{state2} can be expressed as
\beqa
c_n &=& \sum_\pv \Psi_n^\dag(\pv)\big[ u(\pv,\lambda)b_{\pv,\lambda}+ v(-\pv,\lambda)d_{-\pv,\lambda}^\dag\big] \equiv B_{np}b_p+D_{np}d^\dag_p  \label{cndef1}\\[-3mm]
 &&\hspace{10cm} \sum_\pv \equiv \int\frac{d\pv}{(2\pi)^3 2E_p}\sum_\lambda \nn \\[-3mm]
\bar c_n &=& \sum_\pv \big[b_{\pv,\lambda}^\dag u^\dag(\pv,\lambda)+ d_{-\pv,\lambda} v^\dag(-\pv,\lambda)\big]\ovl\Psi_n(\pv)
\equiv \ovl B_{np}b_p^\dag+\ovl D_{np}d_p \label{cndef2}
\eeqa
In the second expressions on the rhs. a sum over the repeated index $p \equiv (\pv,\lambda)$ is implied. In the weak binding limit ($|\pv| \ll m$) the positive energy spinor wave function $\Psi_n$ has only upper components, whereas $\ovl\Psi_n$ has only lower components. Then $\ket{M_n}$ is a single electron state, whereas $\ket{\ovl M_n}$ is a single positron state.

The operators $c_n$ and $\bar c_n$ are related to $b,d$ via the Bogoliubov transformations \eq{cndef1} and \eq{cndef2}. Using the commutation relations \eq{acom2} and the orthonormality of the Dirac wave functions we see that they obey standard anticommutation relations,
\beqa\label{cmnortho}
\acom{c_m}{c_n^\dag} &=& \sum_\pv \Psi_{m,\alpha}^\dag(\pv) \big[u_\alpha(\pv,\lambda)u_\beta^\dag(\pv,\lambda) + v_\alpha(-\pv,\lambda)v_\beta^\dag(-\pv,\lambda)\big]\Psi_{n,\beta}(\pv)
= \int\frac{d\pv}{(2\pi)^3}\Psi_{m,\alpha}^\dag(\pv)\Psi_{n,\alpha}(\pv) =\delta_{mn}\nn \\[2mm]
\acom{\bar c_m}{c_n^\dag} &=&0 \nn \\[2mm]
\acom{\bar c_m}{\bar c_n^\dag} &=& \int\frac{d\pv}{(2\pi)^3}\bar \Psi_{m,\alpha}^\dag(\pv)\bar\Psi_{n,\alpha}(\pv) =\delta_{mn}
\eeqa
Inserting the completeness condition for the Dirac wave functions into the Dirac Hamiltonian \eq{ham1} gives,
\beqa
H_D &=& \sum_n \big[M_n c_n^\dag c_n + \bar M_n \bar c_n^\dag \bar c_n\big] \label{hdiag}
\eeqa

%%%%%%%%%%%%%%%%%%%%%%%%%%
\begin{tcolorbox}
\textit{Exercise \ref{e7}:} Derive \eq{hdiag}.
\end{tcolorbox}
%%%%%%%%%%%%%%%%%%%%%%%%%%

The expression for the vacuum state may be found using the methods in \cite{Blaizot:WjVlxh3f}. $H_D\ket{\Omega}=0$ when, in terms of the $B$ and $D$ coefficients defined in \eq{cndef1} and \eq{cndef2}, 
\beq\label{vac2}
\ket{\Omega} = N_0\exp\Big[-b_q^\dag \big(B^{-1}\big)_{qn}D_{nr}d_r^\dag\Big]\ket{0}
 = N_0\exp\Big[-d_r^\dag \big({\ovl D}^{\,-1})_{rn}{\ovl B}_{nq} b_q^\dag\Big]\ket{0}
\eeq 
Sums over the repeated indices $q,n,r$ are implied in the exponents, and  $N_0$ is a normalization constant. The perturbative vacuum satisfies $b_p\ket{0} = d_p\ket{0} = 0$. The vacuum state $\ket{\Omega}$ describes the distribution of the $e^+e^-$ pairs that arise through perturbative contributions such as \fig{f4}(b). It is a formal expression, involving a sum over all states $n$ and the inverted matrices $\big(B^{-1}\big)_{qn}$ and ${\ovl D}^{\,-1})_{rn}$. In the weak binding limit $D_{nr} \to 0,\ \ovl B_{nq}\to 0$ and $\ket{\Omega}\to\ket{0}$.

The vacuum is ``empty'' in the bound state basis: $c_n\ket{\Omega}=\bar c_n\ket{\Omega} = 0$. The pairs appear only in bases which do not diagonalize the Hamiltonian, such as the free basis generated by the $b^\dag$ and $d^\dag$ operators. 

%%%%%%%%%%%%%%%%%%%%%%%%%%
\begin{tcolorbox}
\textit{Exercise \ref{e8}:} (a) Show the equivalence of the two expressions for $\ket{\Omega}$ in \eq{vac2}. \textit{Hint:} Prove that $B_{mp}\ovl B_{np}+D_{mp}\ovl D_{np}=0$. (b) Prove that $H_D\ket{\Omega}=0$. \textit{Hint:} Note that $b_p$ essentially differentiates the exponents in \eq{vac2}.
\end{tcolorbox}
%%%%%%%%%%%%%%%%%%%%%%%%%%

The Dirac bound states \eq{state1} may be expressed in terms of their electron and positron (``hole'') distributions,
\begin{align} \label{dir18}
\ket{M_n} &= \int\frac{d\pv}{(2\pi)^3 2E_p}\sum_s\big[e_n^-(\pv,s)b^\dag_{\pv s}+e_n^+(\pv,s)d_{-\pv s}\big]\ket{\Omega} \nn\crt
e_n^-(\pv,s) &= u^\dag(\pv,s)\Psi_{n}(\pv) \hspace{2cm}
e_n^+(\pv,s) = v^\dag(-\pv,s)\Psi_{n}(\pv)
\end{align}
with momentum space wave functions $\Psi_{n}(\pv)$ defined as in \eq{fourierdef}.
The corresponding electron and positron densities
\begin{align} \label{dir19}
\rho_n(e^\mp,p) &\equiv \int \frac{d\Omega_p\, p^2}{(2\pi)^3 2E_p} \sum_s|e^\mp(\pv,s,)|^2 
\end{align}
are normalized so that
\begin{align} \label{dir20}
\int_0^\infty dp\big[\rho_n(e^-,p)+\rho_n^+(e^+,p)\big] = 1
\end{align}

%%%%%%%%%%%%%%%%%%%%%%%%%%
\subsection{*\,Dirac wave functions for central $A^0$ potentials} \label{secIII.D}
%%%%%%%%%%%%%%%%%%%%%%%%%%

The wave functions $\Psi(\xv)$ of Dirac bound states in rotationally symmetric potentials $eA^0(\xv)=V(r)$ with $\Av(\xv)=0$ satisfy ($\alv \equiv \gz\gv$)
\begin{align} \label{dircen}
(-i\alv\cdot\nv+m\gz)\Psi(\xv) = \big[M-V(r)\big]\Psi(\xv)
\end{align}
The states may be characterized by their mass $M$, angular momentum $j,\,j^z \equiv \lm$ and parity $\eta_P=\pm 1$. The angular momentum operator in the fermion representation is
\begin{align}\label{dir6}
\bs{\mJ} &= \int d\xv\,\psi^\dag(\xv)\,\bs{J}\,\psi(\xv)
\end{align}
where $\Jv$ is the sum of the orbital $\Lv$ and spin $\Sv$ angular momenta (which are not separately conserved),
\begin{align} \label{dir5a}
\bs{J} = \Lv+\Sv= \xv\times(-i\nv)+\halft\gf\alv
\end{align}
Operating on the states $\ket{M,j\lm}$ in \eq{state1} we get
\begin{align}\label{dir7}
\bs{\mJ}\ket{M,j\lm} &= \int d\xv\,\psi^\dag(\xv)\,\bs{J}\,\Psi_{j\lm}(\xv) \ket{\Omega}
\hspace{2cm} 
\bs{\mJ}^2\ket{M,j\lm} = \int d\xv\,\psi^\dag(\xv)\,\bs{J}^2\,\Psi_{j\lm}(\xv) \ket{\Omega}
\end{align}
The Dirac 4-spinor wave functions are thus required to satisfy
\begin{align}\label{dir8}
\Jv^2\Psi_{j\lm} = j(j+1)\Psi_{j\lm} \hspace{2cm} J^z\Psi_{j\lm} = \lm\Psi_{j\lm}  
\end{align}

The parity operator is defined by
\begin{align} \label{dir9}
\bP \psi^\dag(t,\xv)\bP^\dag &= \psi^\dag(t,-\xv)\gz \nn\crt
\bP\ket{M,j\lm} &= \int d\xv\,\psi^\dag(\xv)\,\gz\,\Psi_{j\lm}(-\xv) \ket{\Omega} = \eta_P \ket{M,j\lm}
\end{align}
Hence the Dirac wave functions of states with parity $\eta_P$ should satisfy
\begin{align} \label{dir9a}
\gz\,\Psi_{j\lm}(-\xv) = \eta_P \Psi_{j\lm}(\xv)
\end{align}

Denoting $\xv = r(\sin\theta\cos\vphi,\sin\theta\sin\vphi,\cos\theta)$ and $\hat\xv \equiv \xv/r$, the angular dependence of $\Psi_{j\lm}(\xv)$ may be expressed using the orthonormalized 2-spinors \cite{Itzykson:1980rh},
\begin{align} \label{dir9b}
\phi_{j\lm+}(\theta,\varphi) = \inv{\sqrt{2j}}&\left(\begin{array}{c}\sqrt{j+\lm}\,Y_{j-\halft}^{\lm-\halft}(\theta,\varphi) \\[4mm] \sqrt{j-\lm}\,Y_{j-\halft}^{\lm+\halft}(\theta,\varphi) \end{array}\right) \nn\crt
\phi_{j\lm-}(\theta,\varphi) = \inv{\sqrt{2(j+1)}}&\left(\begin{array}{c}\sqrt{j-\lm+1}\,Y_{j+\halft}^{\lm-\halft}(\theta,\varphi) \\[4mm] -\sqrt{j+\lm+1}\,Y_{j+\halft}^{\lm+\halft}(\theta,\varphi) \end{array}\right)
= \sv\cdot\hat\xv\,\phi_{j\lm+}(\theta,\varphi)
\end{align}
The $\pm$ notation refers to $j=\ell\pm\halft$, where $\ell$ is the order of the spherical harmonic function $Y_\ell^m(\theta,\varphi)$, which becomes the conserved orbital angular momentum in the non-relativistic limit.
In the standard notation $J^{\pm}=J^x\pm iJ^y$,
\begin{align} \label{dir9c}
\sv\cdot\Lv &= \halft\sigma^+L^- + \halft\sigma^-L^+ +\sigma^z L^z \nn\crt
L^\pm\ket{\ell,\lm} &= \sqrt{(\ell\mp\lm)(\ell\pm\lm+1)}\ket{\ell,\lm\pm 1}
\end{align}
it is straightforward to verify that
\begin{align} \label{dir9d}
\sv\cdot\Lv\,\phi_{j\lm\pm} &= c_{j\pm}\phi_{j\lm\pm} \hspace{2cm}
\left\{\begin{array}{l} c_{j+}=j-\halft \crt  c_{j-}= -(j+\sfrac{3}{2}) \end{array}\right. \nn\crt
(\Lv+\halft\sv)^2\phi_{j\lm\pm} &= j(j+1)\phi_{j\lm\pm}
\end{align}

The 4-spinor Dirac wave functions $\Psi_{j\lm\pm}(\xv)$ describing the states $\ket{M,j\lm\pm}$ of \eq{state1} with $\eta_P=(-1)^{j\mp\halft}$ may now be defined in terms of two radial functions,
\begin{align} \label{dir9e}
\Psi_{j\lm\pm}(\xv) &= \big[F_{j\pm}(r)+i\alv\cdot\hat\xv\, G_{j\pm}(r)\big]\left(\begin{array}{c}\phi_{j\lm\pm}(\theta,\varphi) \crt 0 \end{array}\right)
\end{align}
Since $\com{\Jv}{\alv\cdot\hat\xv}=0$ the angular momentum quantum numbers \eq{dir8} are ensured by \eq{dir9d}. The parity $\eta_P = (-1)^{j\mp1/2}$ follows from $\gz\alv\cdot(-\hat\xv) = \alv\cdot\hat\xv\gz$ and $\phi_{j\lm\pm}(\pi-\theta,\varphi+\pi) = (-1)^{j\mp1/2}\phi_{j\lm\pm}(\theta,\varphi)$.

The eigenvalue condition \eq{hdeigen} determines the bound state equation for $\Psi_{j\lm\pm}(\xv)$. For this we need the relations
\begin{align}\label{dir10}
-i\alv\cdot\nv &= -i(\alv\cdot\hat\xv)\,\partial_r -\inv{r}\,\alv\cdot\hat\xv\times\bs{L} \nn\crt
-i\alv\cdot\nv(i\alv\cdot\hat\xv) &= \frac{2}{r}+ \partial_r+\inv{r}\gf\,\alv\cdot\Lv
\end{align}

%%%%%%%%%%%%%%%%%%%%%%%%%%
\begin{tcolorbox}
\textit{Exercise \ref{e9}:} Derive the identities \eq{dir10}. \textit{Hint:} Use $\alpha^i\alpha^j = \delta^{ij}+i\gf\epsilon^{ijk}\alpha^k$.
\end{tcolorbox}
%%%%%%%%%%%%%%%%%%%%%%%%%%

We may furthermore use
\begin{align} \label{dir10a}
\alv\cdot\hat\xv\times\bs{L}\left(\begin{array}{c}\phi_{j\lm\pm} \crt 0 \end{array}\right) =
\left(\begin{array}{c} 0 \crt \sv\cdot\hat\xv\times\bs{L}\,\phi_{j\lm\pm} \end{array}\right) =
-i\left(\begin{array}{c} 0 \crt (\sv\cdot\hat\xv)(\sv\cdot\Lv)\phi_{j\lm\pm} \end{array}\right) =
-c_{j\pm}\,i\alv\cdot\hat\xv\left(\begin{array}{c} \phi_{j\lm\pm} \crt 0 \end{array}\right)
\end{align}
with $c_{j\pm}$ given in \eq{dir9d}, while $\gf\,\alv\cdot\Lv = 2\Sv\cdot\Lv$ contributes $c_{j\pm}$ with unit Dirac matrix.
The $m\gz$ term in $H_D$ gives $m(F_{j\pm}-i\alv\cdot\hat\xv\, G_{j\pm})$.
Identifying the coefficients in $H_D\ket{M} = M\ket{M}$ of the two Dirac structures in \eq{dir9e}
we get
\begin{align} 
\Big(\frac{j+3/2}{r}+\partial_r\Big)G_{j+} = (M_{j+}-V-m)F_{j+}& \hspace{1cm}
 &\Big(\frac{j-1/2}{r}-\partial_r\Big)F_{j+} = (M_{j+}-V+m)G_{j+} \label{dir10b1}\crt
-\Big(\frac{j+3/2}{r}+\partial_r\Big)F_{j-} = (M_{j-}-V+m)G_{j-}& \hspace{1cm}
 &\hspace{-.3cm}-\Big(\frac{j-1/2}{r}-\partial_r\Big)G_{j-} = (M_{j-}-V-m)F_{j-} \label{dir10b2}
\end{align}

These reduce to second order equations for $F$ and $G$ separately. Suppressing the subscripts $j\pm$, 
\begin{align} \label{dir10c}
F''+\Big(\frac{2}{r}+\frac{V'}{M-V+m}\Big)F'+ \Big[(M-V)^2-m^2-\frac{c(c+1)}{r^2}-\frac{c\,V'}{r(M-V+m)}\Big]F=0 \nn\crt
G''+\Big(\frac{2}{r}+\frac{V'}{M-V-m}\Big)G'+ \Big[(M-V)^2-m^2-\frac{(c+1)(c+2)}{r^2}+\frac{(c+2)V'}{r(M-V-m)}\Big]G=0
\end{align}
At the potentially singular points $M-V\pm m=0$ the solutions behave as $(M-V\pm m=0)^\beta$, with $\beta= 0$ or $\beta=2$, and are thus locally normalizable there.

If $\Psi_{j\lm+}(\xv)$ solves the  Dirac equation \eq{dircen} then $\wt\Psi_{j\lm+}(\xv) \equiv \gf\Psi_{j\lm+}(\xv)$ solves this equation with $m \to -m$ and the same eigenvalue $M_{j+}$. This shows up as a symmetry of the bound state equations. Using $\phi_{j\lm-} = \sv\cdot\hat\xv\,\phi_{j\lm+}$,
\begin{align} \label{dir9f}
\wt\Psi_{j\lm+}(\xv) &= \big[F_{j+}(r)+i\alv\cdot\hat\xv\, G_{j+}(r)\big]\left(\begin{array}{c} 0 \crt \phi_{j\lm+} \end{array}\right) = 
\big[\alv\cdot\hat\xv\, F_{j+}(r)+i\, G_{j+}(r)\big]\left(\begin{array}{c} \sv\cdot\hat\xv\,\phi_{j\lm+} \crt 0 \end{array}\right) \nn\crt
&= i\big[G_{j+}(r)-i\alv\cdot\hat\xv\, F_{j+}(r)\big]\left(\begin{array}{c} \phi_{j\lm-} \crt 0 \end{array}\right) = \Psi_{j\lm-}(\xv)\big[F_{j-}\to iG_{j+},\,G_{j-}\to -iF_{j+}\big]
\end{align}
Eq. \eq{dir10b2} is indeed seen to transform into \eq{dir10b1} when $m \to -m$, $M_{j-}\to M_{j+}$ and the $j-$ radial wave functions are replaced with the $j+$ functions as indicated in \eq{dir9f}. This means that the solution of \eq{dir10b2}, if allowed by the quantum numbers, is given by $F_{j-}(r,m) = G_{j+}(r,-m),\ G_{j-}(r,m) = -F_{j+}(r,-m)$ with the same eigenvalue $M_{j-} = M_{j+}$. "Squaring" the Dirac equation \eq{dircen} by multiplying it with $-i\alv\cdot\nv +m\gz$ gives
\begin{align} \label{pro38}
\big(-\nv^2+m^2\big)\Psi(\xv)=\big[(M-V)^2+i\alv\cdot (\nv V)\big]\Psi(\xv)
\end{align} 
Since this equation depends on $m$ only via $m^2$ the eigenvalue $M$ is independent of the sign of $m$.
The degeneracy $M_{j-} = M_{j+}$ is familiar in the case of a Coulomb potential, to which I turn next.

\subsection{*\,Coulomb potential $V(r)=-\alpha/r$} \label{secIII.E}
%%%%%%%%%%%%%%%%%%%%%%%%%%

There is a standard and elegant method \cite{Itzykson:1980rh} for finding the Dirac spectrum in the case of a Coulomb potential $V(r)=-\alpha/r$. One starts from the squared Dirac equation \eq{pro38}, determining the eigenvalues of the Dirac matrix $i\alv\cdot (\nv V)$. For $V(r)=-\alpha/r$ all terms in \eq{pro38} may then be formally identified, based on their $r$-dependence, with those of the Schr\"odinger equation,
\begin{align} \label{cou0}
\Big(-\inv{2m}\nv^2-\frac{\alpha}{r}\Big)\Psi(\xv) = E_b\Psi(\xv)
\end{align}
The known solution of this equation allows to determine the masses $M$ of the Dirac states,
\begin{align} \label{cou1}
M_{nj} = m\left[1+\Bigg(\frac{\alpha}{n-(j+\halft)+\sqrt{(j+\halft)^2-\alpha^2}}\Bigg)^2\,\right]^{-\halft}
\end{align} 
The principal quantum number $n=1,2,3,\ldots$ and $j \leq n - \halft$. There are two states for each mass, $M_{nj+}=M_{nj-}$, except only $M_{nj+}$ for $j=n-\halft$. For $\alpha\to 0$ we recover the non-relativistic Schr\"odinger result which depends only on $n$, $M_{nj} = m(1-\alpha^2/2n^2)$. 

For $r\to \infty$ \eq{dir10c} reduces to $F''+(M^2-m^2)F=0$, implying $F(r\to\infty) \sim \exp(-r\sqrt{m^2-M^2})$. Hence $M<m$ for normalizable solutions, as in \eq{cou1}. I illustrate using the the radial wave functions $F(r)$ and $G(r)$ of the states with maximal spin $j=n-\halft$, and of the first radial excitation with $j=n-\sfrac{3}{2}$. The wave functions $\Psi_{nj\lm\pm}(\xv)$ are expressed in terms of the radial functions $F_{nj\pm}$ and $G_{nj\pm}$ as in \eq{dir9e}, with the angular functions $\phi_{j\lm\pm}$ given in \eq{dir9b}.

\vspace{.3cm}
\textit{Maximal spin, $j=n-\halft$}
\begin{align} \label{cou2}
M_{j+\inv{2},j,+} &= \frac{m\gamma}{j+\halft} &\gamma\equiv \sqrt{(j+\halft)^2-\alpha^2} \nn\crt
F_{j+\inv{2},j,+}(r) &= N_1\,r^{\gamma-1} \exp(-\mu r) &\mu \equiv \sqrt{m^2-M^2} = \frac{\alpha m}{j+\halft} \nn\crt
G_{j+\inv{2},j,+}(r) &= N_1\,\frac{\mu}{M+m}\,r^{\gamma-1} \exp(-\mu r) & N_1^2 \equiv \frac{(2\mu)^{1+2\gamma}}{\Gamma(1+2\gamma)}\Big[1+\frac{\mu^2}{(M+m)^2}\Big]^{-1}
\end{align}
This state is not degenerate, \ie, there are no radial functions $F_{j+1/2,j,-},\ G_{j+1/2,j,-}$.
In momentum space \eq{fourierdef} the wave functions of the $n=1$ ground state $\ket{M,n=1,j=\halft,\lm,+}$ are, with $\chi_{\inv{2}}=(1\ 0)^{\rm T}$ and $\chi_{-\inv{2}}=(0\ 1)^{\rm T}$,
\begin{align} \label{dir21}
\Psi_{1,1/2,\lm,+}(\pv) &= \big[f(p)+\alv\cdot\hat{\pv}\, g(p)\big] \left(\begin{array}{c} \chi_\lm \crt 0  \end{array}\right) \nn\crt
f(p) &= \sqrt{4\pi}\, N_1\,\Gamma(1+\gamma)\,\frac{\sin[\delta(1+\gamma)/2]}{p(\alpha^2 m^2+p^2)^{(1+\gamma)/2}} \hspace{2cm} \exp(i\delta) \equiv \frac{\alpha m+ip}{\alpha m-ip} \nn\crt
g(p) &= -\frac{\sqrt{4\pi}\,\alpha}{1+\gamma}\,N_1\,\Gamma(\gamma)\,\frac{\partial}{\partial p}\Big[\frac{\sin(\delta\gamma/2)}{p(\alpha^2 m^2+p^2)^{\gamma/2}}\Big]
\end{align}
The electron and positron density distributions \eq{dir19} are then
\begin{align} \label{dir22}
\rho(e^\mp,p) = \frac{p^2}{4\pi^2\,E_p}\big[E_p(f^2+g^2)\pm m(f^2-g^2)\pm 2p\,f\,g\big]
\end{align}

\begin{figure}[h] \centering
\includegraphics[width=0.5\columnwidth]{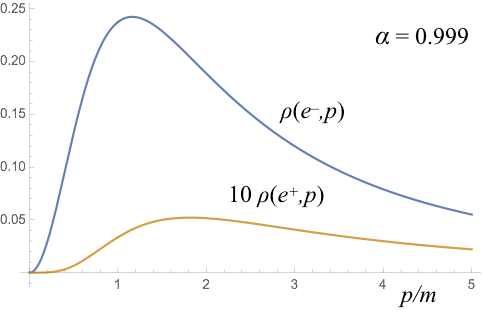}
\caption{Electron and positron densities \eq{dir22} in the $\ket{M,1,\halft,\lm,+}$ Dirac state \eq{dir21} with $V(r)=-\alpha/r$ and $\alpha = 0.999$. The positron density is multiplied by a factor 10. \label{f5}}
\end{figure}

The electron density $\rho(e^-,p)$ is strongly dominant. For $\alpha = 1/137$ the contribution of the positron density to the state normalization is a mere $3.2\cdot 10^{-12}$. Even for $\alpha = 0.999$ (see \fig{f5}) the positron contributes only 3\% to the normalization. The $j=\halft$ eigenvalue $M_{1,\inv{2},+}$ is complex for $\alpha > 1$.

%%%%%%%%%%%%%%%%%%%%%%%%%%%%%%%%%%%%%%%%%%%%%%%%%%%%%%%%
%%%%%%%%%%%%%%%%%%%%%%%%%%%%%%%%%%%%%%%%%%%%%%%%%%%%%%%%%

\vspace{.3cm}
\textit{First radial excitation, $j=n-\sfrac{3}{2}$}
\begin{align} \label{cou3}
M_{j+\frac{3}{2},j,\pm} &= \frac{m(1+\gamma)}{\sqrt{(j+\halft)^2+1+2\gamma}} &\gamma\equiv \sqrt{(j+\halft)^2-\alpha^2} \nn\crt
F_{j+\frac{3}{2},j,+}(r) &= N_{2+}\,r^{\gamma-1} \exp(-\mu r)\Big[r-\frac{\alpha(1+2\gamma)}{(M+m)(j+\halft-\gamma)+\alpha\mu}\Big] &\mu \equiv \sqrt{m^2-M^2} = \frac{\alpha M}{1+\gamma} \nn\crt
G_{j+\frac{3}{2},j,+}(r) &= N_{2+}\,r^{\gamma-1} \exp(-\mu r)\Big[\frac{\mu}{M+m}\,r -\frac{2j+1-2\gamma}{2\alpha}\,\frac{\alpha(1+2\gamma)}{(M+m)(j+\halft-\gamma)+\alpha\mu}\Big]
\end{align}
The degenerate state with the same mass and spin but opposite parity has, as argued at the end of the previous section, radial functions with $F_-(m) = G_+(-m)$ and $G_-(m) = -F_+(-m)$ (up to the normalizations $N_{2\pm}$),
\begin{align} \label{cou4}
F_{j+\frac{3}{2},j,-}(r) &= N_{2-}\,r^{\gamma-1} \exp(-\mu r)\Big[\frac{\mu}{M-m}\,r- \frac{2j+1-2\gamma}{2\alpha}\frac{\alpha(1+2\gamma)}{(M-m)(j+\halft-\gamma)+\alpha\mu}\Big] \nn\crt
G_{j+\frac{3}{2},j,-}(r) &= -N_{2-}\,r^{\gamma-1} \exp(-\mu r)\Big[r -\frac{\alpha(1+2\gamma)}{(M-m)(j+\halft-\gamma)+\alpha\mu}\Big]
\end{align}

\vspace{.3cm}
\textit{Unbound states, $M^2-m^2 > 0$}

States with masses $M >m$ are unbound. The radial equations \eq{dir10b1} imply for all $\ket{M,nj\lm,+}$ states,
\begin{align} \label{cou5}
F_{nj+}(r\to\infty) &= N\, r^\beta \exp(\pm i\mu r)  & \mu = \sqrt{M^2-m^2} \nn\crt
G_{nj+}(r\to\infty) &= N\,\frac{\mp i\mu}{M+m} r^\beta \exp(\pm i\mu r)  & \beta = \mp\frac{i\alpha}{\sqrt{1-m^2/M^2}}-1
\end{align}
In the absence of a normalization condition the mass spectrum is continuous. At large $r$ (where $V \to 0$) the solution is a spherical wave with momentum $p = \pm \mu$, modulated by the phase factor $r^{\beta+1}$. The norm $r^2 |\Psi|^2$ tends to a constant at large $r$.

\subsection{*\,Linear potential $V(r)=V'r$} \label{secIII.F}
%%%%%%%%%%%%%%%%%%%%%%%%%%

Hadron phenomenology, and particularly the description \cite{Eichten:1979ms,Eichten:2007qx} of quarkonia using the Schr\"odinger equation with the Cornell potential \eq{mo3}, motivates studying Dirac states with a linear potential, $eA^0(\xv)=V(|\xv|)=V'r$, $\Av=0$. The solutions of the Dirac equation for polynomial potentials have since the 1930's \cite{Plesset:1930zz} been known to be quite different from those of the Schr\"odinger equation. 

I first recall the solutions of the Schr\"odinger equation \eq{cou0} for a linear potential. The $\ell=0$ wave function $\phi(r)$ satisfies 
\begin{align} \label{lin1}
\Big[-\inv{2m}\Big(\partial_r^2+\frac{2}{r}\partial_r\Big)+V'r\Big]\phi(r) = E_b\phi(r)
\end{align}
The normalizable solutions are given by an Airy function,
\begin{align} \label{lin2}
\phi(r) = \frac{N}{r}\,\textrm{Ai}\big[(2mV')^{1/3}(r-E_b/V')\big]
\end{align}
The discrete values of the binding energy $E_b$ are determined by requiring $\phi(r=0)$ to be regular, which implies $\textrm{Ai}\big[-(2mV')^{1/3}E_b/V')\big]=0$. Since the potential grows linearly with $r$ all states are bound (confined), and their wave functions vanish exponentially for $r \to\infty$.

The Dirac radial functions on the other hand are oscillatory at large $r$, as seen from \eq{dir10b1} and \eq{dir10c},
\begin{align} \label{lin3}
F(r\to\infty) &\simeq N\,r^\beta \exp\big[i(M-V)^2/2V'\big]  & \beta=-\frac{im^2}{2V'}-1 \nn\crt
G(r\to\infty) & \simeq i\,F(r\to\infty)
\end{align}
This result (and its complex conjugate) is independent of the quantum numbers $n,j,\pm$. I retained some non-leading terms in the exponent for ease of notation. The essential feature is that
\begin{align} \label{lin4}
F(r\to\infty) &\sim -iG(r\to\infty) \sim N r^\beta\exp\big[iV'r^2/2\big]
& r^2|F(r\to\infty)|^2 = r^2|G(r\to\infty)|^2 = N^2
\end{align}
Thus the normalization integral diverges even though the potential is confining. In the absence of a normalization constraint the mass spectrum is continuous for all $M$, in contrast to the discrete spectrum of the Schr\"odinger equation.

A Dirac electron state \eq{state1} has Fock components with positrons in the vacuum $\ket{\Omega}$ \eq{vac2}. This is seen perturbatively in time ordered $Z$-diagrams such as in \fig{f4}(b). The distribution of the positrons is traced by the $d$-operator in the state creation operator $c_n^\dag$ \eq{cndef1}, motivating the definitions of the $e^{\mp}(\pv,s)$ probabilities in \eq{dir18}. 

A linear potential confines electrons, limiting their distribution to distances where $V'r \lsim M-m$. The same potential repulses positrons, pushing them to large distances with kinetic energy big enough to cancel their negative potential, $p -V'r \sim M+m$. The exponent $\exp(i r\,V'r/2)$ of $F(r\to\infty)$ in \eq{lin4} implies momenta increasing with $r$ as $p \sim V'r/2$. The relation between the $F$ and $G$ radial functions allows to verify that the $e^+$ distribution indeed dominates at large momenta (equivalent to large $r$),
\begin{align} \label{lin5}
\lim_{|\pv|\to\infty}\,\frac{e^-(\pv,s)}{e^+(\pv,s)} = \lim_{|\pv|\to\infty}\,\frac{u^\dag(\pv,s)\Psi(\pv)}{v^\dag(-\pv,s)\Psi(\pv)} = 0
\end{align}

%%%%%%%%%%%%%%%%%%%%%%%%%%
\begin{tcolorbox}
\textit{Exercise \ref{e10}:} Derive \eq{lin5} for a state with $j=1/2$ and parity $\eta_P = +1$. \textit{Hint:} Calculate the momentum space wave function \eq{fourierdef} for $|\pv| \to \infty$ using the stationary phase approximation.
\end{tcolorbox}
%%%%%%%%%%%%%%%%%%%%%%%%%%

An equivalent interpretation is that the wave function is a superposition of electrons, confined to low $r$, and accelerating/decelerating positrons at large $r$, whose negative kinetic energy balances the positive potential. The spectrum is continuous because the positron energies are continuous.

The tunneling of the $e^+$ to $r \simeq 0$ is exponentially suppressed with growing fermion mass $m$. Hence if the initial condition $G(r=0)/F(r=0)$ of the radial equations \eq{dir10b1} is such as to include a positron contribution (beyond the tunneling rate) the wave function will grow rapidly with $r$ and start oscillating with an amplitude which is exponentially large in $m$. The precise values of $G(r=0)/F(r=0)$ which suppress the positrons at $r=0$ correspond to the discrete bound state masses $M$ of the normalizable solutions of the Schr\"odinger equation. All other values of $M$ give, in the $m \to \infty$ limit, wave functions which grow exponentially with $r$.

These properties were confirmed quantitatively in $D=1+1$ dimensions, using the analytic expression of the wave function in terms of confluent hypergeometric functions \cite{Dietrich:2012un,Hoyer:2016aew}.

\section{Fock expansion of bound states in temporal ($A^0=0$) gauge \label{secIV}}
%%%%%%%%%%%%%%%%%%

\subsection{Definition of the bound state method \label{secIV.A}}
%%%%%%%%%%%%%%%%%%%%%%%%%%

\subsubsection{Considerations \label{secIV.A1}}
%%%%%%%%%%%%%%%%%%%%%%%%%%

Perturbative expansions depend on the choice of a lowest order approximation. The perturbative $S$-matrix expands around free states, which works well for scattering amplitudes. Bound states are stationary in time and thus, in a sense, the very opposites of scattering amplitudes. QED approaches to atoms have been thoroughly considered, with conceptual milestones such as the Bethe-Salpeter equation \cite{Salpeter:1951sz} (1951), the realization that it is not unique \cite{Caswell:1978mt} (1978) and NRQED \cite{Caswell:1985ui} (1986). Even in a first approximation atoms are described by wave functions that are non-polynomial in $\alpha$. NRQED expands around states defined by the Schr\"odinger equation.

Poincar\'e symmetry can be explicitly realized only for generators that mutually commute. Equal-time bound states are defined as eigenstates of the Hamiltonian, which in their rest frame have explicit (kinematic) symmetry under space translations and rotations. The frame dependence of these states is defined by boosts. It is not trivial to determine the boost generators of atoms, which are spatially extended. Alternatively, states with general CM momentum $\Pv$ may be found as eigenstates of the Hamiltonian. Full rotational invariance is lost for $\Pv \neq 0$, but the requirement of a correct $\Pv$-dependence of the energy, $E(\Pv) = \sqrt{\Pv^2+M^2}\,$, is a strong constraint. Field theory ensures covariance, as emphasized by Weinberg in the preface of \cite{Weinberg:1995mt}: 

\textit{``The point of view of this book is that quantum field theory is the way it is because (aside from theories like string theory that have an infinite number of particle types) it is the only way to reconcile the principles of quantum mechanics (including the cluster decomposition property) with those of special relativity.''} 

The examples below will illustrate how subtly Poincar\'e covariance is realized for bound states. Much remains to be understood in this regard. Using ``relativistic wave equations'' is not sufficient, as demonstrated in \cite{Artru:1983gm}.

There are many formally equivalent approaches to bound states. In the following I briefly motivate and define my choice, guided by the properties of atoms and hadrons. Some further comments are given in chapter \ref{secVIII}.

\subsubsection{Choice of approach \label{secIV.A2}}
%%%%%%%%%%%%%%%%%%%%%%%%%%

\textbf{Hamiltonian eigenstates}

Bound states can be identified in two equivalent but distinct ways: As poles in Green functions or as eigenstates of the Hamiltonian. The former involves propagation in time and space, allowing for explicit Poincar\'e invariance as in the Dyson-Schwinger framework. The propagation of bound state constituents is complicated by their state-dependent, mutual interactions. A Hamiltonian framework distinguishes time from space. The eigenstate condition involves no propagation in time, and Poincar\'e invariance emerges dynamically. I shall use the method of Hamiltonian eigenstates, akin to traditional quantum mechanics and NRQED.

\vspace{2mm}
\textbf{Instant time quantization}

Quantum states are traditionally defined at an instant of time $t$ (IT), but relativistic states are also commonly defined  at equal Light-Front (LF) time $t+z$ \cite{Burkardt:1995ct,Brodsky:1997de}. The latter is natural in the description of hard collisions, where a single probe (virtual photon or gluon) interacts with the target at a fixed LF time. LF states are described by boost-invariant wave functions, whereas IT wave functions transform dynamically under boosts. On the other hand, the LF choice of $z$-direction breaks rotational invariance, making angular momentum (other than $J^z$) dynamic even in the rest frame. The so called ``zero modes'' require special attention in LF quantization \cite{Collins:2018aqt,Ji:2020baz,Mannheim:2020rod}. 

A perturbative approach allows to study the frame dependence of IT wave functions at each order of the expansion. Rest frame states can be characterized by their angular momentum $\Jv^2$ and $J^z$. Quantization is simpler at equal ordinary time. For these reasons I choose IT quantization.

\vspace{2mm}
\textbf{Temporal ($\bs{A^0=0}$) gauge}

Gauge theories have a local action, but the gauge may be fixed in all of space at an instant of time. The gauge dependent fields $A^0$ and $\Av_L$ then give rise to an instantaneous potential, such as the Coulomb potential $V(r)=-\alpha/r$. The potential allows to define an initial bound state without the complications of retardation. In temporal gauge ($A^0=0$) the longitudinal electric field $\Ev_L=-\partial_t \Av_L$ is given by a constraint for each physical state. This clearly separates the instantaneous from the propagating fields.

\vspace{2mm}
\textbf{Fock expansion}

States are conventionally defined by their expansion in a complete basis of Fock states. In temporal gauge the Fock state constituents are fermions and transversely polarized photons or gluons. The gauge constraint (Gauss' law) determines the longitudinal electric field within each Fock state.

For strong potentials the number of Fock constituents depends on the basis. The Dirac state \eq{state1} has an infinite number of constituents \eq{vac2} in the free state basis due to $Z$-diagrams, \fig{f4}(b). In the basis of the $c_n$ operators defined by the Bogoliubov transform \eq{cndef1} the same Dirac state has a single constituent $c_n^\dag\ket{0}$. I shall define a fermion Fock state as $\psi^\dag(t,\xv)\ket{0}$, without specifying the expansion of the field in creation and annihilation operators.

\vspace{2mm}
\textbf{Initial state}

I take the valence Fock state as the initial bound state of the perturbative expansion. For Positronium this means $\ket{e^-e^+}$ bound by the $-\alpha/r$ potential. Hadron ($\ket{q\bar q},\ \ket{qqq}$) quantum numbers correspond to their valence quarks.

Higher order corrections in $\alpha$ will involve Fock states with a correspondingly larger number of constituents, as well as loop corrections to Fock states with fewer constituents. At each order of $\alpha$ the usual cancellation of collinear singularities between states with different numbers of constituents should thus be ensured.

\vspace{-.2cm}
\subsection{Quantization in QED \label{secIV.B}}
%%%%%%%%%%%%%%%%%%%%%%%%%%

\vspace{-.3cm}
\subsubsection{Functional integral method \label{secIV.B1}}
%%%%%%%%%%%%%%%%%%%%%%%%%%
\vspace{-.5cm}
Relativistic field theory is commonly defined using functional methods. Green functions are given by a functional integral over the fields weighted by the exponent of the action, $\exp(i\mS/\hbar)$. In QED the photon propagator is thus
\begin{align} \label{eII1aa}
D^{\mu\nu}(x_1,x_2) &= \int \mD(A^\rho)\mD(\bar\psi,\psi)\,e^{i\mS_{QED}/\hbar}\,A^\mu(x_1)A^\nu(x_2) \nn\\
\mS_{QED} &= \int d^4x\big[-\quart F_{\mu\nu}F^{\mu\nu}+\bar\psi(i\slashed{\partial}-m-e\slashed{A})\psi\big]
& F_{\mu\nu} = \partial_\mu A_\nu-\partial_\nu A_\mu
\end{align}
A gauge fixing term $\mS_{GF}$ must be added to the action $\mS_{QED}$ for the integral to be well defined. Explicit Poincar\'e invariance is maintained with
\begin{align} \label{eII1ab}
\mS_{GF}=-\halft\lm \int d^4x\,(\partial_\mu A^\mu)^2
\end{align}
Expanding around free (hence Poincar\'e covariant) states gives the standard perturbative expansion of Green functions in terms of Feynman diagrams.

This approach is well suited for scattering amplitudes. It is less convenient for bound states, for which free states are a poor approximation. As a consequence, the perturbative $S$-matrix \eq{mo2} lacks bound state poles at any finite order. The poles can be generated through the divergence of an infinite sum of Feynman diagrams (or through an equivalent integral equation), as discussed in section \ref{secII.B}. However, it seems unlikely that confinement will be recovered in an expansion starting with free quarks and gluons.

The covariant gauge fixing \eq{eII1ab} introduces a time derivative $\partial_0 A^0$ for the $A^0$ field, which is absent from $\mS_{QED}$. This makes $A^0$ propagate in time like the transverse components $\Av_T$, at the price of introducing a time-dependent kernel in the Bethe-Salpeter equation (section \ref{secII.C}). The $\partial_0 A^0$ term is avoided in Coulomb gauge, $\nv\cdot\Av=0$. The field equation for $A^0$ (Gauss' law) in Coulomb gauge,
\begin{align} \label{eII1b}
\frac{\delta\mS_{QED}}{\delta A^0(x)} = -\nv^2 A^0(x)-e\psi^\dag\psi(x) = 0
\end{align}
defines $A^0$ non-locally in terms of the electron field operator. For Positronium this gives the Coulomb potential $eA^0=-\alpha/r$, which allows an analytic solution of the Schr\"odinger equation and is the leading order interaction. The evaluation of higher order corrections in Coulomb gauge is rather complicated, especially for QCD. See \cite{Feinberg:1977rc} for a study of non-relativistic quarkonia based on the Bethe-Salpeter equation in Coulomb gauge.

\subsubsection{Canonical quantization \label{secIV.B2}}
%%%%%%%%%%%%%%%%%%%%%%%%%%

The conjugate fields $\pi_\alpha$ of the fields $\vphi_\alpha$ in the Lagrangian density $\mL(\vphi,\partial\vphi)$ are defined by
\begin{align} \label{eII1c}
\pi_\alpha(t,\xv) = \frac{\partial\mL(\vphi,\partial\vphi)}{\partial[\partial_0\vphi_\alpha(t,\xv)]}
\end{align}
Equal time (anti)commutation relations are imposed on the (fermion) boson fields,
\begin{align} \label{eII1d}
\com{\vphi_\alpha(t,\xv)}{\pi_\beta(t,\yv)}_\pm = i\delta_{\alpha\beta} \delta^3(\xv-\yv)
\hspace{2cm} \com{\vphi_\alpha(t,\xv)}{\vphi_\beta(t,\yv)}_\pm = \com{\pi_\alpha(t,\xv)}{\pi_\beta(t,\yv)}_\pm =0
\end{align}
and the Hamiltonian is given by
\begin{align} \label{eII1e}
H(t) = \int d^3\xv \Big[\sum_\alpha \pi_\alpha \partial_0\vphi_\alpha(t,\xv)-\mL(\vphi,\partial\vphi)\Big]
\end{align}
In gauge theories the conjugate field of $A^0$ vanishes since $\mL$ is independent of $\partial_0 A^0$. The covariant gauge fixing term \eq{eII1ab} adds $\partial_0 A^0$, giving the conjugate fields
\begin{align}
\pi^0 &= -\lm\, \partial_\mu A^\mu  \label{eII1f}\crt
\pi^i &= -F^{0i} \label{eII1g}
\end{align}
This allows to define covariant commutation relations for the gauge field, the non-vanishing ones being
\begin{align} \label{eII1h}
\com{A_\mu(t,\xv)}{\pi_\nu(t,\yv)}=i\,g_{\mu\nu}\delta^3(\xv-\yv)
\end{align}
The unphysical (gauge) degrees of freedom are removed by constraining physical states not to involve photons with time-like or longitudinal polarizations (Gupta-Bleuler method, see  \cite{Itzykson:1980rh} for details). 

Canonical quantization can be carried out also in Coulomb gauge, $\nv\cdot\Av=0$. Due to the lack of a conjugate $A^0$ field this requires constraints which modify the commutation relations, see \cite{Weinberg:1995mt} for QED. The generalization to QCD is discussed in \cite{Christ:1980ku}, demonstrating how terms related to Faddeev-Popov ghosts arise. The same study also addresses temporal gauge ($A^0=0$), which is an axial gauge without ghosts.

Temporal gauge simplifies canonical quantization since the absence of both $A^0$ and its conjugate allows standard commutation relations for the spatial gauge field components $\Av$. The gauge condition preserves rotational invariance and, most importantly for the present application, Gauss' law is implemented as a constraint on physical states which determines $\Ev_L$, not as an operator relation like \eq{eII1b}. The constraint is trivially satisfied for the vacuum ($\Ev_L=0$), whereas in Coulomb gauge $A^0\ket{0}$ would have an overlap with $\ket{e^-e^+}$. 
I next discuss canonical quantization in temporal gauge for QED, and consider QCD in section \ref{secIV.C}.

\subsubsection{Temporal gauge in QED \label{secIV.B3}}
%%%%%%%%%%%%%%%%%%%%%%%%%%

The canonical quantization of QED in temporal gauge ($A^0=0$) is described in \cite{Willemsen:1977fr,Bjorken:1979hv,Christ:1980ku,Leibbrandt:1987qv,Strocchi:2013awa}. The action \eq{eII1aa} determines the electric field $E^i=F^{i0}=-\partial_0 A^i$ to be conjugate \eq{eII1c} to $A_i \ (i=1,2,3)$, and $i\psi^\dag$ to be conjugate to $\psi$. This gives the canonical commutation relations without constraints, 
\begin{align} \label{eII2}
\com{E^i(t,\xv)}{A^j(t,\yv)} &= i\delta^{ij}\delta(\xv-\yv)  & \acom{\psi^\dag_\alpha(t,\xv)}{\psi_\beta(t,\yv)} = \delta_{\alpha\beta}\,\delta(\xv-\yv)
\end{align}
All other (anti)commutators vanish. The Hamiltonian in temporal gauge is
\begin{align} \label{eII3}
\mH(t) = \int d\xv\big[E^i\partial_0 A_i+i\psi^\dag\partial_0\psi-\mL\big]
= \int d\xv\big[\halft E^iE^i +\quart F^{ij}F^{ij}+\psi^\dag(-i\alpha^i\partial_i-e\alpha^iA^i+m\gz)\psi\big] 
\end{align} 
Gauss' operator is defined as usual by the derivative of the action wrt. $A^0$,
\begin{align} \label{eII4}
G(x) \equiv \frac{\delta\mS_{QED}}{\delta{A^0(x)}} = \partial_i E^{i}(x)-e\psi^\dag\psi(x)
\end{align}
but $G(x)=0$ (Gauss' law) is not an operator relation, since $A^0=0$ is fixed by the gauge condition. The operator relation $\partial_i E^{i}(x)=e\psi^\dag\psi(x)$ would not even be compatible with the commutation relations \eq{eII2}. 

The condition $A^0=0$ does not completely fix the gauge, since it allows time independent gauge transformations parametrized by $\la(\xv)$: $\Av \to \Av+\nv\la(\xv)$. 
Gauss' operator $G(x)$ turns out to generate such transformations. An infinitesimal, time independent gauge transformation $\delta\la(\xv)$ is represented by the unitary operator,
\begin{align} \label{eII5}
U(t) = 1+i\int d\yv\,G(t,\yv)\delta\la(\yv)
\end{align}

%%%%%%%%%%%%%%%%%%%%%%%%%%
\begin{tcolorbox}
\textit{Exercise \ref{e11}:} Show using the commutation relations \eq{eII2} that $U(t)$ \eq{eII5} transforms the $\Av(t,\xv)$ and $\psi(t,\xv)$ fields as required for a time-independent infinitesimal gauge transformation.
\end{tcolorbox}
%%%%%%%%%%%%%%%%%%%%%%%%%%

Constraining the physical states to satisfy
\begin{align} \label{eII7}
G(x)\ket{phys} = \big[\nv\cdot \Ev(x)-e\psi^\dag\psi(x)\big]\ket{phys} = 0
\end{align}
ensures that they are invariant under time-independent gauge transformations. A physical state remains physical under time evolution since, as may be verified, $G(x)$ commutes with the Hamiltonian \eq{eII3}, 
\begin{align} \label{eII4b}
\com{G(t,\xv)}{\mH(t)}=0
\end{align}
The electric field can be separated into its transverse and longitudinal parts, $\Ev = \Ev_T+\Ev_L$, with $\nv\cdot \Ev_T=0$. Gauss constraint \eq{eII7} then allows to solve for $\Ev_L$,
\begin{align} \label{eII7b}
\Ev_L(t,\xv)\ket{phys} &= -\nv_x \int d\yv\, \frac{e}{4\pi|\xv-\yv|}\,\psi^\dag\psi(t,\yv)\ket{phys}
\end{align}
This seems like the instantaneous electric field $-\nv A^0$ in Coulomb gauge, $\nv\cdot \Av=0$. The difference is that Gauss' law is an operator equation in Coulomb gauge, whereas here it is a constraint on the physical states. The constraint specifies $\Ev_L(t,\xv)$ for each state $\ket{phys}$ at all positions $\xv$  at a given time $t$. The electric field of the physical vacuum vanishes in temporal gauge,
\begin{align} \label{eII8}
E_L^i(\xv)\ket{0}= -\partial_i^x \int d\yv \frac{e}{4\pi|\xv-\yv|}\psi^\dag\psi(t,\yv)\ket{0} = 0
\end{align}
since the vacuum state has no net charge at any position.

The Hamiltonian \eq{eII3} has an instantaneous part determined by Gauss' constraint,
\begin{align} \label{eII11}
\mH_V^{QED}\ket{phys} \equiv \halft\int d\xv\, \Ev_L^2(\xv)\ket{phys} 
&= \halft\int d\xv d\yv d\zv\Big[\partial_i^x \frac{e}{4\pi|\xv-\yv|}\psi^\dag\psi(\yv)\Big]
\Big[\partial_i^x \frac{e}{4\pi|\xv-\zv|}\psi^\dag\psi(\zv)\Big]\ket{phys} \nn\crt
&= \halft\int d\xv d\yv\, \frac{e^2}{4\pi|\xv-\yv|}\big[\psi^\dag\psi(\xv)\big]
\big[\psi^\dag\psi(\yv)\big]\ket{phys}
\end{align}
$\mH_V\ket{phys}$ contributes a potential which depends only on the instantaneous positions of the electrons and positrons, regardless of their momenta (which may be relativistic). The other terms of $\mH$ determine the propagation of the transverse photons and fermions in time, as well as the transitions between them.

This method can be applied to atoms in any frame. Given that non-valence Fock states are suppressed by powers of $\alpha$, calculations with a given degree of precision require to include a limited number of terms in the Fock expansion. In section \ref{secV} I illustrate the method by considering several aspects of Positronium at rest and in motion. In section \ref{secVI} I study the strongly bound states of QED in $D=1+1$ dimensions.

\subsection{Temporal gauge in QCD \label{secIV.C}}
%%%%%%%%%%%%%%%%%%%%%%%%%%

\subsubsection{Canonical quantization \label{secIV.C1}}
%%%%%%%%%%%%%%%%%%%%%%%%%%

The canonical quantization of QCD in temporal gauge $A^0_a=0$ proceeds as in QED \cite{Willemsen:1977fr,Bjorken:1979hv,Christ:1980ku,Leibbrandt:1987qv,Strocchi:2013awa}. The QCD action is
\begin{align} \label{eII23}
\mS_{QCD} &= \int d^4x\big[-\quart F_{\mu\nu}^aF^{\mu\nu}_a+\bar\psi(i\slashed{\partial}-m-g\slashed{A}_a T^a)\psi\big]
& F_{\mu\nu}^a = \partial_\mu A_\nu^a-\partial_\nu A_\mu^a-gf_{abc}A_\mu^bA_\nu^c
\end{align}
The electric field $E_a^i=F_a^{i0}=-\partial_0 A_a^i$ is conjugate to $A_i^a=-A_a^i$, giving the equal-time commutation relations
\begin{align} \label{eII24}
\com{E_a^i(t,\xv)}{A_b^j(t,\yv)} &= i\delta_{ab}\delta^{ij}\delta(\xv-\yv)  & \acom{\psi^{A\,\dag}_\alpha(t,\xv)}{\psi_\beta^B(t,\yv)} = \delta^{AB}\delta_{\alpha\beta}\,\delta(\xv-\yv)
\end{align}
The $a,b$ ($A,B$) are color indices in the adjoint (fundamental) representation of SU(3).
The Hamiltonian is
\begin{align} \label{eII25}
\mH_{QCD} &= \int d\xv\big[E_a^i\partial_0 A_i^a+i\psi_A^\dag\partial_0\psi_A-\mL_{QCD}\big]
= \int d\xv\big[\halft E_a^iE_a^i +\quart F_a^{ij}F_a^{ij}+\psi^\dag(-i\alv\cdot\nv+m\gz-g\alv\cdot \Av_a T^a)\psi\big]
\end{align}
where
\begin{align} \label{eII26}
\int d\xv\, \quart F_a^{ij}F_a^{ij} = \int d\xv\big[ \halft A_a^i(-\delta_{ij}\nv^2+\partial_i\partial_j)A_a^j+ gf_{abc}(\partial_i A_a^j)A_b^i A_c^j+ \quart g^2 f_{abc}f_{ade}A_b^i A_c^j A_d^i A_e^j \big]
\end{align}
has both longitudinal and transverse gluon fields.

Gauss' operator
\begin{align} \label{eII27}
G_a(x) \equiv \frac{\delta\mS_{QCD}}{\delta{A_a^0(x)}} = \partial_i E_a^{i}(x)+g f_{abc}A_b^i E_c^i-g\psi^\dag T^a\psi(x)
\end{align}
generates time-independent gauge transformations similarly as in QED \eq{eII5}, which leave the gauge condition $A^0_a=0$ invariant. The longitudinal electric field $\Ev_L^a$ is fixed by constraining physical states to be invariant under the gauge transformations generated by $G_a(x)$,
\begin{align} \label{eII28}
G_a(x)\ket{phys} = 0
\end{align}
This constraint is independent of time since Gauss' operator commutes with the Hamiltonian, $\com{G_a(t,\xv)}{\mH(t)} = 0$. It constrains the longitudinal electric field for physical states,
\begin{align} \label{eII29}
\nv\cdot E_{L}^{a}(\xv)\ket{phys} = g\big[- f_{abc}A_b^i E_c^i+\psi^\dag T^a\psi(\xv)\big]\ket{phys}
\end{align}
We may solve for $\Ev_L^a$ analogously\footnote{At higher orders in $g$ one needs to take into account the contribution of $\Ev_L$ on the rhs. of \eq{eII29}. For large gauge fields this leads to the issue of Gribov copies \cite{Gribov:1977wm}, but they do not appear in a perturbative expansion.} as for QED in section \ref{secIV.B3},
\begin{align} \label{eII31aa}
\Ev_L^a(\xv)\ket{phys} &= -\nv_x \int d\yv\, \frac{g}{4\pi|\xv-\yv|}\,\mE_a(\yv) \ket{phys}\nn\crt
\mE_a(\yv) &= - f_{abc}A_b^i E_c^i(\yv)+\psi^\dag T^a\psi(\yv)
\end{align}
The contribution of the longitudinal electric field to the QCD Hamiltonian \eq{eII25} is then
\begin{align} \label{eII32a}
\mH_V^{QCD}\ket{phys} &\equiv \halft\int d\xv\,\big(\Ev_L^a\big)^2\ket{phys} = \halft\int d\yv d\zv\, \frac{\as}{|\yv-\zv|}\,\mE_a(\yv)\mE_a(\zv)\ket{phys}
\end{align}

\subsubsection{Specification of temporal gauge in QCD \label{secIV.C2}}
%%%%%%%%%%%%%%%%%%%%%%%%%%

There is a relevant difference between QED and QCD which needs to be considered when determining the longitudinal electric field from the QCD gauge constraint \eq{eII29}. To illustrate, compare the expectation value of the field in an $e^-e^+$ Fock component of Positronium and in an analogous color singlet $q\bar q$ component of a meson at $t=0$,
\begin{align}
\ket{e^-e^+} &\equiv \bar\psi_\alpha(\xv_1)\psi_\beta(\xv_2)\ket{0}  \label{diff1a} \crt
\ket{q\,\bar q} &\equiv \sum_A\bar\psi_\alpha^A(\xv_1)\psi_\beta^A(\xv_2)\ket{0}  \label{diff1b}
\end{align}
The Dirac components $\alpha,\beta$ are irrelevant here and will be suppressed. Repeated color indices are summed. Note that ``color singlet'' refers to global gauge transformations\footnote{ The global SU(3) transformations should not be regarded as a subgroup of the local ones, see Ch. 7 of \cite{Strocchi:2013awa}.}, the local temporal gauge being fixed by \eq{eII7b} and \eq{eII31aa}.

The expectation values of the QED \eq{eII7b} and QCD \eq{eII31aa} longitudinal electric fields in these states are, using the canonical commutation relations for the fermions and recalling that $\Ev_L\ket{0}=0$,
\begin{align}
\bra{e^-e^+}\Ev_L(\xv)\ket{e^-e^+} &= -\nv_x \Big(\frac{e}{4\pi|\xv-\xv_1|} - \frac{e}{4\pi|\xv-\xv_2|}\Big)  \bra{e^-e^+}e^-e^+\rangle  \label{diff2a} \crt
\bra{q\,\bar q}\Ev_L^a(\xv)\ket{q\,\bar q} &= -\nv_x \Big(\frac{g}{4\pi|\xv-\xv_1|} - \frac{g}{4\pi|\xv-\xv_2|}\Big)\bra{q\,\bar q} \bar\psi_A(\xv_1)T_{AB}^a\psi_B(\xv_2)\ket{0} \propto \tr T^a =0  \label{diff2b}
\end{align}
In QED the charges of $e^-$ and $e^+$ give rise to the expected dipole electric field, while in QCD the expectation value of an octet field in a singlet state vanishes everywhere. Comparing similarly\footnote{The singular ``self-energy'' contributions $\propto 1/0$ are independent of $\xv_1,\xv_2$ and subtracted.} the instantaneous potentials $\mH_V^{QED}$ \eq{eII11} and $\mH_V^{QCD}$ \eq{eII32a},
\begin{align}
\bra{e^-e^+}\mH_V^{QED}\ket{e^-e^+} &= - \frac{\alpha}{|\xv_1-\xv_2|}\,\bra{e^-e^+}e^-e^+\rangle  \label{diff3a} \crt
\bra{q\,\bar q}\mH_V^{QCD}\ket{q\,\bar q} &= -\frac{\alpha_s}{|\xv_1-\xv_2|}\,\bra{q\,\bar q} \bar\psi_A(\xv_1)T_{AB}^a T_{BC}^a\psi_C(\xv_2)\ket{0} =  -C_F\frac{\alpha_s}{|\xv_1-\xv_2|}\,\bra{q\,\bar q}q\,\bar q\rangle \label{diff3b}
\end{align}
These are the Coulomb potentials of QED and QCD, again as expected. The electron feels only the positron field, and each quark \textit{of a given color} interacts with its antiquark of opposite color. The sum over the potential energies of all color-anticolor components $A$ in \eq{diff1b} gives the Casimir $C_F = (N_c^2-1)/2N_c$ of the fundamental representation.

The solution of the QED gauge constraint \eq{eII7}, the longitudinal electric field \eq{eII7b}, is determined using the physical boundary condition that the electric field vanishes at spatial infinity. This boundary condition is no longer evident for QCD, since the expectation value \eq{diff2b} of the color electric field in any case vanishes at all $\xv$, due to the sum over quark colors. There seems to be no compelling reason to require that the gauge field of each color-anticolor component $A$ of the state \eq{diff1b} should vanish at spatial infinity.

The gauge constraint \eq{eII29} fully determines $\Ev_L^a$ only given a boundary condition at spatial infinity. $\Ev_L^a$ may be specified by the particular solution \eq{eII31aa} and a homogeneous solution $\Ev_H^a$ which satisfies
\begin{align} \label{diff4}
\nv\cdot\Ev_H^a\ket{phys} =0
\end{align}
There is apparently only one homogeneous solution which is invariant under translations and rotations,
\begin{align} \label{eII31a}
\Ev^a_{H}(\xv)\ket{phys} &= -\kappa\,\nv_x \int d\yv \,\xv\cdot\yv\,\mE_a(\yv) \ket{phys}
\end{align}
where $\mE_a(\yv)$ is defined in \eq{eII31aa} and the normalization $\kappa$ is independent of $\xv$, but may depend on the state $\ket{phys}$. The complete longitudinal electric field is then
\begin{align} \label{eII31}
\Ev_{L}^a(\xv)\ket{phys} &= -\nv_x \int d\yv \Big[\kappa\,\xv\cdot\yv + \frac{g}{4\pi|\xv-\yv|}\Big]\mE_a(\yv) \ket{phys}
\end{align}
and its contribution to the Hamiltonian \eq{eII25} is
\begin{align} \label{eII32}
\mH_V &\equiv \halft\int d\xv\,E_{a,L}^i E_{a,L}^i = \halft\int d\xv \Big\{\partial_i^x \int d\yv\Big[\kappa\, \xv\cdot\yv+\frac{g}{4\pi|\xv-\yv|}\Big]\mE_a(\yv)\Big\}
\Big\{\partial_i^x \int d\zv\Big[\kappa\, \xv\cdot\zv+\frac{g}{4\pi|\xv-\zv|}\Big]\mE_a(\zv)\Big\} \nn\crt
&= \int d\yv d\zv\Big\{\,\yv\cdot\zv \Big[\halft\kappa^2\intt d\xv + g\kappa\Big] + \halft \frac{\as}{|\yv-\zv|}\Big\}\mE_a(\yv)\mE_a(\zv)
\end{align} 
where the terms of \order{g\kappa,g^2} were integrated by parts. The term of \order{\kappa^2} is due to an $\xv$-independent field energy density. It is $\propto \int d\xv$ but irrelevant provided it is universal, \ie, the same for all Fock components of all bound states. This determines the normalization $\kappa$ in \eq{eII31} for each state $\ket{phys}$, up to a universal scale $\la$.

The scale $\la$ is unrelated to the coupling $g$, so the $g\kappa$ term in \eq{eII32} may be viewed as an instantaneous \order{\alpha_s^0} potential. All relevant symmetries, in particular exact Poincar\'e invariance, must appear at each order of $\as$. The boost covariance of Positronia in QED is ensured by a combination of the Coulomb potential and \order{\alpha} transverse photon exchange (section \ref{secV}). The boost covariance of QCD bound states must at \order{\alpha_s^0} be achieved by the instantaneous potential alone, akin to QED in $D=1+1$ (section \ref{secVI}). This appears to be satisfied (section \ref{secVII.C}).

\section{Applications to Positronium atoms \label{secV}}
%%%%%%%%%%%%%%%%%%

I now illustrate the approach to QED bound states described above with several applications to Positronium atoms. The expansion starts with valence Fock states, here $\ket{e^-e^+}$, with higher Fock states included perturbatively. The first task is then to define the valence Fock states and determine the constraints on their wave functions imposed by the symmetries of translations and rotations, as well as parity and charge conjugation (section \ref{secV.A}). I express the valence Fock states using field operators (here the electron field $\psi(t,\xv)$), similarly as in the representations \eq{state1}, \eq{state2} of the Dirac bound states. 

In section \ref{secV.B} I determine the wave functions of Para- and Orthopositronium atoms at lowest \order{\alpha^2} in their binding energy $E_b$. The rest frame wave functions then satisfy the Schr\"odinger equation. For atomic CM momentum $\Pv \neq 0$ one needs to include the Fock state with one transverse photon $\ket{e^-e^+\gamma}$ (\ref{secV.C}). The hyperfine splitting between Ortho- and Parapositronium at rest is calculated at \order{\alpha^4} in the rest frame (\ref{secV.D}), taking into account the transverse photon state $\ket{e^-e^+\gamma}$ and Orthopositronium annihilation into a virtual photon, $e^-e^+ \to \gamma \to e^-e^+$. Finally I calculate the electromagnetic form factor of Positronium (\ref{secV.E}) and deep inelastic scattering on Positronia (\ref{secV.F}) in a general frame.

\subsection{The $\ket{e^-e^+}$ Fock states of Para- and Orthopositronium atoms \label{secV.A}}
%%%%%%%%%%%%%%%%%%%%%%%%%%

\subsubsection{Definition of the Fock states \label{secV.A1}}
%%%%%%%%%%%%%%

The $\ket{e^-e^+}$ Fock states of Parapositronium ($J^{PC}=0^{-+}$) and Orthopositronium ($J^{PC}=1^{--}$) atoms, jointly denoted by $\mB$, may be expressed in terms of two electron fields,
\begin{align} \label{fock1}
\ket{e^-e^+;\mB,\Pv} \equiv \int d\xv_1 d\xv_2\,\bar\psi(\xv_1)\lla_+(\xv_1)e^{i\Pv\cdot(\xv_1+\xv_2)/2}\Phi_\mB^{(\Pv)}(\xv_1-\xv_2)\rla_-(\xv_2)\psi(\xv_2)\ket{0}
\end{align}
where $\Phi_\mB^{(\Pv)}$ is a $4\times 4$ wave function of the atom $\mathcal{B}$ with CM momentum $\Pv$. The $\la_\pm$ are Dirac projection operators,
\begin{align} \label{fock2}
\lla_+(\xv_1) &\equiv \inv{2E_1}\Big[E_1-i\alv\cdot\lnab_1 +\gz m\Big] = \big[\lla_+(\xv_1)\big]^2  \nn\crt
\rla_-(\xv_2) &\equiv \inv{2E_2}\Big[E_2 +i\alv\cdot\rnab_2 -\gz m\Big] = \big[\rla_-(\xv_2)\big]^2
\end{align}
where $E=\sqrt{-\nv^2+m^2}$.
The projectors select the $b^\dag$ operator in $\bar\psi(\xv_1)$ and $d^\dag$ in $\psi(\xv_2)$, defined as in \eq{eop},
\begin{align} \label{fock2b}
\bar\psi(\xv)\lla_+(\xv)&=  \int\frac{d\kv}{(2\pi)^32E_k}\sum_\lm \bar u(\kv,\lm)\,e^{-i\kv\cdot\xv}\,b_{\kv,\lm}^\dag \hspace{2cm}
\rla_-(\xv)\psi(\xv)=  \int\frac{d\kv}{(2\pi)^32E_k}\sum_\lm v(\kv,\lm)\,e^{-i\kv\cdot\xv}\,d_{\kv,\lm}^\dag
\end{align}
Since $b\ket{0} = d\ket{0} = 0$ in \eq{fock1} the projectors actually have no effect. However, in operations on the states they allow to use the anticommutation relations $\acom{\psi^\dag_\alpha(t,\xv)}{\psi_\beta(t,\yv)}=\delta_{\alpha,\beta}\,\delta^3(\xv-\yv)$, to which the $b$ and $d$ operators contribute. The projectors ensure that the coefficients of $b$ and $d$ in \eq{fock1} vanish, so that no spurious contributions arise. I assume the normalization
\begin{align} \label{fock3}
\langle{e^-e^+;\mB',\Pv'}\ket{e^-e^+;\mB,\Pv} = 2E_P (2\pi)^3 \delta(\Pv-\Pv')\delta_{\mB,\mB'} \hspace{2cm} E_P=\sqrt{\Pv^2+4m^2}
\end{align}
where $E_P$ is the energy of the atom at \order{\alpha^0}.

The Hamiltonian \eq{eII3} is symmetric under translations, rotations, parity and charge conjugation. The states may be classified by their transformation under those symmetries, giving constraints on the wave functions $\Phipb(\xv)$.

\subsubsection{Translations \label{secV.A2}}
%%%%%%%%%%%%%%

Under space translations $\xv \to \xv+\ellv$ the electron field is transformed by the operator
\beq\label{A2}
U(\bs{\ell}) = \exp[-i\ellv\cdot\bs{\mP}] \hspace{1cm} {\rm where} \hspace{1cm} \bs{\mP} = \int d\xv\,\psi^\dag(\xv)(-i\rnab)\psi(\xv)
\eeq 
The momentum operator satisfies
\begin{align} \label{A3}
\com{\bs{\mP}}{\psi(\xv)} = i\rnab \psi(\xv)  \hspace{2cm}  \com{\bs{\mP}}{\bar\psi(\xv)} = \bar\psi(\xv)i\lnab
\end{align}
With $\bs{\mP}\ket{0}=0$ we have $\bs{\mP}\ket{e^-e^+;\mB,\Pv} = \Pv\ket{e^-e^+;\mB,\Pv}$.

\subsubsection{Rotations \label{secV.A3}}
%%%%%%%%%%%%%%
 
Rotations are generated by the angular momentum operator $\bs{\mJ}$, which was defined already in \eq{dir6} for the Dirac equation. With $\alv \equiv \gz\gv$, 
\begin{align}
\bs{\mJ} &= \int d\xv\,\psi^\dag(\xv)\,\bs{J}\,\psi(\xv) \hspace{2.9cm} \bs{J} = \Lv+\Sv= \xv\times(-i\nv)+\halft\gf\alv \label{A5}
\end{align}
Rest frame states are taken to be eigenstates of $\bs{\mJ}^2$ and $\mJ^z$. For a $\Pv=0$ Positronium state expressed as in \eq{fock1},
\begin{align} \label{fock4}
\bs{\mJ}\ket{e^-e^+;\mB,\Pv=0}=\int d\xv_1 d\xv_2\, \bar\psi(\xv_1)\lla_+(\xv_1) \com{\Jv}{\Phirb(\xv_1-\xv_2)} \rla_-(\xv_2)\psi(\xv_2)\ket{0}
\end{align}

%%%%%%%%%%%%%%%%%%%%%%%%%%
\begin{tcolorbox}
\textit{Exercise \ref{e12a}:} Derive \eq{fock4}. %D5339
\end{tcolorbox}
%%%%%%%%%%%%%%%%%%%%%%%%%%

For the state to have total angular momentum $j$ and $j^z=\lm$ in the rest frame, 
\begin{align} \label{focka5}
\bs{\mJ}^2\ket{e^-e^+;\mB,\Pv=0} = j(j+1)\ket{e^-e^+;\mB,\Pv=0} \hspace{2cm} \mJ^z\ket{e^-e^+;\mB,\Pv=0} = \lm\ket{e^-e^+;\mB,\Pv=0}
\end{align}
the wave function should satisfy
\begin{align}\label{fock5}
\com{J^i}{\com{J^i}{\Phirb(\xv)}} = j(j+1)\Phirb(\xv) \hspace{2cm}
\com{J^z}{\Phirb(\xv)}= \lm\Phirb(\xv)
\end{align}

\subsubsection{Parity $\eta_P$ \label{secV.A4}}
%%%%%%%%%%%%%%

The parity operator $\bP$ transforms the electron field as
\beq\label{A9}
\bP \psi(t,\xv)\bP^\dag = \gz\psi(t,-\xv) \hspace{2cm} \bP \bar\psi(t,\xv)\bP^\dag = \bar\psi(t,-\xv)\gz
\eeq
Changing the integration variables $\xv_{1,2} \to - \xv_{1,2}$ in \eq{fock1} and noting that $\gz\la_\pm(\xv) = \la_\pm(-\xv)\gz$,
\beq\label{A10}
\bP\ket{e^-e^+;\mB,\Pv}=\int d\xv_1 d\xv_2\, \bar\psi(\xv_1) \lla_+(\xv_1)\gz e^{-i\Pv\cdot(\xv_1+\xv_2)/2}\Phi_\mB^{(\Pv)}(-\xv_1+\xv_2)\gz\rla_-(\xv_2)\psi(\xv_2) = \eta_P\ket{e^-e^+;\mB,-\Pv}
\eeq
if the wave function satisfies
\beq\label{A11}
\gz\Phi_\mB^{(\Pv)}(-\xv)\gz = \eta_P \Phi_\mB^{(-\Pv)}(\xv)   \hspace{2cm} (\eta_P = \pm 1)
\eeq
Note that parity reverses the CM momentum $\Pv$ of the state.

\subsubsection{Charge conjugation $\eta_C$ \label{secV.A5}}
%%%%%%%%%%%%%%

The charge conjugation operator $\bC$ transforms particles into antiparticles.
\beq\label{A12}
\bC b(\pv,\lambda)\bC^\dag = d(\pv,\lambda) \hspace{2cm} \bC d(\pv,\lambda)\bC^\dag = b(\pv,\lambda)
\eeq 
In the Dirac representation of the $\gamma$ matrices (here $T$ indicates transpose and $\aly \equiv \gz\gamma^2$)
\beq\label{A13}
\bC \psi(t,\xv)\bC^\dag = -i\gamma^2\psi^*(t,\xv) = i\aly\bar\psi^T(t,\xv) \hspace{2cm} \bC \bar\psi(t,\xv)\bC^\dag = i\psi^T(t,\xv)\aly
\eeq
This implies $v(\kv,\lm) = -i\gamma^2 u^*(\kv,\lm)$ and thus $\bar\chi_\lm = i\sigma_2\chi_\lm$ in \eq{pro6}. 
For a Positronium state to be an eigenstate of charge conjugation,
\beq\label{A14}
\bC\ket{e^-e^+;\mB,\Pv} = \eta_C \ket{e^-e^+;\mB,\Pv}
\eeq
its wave function should satisfy
\beq\label{A15}
\aly\big[\Phi_\mB^{(\Pv)}(-\xv)\big]^T\aly = \eta_C \Phi_\mB^{(\Pv)}(\xv) \hspace{2cm} (\eta_C = \pm 1)
\eeq

%%%%%%%%%%%%%%%%%%%%%%%%%%
\begin{tcolorbox}
\textit{Exercise \ref{e12}:} Derive \eq{A15}. %D5304
\end{tcolorbox}
%%%%%%%%%%%%%%%%%%%%%%%%%%

\subsubsection{Wave functions of Para - and Orthopositronium \label{secV.A6}}
%%%%%%%%%%%%%%
Non-relativistic Para- and Orthopositronium have zero orbital angular momentum in the rest frame, $\comb{\Lv}{\Phirb(\xv)}=0$. Hence their wave functions have no angular dependence. The radial dependence factorizes from the Dirac structure since the spin, parity and charge conjugation constraints are independent of the radial coordinate $r=|\xv|$,
\begin{align} \label{fock5a}
\Phirb(\xv) = N_\mB^{(0)}\, \Gamma_\mB\, F^{(0)}(r)
\end{align}
where $\Gamma_\mB$ is an $\xv$-independent $4\times 4$ Dirac matrix.
Para- and Orthopositronia have the same radial function $F(r)$ and the same binding energy $E_b$ at \order{\alpha^2}. The energy degeneracy holds for all $\Pv$, indicating that the factorization \eq{fock5a} holds in any frame,
\begin{align} \label{fock5c}
\Phipb(\xv) = \Npb\, \Gamma_\mB\, \Fp(\xv) \hspace{2cm} \int d\xv\,|\Fp(\xv)|^2 = E_P
\end{align}
I verify this in section \ref{secV.C}. The boosted radial function $\Fp(\xv)$ is angular dependent for $\Pv \neq 0$ due to Lorentz contraction in the $\Pv$-direction. With its normalization fixed as in \eq{fock5c} the constants $\Npb$ are determined by the normalization \eq{fock3} of the state.

In the following I take $\Pv = (0,0,P)$ in the $z$-direction, and consider Orthopositronium with $j^z=\lm$. The following Dirac structures $\Gamma_\mB$ in \eq{fock5a} give the correct $J^{PC}$ quantum numbers of the Positronia:
\begin{flalign} \label{para}
&\textbf{Parapositronium:}\ J^{PC} = 0^{-+} \hspace{1cm} \Gamma_{Para}= \gf&
\end{flalign}
\begin{itemize}
\item \textit{Spin:} $\com{\Sv}{\gf} = \com{\halft\gf\alv}{\gf} =0$, hence $j=s=0$. 
\item \textit{Parity:} $\gz \gf\gz = -\gf$, hence $\eta_P=-1$.
\item \textit{Charge conjugation:} $\aly \gf^T\aly = \gf$, hence $\eta_C=+1$.
\end{itemize}
\begin{flalign} \label{ortho}
&\textbf{Orthopositronium:}\ J^{PC} = 1^{--} \hspace{1cm} \Gamma_{Ortho}^\lm = \ev_\lm\cdot\alv \hspace{1cm} \ev_{\pm 1} = -\inv{\sqrt{2}}(\pm 1,i,0) \hspace{1cm} \ev_0 = (0,0,1)&
\end{flalign} 
\begin{itemize}
\item \textit{Spin:} $\com{S^z}{\ev_\lm\cdot\alv} = \lm\,\ev_\lm\cdot\alv$, hence $j^z=\lm$, and $\sum_i\com{S^i}{\com{S^i}{\ev_\lm\cdot\alv}}= 2\,\ev_\lm\cdot\alv$, hence $j=1$.
\item \textit{Parity:} $\gz\, \ev_\lm\cdot\alv\,\gz = -\ev_\lm\cdot\alv$, hence $\eta_P=-1$.
\item \textit{Charge conjugation:} $\aly\, \ev_\lm\cdot\alv^T\,\aly = -\ev_\lm\cdot\alv$, hence $\eta_C=-1$.
\end{itemize}

The constants $\Npb$ of \eq{fock5c} are then determined by the state normalization \eq{fock3} to be, at \order{\alpha^0},
\begin{align} \label{fock6}
N_{Para}^{(\Pv)} = N_{Ortho}^{(\Pv)}(\lm=0) = \frac{E_P}{2m} \hspace{2cm} N_{Ortho}^{(\Pv)}(\lm=\pm 1) = 1
\end{align}

%%%%%%%%%%%%%%%%%%%%%%%%%%
\begin{tcolorbox}
\textit{Exercise \ref{e13}:} Verify \eq{fock6}. %D5338
\end{tcolorbox}
%%%%%%%%%%%%%%%%%%%%%%%%%%

\subsection{The Schr\"odinger equation for Positronium at $\Pv=0$ \label{secV.B}}
%%%%%%%%%%%%%%%%%%%%%%%%%%

The Schr\"odinger equation for the rest frame wave function follows from the condition that the (Para- or Ortho-) Positronium state \eq{fock1} be an eigenstate of the Hamiltonian \eq{eII3},
\begin{align} \label{sch1}
\mH\ket{e^-e^+;\mB,\Pv=0}= \big[\mH_0(f)+\mH_V\big]\ket{e^-e^+;\mB,\Pv=0} = (2m+E_b)\ket{e^-e^+;\mB,\Pv=0}
\end{align}
Transverse photons do not contribute to $E_b$ at \order{\alpha^2}. Their coupling to electrons is proportional to the 3-momentum of the electron, which in the rest frame is of \order{\alpha m}.

The free fermion Hamiltonian acting on the electron fields gives (note that $\acom{\psi}{\bar\psi}$ leaves a $\gz$, and $\alv\gz = -\gz\alv$)
\begin{align} \label{sch2}
\comb{\mH_0(f)}{\bar\psi(\xv_1)} &= \int d\xv\, \psi^\dag(\xv)(-i\alv\cdot\rnab+m\gz)\acomb{\psi(\xv)}{\bar\psi(\xv_1)} 
= \bar\psi(\xv_1)(-i\alv\cdot\lnab_1+m\gz) \nn\crt
\comb{\mH_0(f)}{\psi(\xv_2)} &= -(-i\alv\cdot\rnab_2+m\gz)\psi(\xv_2)
\end{align}
Together with the projection operators $\la_\pm$ in $\ket{e^-e^+;\mB,\Pv=0}$ these give energies $E=\sqrt{-\nv^2+m^2}$, 
\begin{align} \label{mov4}
(-i\alv\cdot\lnab_1+m\gz)\inv{2E_1}(E_1-i\alv\cdot\lnab_1+m\gz)&= \inv{2E_1}\big[E_1(-i\alv\cdot\lnab_1+m\gz)-\lnab_1^2+m^2\big]= \la_+(i\lnab_1)E_1 \nn\crt
-\inv{2E_2}(E_2+i\alv\cdot\rnab_2-m\gz)(-i\alv\cdot\rnab_2+m\gz) &= \inv{2E_2}\big[E_2(i\alv\cdot\rnab_2-m\gz)-\rnab_2^2+m^2\big] = E_2\la_-(i\rnab_2) 
\end{align}
Through a partial integration the $\nv^2$ in the energies acts on the wave function. At \order{\alpha^2} we have $E \simeq m-\nv^2/2m$, giving the kinetic contribution of the Schr\"odinger equation with reduced mass $m/2$,
\begin{align} \label{sch3}
\Big(2m-\frac{\nv_1^2}{m}\Big)\Phirb(\xv_1-\xv_2)
\end{align} 

The potential energy arises from the instantaneous part \eq{eII11} of the Hamiltonian,
\begin{align} \label{sch4}
\mH_V\ket{e^-e^+;\mB,\Pv=0} &= \inv{2}\int d\xv d\yv\, \frac{e^2}{4\pi|\xv-\yv|}\big[\psi^\dag\psi(\xv)\big]
\big[\psi^\dag\psi(\yv)\big]\ket{e^-e^+;\mB,\Pv=0}
\end{align}
Because Gauss' law in temporal gauge is imposed as a constraint on the physical states we have $\psi^\dag\psi\ket{0}=0$ as in \eq{eII8}. The effect of $\mH_V$ is then to multiply the wave function by the Coulomb potential,
\begin{align} \label{sch5}
-\frac{\alpha}{|\xv_1-\xv_2|}\,\Phirb(\xv_1-\xv_2)
\end{align} 
Combining \eq{sch3} and \eq{sch5} the eigenstate condition \eq{sch1} implies the Schr\"odinger equation for the wave function,
\begin{align} \label{sch6}
\Big(2m-\frac{\nv^2}{m}-\frac{\alpha}{|\xv|}\Big)\Phirb(\xv) = (2m+E_b)\Phirb(\xv)
\end{align} 

Since the Schr\"odinger equation is a Dirac scalar it gives the usual $\ell=0$ radial equation for $F^{(0)}(r)$ in \eq{fock5a},
\begin{align} \label{fock9}
\inv{m}\Big[{F^{(0)}}''(r)+\frac{2}{r}F^{(0)'}(r)\Big]+\Big[\frac{\alpha}{r}+E_b\Big]F^{(0)}(r) =0
\ \ \ \ \ {\color{red} \Longrightarrow}\ \ \ \ \ 
F^{(0)}(r)=\frac{\alpha^{3/2}\,m^2}{\sqrt{4\pi}}\exp(-\alpha m r/2) \hspace{1cm} E_b = -\quart m\alpha^2
\end{align}
where the normalization of $F^{(0)}(r)$ is determined by \eq{fock5c} ($E_{P=0}= 2m$).

\subsection{Positronium with momentum $\Pv$ \label{secV.C}}
%%%%%%%%%%%%%%%%%%%%%%%%%%

Fock states quantized at equal time transform non-covariantly under boosts since the definition of time is frame dependent. The Poincar\'e invariance of the QED action nevertheless guarantees that measurables will be Lorentz covariant. In this section I demonstrate this for the binding energy of Positronia at \order{\alpha^2}. Lorentz covariance determines the momentum dependence of the binding,
\begin{align} \label{ebp1}
\Delta E(\Pv) \equiv \sqrt{\Pv^2+(2m+E_b)^2} - \sqrt{\Pv^2+4m^2} = \frac{2m E_b}{E_P} +\morder{\alpha^4} \hspace{2cm} \Delta E(\Pv=0) = E_b
\end{align}
where $E_P = \sqrt{\Pv^2+4m^2}$. In section \ref{secV.D} I evaluate the hyperfine splitting between Ortho- and Parapositronia at \order{\alpha^4} for $\Pv=0$.
In sections \ref{secV.E} and \ref{secV.F} I consider the covariance of form factors and deep inelastic scattering. 

The importance of properly taking into account the momentum dependence of bound state wave functions was emphasized in \cite{Brodsky:1968ea}. The frame dependence of atomic wave functions is of general interest, since it shows how the classical concept of Lorentz contraction is realized for quantum bound states. Surprisingly, there appears to be only one study \cite{Jarvinen:2004pi} of atoms with general CM momenta $\Pv$, even at leading order. The following analysis is equivalent to that one, but is formulated in terms of Fock states in temporal gauge.

In atoms with CM momentum $\Pv$ transverse photons contribute at leading order to the binding energy $E_b$, since they couple at \order{\alpha^0} to electrons whose momenta $\propto \Pv$. I consider first the kinetic and potential energies of the $\ket{e^-e^+}$ Fock state \eq{fock1}, and then determine the wave function of the $\ket{e^-e^+\gamma}$ state in Positronium. Only terms which contribute to $E_b$ at \order{\alpha^2} are retained.

\subsubsection{Kinetic and potential energy \label{secV.C1}}
%%%%%%%%%%%%%%%%%%%%%%%%%%

The above relations \eq{sch2} and \eq{mov4} are valid for all $\Pv$. After a partial integration the derivatives in $E_1$ and $E_2$ operate in \eq{fock1} on $\exp\big[i\Pv\cdot(\xv_1+\xv_2)/2\big]\Phipb(\xv_1-\xv_2)$. Let $\rnab_1 = \halft(\rnab_1+\rnab_2)+\halft(\rnab_1-\rnab_2)$, and similarly for $\lnab_2$. Then the first term gives $i\Pv/2$ while the second, denoted $\nv \equiv \halft(\rnab_1-\rnab_2)$, gives \order{\alpha} derivatives of $\Phipb(\xv)$, 
\begin{align} \label{mov5}
E_1 = \sqrt{(\halft \Pv-i\nv)^2+m^2} \hspace{2cm} E_2 = \sqrt{(\halft \Pv+i\nv)^2+m^2}
\end{align}
At \order{\alpha^2} we need to keep two powers of $\nv$. Using $\sqrt{1+x}=1+\halft x-\frac{1}{8}x^2+\morder{x^3}$ and denoting
\begin{align} \label{mov6}
E_P \equiv \sqrt{\Pv^2+4m^2}  \hspace{2cm} \gamma \equiv \frac{E_P}{2m}  \hspace{2cm} \beta \equiv \frac{P}{E_P}
\end{align}
we get
\begin{flalign} \label{mov7}
&\mH_0(f)\ket{e^-e^+;\mB,\Pv}: \hspace{1cm} E_1+E_2 \simeq E_P-\frac{2}{E_P}\Big(\nv_\perp^2-\frac{1}{\gamma^2}\nv_\|^2\Big) = E_P- \inv{m\gamma}\Big(\nv_\perp^2-\frac{1}{\gamma^2}\nv_\|^2\Big)&
\end{flalign}
where $\perp$ and $\|$ refer to the $\Pv$-direction, here taken to be along the $z$-axis.

The potential energy depends only on the instantaneous positions of the fermions and thus gives the same result as in \eq{sch5} for $\Pv=0$, the wave functions is multiplied by
\begin{flalign} \label{mov7a}
&\mHV\ket{e^-e^+;\mB,\Pv}: \hspace{1.5cm} -\frac{\alpha}{|\xv_1-\xv_2|} = -e^2 \int\frac{d\qv}{(2\pi)^3}\,\inv{\qv^2}e^{-i\qv\cdot(\xv_1-\xv_2)}&
\end{flalign}
This already signals the importance of transverse photon exchange, since the potential energy should be commensurate with $\Delta E(\Pv)$ in \eq{ebp1}, which is $\Pv$-dependent.

\subsubsection{The transverse photon Fock state \label{secV.C2}}
%%%%%%%%%%%%%%%%%%%%%%%%%%

For $\Pv \neq 0$ the transverse photon vertices $\propto \Pv$ are of \order{\alpha^0}, so transverse and Coulomb photon exchanges contribute at the same order in $\alpha$. The transverse photon and its conjugate electric field in temporal gauge $(A^0 = 0)$ may at $t=0$ be expanded in photon creation and annihilation operators $a^\dag$ and $a$, with polarization 3-vectors $\vepsv_{\smu}$, $\smu=\pm1$,
\begin{align}\label{hv1a}
&\Av_T(\xv) = \int\frac{d\qv}{(2\pi)^3 2|\qv|} \sum_{\smu=\pm 1}\Big[\vepsv_\smu(\qv)\,e^{i\qv\cdot\xv}a(\qv,\smu) +\vepsv_\smu^*(\qv)\,e^{-i\qv\cdot\xv}a^\dag(\qv,\smu)\Big] \nn\crt
&\Ev_T(\xv) = i\int\frac{d\qv}{2(2\pi)^3} \sum_{\smu=\pm 1}\Big[\vepsv_\smu(\qv)\,e^{i\qv\cdot\xv}a(\qv,\smu) -\vepsv_\smu^*(\qv)\,e^{-i\qv\cdot\xv}a^\dag(\qv,\smu)\Big] \crt
&\com{a(\qv,\smu)}{a^\dag(\qv',\smu')} = (2\pi)^3 2|\qv|\delta(\qv-\qv')\delta_{\smu,\smu'} \nn\crt
&\qv\cdot\vepsv_\smu(\qv)=0 \hspace{1cm}
\vepsv_\smu^*(\qv)\cdot\vepsv_{\smu'}(\qv) = \delta_{\smu,\smu'}  \hspace{1cm}
\sum_{\smu=\pm 1}\veps_\smu^i(\qv){\veps_\smu^j}^*(\qv) = \delta^{ij}-\frac{q^iq^j}{\qv^2}\nn
\end{align}

The interaction between the electron and the transverse photon fields in the Hamiltonian \eq{eII3} is given by
\begin{align} \label{mov39}
\mH_{int}(\Av_T) = -e\int d\xv\, \psi^\dag(\xv)\,\alv\cdot\Av_T(\xv)\psi(\xv)
\end{align}
This creates $\ket{e^-e^+\gamma}$ states with a photon of momentum $\qv$ and polarization $\smu$. At leading order, 
\begin{align} \label{tra1}
&\mH_{int}\ket{e^-e^+;\mB,\Pv} = e\int d\xv_1 d\xv_2\,\bar\psi(\xv_1)\lla_+(\xv_1)e^{i\Pv\cdot(\xv_1+\xv_2)/2}\Phipb(\xv_1-\xv_2)\rla_-(\xv_2)\psi(\xv_2) \nn \crt
&\times \int\frac{d\qv}{(2\pi)^32|\qv|}\sum_\smu \inv{E_P}\Pv\cdot\veps_\smu^*(\qv)\,a^\dag(\qv,\smu)\big(-e^{-i\qv\cdot\xv_1}+e^{-i\qv\cdot\xv_2}\big)\ket{0} 
 \equiv \int\frac{d\qv}{(2\pi)^32|\qv|}\sum_\smu\ket{e^-e^+\gamma;\qv,\smu} 
\end{align}

%%%%%%%%%%%%%%%%%%%%%%%%%%
\begin{tcolorbox}
\textit{Exercise \ref{e14a}:} Derive the expression for $\ket{e^-e^+\gamma;\qv,\smu}$ in \eq{tra1}. %D5342
\end{tcolorbox}
%%%%%%%%%%%%%%%%%%%%%%%%%%

Operating with $\mH_{int}$ a second time and retaining only the terms where the transverse photon is absorbed, giving an $\ket{e^-e^+}$ Fock state,
\begin{align}
\mH_{int}^2\ket{e^-e^+;\mB,\Pv} &= e^2\int d\xv_1 d\xv_2\,\bar\psi(\xv_1)\lla_+(\xv_1)e^{i\Pv\cdot(\xv_1+\xv_2)/2}\Phipb(\xv_1-\xv_2)\rla_-(\xv_2)\psi(\xv_2) \nn \crt
&\times \int\frac{d\qv}{(2\pi)^3 2|\qv|}\sum_\smu \inv{E_P^2}\big[\Pv\cdot\veps_\smu^*(\qv)\big]\big[\Pv\cdot\veps_\smu(\qv)\big]\big|-e^{-i\qv\cdot\xv_1}+e^{-i\qv\cdot\xv_2}\big|^2\ket{0} \label{tra2} \crt 
&\mbox{where}\ \ \ 
\big|-e^{-i\qv\cdot\xv_1}+e^{-i\qv\cdot\xv_2}\big|^2 = 2-e^{-i\qv\cdot(\xv_1-\xv_2)}-e^{i\qv\cdot(\xv_1-\xv_2)} \label{tra3}
\end{align}
The $\xv_{1,2}$-independent term in \eq{tra3} corresponds to the absorption of the photon on the same fermion from which it was emitted. This loop contribution gives a multiplicative renormalization of the state and does not contribute to the eigenstate condition at lowest order. Neglecting this term and summing over the photon polarization $\smu$ using \eq{hv1a} we have
\begin{align} \label{tra4}
\mH_{int}^2\ket{e^-e^+;\mB,\Pv} &= -e^2\int d\xv_1 d\xv_2\,\bar\psi(\xv_1)\lla_+(\xv_1)e^{i\Pv\cdot(\xv_1+\xv_2)/2}\Phipb(\xv_1-\xv_2)\rla_-(\xv_2)\psi(\xv_2) \nn \crt
&\times \int\frac{d\qv}{(2\pi)^3 2|\qv|}\,\beta^2\,\frac{\qv_\perp^2}{\qv^2}\,\big[e^{-i\qv\cdot(\xv_1-\xv_2)}+e^{i\qv\cdot(\xv_1-\xv_2)}\big]\ket{0}
\end{align}
where $\beta = |\Pv|/E_P$ and $\qv_\perp$ is the component of $\qv$ orthogonal to $\Pv$ (\ie, to the $z$-axis).

\subsubsection{The bound state condition \label{secV.C3}}
%%%%%%%%%%%%%%%%%%%%%%%%%%

The Positronium state, including the $\ket{e^-e^+\gamma}$ Fock component, can be expressed as the superposition
\begin{align} \label{tra5}
\ket{\mB,\Pv} = \ket{e^-e^+;\mB,\Pv} + \int\frac{d\qv}{(2\pi)^3 2|\qv|}\sum_\smu C_\gamma(\qv)\ket{e^-e^+\gamma;\qv,\smu} 
\end{align}
where $\ket{e^-e^+\gamma;\qv,\smu}$ is defined in \eq{tra1} and I anticipated that its relative weight $C_\gamma(\qv)$ is independent of $\smu$. $C_\gamma(\qv)$ should be determined so that $\ket{\mB,\Pv}$ is an eigenstate of $\mH$,
\begin{align} \label{tra6}
\mH\ket{\mB,\Pv} &= (\mH_0(f)+\mH_V)\ket{e^-e^+;\mB,\Pv} + \mH_{int}\int\frac{d\qv}{(2\pi)^3 2|\qv|}\sum_\smu C_\gamma(\qv)\ket{e^-e^+\gamma;\qv,\smu} \nn\crt
&+ \int\frac{d\qv}{(2\pi)^3 2|\qv|}\sum_\smu \Big[1+\big(\mH_0(f)+\mH_0(A)\big)C_\gamma(\qv)\Big]\ket{e^-e^+\gamma;\qv,\smu} = E_\mB^{(\Pv)}\ket{\mB,\Pv}
\end{align}
$\mH_0(f)+\mH_V$ modify $\Phipb$ by the factors in \eq{mov7} and \eq{mov7a}. The $\mH_{int}$ term is given by \eq{tra4}, adding the factor $C_\gamma(\qv)$ to the integrand of the $\qv$-integration. The first contribution to $\ket{e^-e^+\gamma;\qv,\smu}$ follows from \eq{tra1}. The action of $\mH_0(f)$ is similar to that on $\ket{e^-e^+;\mB,\Pv}$ in \eq{mov7}, but now the additional $\xv_{1,2}$ dependence in $\exp(-i\qv\cdot\xv_{1,2})$ leaves an \order{\alpha} contribution. Since $\ket{e^-e^+\gamma;\qv,\smu}$ is already suppressed by a factor $e$ we may here neglect the \order{\alpha^2} terms, including its potential energy ($\mH_V$). Recalling that $E_i =$ ${\scriptsize\sqrt{-\nv_i^2+m^2}}$ and separating the \order{\alpha^0} contribution through,
\begin{align} \label{tra7}
(E_1+E_2)e^{i\Pv\cdot(\xv_1+\xv_2)/2} &\simeq e^{i\Pv\cdot(\xv_1+\xv_2)/2}\Big(\sqrt{\quart E_P^2-i\Pv\cdot\nv_1} + \sqrt{\quart E_P^2-i\Pv\cdot\nv_2}\,\Big) \nn\crt
&\simeq e^{i\Pv\cdot(\xv_1+\xv_2)/2}\Big[E_P-\frac{i}{E_P}\,\Pv\cdot(\nv_1+\nv_2)\Big]
\end{align}
gives the sum of the fermion kinetic energies as
\begin{flalign} \label{tra8}
&\mH_0(f)\ket{e^-e^+\gamma;\qv,\smu}: \hspace{1cm} E_1+E_2 = E_P-\beta q_\|+\morder{\alpha^2}&
\end{flalign}
where $\beta=P/E_P$ and $\Pv\cdot\qv = P q_\|$.

The kinetic energy of the photon follows from $\mH_0{(A)} = \int\frac{d\qv}{(2\pi)^3\,2|\qv|}|\qv|\sum_{\smu=\pm1} a^\dag(\qv,\smu)a(\qv,\smu)$,
\begin{flalign} \label{tra9}
&\mH_0(A)\ket{e^-e^+\gamma;\qv,\smu}: \hspace{1cm} E_\gamma = |\qv|&
\end{flalign}
Comparing the coefficients of $\ket{e^-e^+\gamma;\qv,\smu}$ in the eigenstate condition \eq{tra6} gives, since $E_\mB^{(\Pv)}=E_P+\morder{\alpha^2}$,
\begin{align} \label{tra10}
1+C_\gamma(\qv)(E_P+|\qv|-\beta q_\|) = C_\gamma(\qv)E_P +\morder{\alpha^2} \hspace{1cm} {\color{red} \Longrightarrow } \hspace{1cm} 
 C_\gamma(\qv) = -\inv{|\qv|-\beta q_\|} = -\frac{|\qv|+\beta q_\|}{\qv_\perp^2+q_\|^2/\gamma^2}
\end{align}
Including $C_\gamma(\qv)$ in \eq{tra4} we have in \eq{tra6},
\begin{align} \label{tra11}
\mH_{int}\int\frac{d\qv}{(2\pi)^3 2|\qv|}\sum_\smu C_\gamma(\qv)\ket{e^-e^+\gamma;\qv,\smu}
&=\int d\xv_1 d\xv_2\,\bar\psi(\xv_1)\lla_+(\xv_1)e^{i\Pv\cdot(\xv_1+\xv_2)/2}\Phipb(\xv_1-\xv_2)\rla_-(\xv_2)\psi(\xv_2) \nn \crt
&\times e^2\int\frac{d\qv}{(2\pi)^3 2|\qv|}\beta^2\,\frac{\qv_\perp^2}{\qv^2}\frac{|\qv|+\beta q_\|}{\qv_\perp^2+q_\|^2/\gamma^2}\,\big[e^{-i\qv\cdot(\xv_1-\xv_2)}+e^{i\qv\cdot(\xv_1-\xv_2)}\big]\ket{0}
\end{align}
The $\beta q_\|$ term in the numerator does not contribute since the remaining integrand is symmetric under $\qv \to -\qv$. We may then use this symmetry to set $\exp[i\qv\cdot(\xv_1-\xv_2)] \to \exp[-i\qv\cdot(\xv_1-\xv_2)]$ and cancel a factor $2|\qv|$. Combining this transverse photon contribution with the Coulomb one \eq{mov7a} the integral over $\qv$ becomes (with $\xv=\xv_1-\xv_2$),
\begin{align} \label{mov22}
-e^2&\int\frac{d\qv}{(2\pi)^3}\,\inv{\qv^2}\Big[1-\frac{\beta^2q_\perp^2}{\qv_\perp^2+q_\|^2/\gamma^2}
=\frac{\qv^2/\gamma^2}{\qv_\perp^2+q_\|^2/\gamma^2}\Big]e^{-i\qv\cdot\xv}
= -\frac{e^2}{\gamma^2}\int\frac{d\qv}{(2\pi)^3}\,\frac{e^{-i\qv\cdot\xv}}{\qv_\perp^2+q_\|^2/\gamma^2}
= -\frac{\alpha}{\gamma\sqrt{\xv_\perp^2+\gamma^2 x_\|^2}}
\end{align}
Adding the kinetic energy \eq{mov7} the eigenstate condition \eq{tra6} imposes the bound state condition on $\Phipb$, with the required eigenvalue \eq{ebp1},
\begin{align} \label{tra12}
\Big[E_P-\inv{m\gamma}\Big(\nv_\perp^2-\frac{1}{\gamma^2}\nv_\|^2\Big) -\frac{\alpha}{\gamma\sqrt{\xv_\perp^2+\gamma^2 x_\|^2}}\Big]\Phipb(\xv) = \Big(E_P+\inv{\gamma}\,E_b\Big)\Phipb(\xv)
\end{align}
A comparison with the $\Pv=0$ equation \eq{sch6} shows that up to a normalization we have 
\begin{align} \label{tra12a}
\Phipb(\xv)=\Phirb(\xv_\perp,\gamma x_\|)
\end{align} 
\ie, standard Lorentz contraction. The contraction is the same for all Dirac components, justifying the factorization \eq{fock5c}. According to \eq{fock9} the contracted radial function is
\begin{align} \label{tra13}
\Fp(\xv) = \gamma\,F^{(0)}(r_P) = \gamma\,\frac{\alpha^{3/2}m^2}{\sqrt{4\pi}}\,e^{-\alpha m r_P/2} \hspace{2cm} r_P \equiv \sqrt{\xv_\perp^2+\gamma^2 x_\|^2} \hspace{1cm} \gamma = \frac{E_P}{2m} = \frac{\sqrt{\Pv^2+4m^2}}{2m}
\end{align}
The Lorentz contraction of $\Fp(\xv)$ agrees with the classical result. Recall, however, that the $\ket{e^-e^+\gamma;\qv,\smu}$ Fock component \eq{tra5} also contributes. The kinetic energies of its constituents \eq{tra8}, \eq{tra9} are of \order{\alpha}, \ie, large compared to the \order{\alpha^2} binding energy of $\ket{\mB,\Pv}$. By the uncertainty principle $\ket{\mB,\Pv}$ fluctuates into $\ket{e^-e^+\gamma;\qv,\smu}$ only a fraction $\alpha$ of the time. Equivalently, the norm of the $\ket{e^-e^+\gamma;\qv,\smu}$ Fock component is of \order{\alpha}.

\subsection{*\,Hyperfine splitting of Positronium at $\Pv=0$ \label{secV.D}}
%%%%%%%%%%%%%%%%%%%%%%%%%%

Hyperfine splitting (hfs) is defined as the difference between the Ortho- and Parapositronium binding energies, $E_{hfs} = E_b(Ortho)-E_b(Para)$. The hfs \eq{mo1} is known to high accuracy, with current work addressing the \order{\alpha^7 m} contribution using the methods of NRQED. Here I shall illustrate the Fock state method in temporal gauge by evaluating the \order{\alpha^4 m} contribution. At this order the hfs arises from transverse photon exchange between the $e^-$ and $e^+$, as well as from the annihilation contribution  $e^-e^+ \to \gamma \to e^-e^+$ (for Orthopositronium only). In the Fock state approach this means considering the $\ket{e^-e^+\gamma}$ and $\ket{\gamma}$ Fock states, respectively.

\subsubsection{The transverse photon Fock state $\ket{e^-e^+\gamma}$ contribution \label{secV.D1}}
%%%%%%%%%%%%%%%%%%%%%%%%%%

In section \ref{secV.C} I evaluated the transverse photon exchange contribution to Positronium with $\Pv \neq 0$ at leading order. Here I consider transverse photon exchange for Positronium at rest ($\Pv=0$). The two electron-photon vertices are then proportional to \order{\alpha} Bohr momenta, which makes the transverse photon contribution to be of \order{\alpha^4} in the rest frame. I discuss only photon emission from the $e^-$ and absorption by the $e^+$. The converse contribution is identical and is taken into account by a factor 2 in the final result. Photons both emitted and absorbed on the $e^-$ do not contribute to the spin correlation (hfs) between the $e^-$ and $e^+$.

The Positronium state including the Fock state with a transverse photon is \eq{tra5}
\begin{align} \label{v1}
\ket{\mB} = \ket{e^-e^+;\mB} + \ket{e^-e^+\gamma;\mB} 
\end{align}
where $\ket{e^-e^+;\mB}$ is defined in \eq{fock1} with $\Pv=0$ and the wave function $\Phi_\mB(\xv)= \Gamma_\mB\,F(r)$ according to \eq{fock5a} and \eq{fock6}. Its Dirac structures are $\Gamma_{Para} = \gf$, $\Gamma_{Ortho} = \ev_\lm\cdot\alv$ as in \eq{para}, \eq{ortho} and the radial function $F(r)$ is given in \eq{fock9}. The transverse photon state is as in \eq{tra5} with $C_\gamma=-1/|\qv|$ from \eq{tra10} with $\Pv=0$,
\begin{align} \label{v1b}
\ket{e^-e^+\gamma;\mB} = \int\frac{d\qv}{(2\pi)^3 2|\qv|}\sum_\smu \frac{-1}{|\qv|}\ket{e^-e^+\gamma;\qv,\smu}
\end{align}
The $\ket{e^-e^+\gamma;\qv,\smu}$ state created by $\mH_{int}(\Av_T)$ \eq{mov39} acting on $\ket{e^-e^+;\mB}$ with $e^- \to e^-\gamma$ is given in \eq{ex14a.1}.

When the Hamiltonian acts on its eigenstate $\ket{\mB}$ the energy eigenvalue may be read off from the coefficient of the $\ket{e^-e^+;\mB}$ Fock state. Projecting on this state,
\begin{align} \label{v2}
\mH\ket{\mB} = \Big[2m-\quart m\alpha^2+\morder{\alpha^4}+\frac{\bra{e^-e^+;\mB}\mH_{int}\ket{e^-e^+\gamma;\mB}}{\bra{e^-e^+;\mB}e^-e^+;\mB\rangle}\Big] \ket{e^-e^+;\mB} +\ldots
\end{align}
The \order{\alpha^4} term represents higher order corrections to the eigenvalue from $(\mH_0+\mH_V)\ket{e^-e^+;\mB}$, \eg, due to the Taylor expansion of the energies $E_i$ in \eq{mov4}. They are the same for Para- and Orthopositronium and do not affect the hfs. I keep only terms which are spin-, \ie, $\Gamma_\mB$-dependent. 

The absorption of the photon on the $e^+$ is given by $\com{\mH_{int}}{\psi(\xv_2)}$ analogously as the emission from $e^-$ in \eq{ex14a.1},
\begin{align} \label{v3}
\mH_{int}\ket{e^-e^+\gamma;\mB} = -e^2\int\frac{d\qv}{(2\pi)^3 2\qv^2}\sum_\smu \int d\xv_1 d\xv_2\,\bar\psi(\xv_1)&\lla_+(\xv_1) \alv\cdot\vepsv_\smu^*(\qv)e^{-i\qv\cdot\xv_1}\lla_+(\xv_1)\Gamma_\mB F(r) \nn\crt
&\times\rla_-(\xv_2)\alv\cdot\vepsv_\smu(\qv)e^{i\qv\cdot\xv_2}\rla_-(\xv_2)\psi(\xv_2)\ket{0}
\end{align}
where $r=|\xv_1-\xv_2|$ and the extra pair of $\la_\pm$ project $b^\dag$ from $\bar\psi(\xv_1)$ and $d^\dag$ from $\psi(\xv_2)$ \eq{fock2b}. At \order{\alpha^4} two momenta can contribute, one each from the photon vertices. The identities in \eq{ex14a.2} and \eq{ex14a.4} show that $\alv$ bracketed by two projectors becomes a derivative. This contribution reduces the Dirac structure of \eq{v3} so that the overlap in \eq{v2} is independent of $\Gamma_\mB$, \ie, it does not contribute to the hfs. Hence we need only consider the contributions
\begin{align} \label{v4}
e^{-i\qv\cdot\xv_1}\lla(\xv_1) \to -\inv{2m}\alv\cdot\qv\,e^{-i\qv\cdot\xv_1} \hspace{2cm}
\rla(\xv_2)e^{i\qv\cdot\xv_2} \to -\inv{2m}\alv\cdot\qv\,e^{i\qv\cdot\xv_2}
\end{align}
At \order{\alpha^4} the remaining projectors can be set to lowest order, $\la_\pm=(1\pm\gz)/2$. The products of $\alv$-matrices may be reduced through $\alpha_i\alpha_j = \delta_{ij}+i\veps_{ijk}\alpha_k\gf$. The $\delta_{ij}$-function does not contribute here since $\qv\cdot\vepsv_\smu(\qv)=0$,
\begin{align} \label{v5}
{\veps_\smu^i}^*q^j\alpha_i\alpha_j = i\veps_{ijk}{\veps_\smu^i}^*q^j\alpha_k\gf \hspace{2cm}
q^\ell\veps_\smu^m\alpha_\ell\alpha_m = i\veps_{\ell mn}q^\ell\veps_\smu^m\alpha_n\gf 
\end{align}
Since $\com{\gf}{\Gamma_\mB}=0$ the $\gf$'s cancel, $\gf^2=1$. The sum over photon polarizations \eq{hv1a} gives $\sum_\smu {\veps_\smu^i}^*(\qv)\veps_\smu^m(\qv) \to \delta^{im}$, since the $-q^iq^m/\qv^2$ term vanishes. Then $i^2\veps_{ijk}\veps_{in\ell}q^jq^\ell = \qv^2\delta^{kn}-q^kq^n$.

Writing $\alpha_k\Gamma_\mB = \com{\alpha_k}{\Gamma_\mB}+\Gamma_\mB\alpha_k$ the second term does not give an hfs, while the commutator vanishes for $\Gamma_{Para}=\gf$. Hence is suffices to consider the Orthopositronium $(j^z=\lm)$ contribution, $\com{\alpha_k}{\ev_\lm\cdot\alv} = -2i\veps_{jkp}e_\lm^j\alpha_p\gf$. This is multiplied by the $\alpha_n$ in \eq{v5}, giving $-2i\veps_{jkp}e_\lm^j\alpha_p\alpha_n\gf = 2\veps_{jkp}\veps_{nip}e_\lm^j\alpha_i = 2(e_\lm^n\alpha_k-\delta^{kn}\ev_\lm\cdot\alv)$. Combined with the $\qv$-dependence found above we have
\begin{align} \label{v6}
(\qv^2\delta^{kn}-q^kq^n) 2(e_\lm^n\alpha_k-\delta^{kn}\ev_\lm\cdot\alv) = -2\big[\qv^2\ev_\lm\cdot\alv+(\ev_\lm\cdot\qv)(\alv\cdot\qv)\big]
\end{align}
The contribution to \eq{v3} that is relevant for the hfs is thus
\begin{align} \label{v7}
\mH_{int}\ket{e^-e^+\gamma;Ortho} = \frac{e^2}{4m^2} \int\frac{d\qv}{(2\pi)^3 \qv^2} \int d\xv_1 d\xv_2\,\bar\psi(\xv_1) \la_+F(r)e^{-i\qv\cdot(\xv_1-\xv_2)} \big[\qv^2\ev_\lm\cdot\alv+(\ev_\lm\cdot\qv)(\alv\cdot\qv)\big] \la_-\psi(\xv_2)\ket{0}
\end{align}
where $\la_\pm=(1\pm\gz)/2$. In the matrix elements of \eq{v2} both electron fields are annihilated and the integral $\int d(\xv_1+\xv_2)/2 = (2\pi)^3\delta(\bs{0})$ cancels with the norm \eq{fock3} in the denominator up to a factor $4m$. Recalling that we only considered photon emission from $e^-$ and absorption on $e^+$ and so are getting half of the hfs, 
\begin{align} \label{v8}
\halft E_{hfs}^T &= \frac{\bra{e^-e^+;\mB}\mH_{int}\ket{e^-e^+\gamma;\mB}}{\bra{e^-e^+;\mB}e^-e^+;\mB\rangle} \nn\crt
&= \frac{e^2}{16m^3}\int d\xv\, \frac{d\qv}{(2\pi)^3 \qv^2}\,|F(r)|^2 e^{-i\qv\cdot\xv}\tr\big\{ \halft(1-\gz) \ev_\lm^*\cdot\alv \halft(1+\gz)\big[\qv^2\ev_\lm\cdot\alv+\ev_\lm^i\alpha^jq^iq^j\big]\big\}
\end{align}
The factor multiplying $q^iq^j$ in the integrand is symmetric under $q^i \to -q^i,\ x^i \to -x^i$, allowing $q^iq^j \to \sfrac{1}{3}\qv^2\delta^{ij}$. The trace factor becomes $\sfrac{2}{3}\qv^2\tr\big\{(\ev_\lm^*\cdot\alv)(\ev_\lm\cdot\alv)\big\} = \sfrac{8}{3}\qv^2$. With $F(r)$ given by \eq{fock9},
\begin{align} \label{v9}
\halft E_{hfs}^T = \frac{e^2}{6m^3}\int d\xv\,|F(r)|^2 \int\frac{d\qv}{(2\pi)^3} e^{-i\qv\cdot\xv}
= \frac{4\pi\alpha}{6m^3}|F(0)|^2 = \inv{6} m\alpha^4
\end{align}
The contribution to the hfs from transverse photon exchange is thus as expected \cite{Penin:2014bea},
\begin{align} \label{v10}
E_{hfs}^T = \inv{3}m\alpha^4
\end{align}

\subsubsection{Hyperfine splitting from annihilation: $e^-e^+ \to \gamma \to e^-e^+$ \label{secV.D2}}
%%%%%%%%%%%%%%%%%%%%%%%%%%

The $\ket{e^-e^+} \to \ket{\gamma} \to \ket{e^-e^+}$ transition is proportional to the square $|\Phi_\mB(0)|^2$ of the $\Pv=0$ Positronium \eq{fock1}  wave function at the origin. $\Phi_\mB(\xv) = \Gamma_\mB F(r)$ where $\Gamma_\mB$ for Para- and Orthopositronium are in \eq{para}, \eq{ortho} and their common radial function $F(r)$ is in \eq{fock9}. Counting also the vertex couplings the transition is $\propto e^2\,|F(0)|^2 \propto \alpha^4$. Hence we may neglect the \order{\alpha m} relative (Bohr) momenta in evaluating the hfs at \order{\alpha^4}. The projectors $\la_\pm$ in the state \eq{fock1} may then be replaced with $\halft(1\pm \gz)$. Annihilating both the $e^-$ and $e^+$ in the state  \eq{fock1} by $\mH_{int}$ gives
\begin{align} \label{ann1}
\mH_{int}\ket{e^-e^+;\mB} &= -e\int d\xv\,\bar\psi(\xv)\alv\cdot\Av(\xv)\psi(\xv)
\int d\xv_1d\xv_2\,\bar\psi(\xv_1)\halft(1+\gz)\Phi_\mB(\xv_1-\xv_2)\halft(1-\gz)\psi(\xv_2)\ket{0} \crt
&= -e\int d\xv\,\tr\big\{\alv\cdot\Av(\xv)\gz\halft(1+\gz)\Gamma_\mB F(0)\halft(1-\gz)\big\}\ket{0}
=-\halft e F(0)\int d\xv\, \tr[\alpha^i \Gamma_\mB]\,A^i(\xv)\ket{0} \nn
\end{align}
As expected due to charge conjugation invariance, this vanishes for Parapositronium, $\Gamma_{Para}=\gf$. Hence the annihilation contribution to the hfs arises only from Orthopositronium, $\Gamma_{Ortho}^\lm=\ev_\lm\cdot\alv$ for states with $j^z=\lm$:
\begin{align} \label{ann1a}
\mH_{int}\ket{e^-e^+;\mO_\lm} &= -2e F(0) \int d\xv\, \ev_\lm\cdot \Av(\xv)\ket{0} \equiv -2e F(0)\ket{\Av,\lm}
\end{align}
The relevant action of the Hamiltonian \eq{eII3} on this state is given by the canonical commutation relations \eq{eII2},
\begin{align} \label{ann2}
\mH\ket{\Av,\lm} \to \int d\yv\,\halft \Ev^2(\yv)\int d\xv\, \ev_\lm\cdot \Av(\xv)\ket{0} = i\int d\xv\, \ev_\lm\cdot \Ev(\xv)\ket{0} \equiv i\ket{\Ev,\lm}
\end{align}
$\mH\ket{\Ev,\lm}$ has an overlap $C_\mO$ with Orthopositronium. Neglecting the other states which do not contribute here,
\begin{align} \label{ann3}
\mH_{int}\ket{\Ev,\lm} = -e\int d\yv\,\psi^\dag(\yv)\,\alv\cdot\Av(\yv)\psi(\yv)\int d\xv\, \ev_\lm\cdot \Ev(\xv)\ket{0} = ie\int d\xv\, \psi^\dag(\xv)\,\ev_\lm\cdot\alv\,\psi(\xv)\ket{0} = C_\mO\ket{e^-e^+;\mO_\lm} + \ldots
\end{align}
With the normalization \eq{fock3} the overlap $C_\mO$ is
\begin{align} \label{ann4}
C_\mO = \frac{\bra{e^-e^+;\mO_\lm}\mH\ket{\Ev,\lm}}{\bra{e^-e^+;\mO_\lm}e^-e^+;\mO_\lm\rangle}= 
&=\frac{1}{4m(2\pi)^3\delta^3(0)}\bra{0}\int d\xv_1 d\xv_2\,\psi^\dag(\xv_2)\,\ev_\lm^*\cdot\alv\, F^*(|\xv_1-\xv_2|)\halft(1+\gz)\gz\psi(\xv_1) \nn\crt
&\times ie\int d\xv\, \psi^\dag(\xv)\,\ev_\lm\cdot\alv\,\psi(\xv)\ket{0}
=\frac{ie}{2m}\,F^*(0) 
\end{align}

The Orthopositronium state may at \order{\alpha^4} be considered as a superposition of the three states involved,
\begin{align} \label{ann5}
\ket{\mO_\lm} = \ket{e^-e^+;\mO_\lm} + C_A \ket{\Av,\lm} + C_E \ket{\Ev,\lm}
\end{align}
and should be an eigenstate of $\mH$ with eigenvalue $E_\mO$. Using \eq{ann1a}, \eq{ann2} and \eq{ann4},
\begin{align} \label{ann6}
\mH\ket{\mO_\lm} &= (2m-\quart m\alpha^2+C_\mO C_E)\ket{e^-e^+;\mO_\lm} + C_A\,i \ket{\Ev,\lm} -2e F(0) \ket{\Av,\lm} \nn\crt
&= E_\mO\,\Big[\ket{e^-e^+;\mO_\lm}- \frac{2eF(0)}{E_\mO}\ket{\Av,\lm}+\frac{iC_A}{E_\mO}\ket{\Ev,\lm}\Big] = E_\mO\ket{\mO_\lm}
\end{align}
The eigenstate constraint requires (at leading order) $C_A = -2eF(0)/2m$ and $C_E=iC_A/2m = -2ieF(0)/4m^2$. With the value of $F(0)$ in \eq{fock9} this gives the hfs term in $E_\mO$ as
\begin{align} \label{ann6b}
C_\mO C_E = \frac{ie}{2m}\,\frac{-2ie}{4m^2}|F(0)|^2 = \quart m\alpha^4
\end{align}
as quoted in \cite{Penin:2014bea}.

\subsection{*\,Electromagnetic form factor of Positronium atoms in an arbitrary frame} \label{secV.E}
%%%%%%%%%%%%%%%%%%%%%%%%%%

In this section I evaluate the electromagnetic form factors of Positronium. The elastic form factor is evaluated with leading order wave functions in an arbitrary frame, demonstrating the Lorentz covariance of the result. The transition form factor from Para- to Orthopositronium is calculated in the rest frame only. 

The electromagnetic current $j^\mu(z)$ may be translated to the origin ($z=0$) using the four-momentum operator $\hat P$,
\begin{align} \label{ff1}
j^\mu(z) = \bar\psi(z)\gamma^\mu\psi(z) = e^{i\hat P\cdot z}j^\mu(0)e^{-i\hat P\cdot z}
\end{align}
The EM form factor $F_{AB}^\mu$ is the expectation value of the current between atoms $A,B$ of three-momenta $\Pva,\Pvb$ whose four-momenta satisfy $P_A^2=M_A^2$, $P_B^2=M_B^2$,
\begin{align} \label{ff2}
F_{AB}^\mu(z) &= \bra{B, \Pvb}j^\mu(z)\ket{A,\Pva} = e^{i(P_B-P_A)\cdot z}\bra{B, \Pvb}j^\mu(0)\ket{A,\Pva}\nn\crt
F_{AB}^\mu(q) &= \int d^4z\,e^{-iq\cdot z}\,F^\mu_{AB}(z) = (2\pi)^4\delta^4(P_b-P_a-q)G^\mu_{AB}(q)
\end{align}
In the following I consider $G^\mu_{AB}(q)$, keeping in mind the four-momentum constraint.

The Positronium state is defined in \eq{fock1}. 
With a short-hand notation $\Psi_A$ the wave function of the incoming state is
\begin{align} \label{ff3}
\ket{A,\Pv} &= \int d\xv_1d\xv_2\,\bar\psi(\xv_1)\Psi_A(\xv_1,\xv_2)\psi(\xv_2)\ket{0} \nn\crt
\Psi_A(\xv_1,\xv_2) &= \lla_+(\xv_1)e^{i\Pv_A\cdot(\xv_1+\xv_2)/2}\Phi^{(\Pv_A)}_A(\xv_1-\xv_2)\rla_-(\xv_2)
\end{align}
where the projectors $\la_\pm$ are defined in \eq{fock2}. The same notation for the final state $\bra{B,\Pv}$ gives
\begin{align} \label{ff4}
G_{AB}^\mu(q) = \int d\yv_1d\yv_2\, d\xv_1d\xv_2\,\bra{0}\psi^\dag(\yv_2)\Psi_B^\dag(\yv_1,\yv_2)\gz\psi(\yv_1)\,\bar\psi(0)\gamma^\mu\psi(0)\,\bar\psi(\xv_1)\Psi_A(\xv_1,\xv_2)\psi(\xv_2)\ket{0}
\end{align}
The contraction of $\psi(0)$ with $\bar\psi(\xv_1)$ corresponds to $j^\mu$ interacting with the $e^-$. This sets $\xv_1=\yv_1=0$ and $\yv_2=\xv_2$. I denote this contribution $G_{AB}^\mu(q,e^-)$. Interaction with $e^+$ corresponds to $\psi(0)$ contracting with $\psi^\dag(\yv_2)$ and is denoted $G_{AB}^\mu(q,e^+)$. It has a minus sign due to anticommutation and sets $\xv_2=\yv_2=0$ and $\yv_1=\xv_1$. Thus
\begin{align} \label{ff5}
G_{AB}^\mu(q,e^-) &= \int d\xv\,\tr\big\{\Psi_A(0,-\xv)\Psi_B^\dag(0,-\xv)\gamma^\mu\gz\big\}\crt
G_{AB}^\mu(q,e^+) &= -\int d\xv\,\tr\big\{\Psi_B^\dag(-\xv,0)\Psi_A(-\xv,0)\gz\gamma^\mu\big\} \nn
\end{align}
The two contributions are related by charge conjugation. Using \eq{A15} and recalling that $\alpha_2\gamma^\mu\alpha_2 = -(\gamma^\mu)^T$ and $\alpha_2\,\alv\,\alpha_2 = - \alv^T$ we have
\begin{align} \label{ff6}
\aly\Psi^T(\xv_2,\xv_1)\aly = \eta_C\,\Psi(\xv_1,\xv_2)
\end{align}
Multiplying the argument of the $\tr$ in $G_{AB}^\mu(q,e^-)$ by $\aly$ from the left and right and taking its transpose shows that
\begin{align} \label{ff8}
G_{AB}^\mu(q,e^+) = -\eta^A_C \eta^B_C G_{AB}^\mu(q,e^-)
\end{align}
As expected the photon ($\eta_C^\gamma = -1$) requires $\eta^A_C = - \eta^B_C$ when $A$ and $B$ are eigenstates of charge conjugation,
\begin{align} \label{ff9}
G_{AB}^\mu(q) = (1-\eta^A_C \eta^B_C)\int d\xv\,\tr\big\{\Psi_A(0,-\xv)\Psi_B^\dag(0,-\xv)\gamma^\mu\gz\big\}
\end{align}

%%%%%%%%%%%%%%
\subsubsection{Parapositronium form factor \label{secV.E1}}
%%%%%%%%%%%%%%

I take both $A$ and $B$ to be Parapositronium and consider only $G_{AB}^\mu(q,e^-)$. This is relevant for states which are not eigenstates of charge conjugation, \eg, a hypothetical $\mu^- e^+$ atom where the muon and electron have the same mass. Even for standard Positronium $G_{AB}^\mu(q,e^-) \neq 0$  and should have the form required by Lorentz covariance,
\begin{align} \label{ff10}
G^\mu(q,e^-) = (P_A+P_B)^\mu F(q^2)
\end{align}

After partial integrations in the state \eq{ff3} we need consider only the projector derivatives acting on the \order{\alpha^0} phase $\exp[i\Pv\cdot(\xv_1+\xv_2)/2]$, since $\nv\Phip_{A,B}(\xv)$ is of \order{\alpha}. The projectors then become,
\begin{align} \label{ff11}
\la_\pm(\Pv) \equiv \inv{2E_P}\big(E_P\mp \alv\cdot\Pv\pm M\gz \big) = \la_\pm^\dag(\Pv) = \big[\la_\pm(\Pv)\big]^2
\hspace{2cm} \la_+(\Pv)\la_-(\Pv) = 0
\end{align}
where $E_P = \sqrt{\Pv^2+M^2}$ and $M=2m$ at leading order.

The Parapositronium states are relativistically normalized \eq{fock3}, with the Lorentz contracted wave function $\Phip(\xv)$ given in \eq{fock5c}, \eq{para}, \eq{fock6} and \eq{tra13}. Taking $\Pv = (0,0,P) = (0,0,M\sinh\xi)$ along the $z$-axis,
\begin{align} \label{ff12}
\Phip(\xv) = \frac{E_P}{M}\, \gf \Fp(\xv) \hspace{2cm} \Fp(\xv) &= \frac{E_P}{M}\, F^{(0)}(r_P) = \frac{E_P}{M}\, \frac{\alpha^{3/2}m^2}{\sqrt{4\pi}}\,e^{-\alpha m r_P/2} \crt
r_P &\equiv \sqrt{x^2+y^2+z^2\cosh^2\xi}
\end{align}
The photon momentum $\qv$ must be of \order{\alpha} for a leading order overlap between $\Psi_A$ and $\Psi_B$. Hence we may set $\Pv_B=\Pv_A+\qv = \Pv+ \morder{\alpha}$ in $G_{AB}^\mu(q,e^-)$ \eq{ff5}. However, we need to retain the $\qv$-dependence of the $\Psi_A(0,-\xv)\Psi_B^\dag(0,-\xv)$ phase factor $\exp[i(\Pv_B-\Pv_A)\cdot\xv/2] = \exp(i\qv\cdot\xv/2)$, which reflects the photon wave function. The expression for $G_{AB}^\mu(q,e^-)$ is then
\begin{align} \label{ff13}
G^\mu(q,e^-) &= \Big(\frac{E_P}{M}\Big)^2 \int d\xv\,|F^{(\Pv)}(\xv)|^2 e^{i\qv\cdot\xv/2}\tr\big\{\la_+(\Pv)\gf\la_-(\Pv)\la_-(\Pv) \gf\la_+(\Pv)\gamma^\mu\gz\big\}
\end{align}
From the definitions \eq{ff11} of $\la_\pm(\Pv)$ follows that
\begin{align} \label{ff14}
\la_+(\Pv)\gf\la_-(\Pv) = \frac{M}{E_P}\,\la_+(\Pv)\gz\gf
\end{align}
Using this the trace in \eq{ff13} becomes
\begin{align} \label{ff15}
\tr^\mu = \Big(\frac{M}{E_P}\Big)^2\tr\big\{\la_+(\Pv)\gamma^\mu\gz\big\} = \Big(\frac{M}{E_P}\Big)^2 \,\frac{2P^\mu}{E_P}
\end{align}
so that
\begin{align} \label{ff16}
G^\mu(q,e^-) &= \frac{2P^\mu}{E_P}\,\int d\xv\,|F^{(\Pv)}(\xv)|^2 e^{i\qv\cdot\xv/2}
\end{align}
Changing the integration variable to $\xv_R \equiv (x,y,z\cosh\xi)$ gives $d\xv = d\xv_R/\cosh\xi$ and $r_P = |\xv_R|$ in \eq{ff12}. The photon four-momentum $q$ is constrained by kinematics,
\begin{align} \label{ff17}
M^2 = (P+q)^2 = M^2 + 2P\cdot q +\morder{\alpha^2} = M^2
\end{align}
which at \order{\alpha} implies $P\cdot q = E_P q^0-\Pv\cdot\qv =0$. In the Positronium rest frame ($\Pv=0$) this means $q^0_R = 0$. Thus $q^z = q^z_R\cosh\xi$, where $q^z_R$ is the $z$-component of the photon momentum in the rest frame. Hence $\qv\cdot\xv = \qv_R\cdot\xv_R$. Recalling that $\cosh\xi = E_P/M$ and using the expression for $\Fp(\xv)$ in \eq{ff12} gives
\begin{align} \label{ff18}
G^\mu(q,e^-) &= \frac{2P^\mu}{E_P}\,\frac{E_P}{M}\,\frac{\alpha^3m^4}{4\pi}\int d\xv_R\,\exp\big[(-\alpha M|\xv_R|+i\qv_R\cdot\xv_R)/2\big]
\end{align}
The integral is as in the rest frame, \ie, it is $\Pv$-independent, so the result is covariant and agrees with \eq{ff10},
\begin{align} \label{ff19}
G^\mu(q,e^-) &= 2P^\mu\,\frac{(\alpha M)^4}{(Q^2+\alpha^2 M^2)^2}
\end{align}
where $2P^\mu = (P_A+P_B)^\mu +\morder{\alpha}$ and $Q^2 = \qv_R^2 = -q^2$.

%%%%%%%%%%%%%%
\subsubsection{Positronium transition form factor \label{secV.E2}}
%%%%%%%%%%%%%%

The $\gamma^*(q)$ + Parapositronium $\to$ Orthopositronium transition electromagnetic form factor has the structure
\begin{align} \label{ff20}
G_\lm^\mu(q) = i\veps^{\mu\nu\rho\sigma}P_\nu q_\rho e_\sigma^\lm F(q^2)
\end{align}
where $P$ is the four-momentum of one of the Positronia, $e^\lm$ is the polarization vector \eq{ortho} of the Orthopositronium (with $e^\lm_{\sigma=0}=0$) and $q$ is the photon momentum. Symmetries and gauge invariance force the kinematic factor to be of \order{q}, \ie, of \order{\alpha}. This reflects the spin flip, from $S=0$ for Parapositronium to $S=1$ for Orthopositronium.

In section \ref{secV.C} I demonstrated that transverse photon exchange contributes to the binding of Positronium at leading order for $\Pv \neq 0$. The photon exchange may at \order{q} involve a spin flip at one of its vertices, whereupon the transition to Orthopositronium proceeds at \order{\alpha^0}. I shall not here work out the \order{\alpha} corrections to the wave function of Positronium in motion, but limit myself to evaluating the transition form factor in the rest frame of the target, $\Pv_A=0$.

Using the expressions \eq{ff11} for the $\la_\pm(\Pv)$ projectors (with $\Pv_A =0,\  \Pv_B = \qv,\ E_B \simeq M$) the Positronium wave functions given in \eq{fock5c}, \eq{para}, \eq{ortho} and \eq{fock6} are,
\begin{align} \label{ff20a}
\Psi_A(0,-\xv) &= \halft(1+\gz)\gf\,F(r) \nn\crt
\Psi_B(0,-\xv) &= N_\lm \la_+(\qv)\ev_\lm\cdot\alv \la_-(\qv)F(r) e^{-i\qv\cdot\xv/2}
\end{align}
The expression for $\Psi_B(0,-\xv)$ may be simplified using $\la_+(\qv)\la_-(\qv)=0$,
\begin{align} \label{ff20b}
\la_+(\qv)\ev_\lm\cdot\alv\, \la_-(\qv) = \acom{\la_+(\qv)}{\ev_\lm\cdot\alv}\la_-(\qv) = \big(\ev_\lm\cdot\alv-\inv{M}\ev_\lm\cdot\qv\big)\la_-(\qv)
\end{align}
The form factor \eq{ff9} is then in the target rest frame,
\begin{align} \label{ff20c}
G_\lm^\mu(q) &= \frac{{N_\lm}^*}{2 M}\int d\xv\,|F(r)|^2 e^{i\qv\cdot\xv/2} \tr_\lm^\mu \nn\crt
\tr_\lm^\mu &= \tr\big\{(1+\gz)\gf(M+\alv\cdot\qv-M\gz)\big(\ev_\lm^*\cdot\alv-\inv{M}\ev_\lm^*\cdot\qv\big)\gamma^\mu\gz\big\}
= \tr\big\{\gf\,\alv\cdot\qv\,\ev_\lm^*\cdot\alv\,\gamma^\mu\gz\big\} \nn\crt
&= -4i\,\delta^{\mu,i}\veps_{ijk}q^j{e_\lm^k}^*
\end{align}
This agrees with the kinematic factor in \eq{ff20} for $\Pv=0$. The Orthopositronium is transversely polarized because $\ev_{\lm=0}\parallel\Pv=\qv$ gives $\tr_0^\mu = 0$. The normalization $N_{\lm = \pm 1} =1$ \eq{fock6}. The invariant form factor has the same integral as in \eq{ff18},
\begin{align} \label{ff21}
F(q^2) = \frac{2}{M^2}\int d\xv\,|F(r)|^2 e^{i\qv\cdot\xv/2} = \frac{2}{M}\,\frac{(\alpha M)^4}{(Q^2+\alpha^2 M^2)^2}
\end{align}
where $Q^2 = -q^2$.

I leave it as an exercise (without a worked-out solution) to show that the transition form factor agrees with \eq{ff20} in a general frame.

\subsection{*\,Deep inelastic scattering on Parapositronium in a general frame \label{secV.F}}
%%%%%%%%%%%%%%%%%%%%%%%%%%

The target $A$ vertex of Deep Inelastic Scattering (DIS), $\gamma^*(q) + A(P_A) \to X$, is as in a transition form factor \eq{ff2} for each final state $X$, with the Parapositronium state $A$ defined in \eq{ff3}. Now the photon is taken to have an asymptotically large momentum $q$, and the squared vertex is summed over the final states $X$. In the absence of radiative effects we may describe $X$ in the basis of free $e^-e^+$ states,
\begin{align} \label{dis0}
\ket{X} = b^\dag_{\kv_1,\lm_1}\,d^\dag_{\kv_2,\lm_2}\ket{0}
\end{align}
constrained by momentum conservation, $q+P_A = k_1+k_2$.
In the Bjorken limit
\begin{align} \label{dis1}
\xbj = \frac{Q^2}{2P_A\cdot q}
\end{align} 
is fixed as $q \to \infty$, with $Q^2 = -q^2 > 0$. The frame is defined by keeping the target 4-momentum $P_A^\mu$ fixed in the Bj limit, with the three-momenta of $q^\mu=(q^0,0,0,-|\qv|)$ and $P_A^\mu = (E_A,0,0,P_A)$ along the $z$-axis. The target mass $M=2m+\morder{\alpha^2}$.

The amplitude for $\gamma^* A \to X$ corresponding to \eq{ff4} is, suppressing the momentum conserving $\delta$-function as in \eq{ff2},
\begin{align} \label{dis2}
G^\mu_{AX}(q)=\int d\xv_1 d\xv_2\,\bra{0}d_{\kv_2,\lm_2}b_{\kv_1,\lm_1}\,\bar\psi(0)\gamma^\mu\psi(0)\bar\psi(\xv_1)\Psi_A(\xv_1,\xv_2)\psi(\xv_2)\ket{0}
\end{align}
I consider only scattering from the electron, hence the contractions
\begin{align} \label{dis3}
\acom{b_{\kv_1,\lm_1}}{\bar\psi(0)} &= \bar u(\kv_1,\lm_1) \nn\crt
\acom{d_{\kv_2,\lm_2}}{\psi(\xv_2)} &= e^{-i\kv_2\cdot\xv_2} v(\kv_2,\lm_2)
\end{align}
resulting in
\begin{align} \label{dis4}
G^\mu_{AX}(q)&=\int d\xv_2\,\bar u(\kv_1,\lm_1) \gamma^\mu\gz\Psi_A(0,\xv_2) v(\kv_2,\lm_2) e^{-i\kv_2\cdot\xv_2}
\end{align}
According to \eq{fock2b} the $\la_-$ projector in \eq{ff3} reduces to unity when acting on $e^{-i\kv_2\cdot\xv_2} v(\kv_2,\lm_2)$. After a partial integration of the $\gz\la_+$ projector in the target state \eq{ff3} it acts on $\Psi_A(\xv_1,\xv_2)$ and gives $(\slashed{P}_A+M)$ when the \order{\alpha} contribution from differentiating the radial function is neglected. The radial function $F(r)$ is as in \eq{ff12} evaluated with Lorentz contracted argument. Thus
\begin{align} \label{dis4a}
\gz\Psi_A(0,\xv_2) &= \frac{E_A}{2M^2}(\slashed{P}_A+M) e^{i\Pva\cdot\xv_2/2}\gf\,F(|\xv_{2P}|) \nn\crt
\xv_{2P}^\perp &= \xv_{2}^\perp \hspace{2cm} z_{2P} = z_2\cosh\xi = z_2\, E_A/M
\end{align}

The integral over $\xv_{2}$ in $G^\mu_{AX}(q)$ can be done by a change of integration variable, $\xv_2 \to \xv_{2P}$, giving
\begin{align} \label{dis6}
G^\mu_{AX}(q) &= \frac{\sqrt{\pi}\, \alpha^{5/2} M^2}{8\big(\alpha^2 M^2/16+\Pv_{2}^2\big)^2}\,\bar u(\kv_1,\lm_1) \gamma^\mu (\slashed{P}_A+M) \gf\, v(\kv_2,\lm_2) \nn\crt
\Pv_{2} &\equiv \big(-\kv_2^\perp,(\halft P_A^z-k_2^z)/\cosh\xi\big)
\end{align}
The denominator of $G^\mu_{AX}(q)$ shows that $\Pv_{2}$, \ie, $\kv_2^\perp$ and $\halft P_A^z-k_2^z$ must be of \order{\alpha} for a leading order contribution.

Squaring the form factor and summing over the helicities $\lm_1,\lm_2$,
\begin{align} \label{dis7}
\sum_{\lm_1,\lm_2}G^\mu_{AX}(q){G^\nu_{AX}}^\dag(q) &= \frac{\pi\alpha^5 M^4}{64\big(\alpha^2 M^2/16+\Pv_2^2\big)^4}\,\tr^{\mu\nu} \nn\crt
\tr^{\mu\nu} &= \tr\big\{\gamma^\mu(\slashed{P}_A+M)(\slashed{k}_2+m)(\slashed{P}_A+M)\gamma^\nu(\slashed{k}_1+m)\big\} \nn\crt
&=2M^2\tr\big\{\gamma^\mu(\slashed{P}_A+M)\gamma^\nu(\slashed{k}_1+m)\big\}
=8M^2\big[P_A^\mu k_1^\nu+P_A^\nu k_1^\mu-g^{\mu\nu}(P_A\cdot k_1-\halft M^2)\big\}
\end{align}
where I used $k_2 = \halft P_A +\morder{\alpha}$.
We may check gauge invariance at leading order,
\begin{align} \label{dis9}
q_\mu\tr^{\mu\nu} &= 2M^2\tr\big\{(\slashed{k}_1+\slashed{k}_2-\slashed{P}_A)(\slashed{P}_A+M)\gamma^\nu(\slashed{k_1}+m)\big\} = 2M^2\tr\big\{(m-\halft\slashed{P}_A)(\slashed{P}_A+M)\gamma^\nu(\slashed{k_1}+m)\big\}=0
\end{align}

In the notation \cite{Peskin:1995ev} % p. 624-626,
\begin{align} \label{dis10}
\textrm{Im}\,W^{\mu\nu}(P_a,q) &= \inv{2} \int \frac{d\kv_1d\kv_2}{(2\pi)^6 4E_1E_2}\,(2\pi)^4 \delta^4(q+P_a-k_1-k_2) G^\mu(q){G^\nu}^\dag(q) \nn\crt
W^{\mu\nu} &= \Big(-g^{\mu\nu}+\frac{q^\mu q^\nu}{q^2}\Big)W_1 + \Big(P_a^\mu-q^\mu\frac{P_a\cdot q}{q^2}\Big)\Big(P_a^\nu-q^\nu\frac{P_a\cdot q}{q^2}\Big)W_2
\end{align}
Identifying $W_1$ from the coefficient of $-g^{\mu\nu}$ in \eq{dis7} ($P_A\cdot k_1 \gg \halft M^2$) we get for the electron distribution,
\begin{align} \label{dis11}
f_{e/A}(\xbj) &= \inv{\pi}\textrm{Im}\,W_1 = \int\frac{d\kv_2}{(2\pi)^2 4E_1E_2}\delta(q^0+E_A-E_1-E_2) \frac{\alpha^5 M^6\,P_A\cdot k_1}{16\big(\alpha^2 M^2/16+\Pv_2^2\big)^4}
\end{align}

As $q = (q^0,0,0,-|\qv|) \to \infty$ at fixed $\xbj$ and $P_A$,
\begin{align} \label{dis11a}
E_1 = \sqrt{(\qv+\Pv_A-\kv_2)^2+m^2} \simeq \sqrt{\qv^2-2|\qv|(P_A^z-k_2^z)} \simeq |\qv| - (P_A-k_2)^z
\end{align}
Defining the light-front notation by $q^\pm \equiv q^0 \pm q^z$ we have,
\begin{align} \label{dis11b}
Q^2 = -q^+q^- = 2\xbj P_A\cdot q \simeq \xbj P_A^+q^- \ \ \ \ \Longrightarrow \ \ q^+ = -\xbj P_A^+
\end{align}
The energy constraint in \eq{dis11} becomes
\begin{align} \label{dis12}
q^0+E_A-E_1-E_2 &= q^0-|\qv|+(P_A-k_2)^z + E_A-E_2 = q^+ +P_A^+-k_2^+ =  P_A^+(1-\xbj)-k_2^+ = 0
\end{align}
Recalling that $k_2 = \halft P_A$ at \order{\alpha^0} I denote the \order{\alpha} difference $\xbj-\halft$ as
\begin{align} \label{dis13}
\xbj = \halft(1 +\alpha\,\tilde x_{B})  \hspace{2cm} k_2^+ = \halft P_A^+(1-\alpha\tilde x_{B}) = m\,e^\xi(1-\alpha\tilde x_{B})
\end{align}
Neglecting terms of \order{\alpha^2} we have in $\Pv_2$ \eq{dis6},
\begin{align} \label{dis14}
k_2^z-\halft P_a^z = \halft(k_2^+ - k_2^-)-\halft P_a^z = \halft me^\xi(1-\alpha\tilde x_{B})-\frac{m^2 e^{-\xi}}{2m(1-\alpha\tilde x_{B})}-m\sinh\xi = -m \alpha\tilde x_{B}\cosh\xi
\end{align}

With the change of variables $dk_2^z/E_2 = dk_2^+/k_2^+$ the $\delta$-function in \eq{dis11} may be integrated using \eq{dis12}. Substituting $P_A\cdot k_1/E_1k_2^+ \simeq 2$ and the result \eq{dis14} we have the frame independent result
\begin{align} \label{dis15}
f_{e/A}(\xbj) &= \frac{\alpha^5 M^6}{32}\int\frac{d\kv_2^\perp}{(2\pi)^2}\inv{\big({\kv_2^\perp}^2+\alpha^2 M^2/16+\alpha^2 M^2 {\tilde x_{B}}^2/4\big]^4}
= \inv{6\pi\,\alpha}\,\inv{({\tilde x_{B}}^2+\quart)^3} \hspace{2cm} \tilde x_{B} = \frac{2}{\alpha}(\xbj-\halft)
\end{align}
We may check that there is a single $e^-$ in the bound state,
\begin{align} \label{dis16}
\int_0^1 d\xbj\, f_{e/A}(\xbj) = \halft\alpha\int_{-\infty}^\infty \frac{d\tilde x_{B}}{6\pi\alpha({\tilde x_{B}}^2+\quart)^3} =1
\end{align}

%%%%%%%%%%%%%%
\section{QED in $D=1+1$ dimensions \label{secVI}}
%%%%%%%%%%%%%%

In this section I apply the perturbative bound state method described in chapter \ref{secIV} to QED in $D=1+1$ dimensions (QED$_2$), also known as the ``Massive Schwinger model'' \cite{Schwinger:1962tp,Coleman:1975pw,Coleman:1976uz}. QED (and QCD) in two dimensions is often considered as a model for confinement, since the Coulomb potential is linear. The coupling $e$ has dimension of mass, so the dimensionless parameter relevant for dynamics is its ratio $e/m$ to the electron mass. For $e/m \ll 1$ the fermions are weakly bound and their wave function satifies the Schr\"odinger equation. For $e/m \gg 1$ the spectrum is that of weakly interacting bosons: The strong coupling locks the fermion degrees of freedom into compact neutral bound states. In the limit of the massless Schwinger model ($e/m \to\infty$) QED$_2$ reduces to a free theory, with only a pointlike, non-interacting massive ($M=e/\sqrt{\pi}$) boson field.

A perturbative approach requires the coupling to be small, $e/m < 1$. Highly excited states are nevertheless strongly bound due to the linear potential. Several features are similar to those of the Dirac equation in a linear potential discussed in section \ref{secIII.F}. There are also important differences, first of all because translation invariance allows to define the bound state momentum. The peculiar feature of a constant (local) norm for wave functions at large separations of the charges occurs here as well, but now it does not imply a continuous spectrum. Highly excited states have features of duality similar to those observed for hadrons. 
 
In section \ref{secV} we saw that transverse photon exchange contributes to the binding of Positronium atoms even at leading order for non-vanishing atomic momentum $P$. In $D=1+1$ there are no transverse photons. Boost covariance is realized differently, and requires a linear potential. I shall verifty that form factors and deep inelastic scattering are frame independent.

%%%%%%%%%%%%%%
\subsection{QED$_2$ bound states in $A^0=0$ gauge \label{secVI.A}}

%%%%%%%%%%%%%%
\subsubsection{Temporal gauge in $D=1+1$ \label{secVI.A1}}

Quantization in temporal gauge proceeds as in section \ref{secIV.B3}, adapted to $D=1+1$. The QED$_2$ action is
\begin{align} \label{2d1}
\mS = \int d^2x \big[-\halft F_{10}F^{10} + \bar\psi(\slashed{\partial}-m-e\Asl)\psi\big]
\end{align}
The electric field $E^1 = F^{10} = -\partial_0A^1$ is conjugate to the photon field $A_1$, hence
\begin{align} \label{2d2}
\com{E^1(t,x)}{A^1(t,y)} = i\delta(x-y)
\end{align}
The Hamiltonian is
\begin{align} \label{2d3}
\mH &= \int dx \big[E^1 \partial_0 A_1 + i\psi^\dag\partial_0\psi-\mL\big]
= \int dx \big[\halft (E^1)^2 +\psi^\dag(-i\alpha^1\partial_1+m\gz-e\alpha^1A^1)\psi\big] \equiv \mH_V+\mH_0+\mH_{int} \crt
\mH_V(t) &= \int dx\, \halft \big[E^1(t,x)\big]^2 \hspace{1cm} \mH_0(t) = \int dx \big[\bar\psi(-i\alpha^1\partial_1+m\gz\psi) \big] \hspace{1cm} \mH_{int}(t) = -e\int dx \big[\psi^\dag\,\alpha^1 A^1(t,x)\psi \big] \nn
\end{align}
Gauss' operator is
\begin{align} \label{2d4}
G(t,x) \equiv \frac{\delta\mS}{\delta{A^0(t,x)}} = \partial_1 E^{1}(t,x)-e\psi^\dag\psi(t,x)
\end{align}
$G(t,x)=0$ (Gauss' law) is imposed as a constraint on physical states, fixing the remaining gauge degrees of freedom and defining the value of $E^1$ for those states, 
\begin{align} \label{2d5}
G(t,x)\ket{phys} &= \big[\partial_1 E^{1}(x)-e\psi^\dag\psi(x)\big]\ket{phys} = 0 
\end{align}
Solving for $E^1$, using $\partial_x^2 |x-y| = 2\delta(x-y)$,
\begin{align} \label{2d6}
E^1(t,x)\ket{phys} &= \partial_x \int dy\, \halft e|x-y|\psi^\dag\psi(t,y)\ket{phys}
\end{align}
The vacuum $\ket{0}$ is a physical state with locally vanishing charge distribution,
\begin{align} \label{2d7}
E^1(t,x)\ket{0} = 0
\end{align}
The $\mH_V$ part of the Hamiltonian \eq{2d3} generates an instantaneous linear potential,
\begin{align} \label{2d8}
\mH_V\ket{phys} &= \inv{2}\int dx\, \big[E^1(x)\big]^2\ket{phys} 
= \frac{e^2}{8}\int dx dy dz\big[\partial_x |x-y|\psi^\dag\psi(y)\big]
\big[\partial_x |x-z|\psi^\dag\psi(z)\big]\ket{phys} \nn\crt
&= -\frac{e^2}{4}\int dx dy\, \psi^\dag\psi(x)|x-y|\psi^\dag\psi(y)\ket{phys}
\end{align}
 
%%%%%%%%%%%%%%
\subsubsection{States and wave functions of QED$_2$ \label{secVI.A2}}

An $e^-e^+$ valence Fock state with CM momentum $P$ is defined analogously to Positronium \eq{fock1},
\begin{align} \label{2d9}
\ket{M,P} = \int dx_1dx_2\,\bar\psi(x_1) e^{iP(x_1+x_2)/2}\Phip(x_1-x_2)\psi(x_2)\ket{0}
\end{align}
When bound by $\mH_V$ \eq{2d8} this is taken as the lowest order contribution of a bound state expansion, where higher orders are perturbatively generated by $\mH_{int}$. Hence the lowest order wave functions $\Phip(x_1-x_2)$ and energy eigenvalues $E(P)$ are determined by the eigenstate condition
\begin{align} \label{2d10}
(\mH_0+\mH_V)\ket{M,P} = E(P)\ket{M,P}
\end{align}
Each order of the perturbative expansion should be Poincar\'e covariant, which implies $E(P) = \sqrt{P^2+M^2}$. This will be seen to be satisfied, and the $P$-dependence of the wave function $\Phip(x)$ determined. I do not here consider the higher order corrections defined by $\mH_{int}$ \eq{2d3}.

At large values of the linear potential generated by $\mH_V$ the state $\ket{M,P}$ has contributions from virtual $e^\pm$ pairs, as in \eq{cndef1} and \eq{vac2} for the Dirac case. In terms of Feynman diagrams these effects are due to $Z$-diagrams (\fig{f4}(b)), and they give rise to negative energy components of the wave function. The virtual pairs are implicitly included by omitting from \eq{2d9} the energy projectors $\la_\pm$ \eq{fock2} used for Positronium in \eq{fock1}.

Applying the free Hamiltonian to the state \eq{2d9},
\begin{align} \label{2d11a}
\mH_0\ket{M,P} = \int dx_1dx_2\big[&\bar\psi(x_1)(-i\alpha^1\lder_1+m\gz)e^{iP(x_1+x_2)/2}\Phip(x_1-x_2)\psi(x_2) \nn\crt
&-\bar\psi(x_1)e^{iP(x_1+x_2)/2}\Phip(x_1-x_2)(-i\alpha^1\rder_2+m\gz)\psi(x_2)\big]\ket{0}
\end{align}
Partially integrating the derivatives, so that they act on the wave function instead of on the electron fields,
\begin{align} \label{2d12}
\mH_0\ket{M,P} = \int dx_1dx_2\big[&\bar\psi(x_1)e^{iP(x_1+x_2)/2}(i\alpha^1\rder_1-\halft\alpha^1 P+m\gz)\Phip(x_1-x_2)\psi(x_2) \nn\crt
&+\bar\psi(x_1)\Phip(x_1-x_2)(-i\alpha^1\lder_2+\halft\alpha^1 P-m\gz)e^{iP(x_1+x_2)/2}\psi(x_2)\big]\ket{0}
\end{align}
The instantaneous potential generated by $\mH_V$ \eq{2d8} is seen from
\begin{align} \label{2d13}
\mH_V\ket{M,P} = \int dx_1dx_2\,\bar\psi(x_1) e^{iP(x_1+x_2)/2}\halft e^2|x_1-x_2|\Phip(x_1-x_2)\psi(x_2)\ket{0}
\end{align}

In $D=1+1$ the Dirac matrices can be represented by the $2\times 2$ Pauli matrices. I shall use
\begin{align} \label{2d11}
\gz = \sigma_3 \hspace{2cm} \gamma^1 = i\sigma_2 \hspace{2cm} \alpha^1 = \alpha_1 = \gz\gamma^1 = \sigma_1
\end{align}
With this notation the bound state condition \eq{2d10} implies for the wave function
\begin{align} \label{2d14}
i\partial_x \acomb{\sigma_1}{\Phip(x)} - \halft P\comb{\sigma_1}{\Phip(x)} + m\comb{\sigma_3}{\Phip(x)} &= \big[E-V(x)\big]\Phip(x) \nn\crt
V(x) = \halft e^2\,|x| \equiv V'|x| = V'x\ \ \ (x \geq 0)
\end{align}
In the following I assume that $x \geq 0$. The wave function for $x<0$ is then determined by its parity $\eta_P$ as in \eq{A11},
\beq\label{2d14a}
\sigma_3\,\Phip(-x)\sigma_3 = \eta_P \Phi^{(-P)}(x)   \hspace{2cm} (\eta_P = \pm 1)
\eeq 
It can be shown (see Exercise \ref{e21} for the derivation in $D=3+1$) that \eq{2d14} is equivalent to the two coupled equations
\begin{align} \label{eVI7}
\Big[\frac{2}{E-V}\big(i\sigma_1\rder_x+m\sigma_3-\halft\sigma_1 P\big)-1\Big]\Phip &= -\frac{2i}{(E-V)^2}P\partial_x\Phip+\frac{iV'}{(E-V)^2}\comb{\sigma_1}{\Phip}  \nn \crt
\Phip\Big[\big(i\sigma_1\lder_x-m\sigma_3+\halft\sigma_1 P\big)\frac{2}{E-V}-1\Big] &= \frac{2i}{(E-V)^2}P\partial_x\Phip-\frac{iV'}{(E-V)^2}\comb{\sigma_1}{\Phip} 
\end{align}

The $2\times 2$ wave function may be expanded in Pauli matrices,
\begin{align} \label{2d15}
\Phip(x) = \phip_0(x)\,I + \phip_1(x)\,\sigma_1 + \phip_2(x)\,i\sigma_2 + \phip_3(x)\,\sigma_3
\end{align}
where $I$ stands for the unit $2\times 2$ matrix. The coefficients of the Pauli matrices in the bound state equation \eq{2d14} give four conditions,
\begin{align} \label{2d16}
I&: \hspace{1cm} 2i\partial_x\phip_1(x) = (E-V)\phip_0(x) \nn\crt
\sigma_1&: \hspace{1cm} 2i\partial_x\phip_0(x) +2m\phip_2(x) = (E-V)\phip_1(x) \nn\crt
i\sigma_2&: \hspace{1cm} P\phip_3(x) +2m\phip_1(x) = (E-V)\phip_2(x) \nn\crt
\sigma_3&: \hspace{1cm} P\phip_2(x) = (E-V)\phip_3(x)
\end{align}

%%%%%%%%%%%%%%
\subsubsection{Rest frame and non-relativistic limit \label{secVI.A3}}

Consider first the rest frame, $P=0,\ E=M$. The conditions \eq{2d16} give
\begin{align} \label{2d17}
\phir_0(x) = \frac{2i}{M-V}\,\partial_x \phir_1(x)&  \hspace{2cm} 
\phir_2(x) = \frac{2m}{M-V}\,\phir_1(x)  \hspace{2cm} 
\phir_3(x) = 0 \nn\crt
\partial_x^2\phir_1(x) &+ \frac{V'}{M-V}\,\partial_x\phir_1(x) + \big[\quart(M-V)^2-m^2\big]\phir_1(x) = 0
\end{align} %PH 8.5.21: Added \partial_x in 2nd term
In the NR limit, with $e \ll m$ and $V(x) \ll m$, the equation for $\phir_1(x)$ reduces to the Schr\"odinger equation with binding energy $E_b = M-2m$,
\begin{align} \label{2d18}
\Big[-\inv{m}\partial_x^2+ V(x)\Big]\phir_1(x) = E_b\phir_1(x)
\end{align}
The normalizable solution is given by the Airy function,
\begin{align} \label{2d19}
\phir_1(x) = N \mbox{Ai}\big[m(V-E_b)/(mV')^{2/3}\big]\hspace{2cm} (x \geq 0)
\end{align}
The coefficient $N$ may be chosen to be real, with size fixed by the normalization \eq{fock3} of the state. The energy eigenvalues are determined by continuity at $x=0$. From \eq{2d14a} and \eq{2d15} follows $\phir_1(-x) = -\eta_P\phir_1(x)$, so that
\begin{align} \label{2d20}
\phir_1(x=0) = 0\ \ \ (\eta_P=+1) \hspace{2cm} \partial_x\phir_1(x=0) = 0\ \ \ (\eta_P=-1) 
\end{align}

The relations \eq{2d17} reduce in the NR limit to $\phir_2(x)=\phir_1(x)$, $\phir_0(x)=\phir_3(x) =0$. Hence the $2\times 2$ wave function has the structure
\begin{align} \label{2d21}
\wfr_{NR}(x) = (\sigma_1+i\sigma_2)\phir_1(x)
\end{align}
The projectors \eq{fock2} in the NR limit are $\la_\pm = \halft(1\pm\sigma_3)$. The wave function satisfies $\la_+\wfr_{NR}(x)=\wfr_{NR}(x)\la_- = \wfr_{NR}(x)$, showing that it has no negative energy components. Hence there are no virtual $e^-e^+$ pairs in the NR bound state \eq{2d9}.

%%%%%%%%%%%%%%
\subsubsection{Solution for any $M$ and $P$  \label{secVI.A4}}

Consider now the bound state conditions \eq{2d16} for arbitrary momenta $P$, without assuming $V \ll E$.
The last two relations allow to express $\phip_2(x)$ and $\phip_3(x)$ in terms of $\phip_1(x)$,
\begin{align} \label{2d22}
\phip_2(x) = \frac{E-V}{(E-V)^2-P^2}\,2m \phip_1(x) \hspace{2cm}
\phip_3(x) = \frac{P}{(E-V)^2-P^2}\,2m \phip_1(x)
\end{align}
The denominators are the square of the kinetic 2-momentum $\Pi(x) \equiv (E-V,P)$. This motivates changing the variables $x$ into the ``Lorentz invariant'' $\tau(x)$, defined as
\begin{align} \label{2d23}
\tau(x) \equiv \big[(E-V)^2-P^2\big]/V' \hspace{2cm} \partial_x = -2(E-V)\partial_\tau
\end{align}
The relation between $\partial_x$ and $\partial_\tau$ is crucial in the following, and is valid only for a linear potential, $V(x) = V'x$ ($x\geq 0$). When the equations \eq{2d16} for $\phip_0(x)$ and $\phip_1(x)$ are expressed in terms of $\tau$ rather than $x$ they turn out to be frame independent, \ie, no factors of $E$ or $P$ appear \cite{Hoyer:1986ei}. With the shorthand notation $\phi_{0,1}(\tau) \equiv \phip_{0,1}\big[x(\tau)\big]$,
\begin{align} \label{2d24}
\partial_\tau \phi_1(\tau) = \frac{i}{4}\,\phi_0(\tau) \hspace{2cm}
\partial_\tau \phi_0(\tau) = \frac{i}{4}\Big(1-\frac{4m^2}{V'\tau}\Big)\phi_1(\tau)
\end{align}
The superscript $(P)$ on $\phi_{0,1}(\tau)$ is omitted since as functions of $\tau$ they are the same in all frames. The $P$-dependence of $\phip_{0,1}\big[x(\tau)\big]$ as functions of $x$ arises only from the mapping $x(\tau)$ defined by \eq{2d23}. The equivalence of this with the $P$-dependence induced by actually boosting the state was verified in \cite{Dietrich:2012iy} (in $A^1=0$ gauge).

The parity constraint \eq{2d14a} on $\Phip(x)$ implies, in view of the expansion \eq{2d15}, the relations
\begin{align} \label{2d27}
\phip_{0,3}(x) = \eta_P\phi^{\scriptscriptstyle{(-P)}}_{0,3}(-x) \hspace{2cm} 
\phip_{1,2}(x) = -\eta_P\phi^{\scriptscriptstyle{(-P)}}_{1,2}(-x)
\end{align}
Consider first $\phi_0$ and $\phi_1$, which for $x\geq 0$ are functions only of $\tau$ in \eq{2d24}. Since $\tau(x)$ is invariant under $P \to -P$ we need not be concerned with the sign change of $P$ under parity. Continuity at $x=0$ requires for $\eta_P=+1$ that $\phi_1\big[\tau(x=0)\big] = 0$ and for $\eta_P=-1$ that $\phi_0\big[\tau(x=0)\big] = 0$. The relations \eq{2d24} ensure that $\partial_\tau\phi_1(\tau)=0$ when $\phi_0(\tau)=0$ and \textit{vice versa}, as required by the opposite parities of $\phi_0$ and $\phi_1$. 

For $P=0$ \eq{2d23} gives $V'\tau(x=0) = M^2$. Hence the condition $\phi_1(\tau=M^2/V') = 0$ determines the masses $M$ of the bound states with $\eta_P=+1$. Similarly, the zeros of $\phi_0(\tau)$ determine the $\eta_P=-1$ masses. When $P \neq 0$ we have $V'\tau(x=0) = E^2-P^2$, whereas the zeros of the functions $\phi_{0,1}(\tau)$ are independent of $P$. Satisfying the parity constraint for all $P$ then requires the energies $E$ to satisfy $E^2-P^2 = M^2$, as expected from Lorentz covariance. This allows to express $\tau(x)$ in \eq{2d23} as $\tau(x) = \big[M^2-2EV+V^2\big]/V'$.

The function $\phip_2(x)$ given by \eq{2d22} has the same $x \to -x$ symmetry as $\phip_1(x)$ as required by \eq{2d27}. $\phip_3(x)$ is likewise related to $\phip_1(x)$ by a coefficient which is symmetric under $x \to -x$, but antisymmetric under $P \to -P$. Hence $\phip_3(x)$ has opposite parity constraint compared to $\phip_1(x)$, which is again consistent with \eq{2d27}.

Defining the $x$-dependent ``boost parameter'' $\zeta(x)$ by
\begin{align} \label{2d24a}
\cosh\zeta = \frac{E-V}{\sqrt{V'\tau}} \hspace{2cm} \sinh\zeta =\frac{P}{\sqrt{V'\tau}}
\end{align}
the full wave function \eq{2d15} may be expressed using \eq{2d22} as
\begin{align} \label{2d24b}
\Phip = \phi_0 + \phi_1\Big[\sigma_1+ \frac{2m(E-V)}{V'\tau}\,i\sigma_2 + \frac{2mP}{V'\tau}\,\sigma_3\Big]
= e^{-\sigma_1\zeta/2}\Big(\phi_0 + \phi_1\sigma_1+\frac{2m}{\sqrt{V'\tau}}\phi_1\,i\sigma_2\Big)e^{\sigma_1\zeta/2}
\end{align}
In the latter expression the term in (\ ) depends on $\tau$ only, whereas $\zeta$ depends also explicitly on $P$. In the weak coupling limit $(V \ll m)$ $\zeta(x)$ reduces to the standard boost parameter $\xi$,
\begin{align} \label{2d24c}
\cosh\xi = \frac{E}{M} \hspace{2cm} \sinh\xi =\frac{P}{M}
\end{align}

The expression \eq{2d24b} allows to determine the frame dependence of $\Phip(x)$ at constant $x$,
\begin{align} \label{2d24d}
\left.\frac{\partial\Phi^{(P)}(x)}{\partial\xi}\right|_x = \left.\frac{xP}{E-V}\partial_x\Phi^{(P)}\right|_\xi-\frac{E}{2(E-V)}\,\comb{\sigma_1}{\Phi^{(P)}}
\end{align}
where the $x$-derivative on the rhs. is taken at constant $P$.

\vspace{3mm}
%%%%%%%%%%%%%%%%%%%%%%%%%%
\begin{tcolorbox}
\textit{Exercise \ref{e15}:} Derive the expression \eq{2d24d}. \\ %D5382
\textit{Hint:} You may start from $e^{\sigma_1\zeta/2}\Phip e^{-\sigma_1\zeta/2}$, which is a function only of $\tau$.
\end{tcolorbox}
%%%%%%%%%%%%%%%%%%%%%%%%%%

Eliminating $\phi_0(\tau)$ in \eq{2d24} gives
\begin{align} \label{2d25}
\partial_\tau^2 \phi_1(\tau) + \frac{1}{16}\Big(1-\frac{4m^2}{V'\tau}\Big)\phi_1(\tau) = 0
\end{align}
The regular, arbitrarily normalized analytic solutions for $\phi_{0,1}(\tau)$ are\footnote{The factor $\sqrt{V'}$ in the wave function gives it the correct dimension, corresponding to a relativistically normalized state in $D=1+1$.  Analytic solutions are given in \cite{Dietrich:2012un} also for bound states of fermions with unequal masses.} \cite{Dietrich:2012un}
\begin{align} \label{2d26}
\phi_1(\tau) &= \sqrt{V'}\,\tau\,\exp(-i\tau/4)\kum(1-im^2/2V',2,i\tau/2) = \phi_1^*(\tau) \nn\crt
\phi_0(\tau) &= -\phi_1(\tau) -4i\sqrt{V'}\,\exp(-i\tau/4)\kum(1-im^2/2V',1,i\tau/2) = -\phi_0^*(\tau)
\end{align}

%%%%%%%%%%%%%%
\subsubsection{Weak coupling limit  \label{secVI.A5}}

The Positronium states of QED$_4$ were in \eq{fock1} defined with the $\la_\pm$ projectors \eq{fock2}, and the $\xv$-dependence of the $\ket{e^-e^+}$ Fock state wave function $\Phip(\xv)$ \eq{fock5c} was found to Lorentz contract \eq{tra13}. Here I verify that the QED$_2$ wave functions have analogous properties in the weak coupling limit, although there is no transverse photon contribution and the states \eq{2d9} are defined without projecting on the lowest Fock state.

According to \eq{2d16} the Dirac structure \eq{2d15} of the QED$_2$ wave function reduces for $V \ll m$ and $M \simeq 2m$ to
\begin{align} \label{2d28}
\Phip_{NR}(x) = \Big(\sigma_1 + \frac{E}{M}\,i\sigma_2 + \frac{P}{M}\,\sigma_3\Big)\phip_{1,NR}(x)
\end{align}
and the variable $\tau$ of \eq{2d23} simplifies to
\begin{align} \label{2d29}
V'\tau_{NR}(x) = M^2-2EV(x) = M(M-2V'x\cosh\xi) \hspace{2cm} \cosh\xi = \frac{E}{M}
\end{align} 
The dependence on $x\cosh\xi$ means that $\phi_1\big[\tau_{NR}(x)\big]$ Lorentz contracts similarly as $\Fp(\xv)$ in \eq{tra13}.
The leading order expression \eq{ex13.1} of the projectors $\la_\pm$ is in $D=1+1$
\begin{align} \label{2d30}
\la_\pm(P) = \inv{2E}(E\mp \sigma_1 P \pm M\sigma_3)
\end{align}
The Dirac structure of the QED$_2$ wave function is $\sigma_1+i\sigma_2$ for $P=0$ \eq{2d21}. The analogy with QED$_4$ \eq{tra12a} suggests that $\Phip_{NR}(x)$ should for general $P$ be a $2\times 2$ matrix proportional to
\begin{align} \label{2d31}
\la_+(P)(\sigma_1+i\sigma_2)\la_-(P) = \frac{M(E+M)}{2E^2}\Big(\sigma_1 + \frac{E}{M}\,i\sigma_2 + \frac{P}{M}\,\sigma_3\Big)
\end{align}
which agrees with \eq{2d28}: The properties of the weakly bound states in QED$_2$ and QED$_4$ are analogous at all $P$.

The non-relativistic limit of $\phi_1(\tau)$ may be determined from its analytic expression \eq{2d26}. The scaling of the coordinate $x$ in the limit $m \to \infty$ at fixed $V' = \halft e^2$ is given by the Schr\"odinger equation \eq{2d18} for $P=0$: $\partial_x^2 \propto mV'x$, \ie, $x \propto (mV')^{-1/3}$, and $E_b = M-2m \propto (mV')^{2/3}/m$. In this limit \cite{Dietrich:2012un,Hoyer:2014gna}
\begin{align} \label{2d32}
\phi_{1,NR}^{\scriptscriptstyle{(0)}}(x) = \lim_{m\to\infty}\phi_1(\tau)
= 4\sqrt{V'}\Big(\frac{V'}{m^2}\Big)^{1/3}e^{\pi m^2/2V'}\,\mbox{Ai}\big[m(V-E_b)/(mV')^{2/3}\big]\Big[1+{\cal{O}}\Big(m^2/V'\Big)^{-2/3}\Big]
\end{align}
which relates the normalization of the NR solution \eq{2d19} to that of the general solution \eq{2d26}. 

%%%%%%%%%%%%%%
\subsubsection{Large separations between $e^-$ and $e^+$  \label{secVI.A6}}

The variable $\tau(x)$ \eq{2d23} grows with the separation $x$ of the fermions. For $|\tau|\to\infty$ the wave function $\phi_1(\tau)$ \eq{2d26} oscillates with constant amplitude, up to corrections of \order{1/|\tau|}:
\begin{align} \label{2d33}
\phi_1(|\tau|\to\infty) = \frac{4V'}{\sqrt{\pi}\,m}\sqrt{\exp(\pi m^2/V')-1}\;e^{-\theta(-\tau)\pi m^2/2V'}\cos\Big[\quart\tau-(m^2/2V')\log(\halft|\tau|)+\arg\Gamma(1+im^2/2V')-\pi/2\Big]
\end{align} 
where $\theta(x)= 1\ (0)$ for $x>0\ (x<0)$ is the step function. From \eq{2d24} we have $\phi_0(\tau)= -4i\partial_\tau\phi_1(\tau)$, so that
\begin{align} \label{2d34}
\lim_{|\tau|\to\infty}\big[\phi_1(\tau)+\phi_0(\tau)\big]=\lim_{|\tau|\to\infty}\big[\phi_1(\tau)-\phi_0(\tau)\big]^*
&=N\exp\big[i\tau/4-i(m^2/2V')\log(|\tau|/2) + i\arg\Gamma(1+im^2/2V')-i\pi/2\big] \nn\crt
N&= \frac{4V'}{\sqrt{\pi}\,m}\sqrt{\exp(\pi m^2/V')-1}\;e^{-\theta(-\tau)\pi m^2/2V'} \hspace{2cm} 
\end{align}
has an $x$-independent local norm $N^2$. This $x$-dependence of the wave function is made possible by modes with large negative kinetic energy, which balance the linear potential to give a fixed energy eigenvalue. The asymptotic wave function thus describes virtual $e^-e^+$ pairs, illustrated by time-ordered $Z$-diagrams such as in \fig{f4}\,b. The negative energy components created by the $b\,d$ operators in the state \eq{2d9} dominate for $x \to \infty$ \cite{Dietrich:2012un,Hoyer:2014gna}.
The pairs give rise to a sea distribution for $\xbj \to 0$ in deep inelastic scattering (see section \ref{secVI.C} and \fig{f9}). The Dirac radial functions \eq{lin4} similarly have a constant local norm at large values of $r$. 

%%%%%%%%%%%%%%
\subsubsection{Bound state masses and duality \label{secVI.A7}}

The wave function $\phi_1(\tau)$ in \eq{2d26} was chosen to satisfy $\phi_1(\tau=0)=0$, ensuring that $\phi_2(0)$ and $\phi_3(0)$ \eq{2d22} are finite. The general solution of the differential equation \eq{2d25} has $\phi_1(\tau=0) \neq 0$, giving singular $\phi_{2}$ and $\phi_{3}$. The requirement that the wave function is regular at $\tau=0$ implies a discrete QED$_2$ spectrum. This criterion of local normalizability is the relativistic generalization of the requirement of a finite global norm for Schr\"odinger wave functions. In fact, the non-relativistic limit of the general solution for $\phi_1(\tau)$ (with singular $\phi_{2,3}(\tau=0)$) adds an Airy Bi function to \eq{2d32}, which increases exponentially at large $x$ \cite{Dietrich:2012un}.

The bound state masses $M$ of locally normalizable solutions are determined by the parity constraint \eq{2d27} at $x=0$, which requires $\phi_1(\tau=M^2/V')=0$ for $\eta_P=+1$, and $\partial_\tau\phi_1(\tau=M^2/V')=0$ for $\eta_P=-1$. At high masses $M$ we may use the asymptotic expression \eq{2d33} for $\phi_1(\tau\to \infty)$. The squared eigenvalues $M_n^2$ are then given by integers $n$ and lie on asymptotically linear ``Regge trajectories'', 
\begin{align} \label{2d35}
M_n^2 = n\,2\pi V' + \morder{m^2\log n}\hspace{2cm} \eta_P=(-1)^n
\end{align}
For small electron masses $m$ the trajectory is linear down to low excitations \cite{Dietrich:2012un}. In the range of $x$ where $V(x) \ll M$ the state \eq{2d9} is dominated by positive energy $b^\dag\,d^\dag$ contributions, as would be expected for parton-hadron duality \cite{Dietrich:2012un,Hoyer:2014gna}. The states have an overlap with multiple bound states generated via string breaking as shown in \fig{f11}\,a. I return to these issues in the context of QCD in $D=3+1$ dimensions (section \ref{secVII.D}).

\begin{figure}[h] \centering
\includegraphics[width=1.\columnwidth]{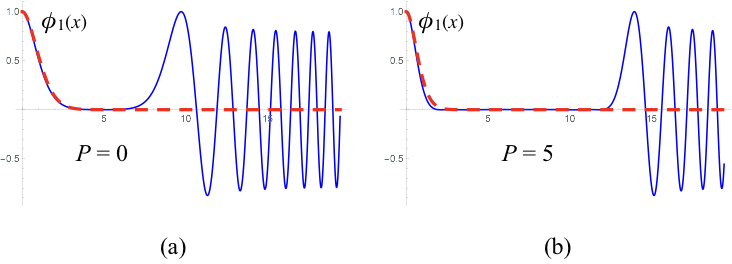}
\caption{The wave function $\phi_1\big[\tau(x)\big]$ of the $\eta_P=-1$ ground state for $m=2\sqrt{V'}$, \ie, $e/m=1/\sqrt{2}$. Blue line: Relativistic expression \eq{2d26}, implying a bound state mass $M=4.86\sqrt{V'}$. Dashed red line: Non-relativistic Airy function \eq{2d32}, requiring $E_b = 0.81\sqrt{V'}$. Both functions are normalized to unity at $x=0$. (a) Rest frame, $P=0$. (b) $P=5\sqrt{V'}$, with the non-relativistic (dashed red) line Lorentz contracted, $x \to xE/M$ as in \eq{2d29}. \label{f6}}
\end{figure}

Consider next a ground state of mass $M$. With increasing $x$ the first virtual $e^-e^+$ pair (string breaking) in the $P=0$ wave function is expected when the potential has reached twice the mass of the bound state, \ie, at $V(x)=2M$. This energetically allows a (virtual) bound state pair to appear (as depicted in \fig{f11}\,b below). I illustrate this with a numerical example in \fig{f6}(a), where $e/m = 0.71$ and $M=4.86\sqrt{V'}$. The dynamics is nearly non-relativistic at low $x$, as evidenced by the agreement of the blue (exact, \eq{2d26}) and red dashed (Schr\"odinger, \eq{2d32}) wave functions. Both are exponentially suppressed with increasing $x$. In the relativistic range, $V(x) \gsim M$, the exact wave function (blue line) begins to increase, reaching a maximum at $V(x)=2M$. The wave function is symmetric around $V'x=M$ because $\tau(x)$ in \eq{2d23} satisfies $\tau(x)=\tau(2M/V'-x)$.\footnote{There is no symmetry for $V'x> 2M$ since the wave function is defined by parity for $x<0$.} For $P = M\sinh\xi > 0$ the virtual pair is expected at $V(x)=2E$, since each bound state has increased energy due to the boost: $M\cosh\xi = E$. This is verified in \fig{f6}(b), and is again due to the symmetry of $\tau(x)$.

%%%%%%%%%%%%%%%%%%%%%%%%%%%%%%%%%%%%%%%%%%%%%%%%%%%%%%%%
%%%%%%%%%%%%%%%%%%%%%%%%%%%%%%%%%%%%%%%%%%%%%%%%%%%%%%%%%

\subsection{*\,Bound state form factors in QED$_2$ \label{secVI.B}} % {\tt (D5026)}
%%%%%%%%%%%%%%%%%%%%%%%%%%

\subsubsection{Form factor definition and symmetry under parity \label{2ffsssec1}}
%%%%%%%%%%%%%%%%%%%%%%%%%%

The electromagnetic form factor is defined as for Positronium in $D=3+1$, see section \ref{secV.E}. The form factor for $\gamma^*+A \to B$ is as in \eq{ff2} and \eq{ff5},
\begin{align} \label{2ff17}
F_{AB}^\mu(q) &= \int d^2z\,e^{-iq\cdot z}\bra{M_B,P_B}\bar\psi(z)\gamma^\mu \psi(z)\ket{M_A,P_A}= (2\pi)^2\delta^2(P_B-P_A-q)G_{AB}^\mu(q) \nn\crt
G_{AB}^\mu(q) &= \int_{-\infty}^\infty dx\,e^{i(P_B-P_A)x/2}\,\tr\big[\Phi_B^\dag(x)\gamma^\mu\gz\Phi_A(x)
 -\Phi_B^\dag(-x)\Phi_A(-x)\gz\gamma^\mu \big]
\end{align} 
where the two contributions to $G_{AB}^\mu(q)$ arise from scattering on $e^-$ and $e^+$, respectively. The wave function $\Phi_A(x) \equiv \Phi_A^{(P_A)}(x)$ determines the state as in \eq{2d9}, and $\gz,\, \gamma^1$ are defined in \eq{2d11}.
The  $e^-$ and $e^+$ contributions can be related using an analogy to charge conjugation in $D=3+1$ \eq{A15}. According to the relations \eq{2d22} the parity relations \eq{2d27} become, when the sign of $P$ is not reversed,
\begin{align} \label{2d27a}
\phip_{0}(x) = \eta_P\phi^{\scriptscriptstyle{(P)}}_{0}(-x) \hspace{2cm} 
\phip_{1,2,3}(x) = -\eta_P\phi^{\scriptscriptstyle{(P)}}_{1,2,3}(-x)
\end{align}
This implies for the wave function expressed as in \eq{2d15}, in analogy to charge conjugation
\begin{align} \label{2d27b}
\sigma_2\big[\Phip(-x)\big]^T\sigma_2 = \eta_P{\Phip}(x) 
\end{align}
Bracketing the trace of the second term in \eq{2ff17} with $\sigma_2$ and transposing it,
\begin{align} \label{2d27c}
-\tr\big[\sigma_2\Phi_B^\dag(-x)\Phi_A(-x)\gz\gamma^\mu \sigma_2\big]^T = -\eta_P^A\eta_P^B\tr\big[\gamma^\mu\gz\Phi_A(x)\Phi_B^\dag(x)\big]
\end{align}
where $\sigma_2(\gz\gamma^\mu)^T\sigma_2 = \gamma^\mu\gz$ for $\mu=0,\,1$. Hence the $e^-$ and $e^+$ contributions are related similarly as in \eq{ff8} and 
\begin{align} \label{2d27d}
G_{AB}^\mu(q) &= (1-\eta_P^A\eta_P^B)\int_{-\infty}^\infty dx\,e^{i(P_B-P_A)x/2}\,\tr\big[\Phi_B^\dag(x)\gamma^\mu\gz\Phi_A(x)\big]
\end{align}
vanishes unless $\eta_P^A=-\eta_P^B$.

\subsubsection{Gauge invariance  \label{2ffsssec2}}
%%%%%%%%%%%%%%%%%%%%%%%%%%

I follow the derivation in \cite{Dietrich:2012un} and consider only scattering from $e^-$, \ie, leave out the factor $1-\eta_P^A\eta_P^B$ of \eq{2d27d}.
Denoting $E_{A,B} = P_{A,B}^{\,0}$ and $P_{A,B} = P_{A,B}^1$, gauge invariance requires that
\begin{align} \label{2ff18}
q_\mu G^\mu_{AB} = (P_B-P_A)_\mu G^\mu_{AB} = \int_{-\infty}^\infty dx\,e^{i(P_B-P_A)x/2}\,\tr\big\{\Phi_B^\dag\big[(E_B-E_A)+ (P_B-P_A)\sigma_1\big]\Phi_A\big\} = 0
\end{align}
The bound state equation \eq{2d14} for $\Phi_A$ and $\Phi_B^\dag$ are, with $V = V'x$ and $x\geq 0$,
\begin{align} \label{2ff19}
(E_A-V)\Phi_A &= i\partial_x\acom{\sigma_1}{\Phi_A}-\halft P_A\com{\sigma_1}{\Phi_A}+m\com{\sigma_3}{\Phi_A} & \Big|\ -\Phi_B^\dag\,\times \nn\crt
\Phi_B^\dag(E_B-V) &= -i\partial_x\acomb{\sigma_1}{\Phi_B^\dag}+\halft P_B\comb{\sigma_1}{\Phi_B^\dag}-m\comb{\sigma_3}{\Phi_B^\dag}  &  \Big|\ \times\,\Phi_A \hspace{.3cm}
\end{align}
Multiplying the equations as indicated in the margin their sum becomes
\begin{align} \label{2ff20}
\Phi_B^\dag(E_B-E_A)\Phi_A =& -i\Phi_B^\dag\partial_x\acom{\sigma_1}{\Phi_A}-i\partial_x\acomb{\sigma_1}{\Phi_B^\dag}\Phi_A+\halft P_B\comb{\sigma_1}{\Phi_B^\dag}\Phi_A+\halft P_A\Phi_B^\dag\com{\sigma_1}{\Phi_A} \nn\crt
&-m\com{\sigma_3}{\Phi_B^\dag}\Phi_A-m\Phi_B^\dag\com{\sigma_3}{\Phi_A}
\end{align}
When the trace is taken the terms with $\partial_x$ form a total derivative,
\begin{align} \label{2ff21}
-i\tr\big\{\Phi_B^\dag(\sigma_1\partial_x\Phi_A+\partial_x\Phi_A\sigma_1) +(\sigma_1\partial_x\Phi_B^\dag+\partial_x\Phi_B^\dag\sigma_1)\Phi_A \big\}
= -i\tr\big\{\partial_x\big(\Phi_B^\dag\sigma_1\Phi_A + \Phi_B^\dag\Phi_A\sigma_1 \big)\big\}
\end{align}
Partially integrating this term in \eq{2ff18} it becomes
\begin{align}\label{2ff22}
-\halft(P_B-P_A)\tr\big\{\Phi_B^\dag\sigma_1\Phi_A + \Phi_B^\dag\Phi_A\sigma_1\big\}
\end{align}
The \order{P} terms in \eq{2ff20} may in the trace be expressed as
\begin{align} \label{2ff23}
\halft P_B\tr\big\{(\sigma_1\Phi_B^\dag-\Phi_B^\dag\sigma_1)\Phi_A\big\}+ \halft P_A\tr\big\{\Phi_B^\dag(\sigma_1\Phi_A-\Phi_A\sigma_1)\big\}
= \halft (P_B-P_A)\tr\big\{\Phi_B^\dag\Phi_A\sigma_1-\Phi_B^\dag\sigma_1\Phi_A\big\}
\end{align}
The \order{m} term vanishes when traced,
\begin{align} \label{2ff24}
-m\tr\big\{(\sigma_3\Phi_B^\dag-\Phi_B^\dag\sigma_3)\Phi_A+ \Phi_B^\dag(\sigma_3\Phi_A-\Phi_A\sigma_3)\big\}=0
\end{align}
The sum of \eq{2ff22} and \eq{2ff23} gives
\begin{align} \label{2ff25}
-(P_B-P_A)\tr\big\{\Phi_B^\dag\sigma_1\Phi_A\big\}
\end{align}
This cancels the second term in \eq{2ff18}, thus ensuring gauge invariance.

\subsubsection{Lorentz covariance} \label{2ffsssec3}
%%%%%%%%%%%%%%%%%%%%%%%%%%

Lorentz covariance requires that the form factor can be written, given the gauge invariance \eq{2ff18}, 
\begin{align} \label{2ff26}
G^\mu_{AB}(q) &= \epsilon^{\mu\nu}(P_B-P_A)_\nu G_{AB}(q^2) & \epsilon^{01}=-\epsilon^{10}=1
\end{align}
With $E_{A,B} = M_{A,B}\cosh\xi$ and $P_{A,B} = M_{A,B}\sinh\xi$ we have
\begin{align} \label{2ff27}
\frac{\delta E_{A,B}}{\delta\xi} = P_{A,B} \hspace{3cm} \frac{\delta P_{A,B}}{\delta\xi} = E_{A,B}
\end{align}
Recalling that $P_0=P^0$ whereas $P_1=-P^1$ we should have
\begin{align} \label{2ff28}
\frac{\delta G_{AB}^0(q)}{\delta\xi} = G_{AB}^1(q) \hspace{3cm} \frac{\delta G_{AB}^1(q)}{\delta\xi} = G_{AB}^0(q)
\end{align}

Subtracting the coupled bound state equations in \eq{eVI7} gives an expression for
\begin{align} \label{2ff29}
\frac{P}{E-V}\partial_x\Phip = -\halft\partial_x\comb{\sigma_1}{\Phip}-\quart iP\acomb{\sigma_1}{\Phip} +\halft im\acomb{\sigma_3}{\Phip} +\frac{V'}{2(E-V)}\comb{\sigma_1}{\Phip}
\end{align}
Using this in \eq{2d24d} gives
\begin{align} \label{2ff30}
\frac{\partial\Phip}{\partial\xi} &=-\halft\partial_x\Big(x\comb{\sigma_1}{\Phip}\Big)+\halft ix\Big(-\halft P\acomb{\sigma_1}{\Phip} +m\acomb{\sigma_3}{\Phip}\Big) \nn\crt
\frac{\partial\Phi^{(P)\dag}}{\partial\xi} &=\halft\partial_x\Big(x\comb{\sigma_1}{\Phi^{(P)\dag}}\Big)+\halft ix\Big(\halft P\acomb{\sigma_1}{\Phi^{(P)\dag}} -m\acomb{\sigma_3}{\Phi^{(P)\dag}}\Big)
\end{align}

According to the expression \eq{2d27d} for $G_{AB}^\mu$ (without the factor $1-\eta_P^A\eta_P^B$),
\begin{align} \label{2ff31}
\frac{\partial G_{AB}^\mu}{\partial\xi} = \int_{-\infty}^\infty dx\,e^{i(P_B-P_A)x/2}\Big(\halft ix(E_B-E_A)\tr\big\{\Phi_B^\dag(x)\gamma^\mu\gz\Phi_A(x)\big\}+\frac{\partial}{\partial\xi}\tr\big\{\Phi_B^\dag(x)\gamma^\mu\gz\Phi_A(x)\big\}\Big)
\end{align}
The second term has the contributions
\begin{align} \label{2ff32}
\tr\Big\{\frac{\partial \Phi_B^\dag}{\partial\xi}\gamma^\mu\gz\Phi_A\Big\}
&=\halft\tr\Big\{\comb{\sigma_1}{\partial_x (x\Phi_B^\dag)}\gamma^\mu\gz\Phi_A+ix\Big(\halft P_B\acomb{\sigma_1}{\Phi_B^\dag}\gamma^\mu\gz\Phi_A-m\acomb{\sigma_3}{\Phi_B^\dag}\Big)\gamma^\mu\gz\Phi_A\Big\} \nn\crt
\tr\Big\{\Phi_B^\dag\gamma^\mu\gz\frac{\partial \Phi_A}{\partial\xi}\Big\}
&=\halft\tr\Big\{-\Phi_B^\dag\gamma^\mu\gz\com{\sigma_1}{\partial_x (x\Phi_A)}+ix\Big(-\halft P_A\Phi_B^\dag\gamma^\mu\gz\acom{\sigma_1}{\Phi_A}+m\Phi_B^\dag\gamma^\mu\gz\acom{\sigma_3}{\Phi_A}\Big)\Big\}
\end{align}

The \order{P} terms may be expressed as
\begin{align} \label{2ff33}
\halft ix\tr\big\{&\halft P_B(\sigma_1\Phi_B^\dag+\Phi_B^\dag\sigma_1)\gamma^\mu\gz\Phi_A - \halft P_A\Phi_B^\dag\gamma^\mu\gz(\sigma_1\Phi_A+\Phi_A\sigma_1)\big\} \nn\crt
&= \halft ix\tr\big\{-\halft P_B\comb{\sigma_1}{\Phi_B^\dag}\gamma^\mu\gz\Phi_A
-\halft P_A\Phi_B^\dag\gamma^\mu\gz\com{\sigma_1}{\Phi_A}\big\}
+\halft ix\tr\big\{(P_B-P_A)\sigma_1\Phi_B^\dag\gamma^\mu\gz\Phi_A\big\}
\end{align}

The \order{m} terms are
\begin{align} \label{2ff34}
\halft ixm\tr\big\{&\Phi_B^\dag\gamma^\mu\gz(\sigma_3\Phi_A+\Phi_A\sigma_3)-(\sigma_3\Phi_B^\dag+\Phi_B^\dag\sigma_3)\gamma^\mu\gz\Phi_A\big\} 
= \halft ixm\tr\big\{ \comb{\sigma_3}{\Phi_B^\dag}\gamma^\mu\gz\Phi_A+ \Phi_B^\dag\gamma^\mu\gz\comb{\sigma_3}{\Phi_A}\big\}
\end{align}

In \eq{2ff31} write $E_B-E_A = (E_B-V)-(E_A-V)$ and add the \order{P} \eq{2ff33} and \order{m} \eq{2ff34}contributions to the respective terms, so as to be able to make use of the bound state equations \eq{2ff19} for $\Phi_A$ and $\Phi_B^\dag$,
\begin{align} \label{2ff35}
\frac{\partial G_{AB}^\mu}{\partial\xi} =& \int_{-\infty}^\infty dx\,e^{i(P_B-P_A)x/2}\Big(\halft\tr\big\{\comb{\sigma_1}{\partial_x (x\Phi_B^\dag)}\gamma^\mu\gz\Phi_A-\Phi_B^\dag\gamma^\mu\gz\com{\sigma_1}{\partial_x (x\Phi_A)}\big\} \nn\crt
&+\halft ix\tr\big\{\big(\Phi_B^\dag(E_B-V)-\halft P_B\comb{\sigma_1}{\Phi_B^\dag}+m\comb{\sigma_3}{\Phi_B^\dag}\big)\gamma^\mu\gz\Phi_A\big\} \nn \crt
&-\halft ix\tr\big\{\Phi_B^\dag\gamma^\mu\gz\big((E_A-V)\Phi_A+\halft P_A\comb{\sigma_1}{\Phi_A}-m\comb{\sigma_3}{\Phi_A}\big)\big\} \nn \crt
&+\halft ix\tr\big\{(P_B-P_A)\sigma_1\Phi_B^\dag\gamma^\mu\gz\Phi_A\big\}\Big)
\end{align}
According to the BSE \eq{2ff19} the expression in (\ ) on the second line equals $-i\partial_x\acomb{\sigma_1}{\Phi_B^\dag}$ and that on the third line equals $i\partial_x\acom{\sigma_1}{\Phi_A}$. Combined with the expression on the first line, noting also the $\partial_x x =1$ contribution we have
\begin{align} \label{2ff36}
\frac{\partial G_{AB}^\mu}{\partial\xi} &= \int_{-\infty}^\infty dx\,e^{i(P_B-P_A)x/2}\Big(\halft\tr\big\{\big(\sigma_1\Phi_B^\dag-\Phi_B^\dag\sigma_1\big)\gamma^\mu\gz\Phi_A-\Phi_B^\dag\gamma^\mu\gz\big(\sigma_1\Phi_A-\Phi_A\sigma_1\big)\big\} \nn\crt
& \hspace{3cm} +x\partial_x\tr\big\{\sigma_1\Phi_B^\dag\gamma^\mu\gz\Phi_A\big\} +\halft ix(P_B-P_A)\tr\big\{\sigma_1\Phi_B^\dag\gamma^\mu\gz\Phi_A\big\}\Big) \nn\crt
&=-\int_{-\infty}^\infty dx\,e^{i(P_B-P_A)x/2}\,\tr\big\{\Phi_B^\dag\sigma_1\gamma^\mu\gz\Phi_A\big\}
\end{align}
I partially integrated the first term on the second line and noted that $\comb{\gamma^\mu\gz}{\sigma_1}=0$. Comparing with the expression \eq{2d27d} for $G_{AB}^\mu$ verifies the Lorentz covariance condition \eq{2ff28}.

%%%%%%%%%%%%%%%%%%%%%%%%%%
\subsection{*\,Deep Inelastic Scattering in D = 1+1} \label{secVI.C} % {\tt [D5105]}
%%%%%%%%%%%%%%%%%%%%%%%%%%

%%%%%%%%%%%%%%%%%%%%%%%%%%
\subsubsection{The Bj limit of the form factor} \label{secVI.C1}

I considered Deep Inelastic Scattering (DIS) $e^-+A \to e^- +X$ on Positronium atoms $A$ of QED$_4$ in section \ref{secV.F}, demonstrating its frame invariance. In that case the final state was taken to be a free $e^-e^+$ pair. In QED$_2$ there are no free electrons, and bound states can have arbitrarily high mass. The target vertex is then described by the form factor $\gamma^* + A \to B$, where $B$ is a bound state whose mass is $\propto Q$ in the Bj limit \eq{dis1}. In $D=1+1$ the mass selects a unique $B$, which represents the inclusive system $X$.

The present approach to DIS in QED$_2$ was previously considered in the Breit frame \cite{Dietrich:2012un}, where $q^0 = E_B-E_A=0$ while $q^1 = -Q \to -\infty$ in the Bj limit. This is a standard frame for QCD$_4$, which allows the target to be described in terms of a parton distribution. It is instructive to repeat the QED$_2$ calculation in a frame where the target momentum is kept fixed in the Bj limit, and to verify that the parton distribution is indeed boost invariant.

The parton distribution was in Eq. (A22) of \cite{Dietrich:2012un} defined in terms of the invariant form factor $G_{AB}(q^2)$ \eq{2ff26} as
\begin{align} \label{2ff37}
f(\xbj)= \inv{16\pi V'm^2}\,\inv{\xbj}|Q^2G_{AB}(q^2)|^2
\end{align} 
I consider the Bj limit where the photon momentum $q^1 \to -\infty$ at fixed $\xbj$, in a frame where the target 2-momentum $P_A$ is fixed. Since $q=P_B-P_A$,
\begin{align} \label{3ff4}
\xbj = \frac{Q^2}{2P_A\cdot q} = -\frac{(P_B-P_A)^2}{2P_A\cdot(P_B-P_A)} \simeq \frac{2P_A\cdot P_B-M_B^2}{2P_A\cdot P_B} = 1-\frac{M_B^2}{2P_A\cdot P_B} \simeq 1-\frac{M_B^2}{2 P_A^+ E_B}
\end{align}
where I used $2P_A\cdot P_B \simeq 2(E_A+P_A^1)E_B \equiv 2 P_A^+ E_B$, neglecting the finite difference between $P_B^0 \equiv E_B$ and $-P_B^1$,
\begin{align} \label{2ff38}
E_B = \sqrt{M_B^2+(P_B^1)^2} \simeq |P_B^1| + \frac{M_B^2}{2|P_B^1|} = -P_B^1+P_A^+(1-\xbj)
\hspace{.5cm} i.e. \hspace{.5cm} P_B^+ = P_A^+(1-\xbj)
\end{align} 

Defining $\gamma^\pm = \gz\pm\go = \sigma_3\pm i\sigma_2$ and with $V'\tau=[(E-V)+P^1][(E-V)-P^1]$ the expression \eq{2d24b} for the bound state wave functions becomes
\begin{align} \label{3ff7}
\Phi = \phi_0 + \phi_1\Big[\sigma_1+ \frac{2m(E-V)}{V'\tau}\,i\sigma_2 + \frac{2mP^1}{V'\tau}\,\sigma_3\Big]
= \phi_0 + \phi_1\Big[\sigma_1+ \frac{m\gamma^+}{E-V-P^1} -\frac{m\gamma^-}{E-V+P^1}\Big]
\end{align}
For $P_B^1 \to -\infty$ this gives using \eq{2ff38}
\begin{align} \label{3ff8}
\Phi_B = \phi_{B0} + \phi_{B1}\Big[\sigma_1- \frac{m\gamma^-}{P_A^+(1-\xbj)-V}\Big]
\end{align}
The invariant form factor $G_{AB}(q^2)$ \eq{2ff26} may be expressed in terms of $G_{AB}^0(q)$ \eq{2d27d}. Using $\gamma^+\gamma^+=0$ and $\gamma^+\gamma^-=2(1-\sigma_1)$ as well as \eq{2d27b} we get for bound states of opposite parities $\eta_A\eta_B=-1$,
\begin{align} \label{3ff9}
G_{AB}(q^2) =& -\inv{q^1}G_{AB}^0(q) \simeq \frac{2}{E_B}\int_{-\infty}^\infty dx\,e^{iq^1x/2}\tr\big[\Phi_B^\dag(x)\Phi_A(x)\big] 
= \frac{4i}{E_B}\int_{0}^\infty dx\,\sin\big(\halft q^1x\big)\tr\big[\Phi_B^\dag(x)\Phi_A(x)\big] \nn\crt
&\halft\tr\big[\Phi_B^\dag(x)\Phi_A(x)\big] = \phi_{B0}^*(\tau_B)\phi_{A0}(\tau_A)+\phi_{B1}^*(\tau_B)\phi_{A1}(\tau_A)\Big[1+ \frac{2m^2}{\big[P_A^+(1-\xbj)-V\big](P_A^+-V)}\Big]
\end{align}

The arguments of the wave functions are, with $V=V'x$:
\begin{align} \label{3ff10}
V'\tau_A &= M_A^2-2E_A V+V^2 \nn\crt
V'\tau_B &= 2E_B\Big(\frac{M_B^2}{2E_B}-V\Big)+V^2 = 2E_B\big[P_A^+(1-\xbj)-V\big]+V^2
\end{align}
At fixed $x$, $\tau_B \to \pm \infty$ for $P_A^+(1-\xbj)
\begin{array}{c}
>\vspace{-2mm}\\<\\ 
\end{array}V$. We may thus use the asymptotic expressions \eq{2d33} and \eq{2d34} for the $\phi_B$ wave functions, see also \eq{3ff11}. For $\eta_B=+1$ and with $n$ an integer,
\begin{align} \label{3ff16}
G_{AB}(q^2) =& (-1)^n\frac{16iV'}{\sqrt{\pi}mE_B}\,\sqrt{e^{\pi m^2/V'}-1}\int_0^\infty dx\, \exp\big[-\theta(-\tau_B)\pi m^2/2V'\big] \nn\crt
&\times\Big\{i\sin\varphi_B(x)\phi_{A0}(\tau_A)+\cos\varphi_B(x)\phi_{A1}(\tau_A)
\Big[1+ \frac{2m^2}{\big[P_A^+(1-\xbj)-V\big](P_A^+-V)}\Big]\Big\} \nn\crt
\varphi_B(x) &\equiv \halft\big[P_A^+(1-\xbj)-P_A^1\big]x + \frac{m^2}{2V'}\log\Big|1-\frac{V'x}{P_A^+(1-\xbj)}\Big|-\quart V'x^2
\end{align}

%%%%%%%%%%%%%%%%%%%%%%%%%%
\begin{tcolorbox}
\textit{Exercise \ref{e16}:} Derive the expression \eq{3ff16}. 
\end{tcolorbox}
%%%%%%%%%%%%%%%%%%%%%%%%%%

So far I arbitrarily normalized the wave functions by adopting the solutions \eq{2d26}. Since $M_B \propto Q$ we need to know the relative normalization of the wave functions $\Phi_B(\tau_B)$ in the Bj limit. This may be determined using duality, as shown in \cite{Dietrich:2012un}. At large $M_B$ and for $V(x) \ll M_B$ the bound state wave functions have the form of free $e^-e^+$ states. The normalization of $\Phi_B(x=0)$ can thus be chosen to agree with that of free $e^-e^+$ (``partonic'') states created by a pointlike current. The result is that the normalization factor is independent of $M_B$ and a function of the electron mass $m$ only, see Eq. (4.16) of \cite{Dietrich:2012un}. Hence the Bj limit and the $\xbj$-dependence of $f(\xbj)$ at fixed $m$ are not affected by this normalization of $\Phi_B$. I do not here consider the normalization of  $\Phi_A$, which affects the magnitude of $f(\xbj)$.

Using $Q^2 = 2\xbj P_A\cdot P_B$ from \eq{3ff4} the electron distribution \eq{2ff37} becomes
\begin{align} \label{3ff18}
f(\xbj) = \frac{(P_A\cdot P_B)^2\xbj}{4\pi V'm^2}|G_{AB}(q^2)|^2
\end{align}
From the boost invariance of $G_{AB}(q^2)$ shown in section \ref{2ffsssec3} follows that also $f(\xbj)$ is invariant. The expression \eq{3ff18} is finite in the Bj limit, given that $2P_A\cdot P_B \simeq P_A^+P_B^- \simeq 2P_A^+E_B$ and that $E_B\,G_{AB}(q^2)$ \eq{3ff16} is finite. I next discuss the numerical evaluation of the $x$-integral in \eq{3ff16}.

%%%%%%%%%%%%%%%%%%%%%%%%%%
\subsubsection{Numerical evaluation of the electron distribution  \label{secVI.C2}} % {\tt [200412\_QED2\_DIS.nb,200416\_QED2\_DIS.nb]}}

The integrand of $G_{AB}(q^2)$ in \eq{3ff16} is regular at $P_A^+-V'x=0$ since $\phi_{A1}(\tau_A=0)=0$. Similarly $\phi_{B1}(\tau_B=0)=0$. In the Bj limit $\tau_B=0$ implies $P_A^+(1-\xbj)-V'x=0$ \eq{3ff10}, \ie, $x=x_0$ with
\begin{align} \label{3ff19}
x_0 = P_A^+(1-\xbj)/V'
\end{align}
The $\tau_B \to \infty$ limit of $\Phi_{B}$ assumed in \eq{3ff16} fails at $x=x_0$. In the asymptotic expression for $\Phi_{B}$ the $1/(x-x_0)$ singularity is regulated by the phase $\varphi_B(x) \propto \log|x-x_0|$ in $\cos[\varphi_B(x)]$. This requires attention in a numerical evaluation of the integral. A Principal Value prescription cannot be used due to the step function $\theta(-\tau_B) \simeq \theta(x-x_0)$. In the following exercise I outline a method for the numerical evaluation of the $x$-integral. 

\begin{figure}[h] 
\includegraphics[width=.5\columnwidth]{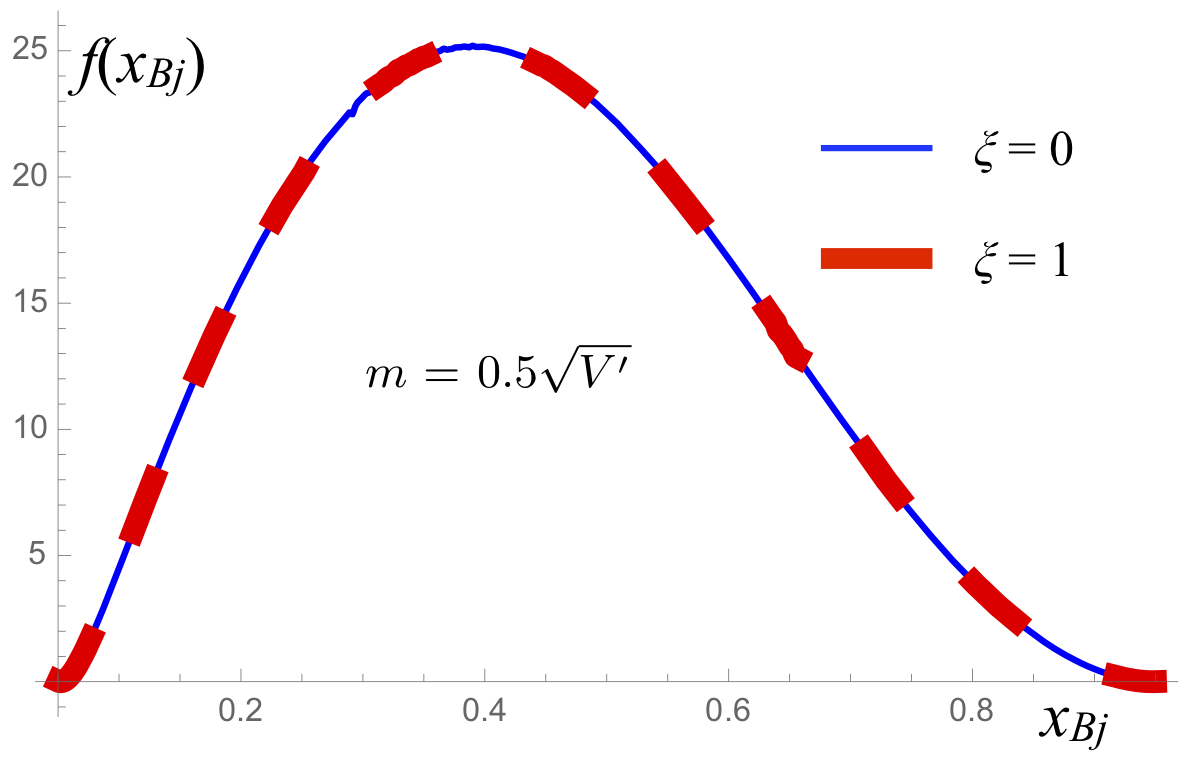}
\caption{QED$_2$ electron distributions \eq{3ff18} for $m=0.5\sqrt{V'}$ and $.05 < \xbj < .95$ of target $A$ ground state $(\eta_A=-1,\ M_A=2.674\sqrt{V'})$ evaluated in the rest frame $(\xi_A=0$, blue) and at $\xi_A=1$ ($E_A=M_A\cosh\xi_A$, thick red dashed line). The two curves agree at \order{10^{-5}}, which indicates the accuracy of the numerical evaluation. The overall normalization is arbitrary. \label{f8}}
\end{figure}

%%%%%%%%%%%%%%%%%%%%%%%%%%
\begin{tcolorbox}
\textit{Exercise \ref{e17}:} Do the $x$-integral in \eq{3ff16} numerically for the parameters in \fig{f8}, and compare the results. 
\end{tcolorbox}
%%%%%%%%%%%%%%%%%%%%%%%%%%

A check of the boost invariance of $f(\xbj)$ \eq{3ff18} is shown in \fig{f8}. The electron distributions obtained in the rest frame and in a frame with boost parameter $\xi =1$ closely agree.

%%%%%%%%%%%%%%%%%%%%%%%%%%%%%%%%%%%%%%%%%%%%%%%%%%%%%%%%
 \parskip.2cm
%%%%%%%%%%%%%%%%%%%%%%%%%%%%%%%%%%%%%%%%%%%%%%%%%%%%%%%%

\begin{figure}[h] 
\includegraphics[width=1.\columnwidth]{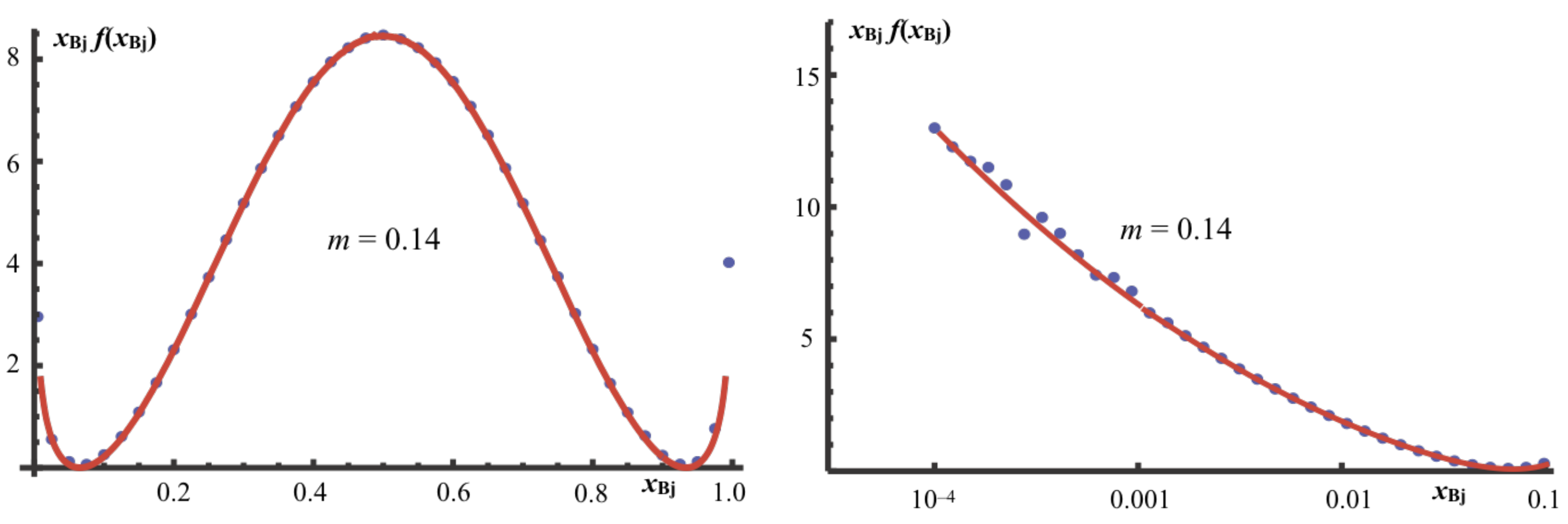}
\caption{QED$_2$ electron distribution \eq{3ff18} for $m=0.14\sqrt{V'}$ ($M_A=2.52\sqrt{V'}$) in the rest frame ($P_A=0$, red curves) compared with the Breit frame result in Fig. 8 of \cite{Dietrich:2012un}, blue dots ($m=0.1\,e = 0.14\,\sqrt{V'}$, see \eq{2d14}). The normalization of the red curve was treated as a free parameter. \label{f9}}
\end{figure}

The electron distribution \eq{3ff18} in the rest frame ($P_A^1=0$) is compared with the one in the Breit frame ($P_A^1=Q/2\xbj \to \infty$ in the Bj limit) in \fig{f9}. The agreement shows the equivalence of the Breit and target rest frames. 

%%%%%%%%%%%%%%%%%%%%%%%%%%%%%%%%%%%%%%%%%%%%%%%%%%%%%%%%
 \parskip.2cm
%%%%%%%%%%%%%%%%%%%%%%%%%%%%%%%%%%%%%%%%%%%%%%%%%%%%%%%%

\subsubsection{$\xbj \to 0$ limit of the electron distribution \label{secVI.C3}}% {\tt [D5122]}} % {\tt 200416\_QED2\_DIS.nb]}}
%%%%%%%%%%%%%%%%%%%%%%%%%%

The electron distribution in \fig{f9} increases for $\xbj \to 0$, analogously to the sea quark distribution for hadrons. In Eq. (6.17) of \cite{Dietrich:2012un} the leading $\xbj$-dependence was found to be (with scale $e^2=2V'$),
\begin{align} \label{3ff20}
\xbj f(\xbj) \sim \cos^2\big[\big(m^2\log\xbj+\halft M_A^2\big)/e^2\big]
\end{align}
States of the form \eq{2d9} appear to have just a single $e^-e^+$ pair, created by the electron fields $\bar\psi$ and $\psi$. However, a strong electric field creates virtual $e^-e^+$ pairs. In time-ordered Feynman diagrams they shows up in ``$Z$''-diagrams like \fig{f4}(b), where an electron scatters into a negative energy state, creating an intermediate state with an additional $e^-e^+$ pair. The mixing of the $b^\dag$ and $d$ operators is explicit for the Dirac states created by the $c_n^\dag$ operator \eq{cndef1}, and the Dirac ground state $\Omega$ \eq{vac2} has an indefinite number of pairs in the free state basis.

The constant norm of the bound state wave functions at large $x$ \eq{2d34} apparently reflects the virtual pairs created by the linear potential $V'x$. The dominant contribution to the form factor \eq{3ff16} for $\xbj\to 0$ comes from the large $x$ part of the integrand, namely from $I_1$ in \eq{3ff24}. More precisely, the leading behavior is due to $I_{1c}$ \eq{3ff25}, for which the angle $\vphi_C$ \eq{3ff26b} depends on $x$ mainly through $\halft P_A^+\xbj\,x$. This allows the integration over $u=x-x_1$ in $I_{1c}$ \eq{3ff28} to contribute over a range $\propto 1/\xbj$. To leading order in the $\xbj \to 0$ limit we may set $x_1=0$.

\begin{figure}[h] \centering
\includegraphics[width=.6\columnwidth]{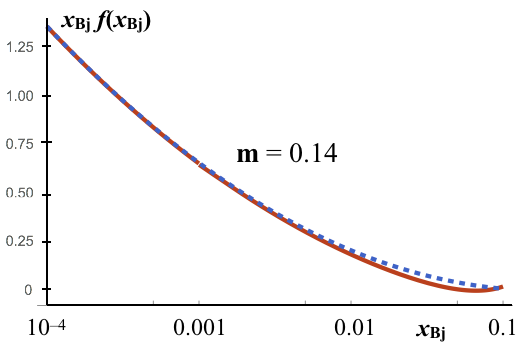}
\caption{Red line: Numerical evaluation of parton distribution \eq{3ff18} using \eq{3ff24} ($P_a=0$ and $m=0.14\,\sqrt{V'}$).\\ Dashed blue line: Analytic approximation for $\xbj \to 0$ given by \eq{3ff34}. \label{f10}}
\end{figure}

The logarithmic terms in $\vphi_C$ are at leading order in the $x\to\infty$ limit,
\begin{align} \label{3ff31}
\log\Big[\frac{x_0\tau_A}{2(x-x_0)}\Big] = \log(\halft P_A^+ x)\,\big[1+\morder{x^{-1}}\big]
\end{align}
Hence
\begin{align} \label{3ff32}
I_{1c} \simeq -\frac{4V'e^{-\pi m^2/4V'}}{\sqrt{\pi}\,m}\sqrt{2\sinh(\pi m^2/2V')}\,\im\int_0^{i\infty}dx\, \exp\Big[\halft iP_A^+ \xbj x+\frac{im^2}{2V'}\log(\halft P_A^+x)-i\frac{M_A^2}{4V'}-i\arg\Gamma\Big(1+\frac{im^2}{2V'}\Big)\Big]
\end{align}
Defining $v=-\halft i P_A^+\xbj x$ we have $dx = 2idv/(P_A^+\xbj)$, $\im \to \re$ and $\log(\halft P_a^+ x) = \log v-\log\xbj+\halft i\pi$. The $v$-integral becomes
\begin{align} \label{3ff33}
\int_0^\infty dv v^{im^2/2V'}e^{-v}= \Gamma(1+im^2/2V') = \frac{\sqrt{\pi}\,m}{\sqrt{2V'\sinh(\pi m^2/2V')}}\,e^{i\arg\Gamma(1+im^2/2V')}
\end{align}
whose modulus and phase cancel the corresponding terms in \eq{3ff32}. This leaves % D5393
\begin{align} \label{3ff34}
I_1 \simeq -\frac{8V'}{\xbj P_a^+}\,e^{-\pi m^2/2V'}\cos\big[(m^2\log\xbj)+\halft M_a^2)/2V'\big] \hspace{2cm} \mbox{for}\ \ \xbj\to 0
\end{align}
Since $I_2$ and $I_3$ give non-leading contributions in the $\xbj\to 0$ limit this defines, via \eq{3ff24}, the frame independent parton distribution \eq{3ff18}. It agrees with the analytic result \eq{3ff20} \cite{Dietrich:2012un}, which was evaluated in the Breit frame. The analytic approximation given by \eq{3ff34} for $f(\xbj\to 0)$ is compared with the numerical evaluation in \fig{f10} for $\xbj \le 0.1$.

%%%%%%%%%%%%%%%%%%%%%%%%%%%%%%%%%%%%%%%%%%%%%%%%%%%%%%%%
 \parskip.2cm
%%%%%%%%%%%%%%%%%%%%%%%%%%%%%%%%%%%%%%%%%%%%%%%%%%%%%%%%

\section{Applications to QCD bound states \label{secVII}}
%%%%%%%%%%%%%%%%%%

\subsection{The instantaneous potential of various Fock states \label{secVII.A}}
%%%%%%%%%%%%%%%%%%

I consider color singlet QCD bound states in temporal gauge $A^0_a = 0$, as described in section \ref{secIV.C}. The scale $\la$ required for confinement is introduced via a boundary condition on the solutions of Gauss' constraint \eq{eII28}, in terms of the homogeneous solution \eq{eII31a}. This affects the longitudinal electric field $\Ev_L^a$ for each color component of a Fock state, whereas the full color singlet state does not generate a color octet field. The gauge invariance condition of electromagnetc form factors is satisfied, and the $q\bar q$ bound state energies have the correct frame dependence.

The longitudinal electric field \eq{eII31} determines the field energy $\mH_V$ \eq{eII32}, which defines an instantaneous potential for each Fock state, 
\begin{align} \label{qcd1}
\mH_V(t=0) &= \int d\yv d\zv\Big[\,\yv\cdot\zv \big(\halft\kappa^2\intt d\xv + g\kappa\big) + \halft \frac{\as}{|\yv-\zv|}\Big]\mE_a(\yv)\mE_a(\zv) \equiv \mH_V^{(0)} + \mH_V^{(1)} \nn\crt
\mE_a(\yv) &= - f_{abc}A_b^i E_c^i(\yv)+\psi_A^\dag T_{AB}^a\psi_B(\yv)
\end{align} 
where a sum over repeated indices is understood. $\mH_V^{(0)}$ is due to the homogeneous solution \eq{eII31a} of Gauss' constraint and generates an \order{\alpha_s^0} potential, while $\mH_V^{(1)}$ gives the standard \order{\as} Coulomb potential. Recall that $\mE_a(\xv)\ket{0}=0$ \eq{eII31a} since Gauss' constraint \eq{eII28} is not an operator condition in temporal gauge. The potentials are independent of the quark Dirac index $\alpha$ and of the gluon Lorentz index $i=1,2,3$. I consider color singlet $q\bar q$ (meson), $qqq$ (baryon), $gg$ (glueball), $q\bar q g$ (higher Fock state of a meson) and $q\bar q\,q\bar q$ (molecular or tetraquark) Fock states.

\subsubsection{The $q\bar q$ potential \label{secVII.A1}}
%%%%%%%%%%%%%%%%%%

A $q\bar q$ Fock state with quarks at $\xv_1,\xv_2$, summed over the colors $A$, is invariant under global color transformations, 
\begin{align} \label{qcd2}
\ket{q\bar q} = \bar\psi_A^\alpha(\xv_1)\psi_A^\beta(\xv_2)\ket{0} \equiv  \bar\psi_A(\xv_1)\psi_A(\xv_2)\ket{0}
\end{align}
I suppress the irrelevant Dirac indices and $t=0$ is understood. The canonical commutation relations \eq{eII24} of the fields in $\mE_a(\xv)$ \eq{qcd1} give
\begin{align} \label{qcd3}
\comb{\mE_a(\xv)}{\bar\psi_A(\xv_1)} &= \bar\psi_{A'}(\xv_1)T_{A'A}^a \delta(\xv-\xv_1) \nn\crt
\comb{\mE_a(\xv)}{\psi_A(\xv_2)} &= -T_{AA'}^a\psi_{A'}(\xv_2) \delta(\xv-\xv_2)
\end{align}
where $T^a$ is the SU(3) generator in the fundamental representation. I shall make use of the following relations for the SU($N_c$) generators\footnote{Useful properties of the SU($N_c$) generators may be found in \cite{Haber:2019sgz}.}
\begin{align} \label{qcd4}
\com{T^a}{T^b} &= if_{abc}T^c \hspace{2cm} \tr\big\{T^aT^b\big\} = \halft\delta^{ab} \hspace{2cm}
T^aT^a = C_F\,I = \frac{N_c^2-1}{2N_c}\,I  \nn\crt
T_{AB}^aT_{CD}^a &= \inv{2}\Big(\delta_{AD}\delta_{BC}-\inv{N_c}\delta_{AB}\delta_{CD}\Big)  \hspace{3.5cm}
T^aT^bT^a = -\inv{2N_c}\,T^b \nn\crt   
f_{abc}f_{abd} &= N_c \delta_{cd} \hspace{6.5cm} f_{abd}T^aT^b = \halft iN_cT^d
\end{align}
The weight $\yv\cdot\zv$ of $\mH_V^{(0)}$ is $\xv_1^2,\,\xv_1\cdot\xv_2$ or $\xv_2^2$ depending on the quark field on which $\mE_a(\yv)$ and $\mE_a(\zv)$ act,
\begin{align} \label{qcd5}
\mH_V^{(0)} \ket{q\bar q} &= \big(\halft\kappa^2\intt d\xv + g\kappa\big)\int d\yv d\zv\;\yv\cdot\zv\,\mE_a(\yv)\mE_a(\zv) \bar\psi_A(\xv_1)\psi_A(\xv_2)\ket{0}  \nn\crt
&= \big(\halft\kappa^2\intt d\xv + g\kappa\big)(\xv_1^2-2\xv_1\cdot\xv_2+\xv_2^2)\bar\psi_{A}(\xv_1)\,T^a_{AB}T^a_{BC}\psi_{C}(\xv_2)\ket{0} \nn\crt
&= \big(\halft\kappa^2\intt d\xv + g\kappa\big)C_F\,(\xv_1-\xv_2)^2\,\ket{q\bar q}
\end{align}
The term $\halft\kappa^2\intt d\xv$ is the contribution of an $\xv$-independent field energy density $E_\la$. Its integral is proportional to the volume of space and irrelevant only if universal, \ie, $E_\la$ must be the same for all Fock states. In particular, $E_\la$ cannot depend on $\xv_1$ or $\xv_2$. This requires to choose the normalization $\kappa$ of the homogeneous solution for this Fock state as
\begin{align} \label{qcd6}
\kappa_{q\bar q} = \frac{\la^2}{gC_F}\frac{1}{|\xv_1-\xv_2|}
\end{align}
which also serves to define the universal constant $\la$. The field energy density is then
\begin{align} \label{qcd7}
E_\la =\frac{\la^4}{2g^2C_F}
\end{align}
This value of $E_\la$ must be imposed on all types of Fock states, \eg, $\ket{qqq}$, and in each case will determine the normalization of the corresponding homogeneous solution. 

Subtracting $E_\la \intt d\xv$ in \eq{qcd5} the remaining $g\kappa$ term gives 
\begin{align} \label{qcd8}
\mH_V^{(0)} \ket{q\bar q} \equiv V_{q\bar q}^{(0)} \ket{q\bar q} \hspace{2cm} V_{q\bar q}^{(0)}(\xv_1,\xv_2) = g\kappa_{q\bar q}\,C_F\,(\xv_1-\xv_2)^2 = \la^2|\xv_1-\xv_2|
\end{align}
The gluon exchange potential due to $\mH_V^{(1)}$ is similarly obtained using \eq{qcd3}. The commutators of $\mE_a(\yv)$ and $\mE_a(\zv)$ with the same quark now gives an infinite, $\sim 1/0$ contribution. This ``self-energy'' is independent of $\xv_1$ and $\xv_2$ and can be subtracted. Altogether,
\begin{align} \label{qcd9}
\mH_V\ket{q\bar q} &=\big[
V_{q\bar q}^{(0)}+ V_{q\bar q}^{(1)}\big]\ket{q\bar q}\hspace{2cm} V_{q\bar q}^{(1)}(\xv_1,\xv_2) =-C_F\frac{\as}{|\xv_1-\xv_2|}
\end{align}
$V_{q\bar q}^{(0)}+ V_{q\bar q}^{(1)}$ agrees with the Cornell potential \eq{mo3} \cite{Eichten:1979ms,Eichten:2007qx}. The first term of the Fock expansion thus gives a good approximation for heavy quarkonia.

\subsubsection{The $qqq$ potential \label{secVII.A2}}
%%%%%%%%%%%%%%%%%%

An SU(3) color singlet $qqq$ Fock state has the form (suppressing the Dirac indices)
\begin{align} \label{qcd10}
\ket{qqq} = \epsilon_{ABC}\,\psi_A^\dag(\xv_1)\psi_B^\dag(\xv_2)\psi_C^\dag(\xv_3)\ket{0}
\end{align}
where $\epsilon_{ABC}$ is the fully antisymmetric tensor with $\epsilon_{123}=1$.
Note that this state is a color singlet of SU($N_c$) only for $N_c=3$. In a global transformation $\psi_A^\dag(\xv) \to \psi_{A'}^\dag(\xv)U_{A'A}^\dag$ the state is invariant: $\epsilon_{ABC}U_{A'A}^\dag U_{B'B}^\dag U_{C'C}^\dag = \epsilon_{A'B'C'}\det(U) = \epsilon_{A'B'C'}$ provided $U$ is a $3\times 3$ matrix with unit determinant.

When $\mH_V^{(0)}$ \eq{qcd1} operates on $\ket{qqq}$ the factor $\yv\cdot\zv$ is $\xv_i\cdot\xv_j$ for the commutator \eq{qcd3} of $\mE_a(\yv)$ with $\psi^\dag(\xv_i)$ and of $\mE_a(\zv)$ with $\psi^\dag(\xv_j)$, or \textit{vice versa}. It suffices to consider the two generic cases,
\begin{flalign} \label{qcd11}
&\xv_1^2: \hspace{1cm} \epsilon_{ABC}\,\psi_{A''}^\dag(\xv_1)\psi_B^\dag(\xv_2)\psi_C^\dag(\xv_3)\ket{0} T_{A''A'}^aT_{A'A}^a = C_F\ket{qqq} = \frac{4}{3}\ket{qqq} \nn \crt
&\xv_1\cdot\xv_2: \hspace{.3cm} 2\epsilon_{ABC}\,\psi_{A'}^\dag(\xv_1)\psi_{B'}^\dag(\xv_2)\psi_C^\dag(\xv_3)\ket{0} T_{A'A}^aT_{B'B}^a =\big(\epsilon_{B'A'C}-\sfrac{1}{N_c}\epsilon_{A'B'C}\big)\,\psi_{A'}^\dag(\xv_1)\psi_{B'}^\dag(\xv_2)\psi_C^\dag(\xv_3)\ket{0} \nn\crt
& \hspace{8.4cm} =-\Big(1+\frac{1}{N_c}\Big)\ket{qqq} = -\frac{4}{3}\ket{qqq} 
\end{flalign} 
where I used \eq{qcd4}. The two eigenvalues are equal and opposite for $N_c=3$, which ensures translation invariance,
\begin{align} \label{qcd12}
\mH_V^{(0)}\ket{qqq} &= \big(\halft\kappa^2\intt d\xv + g\kappa\big)\frac{4}{3}\big[d_{qqq}(\xv_1,\xv_2,\xv_3)\big]^2\ket{qqq} \nn\crt
d_{qqq}(\xv_1,\xv_2,\xv_3) &\equiv \inv{\sqrt{2}}\sqrt{(\xv_1-\xv_2)^2+(\xv_2-\xv_3)^2+(\xv_3-\xv_1)^2}
\end{align}

To ensure the universal value of $E_\la$ in \eq{qcd7}, \ie, the universality of the spatially constant energy density, the homogeneous solution in \eq{eII31} should be normalized for the $\ket{qqq}$ Fock state \eq{qcd10} as 
\begin{align} \label{qcd13}
\kappa_{qqq} = \frac{\la^2}{gC_F}\,\inv{d_{qqq}(\xv_1,\xv_2,\xv_3)}
\end{align}
This gives the \order{\alpha_s^0} potential
\begin{align} \label{qcd14}
V_{qqq}^{(0)}(\xv_1,\xv_2,\xv_3) = g\kappa_{qqq}\,\sfrac{4}{3}\,\big[d_{qqq}(\xv_1,\xv_2,\xv_3)\big]^2 = \la^2\,d_{qqq}(\xv_1,\xv_2,\xv_3)
\end{align}
The \order{\as} gluon exchange potential given by $\mH_V^{(1)}$ in \eq{qcd1} is determined by the eigenvalue of the $\xv_1\cdot\xv_2$ term in \eq{qcd11},
\begin{align} \label{qcd15}
V_{qqq}^{(1)}(\xv_1,\xv_2,\xv_3) = -\frac{2}{3}\,\as\Big(\inv{|\xv_1-\xv_2|}+\inv{|\xv_2-\xv_3|}+\inv{|\xv_3-\xv_1|}\Big)
\end{align}

\subsubsection{The $gg$ potential \label{secVII.A3}}
%%%%%%%%%%%%%%%%%%

A $gg$ Fock state which is invariant under global color SU($N_c$) transformations is expressed in terms of the gluon field in temporal gauge as
\begin{align} \label{qcd16}
\ket{gg} = A_a^i(\xv_1)A_a^j(\xv_2)\ket{0}
\end{align}
To find the action of $\mH_V$ \eq{qcd1} on this state we may use the canonical commutator in \eq{eII24}, 
\begin{align} \label{qcd17}
\com{\mE_a(\yv)}{A_b^i(\xv_1)} &= \com{-f_{acd}A_c^k(\yv)E_d^k(\yv)}{A_b^i(\xv_1)} = if_{abc}A_c^i(\xv_1)\delta(\yv-\xv_1) \nn\crt
\com{\mE_a(\yv)}{A_b^i(\xv_1)A_b^j(\xv_2)} &= if_{abc}A_b^i(\xv_1)A_c^j(\xv_2)\big[-\delta(\yv-\xv_1)+\delta(\yv-\xv_2)\big] \nn\crt
\com{\mE_a(\yv)}{\com{\mE_a(\zv)}{A_b^i(\xv_1)A_b^j(\xv_2)}} &= N_c\,A_a^i(\xv_1)A_a^j(\xv_2)\big[\delta(\yv-\xv_1)-\delta(\yv-\xv_2)\big]\big[\delta(\zv-\xv_1)-\delta(\zv-\xv_2)\big]
\end{align}
Hence
\begin{align} \label{qcd18}
\mH_V^{(0)}\ket{gg} = \big(\halft\kappa^2\intt d\xv + g\kappa\big) N_c (\xv_1-\xv_2)^2\ket{gg}
\end{align}
and to ensure the universal value of $E_\la$ in \eq{qcd7},
\begin{align} \label{qcd19}
\kappa_{gg} = \frac{\la^2}{g\sqrt{C_F N_c}}\,\inv{|\xv_1-\xv_2|}
\end{align}
This gives
\begin{align} \label{qcd20}
\mH_V\ket{gg} = \big(V_{gg}^{(0)}+V_{gg}^{(1)}\big)\ket{gg} = \Big(\sqrt{\frac{N_c}{C_F}}\,\la^2|\xv_1-\xv_2| -N_c\,\frac{\as}{|\xv_1-\xv_2|}\Big) \ket{gg}
\end{align}

\subsubsection{The $q\bar qg$ potential \label{secVII.A4}}
%%%%%%%%%%%%%%%%%%

The Hamiltonian \eq{eII25} creates \order{g} color singlet $\ket{q\bar qg}$ Fock states from $\ket{q\bar q}$ \eq{qcd2}. I consider the instantaneous potential generated by $\mH_V$ \eq{qcd1} for states of the form
\begin{align} \label{qcd21}
\ket{q\bar qg} \equiv \bar\psi_A(\xv_1)\,A_b^i(\xv_g)T^b_{AB}\psi_B(\xv_2)\ket{0}
\end{align}
Proceeding similarly as above gives the potential
\begin{align} \label{qcd22}
V_{q\bar qg}(\xv_1,\xv_2,\xv_g) &= \frac{\la^2}{\sqrt{C_F}}\, d_{q\bar qg}(\xv_1,\xv_2,\xv_g)+\halft\,\as\Big[\inv{N_c}\,\inv{|\xv_1-\xv_2|}-N_c\Big(\inv{|\xv_1-\xv_g|}+\inv{|\xv_2-\xv_g|}\Big)\Big] \nn\crt
d_{q\bar qg}(\xv_1,\xv_2,\xv_g) &\equiv \sqrt{\quart(N_c-\sfrac{2}{N_C})(\xv_1-\xv_2)^2+N_c(\xv_g-\halft\xv_1-\halft\xv_2)^2}
\end{align}
$V_{q\bar qg}$ is a confining potential, as it restricts both $|\xv_1-\xv_2|$ and the distance of $\xv_g$ from the average of $\xv_1+\xv_2$.

%%%%%%%%%%%%%%%%%%%%%%%%%%
\begin{tcolorbox}
\textit{Exercise \ref{e18}:} Derive the $q\bar qg$ potential \eq{qcd22}. 
\end{tcolorbox}
%%%%%%%%%%%%%%%%%%%%%%%%%%

\subsubsection{Limiting values of the $qqq$ and $q\bar qg$ potentials \label{secVII.A5}}
%%%%%%%%%%%%%%%%%%

When two of the quarks in the baryon $\ket{qqq}$ Fock state are close to each other the potential should (for $N_c=3$) reduce to that of the color $3\otimes\bar 3$ meson potential \eq{qcd8}, \eq{qcd9}. Setting $\xv_2=\xv_3$ in $V_{qqq}$ \eq{qcd14} and \eq{qcd15} (and subtracting the infinite Coulomb energy) indeed gives
\begin{align} \label{qcd23}
V_{qqq}(\xv_1,\xv_2,\xv_2) = \la^2 |\xv_1-\xv_2|-\frac{4}{3}\frac{\as}{|\xv_1-\xv_2|} = V_{q\bar q}(\xv_1,\xv_2)
\end{align}

Similarly the potential of the $q\bar qg$ Fock state should reduce to the $3\otimes\bar 3$ meson potential when a quark and gluon coincide. Setting $\xv_g=\xv_2$ in $V_{q\bar qg}$ \eq{qcd22} gives
\begin{align} \label{qcd24}
V_{q\bar qg}(\xv_1,\xv_2,\xv_2) = \la^2 |\xv_1-\xv_2|-C_F\frac{\as}{|\xv_1-\xv_2|} = V_{q\bar q}(\xv_1,\xv_2)
\end{align}
On the other hand, when the quarks in $q\bar qg$ coincide the potential \eq{qcd22} should become the $8\otimes 8$ glueball potential \eq{qcd20}. This is also fulfilled,
\begin{align} \label{qcd25}
V_{q\bar qg}(\xv_1,\xv_g,\xv_1) = \frac{\la^2\sqrt{N_c}}{\sqrt{C_F}}\,|\xv_1-\xv_g| -N_c\,\frac{\as}{|\xv_1-\xv_g|}= V_{gg}(\xv_1,\xv_g) 
\end{align}

\subsubsection{Single quark or gluon Fock states \label{secVII.A6}}
%%%%%%%%%%%%%%%%%%

The above examples indicate that only color singlet Fock states have translation invariant potentials. For the record I confirm this for single quark and gluon states.

In \eq{qcd5} we already saw that
\begin{align} \label{qcd26}
\mH_V^{(0)}\ket{q} \equiv \mH_V^{(0)}\bar\psi_A(\xv_q)\ket{0} = \big(\halft\kappa^2\intt d\xv + g\kappa\big)C_F\,\xv_q^2\,\ket{q}
\end{align} 
For the \order{\kappa^2} term to give $E_\la$ \eq{qcd7} we need
\begin{align} \label{qcd27}
\kappa_q = \frac{\la^2}{gC_F}\,\inv{|\xv_q|}
\end{align} 
This gives the potential
\begin{align} \label{qcd28}
V_q^{(0)} = \la^2|\xv_q|
\end{align} 
which is not invariant under translations.

The case of a single gluon is similar. Using our previous result \eq{qcd18},
\begin{align} \label{qcd29}
\mH_V^{(0)}\ket{g} \equiv \mH_V^{(0)}A_a^i(\xv_g)\ket{0} = \big(\halft\kappa^2\intt d\xv + g\kappa\big) N_c \xv_g^2\ket{g}
\end{align} 
This requires
\begin{align} \label{qcd30}
\kappa_g = \frac{\la^2}{g\sqrt{C_FN_c}}\,\inv{|\xv_g|}
\end{align} 
so that
\begin{align} \label{qcd31}
V_g^{(0)}  = \frac{\la^2\sqrt{N_c}}{\sqrt{C_F}}\,|\xv_g|
\end{align} 
I conclude, without further proof, that only color singlet Fock states are compatible with Poincar\'e invariance.

\subsubsection{The potential of $q\bar q\,q\bar q$ Fock states \label{secVII.A7}}
%%%%%%%%%%%%%%%%%%

As the number of quarks and gluons in a Fock state increases a subset of them may form a color singlet, and thus not be confined. In this final example I show that this is the case for color singlet states $\ket{q\bar q\,q\bar q}$, with two quarks and two antiquarks.

There are two ways to combine the four quarks into a color singlet. In a $q\bar q$ basis we can have 
\begin{align} \label{qcd32}
\ket{1\otimes 1} \equiv \ket{(q_1\bar q_2)_1(q_3\bar q_4)_1} \hspace{1cm} \mbox{and} \hspace{1cm}  \ket{8\otimes 8} \equiv \ket{(q_1\bar q_2)_8(q_3\bar q_4)_8}
\end{align}
where $q_i \equiv q(\xv_i)$.
The two independent configurations in a diquark basis $\ket{(q_1q_3)_{\bar 3}(\bar q_2\bar q_4)_{3}}$ and $\ket{(q_1q_3)_{6}(\bar q_2\bar q_4)_{\bar 6}}$ can be expressed in terms of these. I shall use the $q\bar q$ basis \eq{qcd32} here. Then
\begin{align} \label{qcd33}
\ket{1\otimes 1} &= \bar\psi_A(\xv_1)\psi_A(\xv_2)\,\bar\psi_B(\xv_3)\psi_B(\xv_4)\ket{0} \crt
\ket{8\otimes 8} &= \bar\psi_A(\xv_1)T_{AB}^a\psi_B(\xv_2)\,\bar\psi_C(\xv_3)T_{CD}^a\psi_D(\xv_4)\ket{0} \nn\crt
& = \halft\bar\psi_A(\xv_1)\psi_B(\xv_2)\,\bar\psi_B(\xv_3)\psi_A(\xv_4)\ket{0}
-\sfrac{1}{2N_c}\ket{1\otimes 1} \nn
\end{align}

The coefficients of $\yv\cdot\zv$ in $\mH_V^{(0)}\ket{1\otimes 1}$ are, apart from the common factor $\big(\halft\kappa^2\intt d\xv + g\kappa\big)$:
\begin{align} \label{qcd34}
(\xv_1-\xv_2)^2,\ (\xv_3-\xv_4)^2:&\hspace{1cm} C_F\ket{1\otimes 1} \nn\crt
\xv_1\cdot\xv_3,\ -\xv_1\cdot\xv_4,\ -\xv_2\cdot\xv_3,\ \xv_2\cdot\xv_4:&\hspace{1cm} 2\bar\psi_{A}(\xv_1)T_{AB}^a\psi_B(\xv_2)\bar\psi_{C}(\xv_3)T_{CD}^a\psi_D(\xv_4) = 2\ket{8\otimes 8}
\end{align}
In terms of the relative separations
\begin{align} \label{qcd35}
\dv_{12} = \xv_1-\xv_2 \hspace{2cm} \dv_{34} = \xv_3-\xv_4 \hspace{2cm} \dv=\halft(\xv_1+\xv_2-\xv_3-\xv_4)
\end{align}
and with a similar analysis for $\mH_V^{(0)}\ket{8\otimes 8}$,
\begin{align} \label{qcd36}
\mH_V^{(0)}\ket{1\otimes 1} &= \big(\halft\kappa^2\intt d\xv + g\kappa\big)\big\{\big[C_F(\xv_1-\xv_2)^2+C_F(\xv_3-\xv_4)^2\big]\ket{1\otimes 1} 
+ 2(\xv_1-\xv_2)\cdot(\xv_3-\xv_4)\ket{8\otimes 8}\big\} \nn\crt
&= \big(\halft\kappa^2\intt d\xv + g\kappa\big)\big[C_F(\dv_{12}^2+\dv_{34}^2)\ket{1\otimes 1}
+2\dv_{12}\cdot\dv_{34}\ket{8\otimes 8}\big] \nn\crt
\mH_V^{(0)}\ket{8\otimes 8} &= \big(\halft\kappa^2\intt d\xv + g\kappa\big)\big\{\big[N_c\dv^2+\sfrac{N_c^2-2}{4N_c}(\dv_{12}^2+\dv_{34}^2)+\sfrac{N_c^2-4}{2N_c}\dv_{12}\cdot\dv_{34}\big]\ket{8\otimes 8}+\sfrac{C_F}{N_c}\dv_{12}\cdot\dv_{34}\ket{1\otimes 1}\big\}
\end{align}

%%%%%%%%%%%%%%%%%%%%%%%%%%
\begin{tcolorbox}
\textit{Exercise \ref{e19}:} Derive the expression for $\mH_V^{(0)}\ket{8\otimes 8}$ in \eq{qcd36}. 
\end{tcolorbox}
%%%%%%%%%%%%%%%%%%%%%%%%%%

The expression \eq{qcd1} for $\mH_V$ shows that the color structure of $\mH_V^{(0)}$ and $\mH_V^{(1)}$ is the same. Hence $\mH_V^{(1)}$ is given by the coefficients of $\xv_i\cdot\xv_j$, multiplied by $\halft\as/|\xv_i-\xv_j|$ (for $i \neq j$).

The coefficients of the $\ket{1\otimes 1}$ and $\ket{8\otimes 8}$ states on the rhs. of \eq{qcd36} allow to determine the eigenstates of $\mH_V^{(0)}$ and thus the normalizations $\kappa$ of the corresponding homogeneous solutions. The coefficients, and thus the eigenstates, depend on the separations $\dv,\, \dv_{12}$ and $\dv_{34}$. At large separations of the $q_1\bar q_2$ and $q_3\bar q_4$ pairs, \ie, for $|\dv| \gg |\dv_{12}|, |\dv_{34}|$, $\mH_V\ket{8\otimes 8} \sim N_c\dv^2\ket{8\otimes 8}$. The other eigenstate then approaches $\ket{1\otimes 1}$ with the smaller eigenvalue $\sim C_F(\dv_{12}^2+\dv_{34}^2)$. Since the separation of color octet charges gives a large potential energy it may be expected that eigenstates of the full Hamiltonian are dominated by unconfined $\ket{1\otimes 1}$ color configurations at large separations.

\subsection{Rest frame wave functions of $q \bar q$ bound states \label{secVII.B}}
%%%%%%%%%%%%%%%%%%

In this section I determine the \order{\alpha_s^0} meson eigenstates of the QCD Hamiltonian \eq{eII25} in the rest frame. The valence quark state of the meson is (at $t=0$) defined by its wave function $\Phi_{\alpha\beta}(\xv_1-\xv_2)$,
\begin{align} \label{qcd37}
\ket{M} = \inv{\sqrt{N_c}}\sum_{A,B;\alpha,\beta}\int d\xv_1 d\xv_2\,\bar\psi_\alpha^A(\xv_1)\delta^{AB}\Phi_{\alpha\beta}(\xv_1-\xv_2)\psi_\beta^B(\xv_2)\ket{0}
\end{align}
where $N_c=3$ in QCD and $A,B$ are quark color indices. This is a singlet state under global color SU($N_c$) transformations.
Contributions of higher orders in $\as$ and are neglected here\footnote{The hyperfine splitting of Positronium (section \ref{secV.D}) gives an example of how higher Fock states are taken into account.}.

\subsubsection{Bound state equation for the meson wave function $\Phi_{\alpha\beta}(\xv)$ \label{secVII.B1}}
%%%%%%%%%%%%%%%%%%

The bound state condition for the meson state \eq{qcd37} of mass $M$ is at \order{\alpha_s^0},
\begin{align} \label{qcd38}
\big(\mH_0^{(q)}+\mH_V^{(0)}\big)\ket{M} = M\ket{M}
\end{align}
The quark kinetic energy operator in the Hamiltonian \eq{eII25} is
\begin{gather}
\mH_0^{(q)} = \int d\xv\,\psi_A^\dag(\xv)(-i\alv\cdot\rnab+m\gz)\psi_A(\xv) \label{qcd39} \crt
\com{\mH_0^{(q)}}{\bar\psi(\xv_1)} = \bar\psi(\xv_1)(-i\alv\cdot\lnab_1+m\gz) \hspace{2cm}
\com{\mH_0^{(q)}}{\psi(\xv_2)} = (i\alv\cdot\rnab_2-m\gz)\psi(\xv_2) \label{qcd40}
\end{gather}
The meson is bound by the instantaneous potential generated by $\mH_V^{(0)}$ \eq{qcd1}. Its action on color singlet $q\bar q$ Fock states is given in \eq{qcd8},
\begin{align} \label{qcd41}
\com{\mH_V^{(0)}}{\bar\psi_\alpha^A(\xv_1)\psi_\beta^A(\xv_2)} = V'\,|\xv_1-\xv_2|\,\bar\psi_\alpha^A(\xv_1)\psi_\beta^A(\xv_2) \hspace{2cm} V(\xv)= V'|\xv| = \la^2|\xv|
\end{align}
where a sum over the quark colors $A$ is implied. I neglect the \order{\as} Coulomb energy \eq{qcd9}.

Shifting the derivatives in \eq{qcd40} from the fields onto the wave function by partial integration in \eq{qcd38} the coefficients of each Fock component gives the bound state equation for $\Phi(\xv)$ ($\xv=\xv_1-\xv_2$),
\begin{align} \label{qcd42}
\big(i\alv\cdot\rnab+m\gz\big)\Phi(\xv)+\Phi(\xv)\big(i\alv\cdot\lnab-m\gz\big) = \big[M- V(\xv)\big]\Phi(\xv)
\end{align}
where $V(\xv)=V'|\xv|$. Equivalent forms of this BSE are
\begin{align}
&i\nv\cdot\acom{\alv}{\Phi(\xv)}+m\com{\gz}{\Phi(\xv)} = \big[M- V(\xv)\big]\Phi(\xv)  \label{qcd43} \crt
&\Big[\frac{2}{M-V}(i\alv\cdot\rnab+m\gz)- 1\Big]\Phi(\xv)+\Phi(\xv)\Big[(i\alv\cdot\lnab-m\gz)\frac{2}{M-V}- 1\Big]=0 \label{qcd44}
\end{align}
Introducing the notation
\begin{align} \label{qcd45}
&\rh_\pm \equiv \frac{2}{M-V}(i\alv\cdot\rnab+m\gz)\pm 1
&\lh_\pm \equiv (i\alv\cdot\lnab-m\gz)\frac{2}{M-V}\pm 1
\end{align}
which satisfy $(r=|\xv|)$
\begin{align}\label{qcd46}
\rh_-\rh_+ &= \frac{4}{(M-V)^2}(-\rnab^2+m^2)-1 +\frac{4iV'}{r(M-V)^3}\,\alv\cdot\xv \,(i\alv\cdot\rnab+m\gz) \nn \crt
\lh_+\lh_- &= (-\lnab^2+m^2)\frac{4}{(M-V)^2}-1 + (i\alv\cdot\lnab-m\gz)\, \alv\cdot\xv\,\frac{4iV'}{r(M-V)^3} 
\end{align}
the bound state equation \eq{qcd44} is
\begin{align} \label{qcd47}
\rh_-\Phi(\xv)+\Phi(\xv)\lh_- = 0
\end{align}

The meson states \eq{qcd37} are relativistic (strongly bound) when $m \lsim \la$, and for high excitations at any $m$ due to the linear potential. The example of Dirac states \eq{state1} shows that for strong fields the vacuum \eq{vac2} has $b^\dag d^\dag$ pairs in a free fermion basis, and fermion eigenstates are created by a superposition of the $b^\dag$ and $d$ operators \eq{cndef1}. Analogously, the meson state \eq{qcd37} is a two-particle Fock state only in a Bogoliubov rotated operator basis. In the free basis we may view the pairs as arising from the $Z$-diagrams (\fig{f4}b) of a time-ordered perturbative expansion. For DIS in QED$_2$ the pairs give rise to a sea-like distribution of electrons $\propto 1/\xbj$, as discussed in section \ref{secVI.C3} and shown in \fig{f9}. I return to this issue in section \ref{secVII.D2}.

\subsubsection{Separation of radial and angular variables \label{secVII.B2}}
%%%%%%%%%%%%%%%%%%

The $4\times 4$ wave function $\Phi_{\alpha\beta}(\xv)$ may be expressed as a sum of terms with distinct Dirac structures $\Gamma_{\alpha\beta}^{(i)}(\xv)$, radial functions $F_i(r)$ and angular dependence given by the spherical harmonics $Y_{j\lambda}(\hat\xv)$:
\begin{align} \label{qcd48}
\Phi_{\alpha\beta}^{j\lm}(\xv) = \sum_i \Gamma_{\alpha\beta}^{(i)}F_i(r)Y_{j\lambda}(\hat\xv)
\end{align} 
where $r=|\xv|$ and $\hat\xv=\xv/r$. The 16 independent Dirac structures $\Gamma^{(i)}$ should be rotationally invariant to allow a simple classification of the states according to their angular momentum $j$ and $j^z=\lm$. As discussed in section \ref{secV.A} for Positronium the generator of rotations for the quark fields is
\begin{align} \label{qcd49}
\bs{\mJ} &= \int d\xv\,\psi_A^\dag(\xv)\,\bs{J}\,\psi_A(\xv) \hspace{2.9cm} \bs{J} = \Lv+\Sv= \xv\times(-i\nv)+\halft\gf\alv \nn\crt
\bs{\mJ}\ket{M}&=\int d\xv_1 d\xv_2\, \bar\psi_A(\xv_1) \com{\Jv}{\Phi(\xv_1-\xv_2)}\psi_A(\xv_2)\ket{0}
\end{align}
For example, the rotationally invariant Dirac structure $\alv\cdot\nv$ satisfies $\com{\Jv}{\alv\cdot\nv}=0$ as shown in \eq{ex12a.3} and \eq{ex12a.4}.
When $\com{\Jv}{\Gamma^{(i)}(\xv)}=0$ we get, since $\com{\Lv}{F_i(r)}=\com{\Sv}{F_i(r)}=\com{\Sv}{Y_{j\lambda}(\hat\xv)}=0$, 
\begin{align} \label{qcd50}
\com{\Jv}{\Gamma^{(i)}F_i(r)Y_{j\lambda}(\hat\xv)} = \Gamma^{(i)}F_i(r)\com{\Lv}{Y_{j\lambda}(\hat\xv)}
\end{align}
Because $Y_{j\lambda}(\hat\xv)$ is an eigenfunction of $\Lv^2$ and $L^z$ this ensures that
\begin{align} \label{qcd51}
\bs{\mJ}^2\ket{M} = j(j+1)\ket{M} \hspace{2cm} \mJ^z\ket{M} = \lm\ket{M}
\end{align}
The $\Gamma^{(i)}(\xv)$ need contain at most one power of the Dirac vector $\alv=\gz\gv$ since higher powers may be reduced using 
\begin{align} \label{qcd52}
\alpha_i\alpha_j=\delta_{ij}+i\epsilon_{ijk}\alpha_k\gf
\end{align}
Rotational invariance requires that $\alv$ be dotted into a vector. I shall use the three orthogonal vectors $\xv,\ \Lv=\xv\times(-i\nv)$ and $\xv\times\Lv$. Each of the four Dirac structures $1,\ \alv\cdot\xv,\ \alv\cdot\Lv$ and $\alv\cdot\xv\times\Lv$ can be multiplied by the rotationally invariant Dirac matrices $\gz$ and/or $\gf$. This gives altogether $4\times 2\times 2=16$ possible $\Gamma^{(i)}(\xv)$. Other invariants may be expressed in terms of these, \eg,
\begin{align} \label{qcd53}
i\,\alv\cdot\nv &= (\alv\cdot\xv)\,\inv{r}i\,\partial_r +\inv{r^2}\,\alv\cdot\xv\times\bs{L} \nn\crt
(\alv\cdot\nv)(\alv\cdot\xv) &= 3+r\partial_r +\gf\,\alv\cdot\Lv 
\end{align}

The $\Gamma^{(i)}(\xv)$ may be grouped according to the parity $\eta_P$ \eq{A11} and charge conjugation $\eta_C$ \eq{A15} quantum numbers that they imply for the wave function,
\begin{align} \label{qcd54}
\gz\Phi(-\xv)\gz = \eta_P \Phi(\xv)   \hspace{2cm} \aly\big[\Phi(-\xv)\big]^T\aly = \eta_C \Phi(\xv)
\end{align}
Since $Y_{j\lambda}(-\hat\xv)=(-1)^j Y_{j\lambda}(\hat\xv)$ states of spin $j$ can belong to one of four ``trajectories'', here denoted by the parity and charge conjugation quantum numbers of their $j=0$ member:
\beq\label{qcd55}
\begin{array}{llcl} 0^{-+} \ \mbox{trajectory} & [s=0,\ \ell=j]: & -\eta_P=\eta_C=(-1)^{j} & \gf,\ \gz\gf,\ \gf\,\alv\cdot\xv,\ \gf\,\alv\cdot\xv\times\Lv \\[2mm]
0^{--}\ \mbox{trajectory} & [s=1,\ \ell=j]: & \eta_P= \eta_C=-(-1)^{j} & \gz\gf\,\alv\cdot\xv,\ \gz \gf\,\alv\cdot\xv\times\Lv,\ \alv\cdot\Lv,\ \gz\,\alv\cdot\Lv  \\[2mm]
0^{++}\ \mbox{trajectory} & [s=1,\ \ell=j\pm 1]: & \eta_P= \eta_C=+(-1)^{j} & 1,\ \alv\cdot\xv,\ \gz\alv\cdot\xv,\ \alv\cdot\xv\times\Lv,\ \gz\alv\cdot\xv\times\Lv,\ \gz\gf\,\alv\cdot\Lv  \\[2mm]
0^{+-} \ \mbox{trajectory} & [\mbox{exotic}]: & \eta_P=-\eta_C=(-1)^{j} & \gz,\ \gf\,\alv\cdot\Lv
\vspace{-.4cm}\end{array} 
\eeq

The non-relativistic spin $s$ and orbital angular momentum $\ell$ are indicated in brackets. Relativistic effects mix the $\ell=j\pm 1$ states on the $0^{++}$ trajectory, resulting in a pair of coupled radial equations. The $j=0$ state on the $0^{--}$ trajectory and the entire $0^{+-}$ trajectory are incompatible with the $s,\ell$ assignments and thus exotic in the quark model. They turn out to be missing also in the relativistic case. The bound state equation \eq{qcd43} has no solutions for states on the $0^{+-}$ trajectory ($\Gamma^{(i)} = \gz$ or $\gf\,\alv\cdot\Lv$) since
\begin{align} \label{qcd56}
i\nv\cdot\acom{\alv}{\gz}= i\nv\cdot\acom{\alv}{\gf\,\alv\cdot\Lv}= m\com{\gz}{\gz}=m\com{\gz}{\gf\,\alv\cdot\Lv}=0
\end{align}

%%%%%%%%%%%%%%%%%%%%%%%%%%%%%%%%%%%
\subsubsection{The $0^{-+}$ trajectory: $\eta_P=(-1)^{j+1}, \hspace{.3cm} \eta_C=(-1)^{j}$ \label{secVII.B3}}

According to the classification \eq{qcd55} we expand the wave function $\Phi_{-+}(\xv)$ of the $0^{-+}$ trajectory states as
\begin{align} \label{qcd57}
\Phi_{-+}(\xv) = \Big[F_1(r) + i\,\alv\cdot\xv\,F_2(r) + \alv\cdot\xv\times\Lv\,F_3(r) + \gz\,F_4(r)\Big]\gf\,Y_{j\lambda}(\hat\xv)
\end{align}
Using this in the bound state equation \eq{qcd42}, noting that $i\nv\cdot \xv\times\Lv=\Lv^2$ and collecting terms with the same Dirac structure we get the conditions:
\begin{align}\label{qcd58}
\gf: \hspace{.5cm} & -(3+r\partial_r)F_2+j(j+1)F_3+m F_4=\halft (M-V)F_1 \nn \\
\gf\,\alv\cdot\xv:\hspace{.5cm} & \frac{1}{r}\partial_r F_1= \halft (M-V)F_2 \nn \\
\gf\,\alv\cdot\xv\times\Lv: \hspace{.5cm} & \inv{r^2}F_1 = \halft (M-V)F_3 \nn \\
\gz\gf: \hspace{.5cm} & m F_1 = \halft (M-V)F_4
\end{align}
Expressing $F_2,\ F_3$ and $F_4$ in terms of $F_1$ we find the radial equation (denoting $F_1' \equiv \partial_r F_1$)
\begin{align} \label{qcd59}
F_1''+\Big(\frac{2}{r}+\frac{V'}{M-V}\Big)F_1' + \Big[\quart (M-V)^2-m^2-\frac{j(j+1)}{r^2}\Big]F_1 = 0
\end{align}
in agreement with the corresponding result in Eq. (2.24) of \cite{Geffen:1977bh}.
The wave function \eq{qcd57} may be expressed as
\begin{align}\label{qcd60}
\Phi_{-+}(\xv) &= \Big[\frac{2}{M-V}(i\alv\cdot\rnab+m\gz)+1\Big]\gf\,F_1(r)Y_{j\lambda}(\hat\xv) = F_1(r)Y_{j\lambda}(\hat\xv)\,\gf\Big[(i\alv\cdot\lnab-m\gz)\frac{2}{M-V}+1\Big] \nn \crt
&= \rh_+\gf\,F_1(r)Y_{j\lambda}(\hat\xv) = F_1(r)Y_{j\lambda}(\hat\xv)\,\gf \lh_+
\end{align}

%%%%%%%%%%%%%%%%%%%%%%%%%%
\begin{tcolorbox}
\textit{Exercise \ref{e20}:} Verify that the expression \eq{qcd60} for $\Phi_{-+}(\xv)$ satisfies the bound state equation \eq{qcd47} given the radial equation \eq{qcd59}. \\
\textit{Hint:} The identities \eq{qcd46} are useful.
\end{tcolorbox}
%%%%%%%%%%%%%%%%%%%%%%%%%%

Both the quark and antiquark contributions to the BSE have a spin-dependent ($\Sv=\halft \gf\alv$) interaction which cancels in their sum. The contribution from the quark term is, taking into account the radial equation,
\begin{align} \label{qcd61}
\rh_-\Phi_{-+}(\xv) = \frac{8V'}{r(M-V)^3}\,\Sv\cdot(\rorb\,\gf-im\,\xv\,\gz) \,F_1(r)Y_{j\lambda}(\hat\xv)
\end{align}
 
%%%%%%%%%%%%%%
\subsubsection*{Non-relativistic limit of the $0^{-+}$ trajectory wave functions \label{sIIIC1}}

The non-relativistic (NR) limit in the rest frame is defined by 
\begin{align} \label{qcd62}
\frac{V}{m} \to 0 \hspace{2cm} \frac{\partial}{\partial r} \sim \inv{r} \sim \sqrt{m\,V}
\end{align}
The binding energy $E_b \sim V$ is defined by $M = 2m + E_b$.
In the radial equation \eq{qcd59} we have
\begin{align} \label{qcd63}
\frac{V'}{M-V} = \frac{V}{r(M-V)} \ll \inv{r} \hspace{2cm} \quart (M-V)^2-m^2 \simeq m(E_b-V)
\end{align}
so it becomes the radial Schr\"odinger equation in the NR limit,
\begin{align} \label{qcd64}
F_{1,NR}''+\frac{2}{r}F_{1,NR}' + \Big[m(E_b-V)-\frac{j(j+1)}{r^2}\Big]F_{1,NR} = 0
\end{align}
In the wave function \eq{qcd60} we have at leading order
\begin{align} \label{qcd65}
\rh_+ = \frac{2}{M-V}(i\alv\cdot\nv+m\gz) +1 \simeq 1+\gz
\end{align}
giving
\begin{align} \label{qcd66}
\Phi_{-+}^{NR} = (1+\gz)\gf\,F_{1,NR}(r) \sph(\Omega)
\end{align}

%%%%%%%%%%%%%%%%%%%%%%%%%%%%%%%%%%%
\subsubsection{The $0^{--}$ trajectory: $\eta_P=(-1)^{j+1}, \hspace{.3cm} \eta_C=(-1)^{j+1}$ \label{secVII.B4}}

According to the classification \eq{qcd55} we expand the wave function $\Phi_{--}(\xv)$ of the $0^{--}$ trajectory states as
\begin{align} \label{qcd67}
\Phi_{--}(\xv) = \Big[\gz\,\alv\cdot\Lv\, G_1(r) + i\,\gz\gf\,\alv\cdot\xv\,G_2(r) + \gz\gf\,\alv\cdot\xv\times\Lv\,G_3(r) + m\alv\cdot\Lv\,G_4(r)\Big]Y_{j\lambda}(\hat\xv)
\end{align}
Collecting terms with distinct Dirac structures in the bound state equation \eq{qcd43},
\begin{align}\label{qcd68}
\gz\,\alv\cdot\Lv: \hspace{.5cm} & G_2 -(2+r\partial_r)G_3+m^2G_4=\halft (M-V)G_1 \nn \\
\gz\gf\,\alv\cdot\xv:\hspace{.5cm} & \frac{j(j+1)}{r^2} G_1= \halft (M-V)G_2 \nn \\
\gz\gf\,\alv\cdot\xv\times\Lv: \hspace{.5cm} & \inv{r^2}(1+r\partial_r)G_1 = \halft (M-V)G_3 \nn \\
m\,\alv\cdot\Lv: \hspace{.5cm} & G_1 = \halft (M-V)G_4
\end{align}
Expressing $G_2,\ G_3$ and $G_4$ in terms of $G_1$ we find the radial equation for the $0^{--}$ trajectory,
\begin{align} \label{qcd69}
G_1''+\Big(\frac{2}{r}+\frac{V'}{M-V}\Big)G_1' + \Big[\quart (M-V)^2-m^2-\frac{j(j+1)}{r^2}+\frac{V'}{r(M-V)}\Big]G_1 = 0
\end{align}
in agreement with the corresponding result in Eq. (2.38) of \cite{Geffen:1977bh}. The $0^{--}$ radial equation differs from the $0^{-+}$ one \eq{qcd59} only by the term $\propto \,V'/r(M-V)$. Using
\begin{align} \label{qcd70}
i\alv\cdot\nv\,\gv\cdot\Lv = \gz\gf\,\alv\cdot\xv\,\frac{i\Lv^2}{r^2}+\gz\gf\,\alv\cdot\xv\times\Lv \inv{r^2}(1+r\partial_r)
\end{align}
allows the wave function to be expressed in terms of the projector $\mathfrak{h}_+$ of \eq{qcd45} as,
\begin{align}\label{qcd71}
\Phi_{--}(\xv) &=\rh_+\,\gv\cdot\rorb\,G_1(r)\,Y_{j\lambda}(\hat\xv)
=G_1(r)\,Y_{j\lambda}(\hat\xv)\,\gv\cdot\lorb\,\lh_+
\end{align}
where ${\overset{\lar}{L}}\strut^i = -i{\overset{\lar}{\partial}}\strut_k x^j\veps_{ijk}$.
The $j=0$ state on the $0^{--}$ trajectory is missing since $\Lv\,Y_{00}(\hat\xv)=0$.
The quark contribution to the bound state equation \eq{qcd47} is, with $\bs{S}=\halft \gf\alv$,
\begin{align} \label{qcd72}
\rh_-\Phi_{--}(\xv) = \frac{4V'}{r(M-V)^3}\big[\rorb^2\gz\gf-2m\,\bs{S}\cdot\xv\times\rorb\big]\, G_1(r)\,Y_{j\lambda}(\hat\xv)
\end{align}

%%%%%%%%%%%%%%
\subsubsection*{Non-relativistic limit of the $0^{--}$ trajectory wave functions \label{sIIID1}}

The NR limit of the radial equation \eq{qcd69} reduces as in the $0^{-+}$ case to
\begin{align} \label{qcd73}
G_{1,NR}''+\frac{2}{r}G_{1,NR}' + \Big[m(E_b-V)-\frac{j(j+1)}{r^2}\Big]G_{1,NR} = 0
\end{align}
The equality of the $0^{-+}$ and $0^{--}$ eigenvalues reflects the spin $s$ independence of the NR limit, since $\ell=j$ for both. The wave function is
\begin{align} \label{qcd74}
\Phi_{--}^{NR} = (1+\gz)\alv\cdot\Lv\,G_{1,NR}(r) \sph(\Omega)
\end{align}

%%%%%%%%%%%%%%%%%%%%%%%%%%%%%%%%%%%
\subsubsection{The $0^{++}$ trajectory: $\eta_P=(-1)^{j}, \hspace{.3cm} \eta_C=(-1)^{j}$ \label{secVII.B5}} 

According to the classification \eq{qcd55} we expand the wave function $\Phi_{++}(\xv)$ of the $0^{++}$ trajectory states in terms of six Dirac structures\footnote{The radial functions $F_i$ and $G_i$ are unrelated to those in sections \ref{secVII.B3} and \ref{secVII.B4}.},
\begin{align} \label{qcd75}
\Phi_{++}(\xv) = \left\{ \Big[F_1(r) + i\,\alv\cdot\xv\,F_2(r) + \alv\cdot\xv\times\Lv\,F_3(r)\Big] + \gz\Big[\gf\,\alv\cdot\Lv\,G_1(r)+ i\,\alv\cdot\xv\,G_2(r)+\alv\cdot\xv\times\Lv\,G_3(r)\Big]\right\}Y_{j\lambda}(\hat\xv)
\end{align}
Collecting terms with distinct Dirac structures in the bound state equation \eq{qcd43},
\begin{align} \label{qcd76}
1: \hspace{.5cm} & -(3+r\partial_r)F_2+j(j+1)F_3 = \halft (M-V) F_1 \nn \\
\alv\cdot\xv: \hspace{.5cm} & \frac{1}{r}\partial_r F_1+mG_2 = \halft (M-V) F_2 \nn \\
\alv\cdot\xv\times\Lv: \hspace{.5cm} & \inv{r^2} F_1+mG_3 = \halft (M-V) F_3 \nn \\
\gz\gf\,\alv\cdot\Lv: \hspace{.5cm} & G_2 -(2+r\partial_r)G_3=\halft (M-V)G_1 \nn \\
\gz\,\alv\cdot\xv:\hspace{.5cm} & \frac{1}{r^2}\,j(j+1) G_1 +mF_2= \halft (M-V)G_2 \nn \\
\gz\,\alv\cdot\xv\times\Lv: \hspace{.5cm} & \inv{r^2}(1+r\partial_r) G_1 +mF_3 = \halft (M-V)G_3
\end{align}
It turns out to be convenient to express the above radial functions in terms of two new ones, $H_1(r)$ and $H_2(r)$:
\begin{align} \label{qcd77}
F_1 & =-\frac{2}{(M-V)^2}\big[\quart(M-V)^2-m^2\big]H_1-\frac{4m}{M-V}\partial_r(rH_2) \nn \\
F_2 &= -\frac{1}{r(M-V)}\partial_r H_1 + 2m H_2 \nn \\
F_3 &= -\inv{r^2(M-V)}H_1 \nn \\
G_1 &= 2 H_2 \nn \\
G_2 &= \frac{2}{r}\partial_r\Big[-\frac{m}{(M-V)^2}H_1+\frac{2}{M-V}\partial_r(rH_2)\Big]+(M-V)H_2 \nn \\
G_3 &= \frac{2}{r^2}\Big[-\frac{m}{(M-V)^2}H_1+\frac{2}{M-V}\partial_r(rH_2)\Big]
\end{align}
The bound state conditions \eq{qcd76} are satisfied provided $H_{1,2}$ satisfy the coupled radial equations,
\begin{align}
H_1''+\Big(\frac{2}{r}+\frac{V'}{M-V}\Big)H_1' + \Big[\quart (M-V)^2-m^2-\frac{j(j+1)}{r^2}\Big]H_1 &= 4m(M-V)H_2 \label{qcd78} \crt
H_2''+\Big(\frac{2}{r}+\frac{V'}{M-V}\Big)H_2' + \Big[\quart (M-V)^2-m^2-\frac{j(j+1)}{r^2}+\frac{V'}{r(M-V)}\Big]H_2 &= \frac{mV'}{r(M-V)^2}H_1 \label{qcd79}
\end{align}
These agree with Eqs. (2.48) and (2.49) for $F_2^{GS}$ and $G_1^{GS}$ of \cite{Geffen:1977bh}, when $H_1 = (M-V)F_2^{GS}$ and $H_2=-i\,G_1^{GS}/(M-V)$.

The wave function $\Phi_{++}(\xv)$ \eq{qcd75} can be expressed in terms of the $H_{1,2}(r)$ radial functions  and the $\mathfrak{h}_+$ operators \eq{qcd45} as
\begin{align} \label{qcd80}
\Phi_{++}(\xv) &= 
\rh_+\big[-\halft H_1+2\,\gv\cdot\rorb\,\gf H_2+2im\,\alv\cdot\xv\, H_2\big]Y_{j\lambda}(\hat\xv)
+\frac{m}{M-V}\big[\rh_+\gz H_1+8H_2\big]Y_{j\lambda}(\hat\xv)\crt
&= Y_{j\lambda}(\hat\xv)\big[-\halft H_1-2 H_2\gf\,\gv\cdot\lorb\,+2im H_2\,\alv\cdot\xv\, \big] \lh_+ -Y_{j\lambda}(\hat\xv)\big[H_1\gz \lh_+-8 H_2\big]\frac{m}{M-V} \nn
\end{align}
The quark contribution to the bound state equation \eq{qcd47} is, with $\bs{S}=\halft \gf\alv$,
\begin{align} \label{qcd81}
\rh_- \Phi_{++}(\xv) &= -\frac{4V'}{r(M-V)^3}\Big[\bs{S}\cdot\rorb+\frac{m}{M-V}\gz r\partial_r\Big]H_1(r)Y_{j\lambda}(\hat\xv)+\frac{8V'}{r(M-V)^3}\big[\rorb^2+m^2r^2\big]\gz\,H_2(r)Y_{j\lambda}
\end{align}

When $m=0$ chiral symmetry implies that $\Phi(\xv)$ and $\gf\Phi(\xv)$ define bound states with the same mass $M$, as is apparent from the bound state equation \eq{qcd42}. The radial equations \eq{qcd78} and \eq{qcd79} in fact decouple and coincide with the radial equations of the $0^{-+}$ \eq{qcd59} and $0^{--}$ \eq{qcd69} trajectories, respectively. The $\Phi_{++}$ wave functions correspondingly reduce to $\gf\Phi_{-+}$ and $\gf\Phi_{--}$. I discuss the case of spontaneously broken chiral symmetry in section \ref{secVII.F}.

%%%%%%%%%%%%%%
\subsubsection*{Non-relativistic limit of the $0^{++}$ trajectory wave functions \label{sIIIE1}}

States with the same $j$ on the $0^{-+}$ and $0^{--}$ trajectories are degerate in the NR limit since both have $\ell = j$. States with the same $j$ on the $0^{++}$ trajectory have $\ell = j\pm 1$ and thus unequal binding energies. The radial $0^{++}$ functions $H_1$ \eq{qcd78} and $H_2$ \eq{qcd79} remain coupled in the NR limit,
\begin{align}
H_{1,NR}''+\frac{2}{r}H_{1,NR}' + \Big[m(E_b-V)-\frac{j(j+1)}{r^2}\Big]H_{1,NR} &= 8m^2 H_{2,NR} \label{qcd82} \crt
H_{2,NR}''+\frac{2}{r}H_{2,NR}' + \Big[m(E_b-V)-\frac{j(j+1)}{r^2}\Big]H_{2,NR} &= \frac{V}{4mr^2} H_{1,NR}  \label{qcd83}
\end{align}
The lhs. of both equations scale as $1/r^2 \sim mV$, implying the ratio
\begin{align} \label{qcd84}
\frac{H_{2,NR}}{H_{1,NR}} \sim \frac{V}{m}
\end{align}
In the expression \eq{qcd80} for $\Phi_{++}$ the leading contribution $\propto H_1$ vanishes for $\mathfrak{h}_+\simeq 1+\gz$ \eq{qcd65}. This requires to retain the \order{\sqrt{V/m}} term in $\mathfrak{h}_+$,
\begin{align} \label{qcd85}
\mathfrak{h}_+ \simeq 1+\gz + \frac{i}{m}\alv\cdot\nv
\end{align}
Then the contribution $\sim \sqrt{V/m}\,H_1 \sim \sqrt{m/V}\,H_2$ matches the leading $H_2$ contribution $m\alv\cdot\xv\,H_2 \sim \sqrt{m/V}\,H_2$. The $2\gv\cdot\Lv\,\gf H_2$ term is subdominant, as are the \order{V/m} corrections in $\mathfrak{h}_+$. This gives
\begin{align} \label{qcd86}
\Phi_{++}^{NR} = \frac{i}{2m}(1+\gz)\big[-\alv\cdot\nv H_{1,NR}(r)+4m^2\alv\cdot\xv H_{2,NR}(r)\big]\sph(\Omega)
\end{align}

Orbital angular momentum is conserved in the NR limit, implying
\begin{align} \label{qcd87}
\comb{\Lv^2}{\Phi_{++}^{NR}} = \ell(\ell+1)\Phi_{++}^{NR} \hspace{2cm} \ell = j\pm 1
\end{align}
Using
\begin{align}
\comb{\rorb^2}{\xv} &= 2\big(-\nv\,r^2+\xv\, r\partial_r+3\xv\big) \label{qcd88} \crt
\comb{\rorb^2}{\nv} &= 2\big(\xv\,\nv^2-\nv\,r\partial_r\big)  \label{qcd89}
\end{align}
gives
\begin{align} \label{qcd90}
\comb{\Lv^2}{\Phi_{++}^{NR}} &= i(1+\gz)\,\alv\cdot\nv\,\inv{2m}\big[2r H_{1,NR}'-j(j+1)H_{1,NR}-8m^2r^2 H_{2,NR}\Big]\sph \nn\crt
&+i(1+\gz)\,\alv\cdot\xv\,2m\Big[\inv{2m}(E_b-V)H_{1,NR}+2r H_{2,NR}' +2H_{2,NR} + j(j+1)H_{2,NR}\Big]\sph
\end{align}
Comparing with the Dirac structures in \eq{qcd86} and \eq{qcd87} gives two conditions,
\begin{align}
8m^2 H_{2,NR} &= \frac{2}{r}H_{1,NR}'+\frac{1}{r^2}\big[\ell(\ell+1)-j(j+1)\big]H_{1,NR} = \frac{2}{r}H_{1,NR}'+\frac{1}{r^2}\big[\pm(2j+1)+1\big]H_{1,NR}  \label{qcd91} \crt
m(E_b-V)H_{1,NR} &= -4m^2r H_{2,NR}' + \big[\pm(2j+1)-1\big]2m^2H_{2,NR} \hspace{3cm} \mbox{for}\ \ \ell=j\pm 1  \label{qcd92}
\end{align}
Using the expression \eq{qcd91} for $8m^2 H_{2,NR}$ in the radial equation \eq{qcd82} gives the expected NR radial equation,
\begin{align} \label{qcd93}
H_{1,NR}''+ \Big[m(E_b-V)-\frac{\ell(\ell+1)}{r^2}\Big]H_{1,NR} &= 0
\end{align}
To check the self-consistency of \eq{qcd91} with \eq{qcd92} we may use \eq{qcd91} to express $H_{2,NR}$ and $H_{2,NR}'$ in terms of $H_{1,NR},\, H_{1,NR}'$ and $H_{1,NR}''$ and use this in \eq{qcd92}. The result agrees with \eq{qcd93}.

Using the expression \eq{qcd91} for $H_{2,NR}$ in the wave function \eq{qcd86} we have
\begin{align} \label{qcd94}
\Phi_{++}^{NR} = -\frac{i}{2m}(1+\gz)\Big\{\alv\cdot\nv H_{1,NR}(r)-\alv\cdot\xv\Big[\inv{r}H_{1,NR}' +\frac{1}{2r^2}\big[\pm(2j+1)+1\big]H_{1,NR}\Big]\Big\}\sph
\end{align}
Separating $\nv$ into its radial and angular derivatives,
\begin{align}
\alv\cdot\nv &= (\alv\cdot\xv)\,\inv{r}\partial_r -i\inv{r^2}\,\alv\cdot\xv\times\bs{L} \label{qcd95}
\end{align}
the radial derivative of $H_1$ cancels, so that the $\ell=j\pm 1$ NR wave functions are,
\begin{align} \label{qcd96}
\Phi_{++}^{NR} = \frac{i}{2mr^2}(1+\gz)\Big\{\halft\alv\cdot\xv \big[\pm(2j+1)+1\big]+i\alv\cdot\xv\times\Lv\Big\}H_{1,NR}\sph
\end{align}

\subsection{*\,$q \bar q$ bound states in motion \label{secVII.C}}
%%%%%%%%%%%%%%%%%%

A perturbative expansion for bound states should at each order respect Poincar\'e invariance. The spatial extent and mutual interactions of the constituents make this non-trivial. The bound state energy needs to have the correct dependence on the momentum, $E(\Pv)=\sqrt{M^2+\Pv^2}$, and scattering amplitudes (form factors) should transform covariantly under rotations and boosts.  

For Positronium this requires a frame dependent combination of Coulomb and transverse photon exchange, as discussed in section \ref{secV.C}. However, the \order{\alpha_s^0} instantaneous potential arising from the homogeneous solution of Gauss' law in QCD (section \ref{secVII.A}) must ensure Poincar\'e invariance on its own, without assistance from \order{\as} gluon exchange. This is analogous to $e^+e^-$ bound states in QED$_2$, due to the absence of transverse photons in $D=1+1$ dimensions. In that case it is essential that the potential is linear (section \ref{secVI.A4}). The correct frame dependence of the energy $E(\Pv)$ turns out to be similarly ensured for $q \bar q$ states, due to the linearity of the potential \eq{qcd8}. I have not considered $qqq$ states, which have a different potential \eq{qcd14}.

\subsubsection{The bound state equation \label{secVII.C1}}
%%%%%%%%%%%%%%%%%%

In a general frame the $\Pv=0$ state \eq{qcd37} becomes, at $t=0$,
\begin{align} \label{qcd97}
\ket{M,P} = \inv{\sqrt{N_c}}\sum_{A,B}\int d\xv_1 d\xv_2\,\bar\psi^A(\xv_1)e^{i\Pv\cdot(\xv_1+\xv_2)/2}\delta^{AB}\Phip(\xv_1-\xv_2)\psi^B(\xv_2)\ket{0}
\end{align}
which is an eigenstate of the momentum operator $\bm{\mP}$ \eq{A2} with eigenvalue $\Pv$. In the following I take $\Pv=(0,0,P)$ along the $z$-axis. The derivatives in $\mH_0^{(q)}$ \eq{qcd40} act after the partial integration also on $\exp[i\Pv\cdot(\xv_1+\xv_2)/2]$, giving rise to a new term in the bound state equation \eq{qcd43},
\begin{align} \label{qcd98}
i\nv\cdot\acomb{\alv}{\Phip(\xv)}-\halft \Pv\cdot \comb{\alv}{\Phip(\xv)}+m\comb{\gz}{\Phip(\xv)} &= \big[E-V(\xv)\big]\Phip(\xv)
\end{align}
The potential $V(\xv) = V'|\xv|$ is independent of the bound state momentum $\Pv$, being determined by the instantaneous positions $\xv_{1,2}$ of the quarks. An alternative form of this BSE is
\begin{align} \label{qcd99}
\Big[i\rnab\cdot\alv -\halft\big(E-V+ P\alz\big)+m\gz\Big]\Phip(\xv)  + \Phip(\xv) \Big[i\lnab\cdot\alv -\halft\big(E-V- P\alz\big)-m\gz\Big]=0
\end{align} 
It is possible to express the BSE equivalently as two coupled equations,
\begin{align} \label{qcd100}
\Big[\frac{2}{E-V}\big(i\alv\cdot\nv+m\gz-\halft\alv\cdot\Pv\big)-1\Big]\Phip &= -\frac{2i}{(E-V)^2}\Pv\cdot\nv\Phip+\frac{V'}{r(E-V)^2}\com{i\alv\cdot\xv}{\Phip}  \nn \crt
\Phip\Big[\big(i\alv\cdot\lnab-m\gz+\halft\alv\cdot\Pv\big)\frac{2}{E-V}-1\Big] &= \frac{2i}{(E-V)^2}\Pv\cdot\nv\Phip-\frac{V'}{r(E-V)^2}\com{i\alv\cdot\xv}{\Phip} 
\end{align}

%%%%%%%%%%%%%%%%%%%%%%%%%%
\begin{tcolorbox}
\textit{Exercise \ref{e21}:} Derive the coupled equations \eq{qcd100} from the bound state equation \eq{qcd98}. \\
\textit{Note:} Can you find a simpler derivation than the one presented in \ref{e21}?
\end{tcolorbox}
%%%%%%%%%%%%%%%%%%%%%%%%%%

Since $\Pv$ breaks rotational symmetry in \eq{qcd98} (except for rotations around the $z$-axis) the radial and angular variables do not separate as in \eq{qcd48}. This makes a solution of the BSE more challenging. In QED$_2$ $\Phip(x)$ can be expressed \eq{2d24b} in terms of the rest frame wave function evaluated at a ``boost invariant'' variable $\tau(x)$ \eq{2d23}. A similar relation works here as well, but only at $\xtr=(x,y)=0$. $\Phip(0,0,z)$ then serves as a boundary condition on the BSE \eq{qcd98}.

Before turning to the expression for $\Phip(0,0,z)$ I consider the case of a vanishing potential. The exact solution can be found for $V=0$, at any $P$ and $\xv$. This provides a boundary condition for the $V \neq 0$ BSE in the limit $r \to 0$, in which $E-V'r \to E$ on the rhs. of the BSE \eq{qcd98}. Solving the partial differential equation with boundary conditions at $\xtr=0$ and $r \to 0$ should determine the wave function for all $\xv$ (but this remains to be demonstrated).

\subsubsection{Solution of the $P \neq 0$ bound state equation for $V(\xv)=0$ \label{secVII.C2}}
%%%%%%%%%%%%%%%%%%

The free solution of \eq{qcd98} is, for $\Pv=(0,0,P)$,
\begin{align} \label{qcd101}
\Phip_{V=0}(\xv) &= \exp(-\halft \xi\alz)\,\Phi_{V=0}^{(0)}(\xv_R)\exp(\halft \xi\alz) \crt
\xv_R = (x,y,z\cosh\xi)& \hspace{2cm} E = M\cosh\xi \hspace{2cm} P = M\sinh\xi \nn
\end{align}
where $\Phi_{V=0}^{(0)}(\xv)$ is the solution of the BSE with $V=0$ in the rest frame. Its relation to $\Phip_{V=0}(\xv)$ corresponds to standard Lorentz contraction, with the $j=1/2$ boost representations $\exp(\pm\halft \xi\alz)$ familiar from the Dirac equation.

I denote by $B(\xv)$ the lhs. of the BSE \eq{qcd99} with $V=0$ and $\Phip(\xv) = \Phip_{V=0}(\xv)$ given by \eq{qcd101}. Thus $B=0$ is required for \eq{qcd101} to be a solution. Multiplying by $e^{\xi\alz/2}$ from the left and $e^{-\xi\alz/2}$ from the right,
\begin{align} \label{qcd102}
e^{\xi\alz/2}B(\xv)e^{-\xi\alz/2} &= e^{\xi\alz/2}\big[i\rnab\cdot\alv -\halft\big(E+ P\alz\big)+m\gz\big]e^{-\xi\alz/2}\,\Phi_{V=0}^{(0)}(\xv_R)  \nn\crt
&+ \Phi_{V=0}^{(0)}(\xv_R)e^{\xi\alz/2} \big[i\lnab\cdot\alv -\halft\big(E- P\alz\big)-m\gz\big]e^{-\xi\alz/2}
\end{align}
Since $z_R = z\cosh\xi$ we have $\partial_z = \cosh\xi\,\partial_{z_R}$, and
\begin{align} \label{qcd103}
i\alz\rder_z &= e^{\xi\alz}\alz i\rder_{z_R} - \sinh\xi\,i\rder_{z_R} \nn\crt
i\alz\lder_z &= i\lder_{z_R}\,\alz e^{-\xi\alz} +i\lder_{z_R} \sinh\xi
\end{align}
The terms $\propto \sinh\xi$ give Dirac scalar contributions to \eq{qcd102}, and cancel each other. Using $E\pm P\alz = M\exp(\pm\xi\alz)$, $i\ntr\cdot\atr\,\exp(-\xi\alz/2) = \exp(\xi\alz/2)i\ntr\cdot\atr$ and similarly for the $m\gz$ terms we get,
\begin{align} \label{qcd104}
e^{\xi\alz/2}B(\xv)e^{-\xi\alz/2} = e^{\xi\alz}\big[i\rnab_R\cdot\alv -\halft M+m\gz\big]\Phi_{V=0}^{(0)}(\xv_R) 
+ \Phi_{V=0}^{(0)}(\xv_R) \big[i\lnab_R\cdot\alv -\halft M-m\gz\big]e^{-\xi\alz}
\end{align}
Expressing $\exp(\pm\xi\alz) = \cosh\xi \pm \alz\sinh\xi$ the coefficent of $\cosh\xi$ is the rest frame BSE at $\xv=\xv_R$, which $\Phi_{V=0}^{(0)}$ satisfies by definition. The BSE allows to relate the coefficients of $\alz\sinh\xi$, leaving the anticommutator with $\alz$,
\begin{align} \label{qcd105}
e^{\xi\alz/2}B(\xv)e^{-\xi\alz/2} = \sinh\xi\acomb{\alz}{\big(i\rnab_R\cdot\alv -\halft M+m\gz\big)\Phi_{V=0}^{(0)}(\xv_R)} = \halft M\sinh\xi\acomb{\alz}{\rh_-\Phi_{V=0}^{(0)}}
\end{align}
where $\rh_-$ is defined in \eq{qcd45} and evaluated at $V=0$. The explicit expressions for $\rh_-\Phi(\xv)$ in \eq{qcd61}, \eq{qcd72} and \eq{qcd81} are all $\propto V'$ and thus vanish for $V=0$. Hence $B(\xv)=0$ and $\Phip_{V=0}(\xv)$ of \eq{qcd101} solves the BSE for all $P$.

\subsubsection{Boost of the state $\ket{M,P}$ for $V(\xv)=0$ \label{secVII.C3}}
%%%%%%%%%%%%%%%%%%

Instead of solving the BSE at a finite momentum $\Pv$ we may boost the rest frame state. This is feasible for $V=0$ using the boost generator of free quarks. Suppressing the irrelevant color indices,
\begin{align} \label{add1}
\bs{\mK_0}(t) = t\bs{\mP}+\int d\xv\,\psi^\dag(\xv)\big[-\xv H_0+\halft i\alv\big]\psi(\xv)
\end{align} 
The expressions for the generators of translations $\bs{\mP}$ \eq{A2} in space and $\mH_0$ \eq{qcd39} in time (the free Hamiltonian) are,
\begin{align} \label{add2}
\bs{\mP} &= \int d\xv\,\psi^\dag(\xv)(-i\nv)\psi(\xv) \nn\crt
\mH_0 &= \int d\xv\,\psi^\dag(\xv)H_0\psi(\xv) \equiv \int d\xv\,\psi^\dag(\xv)(-i\alv\cdot\nv+m\gz)\psi(\xv)
\end{align} 

These operators satisfy the Lie algebra of the Poincar\'e group (I do not here consider rotations, and set $t=0$). The commutators of local operators $\mathcal{O} = \int d\xv\,\psi^\dag(\xv) O(\xv)\psi(\xv)$ satisfy
\begin{align} \label{add3}
\com{\mathcal{O}_i}{\mathcal{O}_j} = \int d\xv\,\psi^\dag(\xv) \com{O_i}{O_j}\psi(\xv)
\end{align}
This allows to verify the Lie algebra in terms of the structures $O_i$ (here $P^i=-i\partial_i$):
\begin{align} \label{add4}
\com{P^i}{P^j} &=0 \hspace{2cm} \mbox{since}\ \partial_i\partial_j = \partial_j\partial_i \nn\crt
\com{P^i}{K_0^j} &= \com{-i\partial_i}{-x^j(-i\alv\cdot\nv + m\gz)+\halft i\alpha_j} = i\delta^{ij}H_0 \nn\crt
\com{H}{K_0^i} &= \com{-i\alv\cdot\nv+m\gz}{-x^i H_0+\halft i\alpha_i} = i\alpha_i H_0+ \halft i\com{H_0}{\alpha_i} 
=\halft i \acom{\alpha_i}{H_0} = iP^i
\end{align}
An infinitesimal boost in the $z$-direction of the (non-interacting) state $\ket{M,P}$ with $\Pv = (0,0,P)$ is generated by the operator $1-id\xi\mK_0^z$, as verified by the eigenvalues,
\begin{align} \label{add5}
\mP^z(1-id\xi\mK_0^z)\ket{M,P} &= (1-id\xi\mK_0^z)\mP^z\ket{M,P}-id\xi\com{\mP^z}{\mK_0^z}\ket{M,P} = (P+d\xi E)(1-id\xi\mK_0^z)\ket{M,P} \nn\crt
\mH_0(1-id\xi\mK_0^z)\ket{M,P} &= (1-id\xi\mK_0^z)\mH_0\ket{M,P}-id\xi\com{\mH_0}{\mK_0^z}\ket{M,P} = (E+d\xi P)(1-id\xi\mK_0^z)\ket{M,P}
\end{align}
The expression \eq{qcd97} for $\ket{M,P}$ in terms of the wave function $\Phip$ allows to determine the wave function for $(1-id\xi\mK_0^z)\ket{M,P}$, and thus to deduce its frame dependence \eq{qcd101}.  

%%%%%%%%%%%%%%%%%%%%%%%%%%
\begin{tcolorbox}
\textit{Exercise \ref{e21b}:} Derive the frame dependence  \eq{qcd101} of $\Phip_{V=0}(\xv)$ using the boost generator $\mK_0^z$. \\
\textit{Hint:} Use the bound state equation in the form of \eq{qcd100} (with $V=0$).
\end{tcolorbox}
%%%%%%%%%%%%%%%%%%%%%%%%%%

The boost demonstrates that the relative normalizations of wave functions with different momenta $P$ is correctly given by \eq{qcd101}. This applies also to the interacting case ($V \neq 0$) considered next, since the $P$-dependence of the component $\Phip(\xv=0)$ is given by $V=0$.

\subsubsection{Solution of the $P \neq 0$ bound state equation at $\xtr=0$ \label{secVII.C4}}
%%%%%%%%%%%%%%%%%%

Apart from $\xtr=0$ the following requires $E=\sqrt{M^2+P^2}$ and a linear potential $V=V'z$ with $z>0$. The wave function for $z<0$ may be determined using parity or charge conjugation \eq{qcd54}. As in the $D=1+1$ case (section \ref{secVI.A4}) the coordinate $z$ is transformed into the variable $\tau(z)$
\begin{align} \label{qcd106}
\tau(z) \equiv \big[(E-V)^2-P^2\big]/V' = (M^2-2EV+V^2)/V' \hspace{2cm} (\xv_\perp=0)
\end{align}
Since $\tau(z)$ depends on $E$ the transformation $z \to \tau$ is different for the rest frame wave function $\Phi^{(0)}(0,0,z)$ compared to that for $\Phip(0,0,z)$. These two wave functions will be related at the same value of $\tau$, and therefore at different values of $z$. For $V \ll E$ (weak binding) $\tau(z) \simeq M^2/V'-2Mz\cosh\xi$ and the transformation is equivalent to $z \to z_R$ as in \eq{qcd101} (standard Lorentz contraction). I shall somewhat sloppily denote the wave functions expressed in terms of $\tau$ using the same symbols, $\Phi^{(0)}(\tau)$ and $\Phip(\tau)$. It should be kept in mind that these are related to the original wave functions at $\xv=(0,0,z)$ through the $P$-dependent transformation \eq{qcd106}.

The variable $\zeta(z)$ takes the place of the boost parameter $\xi$,  
\begin{align} \label{qcd107}
\cosh\zeta = \frac{E-V}{\sqrt{V'\tau}} = \sqrt{1+\frac{P^2}{V'\tau}} \hspace{2cm} \sinh\zeta = \frac{P}{\sqrt{V'\tau}}
\end{align}
$\zeta(z)$ depends on $P$ as well as $\tau$. The definition \eq{qcd106} shows that $V'\tau \geq -P^2$ for real values of $z$. Hence when $P \neq 0$ there is a range of $z$ for which $V'\tau <0$. To avoid considering complex values of $\zeta$ I shall assume values of $z$ and $P$ such that $\tau >0$. I discuss below how to determine the $\xtr=0$ wave function in the range where $\tau < 0$.

The solution of the BSE \eq{qcd99} at $\xtr=0$ is related to the rest frame wave function through
\begin{align} \label{qcd108}
\Phip(\tau) &= \exp(-\halft \zeta\alz)\,\Phi^{(0)}(\tau)\exp(\halft \zeta\alz)
\end{align}
The same relation holds also for $\ntr\Phip(\tau)$. This requires $\ntr \zeta =0$, which follows from $\ntr V(\xv) =0$ at $\xtr=0$. By construction, $\Phi^{(0)}(\tau)$ depends only on $\tau$, whereas $\Phip(\tau)$ has an explicit $P$-dependence through $\zeta$ \eq{qcd107}.

%%%%%%%%%%%%%%%%%%%%%%%%%%
\begin{tcolorbox}
\textit{Exercise \ref{e22}:} Show that $\Phip(\tau)$ given by \eq{qcd108} satisfies the BSE \eq{qcd99} at $\xtr=0$. \\
\textit{Hint:} Follow the proof of section \ref{secVII.C2} for  the $V=0$ case. Pay attention to derivatives of $\zeta$.
\end{tcolorbox}
%%%%%%%%%%%%%%%%%%%%%%%%%%

As seen in section \ref{secVII.B} the wave functions of all rest frame $q\bar q$ states are found by solving radial equations, which are ordinary differential equations in $r$. The relation \eq{qcd108} then determines $\Phip(0,0,z)$ and $\ntr\Phip(0,0,z)$ in all frames, when the $q\bar q$ pairs are aligned with $\Pv$. This boundary condition on the BSE (together with the one for $r \to 0$ based on \eq{qcd101}) should allow to determine $\Phip(\xv)$ at all $\xv$ by solving the partial differential equation \eq{qcd98} in $(x_\perp,z)$. This remains to be demonstrated.

For a rest frame wave function $\tau=(M-V'z)^2/V' \geq 0$ \eq{qcd106}, whereas in general $\tau \geq -P^2$. This leaves a gap $-P^2 \leq \tau < 0$ in the boundary condition \eq{qcd108}. In the $D=1+1$ case the analytic functions \eq{2d26} determine the solution for all $\tau$. Here we may use the analog of the expression \eq{2d24d}, 
\begin{align} \label{qcd109}
\left.(E-V)\frac{\partial\Phip(\xv)}{\partial P}\right|_z = \left.\frac{zP}{E}\partial_z\Phip(\xv)\right|_P-\halft\,\comb{\alz}{\Phip(\xv)} \hspace{1cm} \mbox{at}\ \ \ \xv=(0,0,z)
\end{align}
which is a consequence of \eq{qcd108}. On the lhs. the $|_z$ indicates that the $P$-derivative is to be taken at fixed $z$, while on the rhs. the $z$-derivative is at fixed $P$. The derivation is the same as in the $D=1+1$ case, see Exercise \ref{e15}. This equation determines $\Phip(0,0,z)$ for all $P$ and $z$, with the rest frame wave function $\Phi^{(0)}(0,0,z)$ serving as boundary condition at $P=0$. In particular, the solution covers the gap $-P^2\leq\tau < 0$.

\subsection{Properties of the $q \bar q$ bound states \label{secVII.D}}
%%%%%%%%%%%%%%%%%%

The $q \bar q$ wave functions have novel features at large values of the linear potential \eq{qcd8}. There is little \textit{ab initio} knowledge of strongly bound states since they are usually associated with large values of the coupling, \ie, non-perturbative dynamics. Here the confining potential is of \order{\alpha_s^0}, so relativistic binding is compatible with a perturbative expansion in $\as$. It is essential that the lowest order bound states, determined by the relativistic solutions $\Phip(\xv)$ of the BSE \eq{qcd98}, provide a reasonable approximation of the true states.

\subsubsection{String breaking and duality \label{secVII.D1}}
%%%%%%%%%%%%%%%%%%

A first issue is why we should trust the linear potential $V(r)=V'r$ at large $r$. ``String breaking'' will prevent the potential from reaching large values. I touched upon this already in section \ref{secVI.A7}, for QED in $D=1+1$ dimensions. The key appears to be quark-hadron duality, which is a pervasive feature of hadron data and poorly understood theoretically, see the review \cite{Melnitchouk:2005zr} and conference\footnote{``First Workshop on Quark-Hadron Duality and the Transition to pQCD'', \tt{http://www.lnf.infn.it/conference/duality05/} .}. 

Bound state solutions of the BSE with different eigenvalues $E_A \neq E_B$ are orthogonal,
\begin{align} \label{qcd110}
\bra{M_B,\Pv_B}M_A,\Pv_A\rangle \propto \delta(\Pv_A-\Pv_B)\delta_{A,B}
\end{align} 

%%%%%%%%%%%%%%%%%%%%%%%%%%
\begin{tcolorbox}
\textit{Exercise \ref{e23}:} Prove \eq{qcd110} for states with wave functions satisfying the BSE \eq{qcd98}. \\
\textit{Hint:} The proof is analogous to the standard one for non-relativistic systems.
\end{tcolorbox}
%%%%%%%%%%%%%%%%%%%%%%%%%%

\begin{figure}[h] \centering
\includegraphics[width=.6\columnwidth]{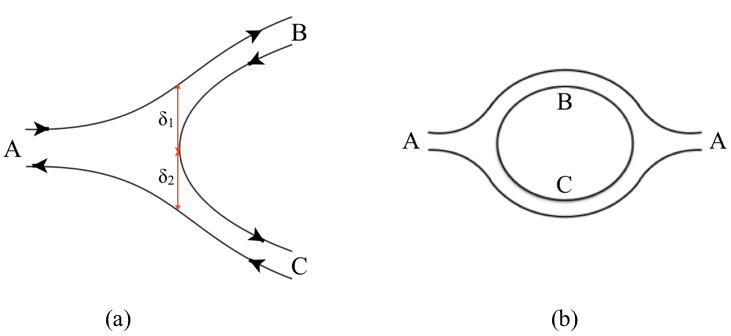}
\caption{(a) The overlap $\bra{B,C}A\rangle$ arises through the tunneling of a $q\bar q$ pair in the instantaneous field, ``string breaking''. Since the potential is linear the field energy is nearly the same before and after the split. (b) Hadron loop correction through the creation and annihilation of a $q\bar q$ pair, which may be important for unitarity. \label{f11}}
\end{figure}

However, the $q \bar q$ states are not orthogonal to $q \bar q\,q \bar q$ states. Contracting the quark fields as in \fig{f11}(a) gives
\beqa\label{qcd111}
\bra{B,C}A\rangle &=& \inv{\sqrt{N_C}}\int\Big[\prod_{k=A,B,C}d\xv_{1k}d\xv_{2k}\Big]
\exp\big\{i\halft\big[(\xv_{1A}+\xv_{2A})\cdot \Pv_A-(\xv_{1B}+\xv_{2B})\cdot \Pv_B-(\xv_{1C}+\xv_{2C})\cdot \Pv_C\big]\big\}\nn\\
&\times& \bra{0}\big[\psi^\dag(\xv_{2B})\Phi_B^\dag\gz\psi(\xv_{1B})\big]
\big[\psi^\dag(\xv_{2C})\Phi_C^\dag\gz\psi(\xv_{1C})\big]
\big[\bar\psi(\xv_{1A})\Phi_A\psi(\xv_{2A})\big]\ket{0} \\[2mm]
&=& -\frac{(2\pi)^3}{\sqrt{N_C}}\,\delta^3(\Pv_A-\Pv_B-\Pv_C)\int d\delv_1 d\delv_2\,e^{i\delv_1\cdot\Pv_C/2-i\delv_2\cdot\Pv_B/2}\tr\big[\gz\Phi_B^\dag(\delv_1)\Phi_A(\delv_1+\delv_2)\Phi_C^\dag(\delv_2)\big] \nn
\eeqa
where $\delv_1=\xv_{1B}-\xv_{2B}$ and $\delv_2=\xv_{1C}-\xv_{2C}$. Note that the process $A \to B+C$ is not mediated by a Hamiltonian interaction. It expresses that state $A$ has an overlap with $B+C$\,\footnote{I thank Yiannis Makris for a helpful comment concerning this.}. For example, Parapositronium has no overlap with $\ket{\gamma\gamma}$, but decays into this state do occur through the action of $\mH_{int}$. A bound state can overlap two bound states only for relativistic binding, so this phenomenon has no precedent for atoms. 

Quark-hadron duality in $e^+e^- \to hadrons$ means that the final state is described, in an average sense, by the $q \bar q$ state first created by the virtual photon. This holds also for individual resonances in the direct channel, and is consistent with an overlap between the $q \bar q$ and final hadron states. I show below (\ref{secVII.D4}) how the wave functions of highly excited bound states indeed reduce to those of free $q \bar q$ states. 

Another example of duality is the observation that the inclusive momentum distribution of hadrons produced in hard processes agrees with the perturbatively calculated gluon distribution \cite{Dokshitzer:2010zza}. This ``Local Parton Hadron Duality'' works down to low momenta. It shows that the transition from parton to hadron occurs with a minimal change of momentum, in the spirit of an overlap between states.

I shall assume that highly excited bound states give a gross description of the final multi-hadron states, similarly as high energy quarks and gluons describe hadron jets. This can and should be quantified by evaluating matrix elements like \eq{qcd111}.

%%%%%%%%%%%%%%%%%%%%%%%%%%%%%%%%%%%%%%%%%%%%%%%%%%%%%%%%
%%%%%%%%%%%%%%%%%%%%%%%%%%%%%%%%%%%%%%%%%%%%%%%%%%%%%%%%%

\subsubsection{Properties of the wave function at large separations $r$ \label{secVII.D2}}

\vspace{-.3cm}
Non-relativistic (Schr\"odinger) wave functions describe the distribution of a fixed number of bound state constituents. The single particle probability constraint $\int d\xv\, |\Phi|^2=1$ (global norm) determines the energy eigenvalues. The Dirac equation also describes a single electron, but in a strong potential ($V \agt m$) the wave function has negative energy components. In time-ordered dynamics these $E<0$ components correspond to $e^+e^-$ pairs in the wave function. This is related to the Klein paradox \cite{Klein:1929zz} and illustrated by the perturbative $Z$-diagram of \fig{f4}(b). For a linear potential the local norm of the Dirac wave function approaches a constant at large $r$ \eq{lin4} \cite{Plesset:1930zz}, where it is dominated by the $E<0$ components \eq{lin5}. This may be interpreted as positrons, which due to their positive charge are repelled by the potential, and accelerated to high momenta at large $r$.

\begin{figure}[h] \centering
\includegraphics[width=.6\columnwidth]{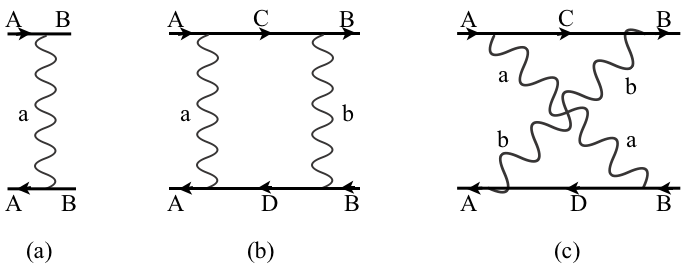}
\caption{Color structure of QCD Feynman diagrams. (a) Single gluon exchange. Planar (b) and non-planar (c) two-gluon exchange. The initial and final states are color singlets, implying a sum over the quark colors $A$ and $B$. \label{f12}}
\end{figure}

In section \ref{secII.B} we saw that the crossed ladder diagram of \fig{f2}(c) does not contribute to atomic bound states at lowest order in $\alpha$, \ie, to non-relativistic dynamics. When time ordered this Feynman diagram is the same as the pair-producing $Z$-diagram of \fig{f4}(b) (after the addition of the antifermion line). In QCD the uncrossed and crossed diagrams are distinguished also by their dependence on the number $N_c$ of colors. This is illustrated in \fig{f12} for $q\bar q \to q\bar q$ with single and double gluon exchange. The initial and final states are taken to be SU($N_c$) singlets,  implying a sum over the colors $A$ and $B$, each normalized by $1/\sqrt{N_c}$. 
\vspace{-.2cm}
%
%%%%%%%%%%%%%%%%%%%%%%%%%%%%%%%%%%%%%%%%%%%%%%%%%%%%%%%%
%%%%%%%%%%%%%%%%%%%%%%%%%%%%%%%%%%%%%%%%%%%%%%%%%%%%%%%%%

The color factors are thus, including also their behavior for $N_c\to\infty$:
\begin{align} \label{qcd112}
C(a) &= \frac{g^2}{(\sqrt{N_c})^2} \sum_{A,B}\sum_a T_{BA}^aT_{AB}^a = \frac{g^2}{N_c}\tr(T^aT^a) = g^2\,\frac{N_c^2-1}{2N_c} \sim \halft\, g^2 N_c \nn\crt
C(b) &= \frac{g^4}{N_c}\tr(T^bT^aT^aT^b) = g^4\,\Big(\frac{N_c^2-1}{2N_c}\Big)^2 \sim \quart\, g^4N_c^2 \nn\crt
C(c) &= \frac{g^4}{N_c}\tr(T^bT^aT^bT^a) = -\frac{g^4}{2N_c^2}\tr(T^aT^a) = -g^4\,\frac{N_c^2-1}{4N_c^2} \sim -\quart\, g^4
\end{align}
The suppression of $C(c)$ compared to $C(b)$ for $N_c\to\infty$ is a general feature of non-planar diagrams \cite{tHooft:1973alw,Witten:1979pi,Coleman:1980nk}. Due to the relation of diagram (c) in \fig{f12} with the $Z$-diagram of \fig{f4}(b) there are no such pair contributions to $q\bar q$ states in the $N_c\to\infty$ limit, despite the relativistic binding. In particular, there are no sea quarks in the 't Hooft model of QCD$_2$ \cite{tHooft:1974pnl}, where $N_c \to \infty$.
The connection between the sea quark distribution at low $\xbj$ and the behavior of the wave function at large quark separations (the virtual pairs) is apparent for QED$_2$ (section \ref{secVI.C3}). 

The normalizing integral of the $0^{-+}$ trajectory wave functions \eq{qcd60} can for all angular momenta $j$ be expressed as,
\begin{align}\label{qcd113}
\int d\xv\,\tr\Big[\Phi_{-+}^\dag(\xv)\Phi_{-+}(\xv)\Big] = 8\int_0^\infty dr\,r^2F_1^*(r)\Big[1-\frac{2V'}{(M-V)^3}\partial_r\Big]F_1(r)
\end{align}

%%%%%%%%%%%%%%%%%%%%%%%%%%
\begin{tcolorbox}
\textit{Exercise \ref{e24}:} Verify the expression \eq{qcd113} for the global norm of $\Phi_{-+}(\xv)$ in terms of $F_1(r)$.
\end{tcolorbox}
%%%%%%%%%%%%%%%%%%%%%%%%%%

For $r \to \infty$ (at fixed $j$) the asymptotic radial wave function $F_1(r)$,
\begin{align} \label{qcd114}
F_1(r) \simeq \inv{r}\,\exp\big[i(M-V)^2/4V'\big] \ \ \mbox{and} \ \ c.c. \ \ \ (r \to \infty)
\end{align} 
satisfies the radial wave equation \eq{qcd59} up to terms of \order{r^0}. 
The integrand (local norm) in \eq{qcd113} tends to $r^2|F_1(r)|^2$ for large $r$, and is thus independent of $r$. This feature is common to states of all quantum numbers. The probability density similarly tends to a constant also in lower spatial dimensions ($D=1+1$ \eq{2d34} and $D=2+1$), as well as in the Dirac equation for a linear potential \eq{lin4}. 

At large $r$ the radial derivative dominates in the expression \eq{qcd60} for $\Phi_{-+}(\xv)$: $\alv\cdot\nv \simeq \alv\cdot\xvh\,\partial_r $ where $\xvh = \xv/r$. Using also $F'_1(r\to\infty) \simeq \halft iV'r\,F_1(r\to\infty)$, 
\begin{align}\label{qcd115}
\Phi_{-+}(r\to\infty) &\simeq \Big[-\frac{2}{V'r}(i\alv\cdot\xvh\,\partial_r+m\gz)+1\Big]\gf\,F_1(r)Y_{j\lambda}(\Omega) = (1+\alv\cdot\xvh)\gf\,F_1(r)Y_{j\lambda}(\xvh)
\end{align}
The projector $\la_+$ which selects the $b^\dag$ operator in $\bar\psi(\xv)$ is as in \eq{fock2b}, 
\begin{align} \label{qcd116}
\bar\psi(\xv)\lla_+(\xv)&=  \int\frac{d\kv}{(2\pi)^32E_k}\sum_\lm \bar u(\kv,\lm)\,e^{-i\kv\cdot\xv}\,b_{\kv,\lm}^\dag \hspace{2cm}
\lla_+(\xv) \equiv \inv{2E}\Big[E-i\alv\cdot\lnab +\gz m\Big]
\end{align}
A partial integration in the $b_{\kv,\lm}^\dag$ component of the state \eq{qcd37} makes $\la_+$ operate on the wave function. Noting that
\begin{align} \label{qcd117}
\nv^2\Phi_{-+}(r\to\infty) \simeq \partial_r^2\Phi_{-+} \simeq -\quart(V'r)^2\Phi_{-+} \hspace{2cm}
E\,\Phi_{-+}\equiv\sqrt{-\nv^2+m^2}\,\Phi_{-+} \simeq \halft V'r\,\Phi_{-+}
\end{align}
we see that $\rla_+$ annihilates the asymptotic wave function,
\begin{align} \label{qcd118}
\rla_+\Phi_{-+}(r\to\infty) &= \inv{2E}\Big[E+i\alv\cdot\rnab +\gz m\Big]\Phi_{-+}
\simeq \inv{2}\Big[1+\frac{i\alv\cdot\xvh\,\partial_r}{E} \Big]\Phi_{-+} \nn\crt
&\simeq \halft(1-\alv\cdot\xvh)(1+\alv\cdot\xvh)\gf\,F_1(r\to\infty)Y_{j\lambda}(\xvh) =0
\end{align}
Consequently the state has no $b^\dag$ contribution at large $r$. Similarly $d^\dag$ does not contribute, leaving only the negative energy component $d\,b$. This is characteristic of the pair (sea quark) contributions from $Z$-diagrams. The negative kinetic energy is cancelled by the positive potential energy at large $r$, leaving a finite eigenvalue $M$. 

My tentative interpretation of this is as follows. The delicate cancellation between large negative kinetic and positive potential energies means that bound state components with $V(r) \gg M$ will not affect physical processes with resolution below those energies. Hadron loop corrections like in \fig{f11}\,b may change the wave function considerably. Nevertheless, since such corrections are due to overlaps the original wave function still approximately describes physical processes. Deep inelastic processes reveal the $\xbj \to 0$ sea quark distribution as far as its resolution permits.

\subsubsection{Discrete mass spectrum \label{secVII.D3}}

Due to the infinite sea one cannot impose a global normalization condition on the bound state wave functions. There is, however, a new local normalization condition. The solutions of the $q\bar q$ bound state equation \eq{qcd42} are generally singular at $M-V(r)=0$, as indicated by the coefficients $\propto 1/(M-V)$ in the radial equations \eq{qcd59}, \eq{qcd69} and \eq{qcd78}-\eq{qcd79}. A physical wave function should be locally normalizable at $M-V=0$, in line with the standard requirement of local normalizability at $r=0$.

The radial equation \eq{qcd59} of the $0^{-+}$ trajectory allows $F_1(r) \sim (M-V)^\gamma$ with $\gamma=0$ and $\gamma=2$ as $M-V(r) \to 0$. The integrand (local norm) in \eq{qcd113} is finite at $M-V=0$ only if $\gamma=2$. For $r \to 0$ we have as usual $F_1(r) \sim r^\beta$, with $\beta= j$ or $\beta= -j-1$. Only $\beta = j$ makes the integrand in \eq{qcd113} finite at $r=0$. The two constraints, at $M-V(r) = 0$ and $r=0$, determine the bound state masses, but leave the magnitude of the wave function unconstrained. This is a general feature, valid for states of all quantum numbers.

The vanishing of the radial wave functions at $M-V(r) = 0$ generalizes the vanishing for $r \to \infty$ of non-relativistic wave functions, which are defined only for $V(r) \ll M$. In the limit of non-relativistic dynamics ($m \to \infty$) the wave functions which are regular at $M-V=0$ become globally normalizable \cite{Dietrich:2012un}. 

The Dirac equation may be viewed as the limit of a two-particle equation where the mass $m_2$ of one particle tends to infinity, turning it into a static source. The point $V(r) = M$ (where $M$ includes $m_2$) recedes to $r=\infty$ as $m_2 \to \infty$. Hence there is no condition on the Dirac wave function at $M-V=0$, and the spectrum is continuous \cite{Plesset:1930zz}.  

\begin{figure}[h] \centering
\includegraphics[width=1.\columnwidth]{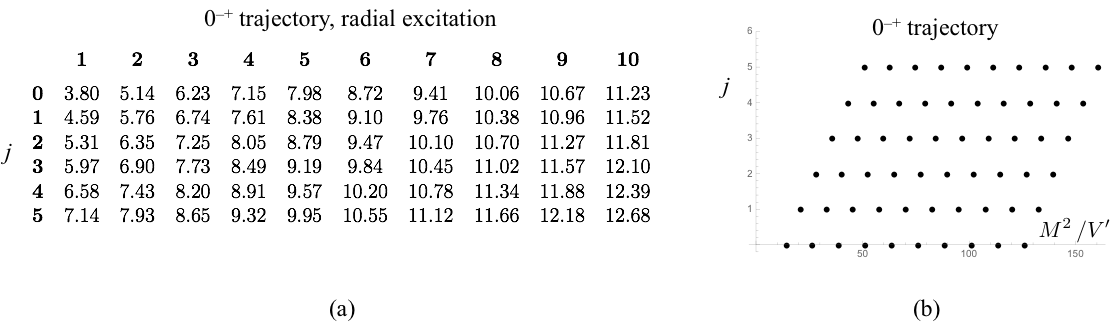}
\caption{(a) Masses $M$ of the mesons on the $0^{-+}$ trajectory for $m=0$, in units of $\sqrt{V'}$. (b) Plot of the spin $j$ {\em vs.} $M^2/V'$ for the states listed in (a). Figure taken from \cite{Hoyer:2016aew}. \label{f13}}
\end{figure}

The radial equation \eq{qcd59} can readily be solved numerically, subject to the boundary conditions $F_1(r\to 0) \sim r^j$ and $F_1(r\to M/V') \sim (M-V)^2$. As seen in \fig{f13}, for the linear potential $V(r)=V'r$ and quark mass $m=0$ the states lie on linear Regge trajectories and their parallel daughter trajectories. The mass spectra of the $0^{--}$ and $0^{++}$ trajectories are similar  \cite{Hoyer:2016aew}.

%%%%%%%%%%%%%%%%%%%%%%%%%%%%%%%%%%%%%%%%%%%%%%%%%%%%%%%%
%%%%%%%%%%%%%%%%%%%%%%%%%%%%%%%%%%%%%%%%%%%%%%%%%%%%%%%%%

%%%%%%%%%%%%%%
\subsubsection{Parton picture for $M \gg V(r)$ \label{secVII.D4}}

Duality in $e^+e^- \to hadrons$ implies that the perturbative process $e^+e^- \to  q\bar q$ gives an average of the direct channel resonances \cite{Melnitchouk:2005zr}. More generally, PQCD describes inclusive cross sections and hadron distributions (jets) at high energies in terms of unconfined quarks and gluons. Confinement (hadronization) is expected to become important as the color charges separate beyond the hadronic scale.

Duality implies that the wave functions of highly excited bound states $(M \gg 2m)$ should agree with the parton model, \ie, be compatible with the perturbative quark and gluon distribution for $V = V'r \ll M$. This is already indicated by the form of the bound state equation \eq{qcd42}, where $M$ and $V$ appear only in the combination $M-V$. When $V \ll M$ this is a free particle equation, with no negative energy ($d,\,b$) contributions from $Z$-diagrams. It is instructive to see how this limit emerges from the wave functions. 

I shall again use the $0^{-+}$ trajectory states to illustrate the emergence of the parton picture. The asymptotic expression \eq{qcd114} for $F_1(r)$ at large $r$ satisfies the radial equation also when $M\to\infty$ at finite $r$, up to terms of \order{M^0}.\footnote{In $D=1+1$ the $x\to\infty$ and $M\to\infty$ limits are similarly related. The QED$_2$ wave functions $\phi_0$ and $\phi_1$ \eq{2d26} depend on $x$ and $M$ only through the variable $\tau$ \eq{2d23}, and $\tau\to\infty$ in both limits.}
In the expression \eq{qcd60} for the wave function the radial derivative then dominates, $F_1' \simeq -i\halft M F_1$, so
\begin{align} \label{qcd119}
\Phi_{-+}(M\to\infty) \simeq \Big[\frac{2}{M}(i\alv\cdot\xvh\,\partial_r+m\gz)+1\Big]\gf\,F_1(r)Y_{j\lm}(\Omega)
\simeq (1+\alv\cdot\xvh)\gf\,F_1Y_{j\lm}
\end{align}
Consider again the projector $\la_+$ \eq{qcd116} which projects out the $b^\dag$ term in the $\bar\psi(\xv_1)$ field of the state \eq{qcd37}. After a partial integration it operates on $\Phi_{-+}(\xv)$, with the quark energy now giving
\begin{align} \label{qcd120}
E\,\Phi_{-+}(M\to\infty)\equiv\sqrt{-\nv^2+m^2}\,\Phi_{-+} \simeq \halft M\,\Phi_{-+}
\end{align}
The dominance of the radial derivative in $\rla_+$ implies,
\begin{align} \label{qcd121}
\rla_+\Phi_{-+}(M\to\infty) \simeq \inv{2}\Big(1+\frac{i\alv\cdot\xvh\,\partial_r + \gz m}{E} \Big)\Phi_{-+}
\simeq \halft(1+\alv\cdot\xvh)(1+\alv\cdot\xvh)\gf\,F_1(r)Y_{j\lambda} = \Phi_{-+}
\end{align}
The result is opposite to that of \eq{qcd118}: Now only the valence quark operators $b^\dag,\,d^\dag$ contribute to the state,
\begin{align} \label{qcd122}
\ket{M}_{V\ll M} \simeq \int d\xv_1 d\xv_2 \int\frac{d\kv_1d\kv_2}{(2\pi)^6 4E_1E_2}\sum_{\lambda_1,\lambda_2} e^{-i(\kv_1\cdot\xv_1+\kv_2\cdot\xv_2)}\,F_1Y_{j\lambda}\big[\bar u(\kv_1,\lambda_1)\gf v(\kv_2,\lambda_2)\big] \,b_{\kv_1,\lambda_1}^\dag \,d_{\kv_2,\lambda_2}^\dag \ket{0}
\end{align}
Since $F_1Y_{j\lambda}$ depends only on $\xv_1-\xv_2$, and
\begin{align} \label{qcd123}
\kv_1\cdot\xv_1+\kv_2\cdot\xv_2 = \halft(\kv_1+\kv_2)(\xv_1+\xv_2) + \halft(\kv_1-\kv_2)(\xv_1-\xv_2)
\end{align}
the integral over $(\xv_1+\xv_2)/2$ sets $\kv_1=-\kv_2 \equiv \kv$. Denoting $\xv \equiv \xv_1-\xv_2$,
\begin{align} \label{qcd124}
\ket{M}_{V\ll M} \simeq \int d\xv \int\frac{d\kv}{(2\pi)^3 4E_k^2}\sum_{\lambda_1,\lambda_2} e^{-i\kv\cdot\xv}\big[\bar u(\kv,\lambda_1)\gf v(-\kv,\lambda_2)\big]\,F_1(r)Y_{j\lambda}(\hat\xv) \,b_{\kv,\lambda_1}^\dag \,d_{-\kv,\lambda_2}^\dag \ket{0}
\end{align}
For $j=0$ the angular integral over $\xvh$ gives, with $\xv\cdot\kv = kr\cos\theta$,
\begin{align} \label{qcd125}
\int_{-1}^1 dcos\theta\int_0^{2\pi} d\vphi\, e^{-ikr\cos\theta} \inv{\sqrt{4\pi}} = -\frac{i\sqrt{\pi}}{kr}\big(e^{ikr}-e^{-ikr}\big)
\end{align}
At large $M \gg V$ the phase $\phi(r)$ of $F_1(r)$ \eq{qcd114} changes quickly with $r$, allowing to estimate the $r$-integral using the method of stationary phase,
\begin{align} \label{qcd126}
\phi(r) = \pm kr+(M-V)^2/4V' \hspace{1cm} \partial_r\phi(r_s) = 0 \hspace{1cm} V'r_s = M-2k \hspace{1cm} \phi(r_s) = k(M-k)/V'
\end{align}
The fact that only the $\exp(+ikr)$ term in \eq{qcd125} contributes sets $\cos\theta = -1$, \ie, $\xv$ and $\kv$ are antiparallel (with my choice of asymptotic solution in \eq{qcd114}). The stationary phase method may again be used in the $k$-integration, since $\phi(r_s)$ depends sensitively on $k$, 
\begin{align} \label{qcd127}
\partial_k\phi(r_s) = 0 \hspace{2cm} k_s = \halft M
\end{align}
Thus the quark and antiquark momenta are each half the resonance mass, as expected by energy conservation. With $k=M/2$ in \eq{qcd126} we have $V'r_s \ll M$, \ie, the  stationary value $r_s$ is compatible with the limit assumed in \eq{qcd119}.

A similar result will be obtained for any (fixed) angular momentum $j$, since each term in the angular integral corresponding to \eq{qcd125} will be $\propto \exp(\pm kr)$, multiplied by powers of $kr$ which do not affect the stationary phase approximation. Thus the wave functions of highly excited states on the $0^{-+}$ trajectory \eq{qcd55} take the form
\begin{align} \label{qcd128}
\ket{M}_{V\ll M} \propto \int d\hat\kv \sum_{\lambda_1,\lambda_2} \big[\bar u(\kv,\lambda_1)\gf v(-\kv,\lambda_2)\big]\,Y_{j\lambda}(\hat\kv) \,b_{\kv,\lambda_1}^\dag \,d_{-\kv,\lambda_2}^\dag \ket{0} \hspace{2cm} |\kv| = \halft M
\end{align}
The bound state wave function thus agrees with that of free $q\bar q$ states perturbatively produced by a current with corresponding $J^{PC}$ quantum numbers, as expected by duality. This may be used to determine the absolute normalization of highly excited bound states, as discussed in section IV of \cite{Dietrich:2012un}.

%%%%%%%%%%%%%%
\subsection{*\,Glueballs in the rest frame \label{secVII.E}}

I consider states of two transversely polarized gluons $\ket{gg}$, bound by the instantaneous linear potential $V_{gg}^{(0)}$ \eq{qcd20},
\begin{align}\label{qcd129}
V_{gg}(r) = \sqrt{\frac{N}{C_F}}\, \la^2\,r = \frac{3}{2}\, \la^2\,r \equiv V_g'r
\end{align}
The \order{\as} instantaneous gluon exchange $V_{gg}^{(1)}$ in \eq{qcd20} as well as higher Fock components ($\ket{ggg}$, $\ket{ggq\bar q}\ldots$) are ignored. Hence the Hamiltonian \eq{eII25} is approximated as $\mH=\mH_0+\mH_V$, where $\mH_V$ \eq{eII32} generates the linear potential and the kinetic term in \eq{eII25},
\begin{align} \label{qcd130}
\mH_0 &= \int d\xv\big[\halft E_{a,T}^iE_{a,T}^i +\halft A_{a,T}^i(-\nv^2)A_{a,T}^i\big]
\end{align}
involves only transversely polarized gluons $A^i_{a,T}$, which satisfy $\nv\cdot \Av_{a,T}=0$, and their conjugate electric fields $-E_{a,T}^i$. The canonical commutation relations \eq{eII24} imply
\begin{align} \label{qcd131}
\com{\mH_0}{A^i_{a,T}(\xv)} &= iE^i_{a,T}(\xv)  &\com{\mH_0}{E^i_{a,T}(\xv)} = i\nv^2 A^i_{a,T}(\xv)
\end{align}
Consequently the bound state condition
\begin{align} \label{qcd132}
(\mH_0+\mH_V)\ket{gg} = M\ket{gg}
\end{align}
requires $\ket{gg}$ to have both $A$ and $E$ components. In terms of the wave functions $\Phi^{ij}(\xv_1-\xv_2)$,
\begin{align} \label{qcd133}
\ket{gg} &\equiv \int d\xv_1 d\xv_2 \big[A^i_{a,T}(\xv_1)A^j_{a,T}(\xv_2)\Phi^{ij}_{AA}(\xv_1-\xv_2)+A^i_{a,T}E^j_{a,T}\Phi^{ij}_{AE}+E^i_{a,T}A^j_{a,T}\Phi^{ij}_{EA}+E^i_{a,T}E^j_{a,T}\Phi^{ij}_{EE}\big]\ket{0} 
\end{align}
where sums over the color $a$ and 3-vector indices $i,j$ are understood (here $A$ is not a color index!). The constituent $\Av$ and $\Ev$ fields are assumed to be normal ordered (their mutual commutators are subtracted).

As shown in section \ref{secVII.A3} the action of $\mH_V$ on $A^i_{a,T}(\xv_1)A^j_{a,T}(\xv_2)\ket{0}$ gives the potential \eq{qcd129}.
Since $\mE_a(\yv)$ \eq{eII31aa} has similar commutators with the $A$ and $E$ fields,
\begin{align} \label{qcd134}
\com{\mE_a(\yv)}{A_d^i(\xv)} &= -i\,f_{abd}A_b^i(\xv)\delta(\xv-\yv) \nn\crt
\com{\mE_a(\yv)}{E_d^i(\xv)} &= -i\,f_{abd}E_b^i(\xv)\delta(\xv-\yv)
\end{align} 
the same potential \eq{qcd129} is obtained for all four components of $\ket{gg}$ in \eq{qcd133},
\begin{align} \label{qcd135}
\mH_V\ket{gg} = \int d\xv_1 d\xv_2\,V_{gg}(|\xv_1-\xv_2|) \big[A_{a}(\xv_1)A_{a}(\xv_2)\Phi_{AA}(\xv_1-\xv_2) +A_{a}E_{a}\Phi_{AE} +E_{a}A_{a}\Phi_{EA} +E_{a}E_{a}\Phi_{EE}\big]\ket{0}
\end{align}
where I suppressed the 3-vector indices $i,j$ and the label $T$ of the transverse fields, which are unaffected by $\mH_0$ and $\mH_V$. Using the commutation relations \eq{qcd131},
\begin{align} \label{qcd136}
\mH_0\ket{gg} = i\int d\xv_1 d\xv_2\,\Big\{&\big[E_a(\xv_1)A_a(\xv_2)+A_a(\xv_1)E_a(\xv_2)\big]\Phi_{AA}(\xv_1-\xv_2)
+\big[E_aE_a+A_aA_a\nv^2\big]\Phi_{AE} \nn\crt
&+\big[A_aA_a\nv^2+E_aE_a\big]\Phi_{EA}+ \big[A_aE_a+E_aA_a\big]\nv^2\Phi_{EE}\Big\}\ket{0}
\end{align}
where $\nv$ differentiates $\Phi(\xv_1-\xv_2)$ wrt. $\xv_1-\xv_2$. 

The stationarity condition \eq{qcd132} implies the following relations between the wave functions $\Phi(\xv)$:
\begin{align} \label{qcd137}
\nv^2(\Phi_{AE}+\Phi_{EA}) &= -i(M-V)\Phi_{AA} \nn\crt
\Phi_{AA}+\nv^2\Phi_{EE} &= -i(M-V)\Phi_{AE} \nn\crt
\Phi_{AA}+\nv^2\Phi_{EE} &= -i(M-V)\Phi_{EA} \nn\crt
\Phi_{AE}+\Phi_{EA} &= -i(M-V)\Phi_{EE}
\end{align}
where $V = V_{g}'|\xv| = V'_g r$ as in \eq{qcd129}. This implies
\begin{align} \label{qcd138}
\Phi_{AE} &= \Phi_{EA} = -\halft i(M-V)\Phi_{EE} \nn\crt
\Phi_{AA} &= \inv{M-V}\,\nv^2\big[(M-V)\Phi_{EE}\big] \nn\crt
\inv{M-V}\,\nv^2&\big[(M-V)\Phi_{EE}\big] + \nv^2\Phi_{EE} = -\halft (M-V)^2 \Phi_{EE}
\end{align}
The last equation is
\begin{align} \label{qcd139}
\nv^2\Phi_{EE}(\xv)-\frac{V_g'}{M-V}\partial_r\Phi_{EE}(\xv)-\frac{V_g'}{r(M-V)}\Phi_{EE}(\xv)+\quart (M-V)^2 \Phi_{EE}(\xv) = 0
\end{align}
Separating the radial and angular dependence according to
\begin{align} \label{qcd140}
\Phi_{EE}(\xv) = F(r)Y_{\ell\lambda}(\Omega)
\end{align}
where $Y_{\ell\lambda}$ is the standard spherical harmonic function, the radial equation becomes
\begin{align} \label{qcd141}
F''(r) + \Big(\frac{2}{r}-\frac{V_g'}{M-V}\Big)F'(r) +\Big[\quart(M-V)^2-\frac{V_g'}{r(M-V)}-\frac{\ell(\ell+1)}{r^2}\Big]F(r) = 0
\end{align}

There is a single dimensionful parameter $V_g'$. Scaling $r = R/\sqrt{V_g'}$ and $M = \mM\sqrt{V_g'}$ the bound state equation in terms of the dimensionless variables $R, \mM$ becomes
\begin{align} \label{qcd142}
\partial_R^2 F(R) + \Big(\frac{2}{R}-\frac{1}{\mM-R}\Big)\partial_R F(R) +\Big[\quart(\mM-R)^2-\frac{1}{R(\mM-R)}-\frac{\ell(\ell+1)}{R^2}\Big]F(R) = 0
\end{align}

%%%%%%%%%%%%%%%%%%%%%%%%%%%%%%%%%%%%%%%%%%%%%%%%%%%%%%%%
\begin{figure}[h] \centering
\includegraphics[width=.6\columnwidth]{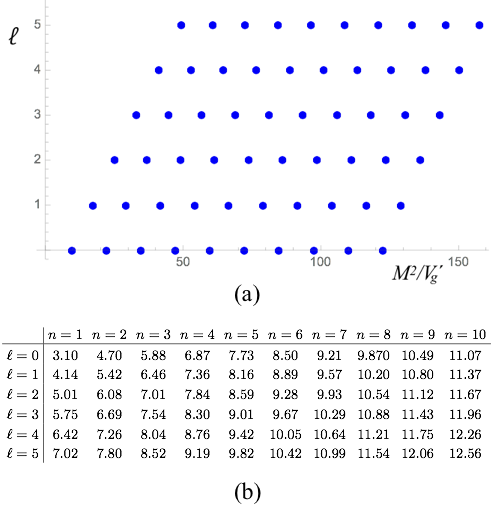}
\caption{(a) Glueball spectrum: Orbital angular momentum $\ell$ versus $M^2/V'_g$. (b) Glueball masses $\mM=M/\sqrt{V'_g}$ of the radial equation \eq{qcd142}. \label{f14}}
\end{figure} \parskip.2cm
%%%%%%%%%%%%%%%%%%%%%%%%%%%%%%%%%%%%%%%%%%%%%%%%%%%%%%%%

For $R\to 0$ we have the standard behaviors $F \sim R^\alpha$, with $\alpha = \ell$ or $\alpha = -\ell-1$. Since $\Phi_{AA} \sim \partial_R^2 \Phi_{EE}$ only the $\alpha = \ell$ solution gives a locally finite norm at $R=0$. For $\mM-R\to 0$ with $F \sim (\mM-R)^\beta$ we have $\beta =0$, and a second solution $F \sim \log(\mM-R)$. Only the $\beta = 0$ solution gives a locally finite norm at $\mM-R = 0$. These constraints on the solutions of \eq{qcd142} at $R=0$ and $R=\mM$ determine the allowed masses $\mM$. 
The glueball states lie on approximately linear Regge and daughter trajectories (\fig{f14}). At \order{\alpha_s^0} the spectrum is independent of the vector indices $i,j$ of the wave function in \eq{qcd133}.
%We may estimate the glueball masses in GeV using $\la^2 \simeq 0.18$ GeV$^2$ according to \eq{eI1}, giving $V_g' = 1.5\,\la^2 = 0.27$ GeV$^2$. Then the mass of the lowest state $M(\ell =0, n=1) = 3.10 \sqrt{V_g'} \simeq 1.6$ GeV.

%\begin{table}[h]
%$\begin{array}{c|cccccccccc}
%& n=1 & n=2 & n=3 & n=4 & n=5 & n=6 & n=7 & n=8 & n=9 & n=10 \\
%\hline
%\ell= 0 & 3.10 & 4.70 & 5.88 & 6.87 & 7.73 & 8.50 & 9.21 & 9.870 & 10.49 & 11.07 \\
%\ell= 1 &  4.14 & 5.42 & 6.46 & 7.36 & 8.16 & 8.89 & 9.57 & 10.20 & 10.80 & 11.37 \\
%\ell= 2 & 5.01 & 6.08 & 7.01 & 7.84 & 8.59 & 9.28 & 9.93 & 10.54 & 11.12 & 11.67 \\
%\ell= 3 & 5.75 & 6.69 & 7.54 & 8.30 & 9.01 & 9.67 & 10.29 & 10.88 & 11.43 & 11.96 \\
%\ell= 4 & 6.42 & 7.26 & 8.04 & 8.76 & 9.42 & 10.05 & 10.64 & 11.21 & 11.75 & 12.26 \\
% \ell=5 & 7.02 & 7.80 & 8.52 & 9.19 & 9.82 & 10.42 & 10.99 & 11.54 & 12.06 & 12.56 \\
%\end{array}$
%\caption{Glueball masses $\mM=M/\sqrt{V'_g}$ of the radial equation \eq{qcd142}. \label{t1}}
%\end{table}

%%%%%%%%%%%%%%
\subsection{Spontaneous breaking of chiral symmetry \label{secVII.F}}

Let us now return to the bound state equation \eq{qcd43} of a meson ($q\bar q$) state in the rest frame,
\begin{align} \label{qcd143}
&i\nv\cdot\acom{\alv}{\Phi(\xv)}+m\com{\gz}{\Phi(\xv)} = \big[M- V(\xv)\big]\Phi(\xv)
\end{align}
where the \order{\alpha_s^0} potential is linear $V=V'|\xv|$ \eq{qcd8}. In section \ref{secVII.D3} we saw that the bound state mass spectrum is determined by the requirement that $|\Phi(\xv)|^2$ is integrable at $r = |\xv| =0$ and at $M-V'r=0$. The latter condition is the generalization of the standard condition that the solutions of the Schr\"odinger equation are normalizable.

So far I did not discuss the special case of $M=0$, in which case the constraints at $r=0$ and $M-V'r=0$ coincide  \cite{Dietrich:2012un,Hoyer:2018hdj}. Such solutions have vanishing four-momentum in all frames. This allows the $J^{PC} = 0^{++}$ state to mix with the ground state (vacuum) without violating Poincar\'e invariance. The chiral symmetry of the QCD action for massless quarks is then not manifest in the states, as in a spontaneous breakdown of chiral invariance. 

In this preliminary study I consider two degenerate flavors, $m_u=m_d=m$. Suppressing color and Dirac indices the states \eq{qcd37} are
\begin{align} \label{qcd144}
\ket{M,i} = \int d\xv_1 d\xv_2\,\bar q(\xv_1)\Phi(\xv_1-\xv_2)\tau^i\, q(\xv_2)\ket{0} \hspace{2cm} q = \left(\begin{array}{c} u \crt d \end{array} \right)
\end{align}
where the $\tau^i$ are Pauli matrices for isospin $I=1$ states and $\tau^i \to 1$ for $I=0$.

%%%%%%%%%%%%%%
\subsubsection{$M=0$ states with vanishing quark mass $m=0$ \label{secVII.F1}}
%%%%%%%%%%%%%%

The $J^{PC} = 0^{++}$, $I=0$ ``$\sigma$'' wave function may be expressed in the general form \eq{qcd48}, with three Dirac structures\footnote{The present radial functions $H_i$ ($i=1,2,3$) are distinct from the functions $H_1,H_2$ in \eq{qcd78} and \eq{qcd79}.} allowed by \eq{qcd55} for $j=0$,
\begin{align} \label{qcd145}
\Phi_\sigma(\xv) = H_1(r) + i\,\alv\cdot\hat\xv\,H_2(r) + i\,\gz\alv\cdot\hat\xv\,H_3(r)
\end{align}
where $\hat\xv = \xv/r$. For $M=m=0$ the BSE \eq{qcd143} is $i\nv\cdot\acom{\alv}{\Phi_\sigma}+V'r\,\Phi_\sigma=0$, and requires\footnote{A more detailed derivation is given in Exercise \ref{e25} for $m \neq 0$.}
\begin{align} \label{qcd146}
i\alv\cdot\hat\xv H_1'-\frac{2}{r}H_2-H_2'+\halft V'r\,\Phi_\sigma=0
\end{align}
The coefficients of the three Dirac structures impose $H_2 = -(2/V'r)H_1',\ H_3=0$ and
\begin{align} \label{qcd147}
H_1''(r)+ \inv{r}H_1'(r)+\quart(V'r)^2 H_1(r)=0
\end{align}
This differential equation has an analytic solution, giving the wave function
\begin{align} \label{qcd148}
\Phi_\sigma(\xv) &= N\big[J_0(\quart V'r^2)+ i\,\alv\cdot\hat\xv\,J_1(\quart V'r^2)\big] \hspace{2cm} (m=M=0) 
\end{align}
where $N$ is a normalization constant and $J_0,\,J_1$ are Bessel functions. The $\sigma$ state \eq{qcd144},
\begin{align} \label{qcd149}
\ket{\sigma} = \hat\sigma\ket{0} \hspace{2cm} \hat\sigma = \int d\xv_1 d\xv_2\,\bar q(\xv_1)\Phi_\sigma(\xv_1-\xv_2)q(\xv_2)
\end{align}
has $M=\Pv=0$, \ie, vanishing 4-momentum in all frames. When mixed with the vacuum it causes a spontaneous breaking of chiral symmetry. The $I=1$ (non-anomalous) chiral transformations are generated by
\begin{align} \label{qcd150}
Q_{5i} = \int d\xv q^\dag(\xv)\gf\halft\tau^i q(\xv)
\end{align}
which transform the $\sigma$ state into a Goldstone boson (pion)
\begin{align} \label{qcd151}
i\com{Q_{5i}}{\hat\sigma} &= \hat\pi_i \hspace{2.7cm} \hat\pi_i = \int d\xv_1 d\xv_2\,\bar q(\xv_1)\Phi_\pi(\xv_1-\xv_2)\tau^i q(\xv_2) \crt
i\com{Q_{5i}}{\hat\pi_j} &= -\delta_{ij}\,\hat\sigma \hspace{2cm} \Phi_\pi(\xv) = -i\Phi_\sigma(\xv)\gf \nn
\end{align}
Like $\hat\sigma\ket{0}$, the state $\hat\pi_i\ket{0}$ is an eigenstate of the \order{\alpha_s^0} Hamiltonian with vanishing 4-momentum in all frames. Chiral transformations thus transform a vacuum state with a $\sigma$ condensate into another one with a mixture of zero-mass pions.

It is common to describe the pion in terms of a local field $\vphi_{\pi,i}$, which is a good approximation at low momentum transfers. Then, recalling that $\ket{\pi_j} = \hat\pi_j\ket{0}$ is time independent and normalized as in \eq{qcd148} and \eq{qcd151},
\begin{align} \label{qcd152}
\bra{0}\vphi_{\pi,i}(x)\ket{\pi_j} &= \delta_{ij} \nn\crt
\vphi_{\pi,i}(x) &= -\frac{i}{4N}\bar q(x)\gf\halft\tau^i q(x)
\end{align}

Spontaneous chiral symmetry breaking implies that the Goldstone bosons $\ket{\pi_j}$ are annihilated by the axial vector current $j^\mu_{5i}(x)=\bar q(x)\gamma^\mu\gf\halft\tau^i q(x)$ and by its divergence $\partial_\mu j_{5i}^\mu(x)= 2im\, \bar q(x)\gf\halft\tau^i q(x)$. With $f_\pi \simeq 93$ MeV,
\begin{align}
\bra{0}\bar q(x)\gamma^\mu\gf\halft\tau^i q(x)\,\ket{\pi_j,P} &= i\,\delta_{ij}P^\mu\, f_\pi\,e^{-iP\cdot x} \label{qcd153} \crt
\bra{0}\bar q(x)\gf\halft\tau^i q(x)\,\ket{\pi_j,P} &= -i\,\delta_{ij}\frac{M_\pi^2}{2m}\,f_\pi\,e^{-iP\cdot x} \label{qcd154}
\end{align}
Since the Goldstone pion has $P^\mu=0$ in all reference frames the rhs. of \eq{qcd153} vanishes. The lhs. also vanishes: $-iN\tr(\gamma^\mu\gf\halft\tau^i \gz J_0(0)\gf\tau^j \gz) = iN\delta^{ij}\tr(\gamma^\mu)=0$. The rhs. of \eq{qcd154} is finite in the $m \to 0$ limit, since $M_\pi^2 \propto m$ \cite{GellMann:1968rz}. We have then
\begin{align}
\bra{0}\bar q(x)\gf\halft\tau^i q(x)\int d\xv_1 d\xv_2\,\bar q(\xv_1)\Phi_\pi(\xv_1-\xv_2)\tau^j q(\xv_2)\ket{0} &=
-iN\tr(\gf\halft\tau^i\gz\gf\tau^j\gz) = 4iN\,\delta^{ij} = -i\frac{M_\pi^2}{2m}\,f_\pi\,\delta^{ij} \nn
\end{align}
This determines the normalization of the pion wave function,
\begin{align} \label{qcd155}
\Phi_\pi(\xv) = i\,\frac{M_\pi^2}{8m}\,f_\pi \big[J_0(\quart V'r^2) + i\,\alv\cdot\hat\xv\,J_1(\quart V'r^2) \big]\gf
\end{align}

%%%%%%%%%%%%%%
\subsubsection{Finite quark mass $m_u=m_d=m \neq 0$  \label{secVII.F2}}
%%%%%%%%%%%%%%

The pion becomes massive ($M_\pi > 0$) through the explicit breaking of chiral symmetry by a non-vanishing quark mass $m \neq 0$. The $0^{++}$ $\sigma$ must remain massless ($M_\sigma=0$) to ensure that its mixing with the vacuum does not break Poincar\'e invariance. The wave function $\Phi_\sigma(\xv)$ which solves the BSE \eq{qcd143} with $M=0$ but $m \neq 0$ has the structure of \eq{qcd145} with
\begin{align} \label{qcd156}
H_1(r) &= N e^{-iV'r^2/4}\Big\{\Big[1-\frac{2m^2}{(V'r)^2}\Big]L_{\sfrac{im^2}{2V'}-\sfrac{1}{2}}\big(\halft iV'r^2\big)+\frac{2m^2}{(V'r)^2}L_{\sfrac{im^2}{2V'}+\sfrac{1}{2}}\big(\halft iV'r^2\big)\Big\} = N\big[ J_0(\quart V' r^2) + \morder{m^2}\big] \nn\crt
H_2(r) &= N e^{-iV' r^2/4}\Big\{\Big[\frac{2}{V'r^2}-i\Big]L_{\sfrac{im^2}{2V'}-\sfrac{1}{2}}\big(\halft iV'r^2\big)-\frac{2}{V'r^2}L_{\sfrac{im^2}{2V'}+\sfrac{1}{2}}\big(\halft iV'r^2\big)\Big\} = N\big[ J_1(\quart  V' r^2) + \morder{m^2}\big] \nn\crt
H_3(r) &= -\frac{2m}{V'r}\,H_2(r) 
\end{align}
where the $L_n(x)$ are Laguerre functions.

%%%%%%%%%%%%%%%%%%%%%%%%%%
\begin{tcolorbox}
\textit{Exercise \ref{e25}:} Verify the expressions \eq{qcd156} for radial functions $H_1(r),H_2(r)$ and $H_3(r)$.
\end{tcolorbox}
%%%%%%%%%%%%%%%%%%%%%%%%%%

The pion state in the rest frame is
\begin{align} \label{qcd157}
\ket{\pi,i} = \hat\pi_i\ket{0} = \int d\xv_1 d\xv_2\,\bar q(\xv_1)\Phi_\pi(\xv_1-\xv_2)\tau^i q(\xv_2)\ket{0}
\end{align}
The pion wave function has the form given in \eq{qcd57}, where $F_3=0$ for $j=0$. With the notational change $F_2 \to F_2/r$ this means
\begin{align} \label{qcd158}
\Phi_{\pi}(\xv) = \Big[F_1(r) + i\,\alv\cdot\hat\xv\,F_2(r) + \gz\,F_4(r)\Big]\gf
\end{align}
Using this in the bound state equation \eq{qcd143} and collecting terms with the same Dirac structure we get the conditions:
\begin{align}\label{qcd159}
\gf: \hspace{.5cm} & -\frac{2}{r}F_2-F_2'+m F_4=\halft (M_\pi-V)F_1 \nn \\
i\,\alv\cdot\hat\xv\,\gf:\hspace{.5cm} & F_1'= \halft (M_\pi-V)F_2 \nn \\
\gz\gf: \hspace{.5cm} & m F_1 = \halft (M_\pi-V)F_4
\end{align}
Eliminating $F_2$ and $F_4$ gives the radial equation
\begin{align} \label{qcd160}
F_1''+\Big(\frac{2}{r}+\frac{V'}{M_\pi-V}\Big)F_1' + \big[\quart (M_\pi-V)^2-m^2\big]F_1 = 0
\end{align}
For the regular solution $F_1(r\to 0) \sim r^0[1+ \morder{r^2}]$, $F_2(r\to 0) \sim r$ and $F_4(0)= F_1(0)\,2m/M_\pi$. Thus
\begin{align} \label{qcd161}
\Phi_\pi^{(0)}(\xv=0) = F_1(0)\Big(1+\frac{2m}{M_\pi}\gz\Big)\gf
\end{align}
The superscript on $\Phi_\pi$ reminds that this is the rest frame wave function. In a frame with $\Pv\neq 0$ the pion state takes the form \eq{qcd97} at $t=0$. For a general time, suppressing color and Dirac indices,
\begin{align} \label{qcd161a}
\ket{M_\pi,i,\Pv,t} = e^{-iP^0t}\int d\xv_1 d\xv_2\,\bar q(t,\xv_1)e^{i\Pv\cdot(\xv_1+\xv_2)/2}\Phi_\pi^{(\Pv)}(\xv_1-\xv_2)\tau^i\, q(t,\xv_2)\ket{0}
\end{align}
where $P^0=\sqrt{\Pv^2+M_\pi^2}$. The axial current identities \eq{qcd153} and \eq{qcd154} probe the pion state at $t=x^0$ and $\xv_1=\xv_2=\xv$, involving $\Phi_\pi^{(\Pv)}(0)$. Since $V'r = 0$ at $r=0$ the frame dependence of the wave function at the origin is that of a non-interacting state, given by \eq{qcd101}. The wave function of a pion with momentum $\Pv = {\bs{\hat\xi}}\, M_\pi \sinh|\bs{\xi}|$ is then, at the origin,
\begin{align} \label{qcd162}
\Phi_\pi^{(\Pv)}(\xv=0) &= e^{-\bs{\xi}\cdot\alv/2}\Phi_\pi^{(0)}(\xv=0)e^{\bs{\xi}\cdot\alv/2}
=F_1(0)\Big(1+\frac{2m}{M_\pi}\gz e^{\bs{\xi}\cdot\alv}\Big)\gf
=F_1(0)\Big[1+\frac{2m}{M_\pi^2}(\gz P^0 + \Pv\cdot\gv)\Big]\gf \nn\crt
\gz\Phi_\pi^{(\Pv)}(0)\gz &= -F_1(0)\Big[1+\frac{2m}{M_\pi^2}\slashed{P}\Big]\gf
\end{align}

The CSB matrix elements become
\begin{align} \label{qcd163}
\bra{0}\bar q(x)\gamma^\mu\gf\halft\tau^i q(x)\,\ket{M_\pi,j,\Pv,x^0} &= \delta_{ij}e^{-iP\cdot x}\,\tr\big\{\gamma^\mu\gf\gz\Phi_\pi^{(\Pv)}(0)\gz\big\}
= \delta_{ij}e^{-iP\cdot x}F_1(0)\,\frac{2m}{M_\pi^2}\,4P^\mu \nn\crt
\bra{0}\bar q(x)\gf\halft\tau^i q(x)\,\ket{M_\pi,j,\Pv,x^0}&= \delta_{ij}e^{-iP\cdot x}\,\tr\big\{\gf\gz\Phi_\pi^{(\Pv)}(0)\gz\big\} = \delta_{ij}e^{-iP\cdot x}(-4)F_1(0)
\end{align}
Comparing with the rhs. of relations \eq{qcd153} and \eq{qcd154} gives in both cases
\begin{align} \label{qcd164}
F_1(0) = i\,\frac{M_\pi^2}{8m}\,f_\pi
\end{align}
in agreement with our previous result \eq{qcd155} for $m=0$.

I leave to the future a more comprehensive study of spontaneous chiral symmetry breaking in the present context.

\vspace{3cm}

\section{Bound state epilogue} \label{secVIII}
 
I conclude with several remarks on bound states. The more subjective ones may serve to stimulate further discussion on these topics.

Confinement is an essential aspect of hadrons, and has been demonstrated in lattice QCD. Analytic approaches to confinement are often formulated in terms of quark and gluon Green functions, aiming to show that colored states are unphysical. This deals directly with the fundamental fields of the theory, and benefits from the accumulated experience with (non-)perturbative methods for local fields. A drawback is that one needs to prove that something does not exist, with little guidance from data.

Here I try to approach confinement from the opposite direction, in terms of the color singlet bound states that are the asymptotic states of QCD. A main challenge is that bound states are extended objects, and more difficult to deal with than the pointlike quarks and gluons. The experience gained from QED atoms is valuable, even though it does not address confinement. Experts regard atoms as ``non-perturbative'' \cite{RevModPhys.57.723}, yet use PQED for precise evaluations of binding energies. This cautions not to rush to judgement based on the non-perturbative nature of hadrons. 

The hadron spectrum has surprising ``atomic'' features, despite their large binding energies. Why can hadrons be classified by their valence quarks and $J^{PC}$ quantum numbers only \cite{Zyla:2020zbs}? Their gluon and sea quark constituents do not feature in the spectrum as clearly as they do in deep inelastic scattering. Why do hadron decays obey the OZI rule \cite{Okubo:1963fa,Zweig:570209,Iizuka:1966fk}, \eg, favor $\phi(1020) \to K\bar K$ over $\phi(1020) \to \pi\pi\pi$? What causes quark-hadron duality, which in various guises pervades hadron dynamics? These features are pictured by dual diagrams (\cite{Harari:1981nn,Rosner:1981np,Zweig:2015gpa} and \fig{f11}\,a) which show only valence quarks, no gluons. We lack understanding based on QCD.

Simple features are precious: experience shows that they often have correspondingly simple explanations. Taking the data at face value limits the options in choosing the approach. An explanation based on QCD needs to use perturbation theory, which is our only general analytic method. Perturbative methods are successful for the atomic spectrum as well as for hard scattering in QCD. Addressing bound states in motion with a perturbative expansion is possible in QED and conceivable in QCD. Imposing restrictive conditions on an explicit yet formally exact method may reveal the QCD solution, or the theoretical inconsistency of such conditions.

Quarks and gluons do not move faster than light. This implies retarded interactions between bound state constituents. In a Fock state expansion the retardation is described by higher Fock states, with gluons ``on their way'' between valence quarks. For the non-relativistic constituents of atoms photon exchange is almost instantaneous, so retardation is a higher order, relativistic correction. For light, relativistic quarks retarded interactions would be expected to be prominent and higher Fock states significant. Yet data suggests that most hadrons may be viewed as $q\bar q$ and $qqq$ states. Is it conceivable that the valence quark Fock states dominate also for light hadrons?

Gauge theories can, depending on the choice of gauge, have instantaneous interactions. The absence of the $\partial_t A^0$ and $\nv\cdot\Av_L$ terms in the action means that $A^0$ and the longitudinal $\Av_L$ fields do not propagate in time and space. Their values are determined by the gauge, which can be fixed over all space at an instant of time. This is illustrated by a comparison of the $A^0$ propagator in the Coulomb and Landau gauges, 
\begin{align} \label{bs1}
D_C^{00}(q^0,\qv) &= i\,\frac{1}{\qv^2} \hspace{2.6cm} \mbox{Coulomb gauge:}\ \nv\cdot \Av =0 \nn\crt
D_L^{00}(q^0,\qv) &= -i\frac{1-(q^0)^2/q^2}{q^2+\ieps}\hspace{1cm} \mbox{Landau gauge:}\ \ \partial_\mu A^\mu = 0
\end{align}
The Coulomb gauge propagator is independent of $q^0$ and thus $\propto \delta(t)$ after a Fourier transform $q^0 \to t$. The gauge fixing term $\propto (\partial_\mu A^\mu)^2$ of Landau gauge adds the missing derivatives $\partial_t A^0$ and $\nv\cdot\Av_L$ to the action, allowing all components of $A^\mu$ to propagate. The free field boundary condition of the perturbative $S$-matrix at $t=\pm\infty$ is also covariant, resulting in an explicitly Poincar\'e invariant expansion of scattering amplitudes in terms of Feynman diagrams in Landau gauge.

Bound states defined at an instant of time retain explicit symmetry only under space translations and rotations (in the rest frame). Coulomb gauge maintains these symmetries, sets $\Av_L=0$ and determines $A^0$ through Gauss' law,
\begin{align} \label{bs2}
\frac{\delta\mS_{QED}}{\delta A^0(x)} = -\nv^2 A^0(t,\xv)-e\psi^\dag\psi(t,\xv) = 0
\end{align}
The positions of the charges at any time $t$ determine $A^0$ instantaneously at all positions $\xv$. This gives the Coulomb potential $V(r)=-\alpha/r$ for Positronium, which is the dominant interaction in atomic rest frames. The $e^-e^+$ Fock state wave function is determined by the Schr\"odinger equation, and other Fock states are suppressed by powers of $\alpha$. These features make Coulomb gauge a common choice for bound state calculations.

Quantization in Coulomb gauge is complicated by the absence of a conjugate field for $A^0$, see \cite{Feinberg:1977rc,Christ:1980ku,Weinberg:1995mt}. Gauss' law \eq{bs2} is an operator relation which defines $A^0$ in terms of $\psi^\dag\psi$ as a non-local quantum field. The classical potential of the Schr\"odinger equation given by $A^0$ in Coulomb gauge is more simply realized in temporal gauge, $A^0=0$. Quantization is straightforward in temporal gauge, since $\Av_L$ does have a conjugate field, $\Ev_L$. Gauss' law is no longer an operator equation of motion since $A^0$ is fixed. Physical Fock states are defined to be invariant under time independent gauge transformations (which preserve $A^0=0$). This constrains the value of $\Ev_L$ for each physical state to be such that Gauss' law is satisfied (\cite{Willemsen:1977fr,Bjorken:1979hv,Christ:1980ku,Leibbrandt:1987qv,Strocchi:2013awa} and section \ref{secIV}), giving rise to the classical potential.

In a perturbative approach the classical potential is weak, being proportional to $\alpha$ or $\as$. The QCD action has no parameter like $\lqcd \sim 1$ fm$^{-1}$ for the confinement scale. However, Gauss' law determines $\Ev_L^a$ only up to a boundary condition. In section \ref{secIV.C2} I consider adding a homogeneous solution \eq{eII31a} to the gluon exchange term. This gives rise to a spatially constant gluon field energy density \eq{qcd7} for each color component of a Fock state. Being a color octet, $\Ev_L^a$ cancels in the sum over the color components of singlet Fock states, avoiding long range effects.

The homogeneous solution generates a confining potential of \order{\alpha_s^0}. In section \ref{secVII} I determined the potential for various Fock states and made first checks of the \order{\alpha_s^0} dynamics. Essential and non-trivial features include the gauge invariance of electromagnetic form factors and the correct dependence of the bound state energies on the CM momentum, $E(\Pv)$. String breaking and duality can arise through an overlap between single and multiple bound states as suggested by dual diagrams (\fig{f11}\,a). There is a $J^{PC}= 0^{++}$ solution with $P^\mu = 0$ in all frames whose mixing with the vacuum allows solutions with spontaneously broken chiral symmetry.

 A non-vanishing field energy density would explain why confinement is not seen in Feynman diagrams, which expand around free states. However, the homogeneous solution \eq{eII31a} represents a major departure from previous experience. It requires much further study, both of theoretical consistency and phenomenological relevance.

\begin{acknowledgments}
During my work on the topics presented here I have benefited from the collaboration and advice of many colleagues. Particular thanks are due to Jean-Paul Blaizot, Stan Brodsky, Dennis D.~Dietrich, Matti J\"arvinen, Stephane Peign\'e and Johan Rathsman. I am grateful for the hospitality of the University of Pavia, and the stimulating response to my lectures there in early 2020. During earlier stages of this project I have enjoyed visits of a month or more to ECT* (Trento), CERN-TH (Geneva), CP$^3$ (Odense), NIKHEF (Amsterdam), IPhT (Saclay) and GSI (Darmstadt). I have the privileges of Professor Emeritus at the Physics Department of Helsinki University. Travel grants from the Magnus Ehrnrooth Foundation have allowed me to maintain contacts and present my research to colleagues.
\end{acknowledgments}

\appendix
\renewcommand{\theequation}{\thesection.\arabic{equation}}

%%%%%%%%%%%%%%
\section{Solutions to exercises \label{exa}}
%%%%%%%%%%%%%%

%%%%%%%%%%%
\subsection{Order of box diagram \label{e1}}
The leading contribution comes from the range of the loop integral $\int d^4\ell$ where $\ell^0 \sim \alpha^2 m$ and $|\ellv| \sim \alpha m$. Thus $\int d^4\ell \sim \alpha^5$. The fermions are off-shell similarly to $E_{p_1}-m \sim \alpha^2$, as are the photons, $q^2 = (q^0)^2-\qv^2 \simeq -\qv^2 \sim \alpha^2$. Considering also the factor $e^4\sim\alpha^2$ from the vertices the power of $\alpha$ is altogether $5+4\times(-2)+2=-1$.

%%%%%%%%%%%
\subsection{Contribution of the diagrams in \fig{f2}(b,c) \label{e2}}
As in Ex. \ref{e1} the leading contribution comes from the range of the loop integral $\int d^4\ell$ where $\ell$ is of the order of the bound state momenta. Then the Dirac structures simplify as in $A_1$ of \eq{pro7}, giving $(2m)^4$. The two photon propagators reduce to the $\ell^0$-independent Coulomb exchanges contained in $A_1(\pv,\ellv)$ and $V(\ellv-\qv)$ of \eq{pro8}. The fermion propagators with momenta $\ell$ and $\ell-p_1-p_2 \equiv \ell-P$ may be expressed as in \eq{pro5}, keeping only the terms with the electron and positron poles. The relevant part of the $\ell^0$ integral is then
\begin{align} \label{ex2.1}
\int\frac{d\ell^0}{2\pi}\,\inv{\ell^0-E_\ell+\ieps}\,\inv{\ell^0-P^0+E_\ell-\ieps}
=\frac{i}{P^0-2E_\ell+\ieps}
\end{align}
Noting that the factor of $1/2E_\ell$ in each fermion propagator \eq{pro5} gives $1/(2m)^2$ we arrive at \eq{pro8}.

In the crossed diagram (c) of \fig{f2} the second factor in \eq{ex2.1} is $1/(q_1^0-p_2^0-\ell^0+E-\ieps)$. The pole in $\ell^0$ now has $\im\, \ell^0 <0$ as in the first factor of \eq{ex2.1}. This allows closing the contour in the $\im \,\ell^0 >0$ hemisphere giving no leading order contribution. In fact only the ladder diagrams shown in \fig{f1} contribute at leading \order{1/\alpha}. They generate the classical field of the bound state. 

%%%%%%%%%%%
\subsection{Derivation of \eq{pro36} \label{e6}}

For any (finite) momentum exchange the antifermion energy in the loops of \fig{f2}(b,c) is $\bar E= m_T+\morder{1/m_T}$. Since $p_2^0 = m_T$ we need only retain the negative energy pole term in the antifermion propagators (\cf\ \eq{pro5}), in which $-p_2^0+\bar E$ is of \order{1/m_T}. For the electron lines to go on-shell requires non-vanishing energy transfer $\ell^0-p_1^0 \neq 0$, which causes the antifermion propagator to be off-shell by \order{m_T}. The photon propagators are thus independent of $\ell^0$, \eg, $D^{00}(\ell-p_1)=i/(\ellv-\pv_1)^2$. The only relevant poles of the $\ell^0$ integration are in the antifermion propagators,
\begin{align} \label{ex6.1}
\int\frac{d\ell^0}{2\pi}\Big[ \frac{i}{\ell^0-p_1^0-\ieps}+\frac{i}{q_1^0-\ell^0-\ieps}\Big] = i\int\frac{d\ell^0}{2\pi}\,2\pi i\delta(\ell^0-p_1^0) = -1
\end{align}
The factor $-i$ at the antifermion vertices cancels with the $i$ of the Coulomb photon propagators. The standard rules for the electron line then gives \eq{pro36}.

In the diagram with three uncrossed photon exchanges with momenta $\ell_1-p_1,\ \ell_2-\ell_1$ and $q_1-\ell_2$ the two antifermion propagators similarly give
\begin{align} \label{ex6.2}
\int\frac{d\ell_1^0\,d\ell_2^0}{(2\pi)^2}\,\frac{i^2}{(\ell_1^0-p_1^0-\ieps)(\ell_2^0-q_1^0-\ieps)}
\end{align}
The five other diagrams with crossed photons ensure the convergence of the integrals similarly as in \eq{ex6.1}, and do not contribute when the integration contours of $\ell_1^0$ and $\ell_2^0$ are closed in the upper half plane. The full result is then due to \eq{ex6.2}, which equals 1 (given the convergence). If the contours are chosen differently then the same result will come from diagrams with crossed photons. The factors associated with the electron line are again given by the standard rules, making the three photon contribution equal to scattering from an external potential.

%%%%%%%%%%%
\subsection{Derivation of \eq{hdiag} \label{e7}}

Inserting the completeness condition for the Dirac wave functions,
\beq\label{ex7.1}
\sum_n \big[\Psi_{n,\alpha}(\xv)\Psi_{n,\beta}^\dag(\yv) +
\ovl\Psi_{n,\alpha}(\xv)\ovl\Psi_{n,\beta}^\dag(\yv)\big]=\delta_{\alpha\beta}\delta^3(\xv-\yv)
\eeq
into the Dirac Hamiltonian \eq{ham1} gives, recalling that the wave functions satisfy \eq{dir1} and \eq{dir2},
\beqa
H_D &=& \int d\xv\,d\yv\,\bar\psi_{\alpha'}(\xv)\big[-i\nv_\xv\cdot\gv+m+e\slashed{A}(\xv)\big]_{\alpha'\alpha}
\sum_n \big[\Psi_{n,\alpha}(\xv)\Psi_{n,\beta}^\dag(\yv) +
\ovl\Psi_{n,\alpha}(\xv)\ovl\Psi_{n,\beta}^\dag(\yv)\big]\psi_\beta(\yv)
\nn\\[2mm]
&=& \sum_n\int d\xv\,d\yv\,\psi_\alpha^\dag(\xv)\big[M_n \Psi_{n,\alpha}(\xv)\Psi_{n,\beta}^\dag(\yv)-\bar M_n \ovl\Psi_{n,\alpha}(\xv)\ovl\Psi_{n,\beta}^\dag(\yv)\big]\psi_\beta(\yv)\nn\\[2mm]
&=& \sum_n \big[M_n c_n^\dag c_n - \bar M_n \bar c_n \bar c_n^\dag\big] \to
\sum_n \big[M_n c_n^\dag c_n + \bar M_n \bar c_n^\dag \bar c_n\big] \label{ex7.2}
\eeqa
In the last step I normal-ordered the operators, neglecting the zero-point energies according to \eq{vac1}.

%%%%%%%%%%%
\subsection{The expressions \eq{vac2} for vacuum state \label{e8}}

(a) The $B$ and $D$ coefficients defined in \eq{cndef1} and \eq{cndef2} satisfy
\beq\label{ex8.1}
B_{mp}\ovl B_{np}+D_{mp}\ovl D_{np}=\sum_\pv \Psi_{m,\alpha}^\dag(\pv)\big[u_\alpha(\pv,\lambda)u_\beta^\dag(\pv,\lambda) + v_\alpha(-\pv,\lambda)v_\beta^\dag(-\pv,\lambda)\big]\ovl\Psi_{n,\beta}(\pv)
= \int\frac{d\pv}{(2\pi)^3}\Psi_{m,\alpha}^\dag(\pv)\ovl\Psi_{n,\alpha}(\pv) =0
\eeq
Multiplying by $(B^{-1})_{qm}({\ovl D}^{\,-1})_{rn}$ and summing over $m,n$ gives
\beq\label{ex8.2}
-(B^{-1})_{qm}D_{mr} = ({\ovl D}^{\,-1})_{rn}{\ovl B}_{nq}
\eeq
Using also $b^\dag_q d^\dag_r = -d^\dag_r b^\dag_q$ shows the equivalence of the two expressions for the vacuum state in \eq{vac2}.

(b) 
In order to verify that $c_n\ket{\Omega}=0$ we note that since $b_p$ essentially differentiates the exponent in $\ket{\Omega}$,
\beq\label{ex8.3}
B_{nq}b_q\ket{\Omega}=-B_{nq}\big(B^{-1}\big)_{qm}D_{mr}d_r^\dag\ket{\Omega} = -D_{nr}d_r^\dag\ket{\Omega}
\eeq
This cancels the contribution of the second term in the definition \eq{cndef1} of $c_n$. 
The demonstration that $\bar c_n$ annihilates the vacuum is similar. Thus
\beq\label{ex8.4}
c_n\ket{\Omega}=\bar c_n\ket{\Omega}=H_D\ket{\Omega}=0
\eeq

%%%%%%%%%%%
\subsection{Derivation of the identities \eq{dir10} \label{e9}}

We may start by evaluating (here I make no difference between lower and upper indices)
\begin{align} \label{ex9.1}
(\hat\xv\times\Lv)^i = \epsilon_{ijk} \hat x^j \epsilon_{kln} x^l(-i\partial_n) = (\delta_{il}\delta_{jn}-\delta_{in}\delta_{jl})\hat x^j x^l(-i\partial_n) =-i\hat x^i r\partial_r +ir \partial_i
\end{align}
Multiplying by $\alpha^i/r$ we have the first relation,
\begin{align} \label{ex9.2}
\inv{r}\alv\cdot\hat\xv\times\Lv = -i\alv\cdot\hat\xv\,\partial_r + i\alv\cdot\nv
\end{align}
The second identity:
\begin{align}  \label{ex9.3}
-i(\alv\cdot\nv)i(\alv\cdot\hat\xv) = \alpha^i\alpha^j \partial_i\frac{x^j}{r} = (\delta_{ij}+i\gf\epsilon_{ijk}\alpha^k)\Big(\frac{\delta^{ij}}{r}-\frac{x^ix^j}{r^3}+\frac{x^j}{r}\partial_i\Big)
= \frac{2}{r}+\partial_r +\inv{r}\gf\alv\cdot\Lv
\end{align}

%%%%%%%%%%%
\subsection{Derivation of \eq{lin5} \label{e10}}

The momentum space wave function \eq{fourierdef} for $j=\halft$ and positive parity is
\begin{align} \label{ex10.1}
\Psi_{1/2,\lm,+} = 2\pi\int_0^\infty dr \int_{-1}^1 d\cos\theta\,r^2 e^{-ipr\cos\theta}\big[F(r)+iG(r)\,\alpha\cdot\hat\pv\,\cos\theta\big]\left(\begin{array}{c}\chi_\lm \crt 0 \end{array} \right)
\end{align}
I chose the $z$-axis of the $\xv$-integration along $\pv$. The integration over the azimuthal angle $\vphi$ leaves only the $\alpha_z$ component of $\alv\cdot\xv$. Expressing the factor $\cos\theta$ as a derivative of $\exp(-ipr\cos\theta)$ and using $G=iF$ (which holds at large $r$, where the stationary point is located for large $p$) we have
\begin{align} \label{ex10.2}
\Psi_{1/2,\lm,+} = 2\pi\int_0^\infty dr \,r^2\,F(r)\big(1-\frac{i}{r}\,\alpha\cdot\hat\pv\,\partial_p\big)\frac{i}{pr}\big(e^{-ipr}-e^{ipr}\big)\left(\begin{array}{c}\chi_\lm \crt 0 \end{array} \right)
\end{align}
For $p\to \infty$ the phase of $e^{\pm ipr}$ is rapidly oscillating, so the leading contribution comes from the region of the $r$-integration where the phase of the integrand is stationary. The stationary phase approximation is
\begin{align} \label{ex10.3}
\int dr\,f(r)e^{i\vphi(r)} \simeq e^{\veps(\vphi''(r_s))i\pi/4}\,f(r_s)\,e^{i\vphi(r_s)}
\end{align}
The function $f(r)$ is assumed to be varying slowly compared to the phase $\exp[i\vphi(r)]$. The phase is stationary at $\vphi'(r_s)=0$ and $\veps(x) = 1\ (-1)$ for $x>0\ (x<0)$. According to \eq{lin4} we have in the integral of \eq{ex10.2} $\vphi(r)= V'r^2/2\mp pr$. There is a stationary phase for $r>0$ only with the $\exp(-ipr)$ term, giving $r_s = p/V'$, $\vphi(r_s)=-p^2/2V'$ and $\vphi''(r_s)>0$. From \eq{lin4} the contribution proportional to the unit Dirac matrix in \eq{ex10.2} is then
\begin{align} \label{ex10.4}
\big(\Psi_{1/2,\lm,+}\big)_F = \frac{2\pi iN}{p}\Big(\frac{p}{V'}\Big)^{\beta+1}\,e^{-ip^2/2V'}\left(\begin{array}{c}\chi_\lm \crt 0 \end{array} \right)
\end{align}
The leading term in the contribution  $\propto \alpha\cdot\hat\pv$ has $\partial_p \to -ip/V' = -ir_s$, giving the full result
\begin{align} \label{ex10.5}
\Psi_{1/2,\lm,+} = \frac{2\pi iN}{p}\Big(\frac{p}{V'}\Big)^{\beta+1}(1-\alpha\cdot\hat\pv)\,e^{-ip^2/2V'}\left(\begin{array}{c}\chi_\lm \crt 0 \end{array} \right)
\end{align}
Consider now the expressions for the $u$ and $v$ spinors for $|\pv| \to \infty$,
\begin{align} \label{ex10.6}
u(\pv,\lm) &\equiv \frac{\psl+m}{\sqrt{E_p+m}}\left(\begin{array}{c}\chi_\lm \crt 0 \end{array} \right)
\simeq \sqrt{|\pv|}\,(1+\alpha\cdot\hat\pv) \left(\begin{array}{c}\chi_\lm \crt 0 \end{array} \right) \ \ \ (|\pv| \to \infty) \nn\crt
v(\pv,\lm) &\equiv \frac{-\psl+m}{\sqrt{E_p+m}}\left(\begin{array}{c}0\crt \bar\chi_\lm  \end{array} \right)
\simeq \sqrt{|\pv|}\,(1+\alpha\cdot\hat\pv) \left(\begin{array}{c}0\crt \bar\chi_\lm  \end{array} \right) \ \ \ (|\pv| \to \infty)
\end{align}
Consequently, omitting a common factor in the distributions $e^\pm(\pv,s)$ \eq{dir18},
\begin{align} \label{ex10.7}
e^-(p,s) &= u^\dag(\pv,s)\Psi_{1/2,\lm,+} \sim \left(\begin{array}{cc}\chi_s^\dag & 0 \end{array} \right) (1+\alpha\cdot\hat\pv)(1-\alpha\cdot\hat\pv)\left(\begin{array}{c}\chi_\lm \crt 0 \end{array} \right) =0 \nn\crt
e^+(p,s) &= v^\dag(-\pv,s)\Psi_{1/2,\lm,+} \sim \left(\begin{array}{cc} 0 & \bar\chi_s^\dag \end{array} \right) (1-\alpha\cdot\hat\pv)(1-\alpha\cdot\hat\pv)\left(\begin{array}{c}\chi_\lm \crt 0 \end{array}\right) = -2\bar\chi_s^\dag\, \sv\cdot\hat\pv\,\chi_\lm
\end{align}
which establishes \eq{lin5}.

%%%%%%%%%%%
\subsection{Gauge transformations generated by Gauss operator \label{e11}}

The unitary operator defined in \eq{eII5},
\begin{align} \label{tgqed17}
U(t) = 1+i\int d\yv\,G(t,\yv)\delta\la(\yv) = 1+i\int d\yv\,\big[\partial_i E^i(t,\yv)-e\psi^\dag\psi(t,\yv)\big]\delta\la(\yv)
\end{align}
transforms $A^j(t,\xv)$ as,
\begin{align} \label{tgqed18}
\delta A^j(t,\xv) &\equiv U(t)A^j(t,\xv)U^{-1}(t)-A^j(t,\xv) = i\com{\int d\yv\,\big[\partial_i E^i(t,\yv)-e\psi^\dag\psi(t,\yv)\big]\delta\la(\yv)}{A^j(t,\xv)}\nn\crt
&= -i\com{\int d\yv\,E^i(t,\yv)\partial_i^y \delta\la(\yv)}{A^j(t,\xv)} 
= \partial_j \delta\la(\xv) \nn\crt
\end{align}
Similarly for the electron field,
\begin{align} \label{tgqed18b}
\delta \psi(t,\xv) &\equiv U(t)\psi(t,\xv)U^{-1}(t)-\psi(t,\xv) = -ie\com{\int d\yv\,\psi^\dag\psi(t,\yv)\delta\la(\yv)}{\psi(t,\xv)} = ie\,\delta\la(\xv)\psi(t,\xv)
\end{align}

%%%%%%%%%%%
\subsection{ Derive \eq{fock4}. \label{e12a}}

The anticommutation relation of the electron fields gives
\begin{align} \label{ex12a.1}
\bs{\mJ}\ket{e^-e^+;\mB,\Pv=0}=\int d\xv_1 d\xv_2\, \bar\psi(\xv_1)\Big[\Jv\lla_+(\xv_1)\Phirb(\xv_1-\xv_2) \rla_-(\xv_2)-\lla_+(\xv_1) \Phirb(\xv_1-\xv_2) \rla_-(\xv_2)\Jv\Big]\psi(\xv_2)\ket{0}
\end{align}
$\Jv$ commutes with the $\la_\pm$ projectors \eq{fock2}, which have a rotationally invariant form. It is instructive to see how this works out explicitly. The commutator $\comb{\rla_-}{\Lv+\Sv}$ gets contributions from the derivatives in $\rla_-$ differentiating $\xv$ in $\Lv=-i\xv\times\nv$, and from the commutators of the Dirac matrices in $\rla_-$ with $\Sv=\halft\gf\alv$. 

The $\nv^2$ in $E=\sqrt{-\nv^2+m^2}$ of $\rla_-(\xv)$ commutes with $L^i$,
\begin{align}\label{ex12a.2}
\com{\partial_j\partial_j}{\veps_{ik\ell}x^k\partial_\ell} = \partial_j \veps_{ij\ell}\partial_\ell+ \veps_{ij\ell}\partial_\ell\partial_j=0
\end{align}
since $\partial_j\partial_\ell = \partial_\ell\partial_j$, whereas $\veps_{ij\ell} =-\veps_{i\ell j}$. The commutator of $i\alv\cdot\nv$ with $L^i$ (in my notation $\alpha_j=\alpha^j$),
\begin{align} \label{ex12a.3}
\com{i\alpha_j\partial_j}{-i\veps_{ik\ell}x^k\partial_\ell} = \veps_{ij\ell}\alpha_j\partial_\ell
\end{align}
is cancelled by its commutator with $S^i$. Using $\alpha_i\alpha_j = \delta_{ij}+i\veps_{ijk}\alpha_k\gf$,
\begin{align} \label{ex12a.4}
\com{i\alpha_j\partial_j}{\halft\gf\alpha_i} = -\veps_{jik}\alpha_k\partial_j
\end{align}
Finally, $\com{\gz}{\Sv}=0$. We see that the commutator $\comb{\rla_-}{\Jv}=0$ requires contributions from both $\Lv$ and $\Sv$. Similarly $\Jv$ commutes with other rotationally invariant structures. Bringing $\Jv$ through the $\la_\pm$ projectors gives \eq{fock4}.

The commutator with the orbital angular momentum $\Lv$ in \eq{fock4} arises from
\begin{align} \label{fock4b}
\xv_1\times(-i\rnab_1)\Phirb(\xv_1-\xv_2)-\Phirb(\xv_1-\xv_2)\xv_2\times(i\lnab_2) = (\xv_1-\xv_2)\times(-i\rnab_1)\Phirb(\xv_1-\xv_2)
\end{align}
Hence in $\com{\Lv}{\Phirb(\xv)} = \xv\times[-i\nv\Phirb(\xv)]$ the $\xv$-derivatives of $\Lv$ apply only to $\Phirb(\xv)$.

%%%%%%%%%%%
\subsection{Derivation of \eq{A15} \label{e12}}

The transformation \eq{A13} of the fields under charge conjugation gives
\begin{align}\label{ex12.1}
\mC\ket{e^-e^+;\mB,\Pv} = -\int d\xv_1 d\xv_2\,\psi(\xv_1)^T\aly\lla_+(\xv_1)\Phipb(\xv_1-\xv_2)\rla_-(\xv_2)\aly\bar\psi^T(\xv_2)\ket{0}
\end{align}
Take the transpose on the rhs., recalling that the anticommutation of the fields gives a minus sign,
\begin{align}\label{ex12.2}
\mC\ket{e^-e^+;\mB,\Pv} &= \int d\xv_1 d\xv_2\,\bar\psi(\xv_2)\aly^T\inv{2E_2}(E_2+i\alv^T\cdot\lnab_2-\gz m){\Phipb}^T(\xv_1-\xv_2) \nn\crt
&\times\inv{2E_1}(E_1-i\alv^T\cdot\rnab_1+\gz m)\aly^T\psi(\xv_1)\ket{0}
\end{align}
Recalling that $\aly \alv^T\aly=-\alv$ and changing integration variables $\xv_1 \lra \xv_2$ we get
\begin{align}\label{ex12.3}
\mC\ket{e^-e^+;\mB,\Pv} = \int d\xv_1 d\xv_2\,\bar\psi(\xv_1)\lla_+(\xv_1)\aly{\Phipb}^T(\xv_2-\xv_1)\aly\rla_-(\xv_2)\psi(\xv_2)\ket{0}
\end{align}
Comparing with the definition \eq{fock1} of $\ket{e^-e^+;\mB,\Pv}$ we see that \eq{A14} implies \eq{A15}.

%%%%%%%%%%%
\subsection{ Verify \eq{fock6}.} \label{e13}

At lowest order in $\alpha$ the projectors $\la_\pm$ in the definition \eq{fock1} of the Positronium state may be
expressed in terms of the CM momentum $\Pv$, by partial integration and ignoring the \order{\alpha} contributions from differentiating $\Phipb(\xv_1-\xv_2)$,
\begin{align} \label{ex13.1}
\la_\pm(\Pv) = \inv{2E_P}\big(E_P\mp \alv\cdot\Pv\pm 2m\gz \big) = \la_\pm^\dag(\Pv) = \big[\la_\pm(\Pv)\big]^2 \hspace{2cm} \la_+(\Pv)\la_-(\Pv) = 0
\end{align}
Anticommutating the fields in $\bra{e^-e^+;\mB',\Pv'}$ with those in $\ket{e^-e^+;\mB,\Pv}$ gives using \eq{fock5c},
\begin{align} \label{ex13.2}
\langle{e^-e^+;\mB',\Pv'}\ket{e^-e^+;\mB,\Pv} &= N_{\mB'}^{(\Pv')*}\Npb\int d\xv_1 d\xv_2\, e^{i(\Pv-\Pv')\cdot(\xv_1+\xv_2)/2} {F^{(\Pv')}}^*(\xv_1-\xv_2)\Fp(\xv_1-\xv_2) \tr_{\mB,\mB'} \nn\crt
&= \big|\Npb\big|^2 (2\pi)^3 \delta(\Pv-\Pv')\int d\xv\,|\Fp(\xv)|^2\, \tr_{\mB,\mB'} \crt
\tr_{\mB,\mB'}&= \tr\big\{\Gamma_{\mB'}^\dag \la_+(\Pv)\Gamma_{\mB}\la_-(\Pv) \big\} \nn
\end{align}
Commuting $\la_-(\Pv)$ through $\Gamma_{\mB}$ gives for $\Gamma_{\mB} = \gf$ (Parapositronium) and $\Gamma_{\mB} = \alz$ (Orthopositronium with $\lm = 0$),
\begin{align} \label{ex13.3}
\tr_{\mB,\mB'} = \frac{2m}{E_P}\, \tr\big\{\Gamma_{\mB'}^\dag \la_+(\Pv)\gz\Gamma_{\mB} \big\}
= \frac{8m^2}{E_P^2}\delta_{\mB,\mB'}
\end{align}
while for $\Gamma_{\mB} = \ev_\pm\cdot\alv$ (Orthopositronium with $\lm = \pm 1$),
\begin{align} \label{ex13.4}
\tr_{\mB,\mB'} = \tr\big\{\Gamma_{\mB'}^\dag \la_+(\Pv)\Gamma_{\mB} \big\} = 2 \delta_{\mB,\mB'}
\end{align}
Using these expressions for $\tr_{\mB,\mB'}$ and the normalization \eq{fock5c} of $\Fp(\xv)$ in \eq{ex13.2} the normalization \eq{fock3} of the state implies \eq{fock6}.

%%%%%%%%%%%
\subsection{Derive the expression for $\ket{e^-e^+\gamma;\qv,\smu}$ in \eq{tra1}.} \label{e14a}

The contribution to $\ket{e^-e^+\gamma;\qv,\smu}$ from $\com{\mH_{int}}{\bar\psi(\xv_1)}$ is
\begin{align} \label{ex14a.1}
e\int d\xv_1 d\xv_2\,\bar\psi(\xv_1)\lla_+(\xv_1) \alv\cdot\vepsv_\smu^*(\qv)e^{-i\qv\cdot\xv_1}a^\dag(\qv,\smu)\lla_+(\xv_1)e^{i\Pv\cdot(\xv_1+\xv_2)/2}\Phipb(\xv_1-\xv_2)\rla_-(\xv_2)\psi(\xv_2)\ket{0}
\end{align}
where I inserted $\lla_+(\xv_1)$ to select the $b^\dag$ contribution in $\bar\psi(\xv_1)$ as in \eq{fock2b}. 
We have then
\begin{align} \label{ex14a.2}
\lla_+(\xv_1)\alpha^j \lla_+(\xv_1) &=
\lla_+(\xv_1)\inv{2E_1}\Big[(E_1+i\alv\cdot\lnab_1-m\gz)\alpha^j + \acomb{\alpha^j}{-i\alv\cdot\lnab_1}\Big] = \lla_+(\xv_1)\frac{-2i\lder_{1,j}}{2E_1} \nn\crt
&\to \lla_+(\xv_1)\frac{-P^j}{E_P}
\end{align}
The first term in the square bracket vanishes when multiplied by $\lla_+(\xv_1)$. The final result follows after partial integration, with the leading order term due to $i\rder_{1,j}\exp[i\Pv\cdot(\xv_1+\xv_2)/2]$ in \eq{ex14a.1}.

The contribution to $\ket{e^-e^+\gamma;\qv,\smu}$ from $\com{\mH_{int}}{\psi(\xv_2)}$ is similarly
\begin{align} \label{ex14a.3}
e\int d\xv_1 d\xv_2\,\bar\psi(\xv_1)\lla_+(\xv_1) e^{i\Pv\cdot(\xv_1+\xv_2)/2}\Phipb(\xv_1-\xv_2)\rla_-(\xv_2)\alv\cdot\vepsv_\smu^*(\qv)e^{-\qv\cdot\xv_2}a^\dag(\qv,\smu)\rla_-(\xv_2)\psi(\xv_2)\ket{0}
\end{align}
where now
\begin{align} \label{ex14a.4}
\rla_-(\xv_2)\alpha^j \rla_-(\xv_2) &=
\alpha^j\inv{2E_2}\big[(E_2-i\alv\cdot\rnab_2+m\gz) + \acomb{\alpha^j}{i\alv\cdot\rnab_2}\big]\rla_-(\xv_2) = \frac{2i\rder_{2,j}}{2E_2}\rla_-(\xv_2) \nn\crt
 &\to \frac{P^j}{E_P}\rla_-(\xv_2)
\end{align}
Using \eq{ex14a.2} in \eq{ex14a.1} and \eq{ex14a.4} in \eq{ex14a.3} and adding the two contributions gives \eq{tra1}.

%%%%%%%%%%%
\subsection{Derive the expression \eq{2d24d}.} \label{e15}

The frame dependence of functions like $\phi_0(\tau)$ and $\phi_1(\tau)$ that do not explicitly depend on $P$ or $E$ arises only due to the $P$-dependence of $\tau(x)$.
\begin{align} \label{ex15.1}
\left.\frac{\partial \tau}{\partial P}\right|_x = \frac{\partial}{\partial P}\big[M^2-2EV+V^2\big]/V' = -\frac{2xP}{E}
\end{align}
Recalling also that for functions that depend on $x$ only via $\tau$,
\begin{align} \label{ex15.2}
\left.\frac{\partial}{\partial x}\right|_P = \left.\frac{\partial \tau}{\partial x}\right|_P  \frac{\partial}{\partial \tau} = -2(E-V)\frac{\partial}{\partial \tau}
\end{align}
we have
\begin{align} \label{ex15.3}
\left.\frac{\partial \phi_{0,1}}{\partial \xi}\right|_x = \left. E \frac{\partial \phi_{0,1}}{\partial P}\right|_x = \left.\frac{xp}{E-V}\partial_x \phi_{0,1}\right|_P
\end{align}
Applying this to $e^{\sigma_1\zeta/2}\Phip e^{-\sigma_1\zeta/2}$ gives
\begin{align} \label{ex15.4}
\left.\frac{\partial}{\partial \xi}\Big[e^{\sigma_1\zeta/2}\Phip e^{-\sigma_1\zeta/2}\Big]\right|_x
&= \left.e^{\sigma_1\zeta/2}\bigg\{\inv{2}\,\frac{\partial \zeta}{\partial \xi}\right|_x\comb{\sigma_1}{\Phip}  + \left.\frac{\partial \Phip}{\partial \xi}\right|_x \bigg\} e^{-\sigma_1\zeta/2} \crt
&= \frac{xp}{E-V}\,\partial_x \Big[e^{\sigma_1\zeta/2}\Phip e^{-\sigma_1\zeta/2}\Big]\Big|_\xi 
= \left.\frac{xp}{E-V}\,e^{\sigma_1\zeta/2}\bigg\{\inv{2}\,\frac{\partial \zeta}{\partial x}\right|_\xi\comb{\sigma_1}{\Phip}  + \left.\frac{\partial \Phip}{\partial x}\right|_\xi \bigg\} e^{-\sigma_1\zeta/2}  \nn
\end{align}
which implies
\begin{align} \label{ex15.5}
\frac{\partial \Phip}{\partial \xi}\Big|_x = \frac{xp}{E-V}\,\frac{\partial \Phip}{\partial x}\Big|_\xi 
+\inv{2}\Big(\frac{xp}{E-V}\,\frac{\partial \zeta}{\partial x}\Big|_\xi-\frac{\partial \zeta}{\partial \xi}\Big|_x \Big)\comb{\sigma_1}{\Phip} 
\end{align}
It remains to work out the derivatives of $\zeta$. From its definition \eq{2d24a},
\begin{align}\label{ex15.6}
\partial_x(\sinh\zeta)\big|_\xi &= \partial_x\zeta\big|_\xi\,\cosh\zeta = \frac{V'P(E-V)}{(V'\tau)^{3/2}}
 \hspace{2.15cm} \Longrightarrow \hspace{1cm} \frac{\partial\zeta}{\partial x}\Big|_\xi = \frac{P}{\tau} \nn\crt
 \partial_\xi(\sinh\zeta)\big|_x &= \partial_\xi\zeta\big|_x\,\cosh\zeta\, = \frac{(E-V)(M^2-EV)}{(V'\tau)^{3/2}}
 \hspace{1cm} \Longrightarrow \hspace{1cm} \frac{\partial\zeta}{\partial \xi}\Big|_x = \frac{M^2-EV}{V'\tau}
\end{align}
Using these in \eq{ex15.5} gives \eq{2d24d}.

%%%%%%%%%%%
\subsection{Derive the expression \eq{3ff16}.} \label{e16}

The expressions \eq{2d33} for $\phi_1(\tau)$ and $\phi_0(\tau)$ at large $\tau$ are,
\begin{align} \label{3ff11}
\phi_1(|\tau|\to\infty) &= \frac{4V'}{\sqrt{\pi}\, m}\,\sqrt{e^{\pi m^2/V'}-1}\,e^{-\theta(-\tau)\pi m^2/2V'}
\sin\big[\quart\tau-\halft m^2/V'\log(\halft |\tau|)+\arg\Gamma(1+im^2/2V')\big] \nn\crt
\phi_0(|\tau|\to\infty) &= -i\frac{4V'}{\sqrt{\pi}\, m}\,\sqrt{e^{\pi m^2/V'}-1}\,e^{-\theta(-\tau)\pi m^2/2V'}
\cos\big[\quart\tau-\halft m^2/V'\log(\halft |\tau|)+\arg\Gamma(1+im^2/2V')\big]
\end{align}
Since state $B$ has positive parity $\phi_{B1}[\tau_B(x=0)]= \phi_{B1}(\tau_B=M_B^2/V')=0$ according to \eq{2d27}. This determines the masses $M_{Bn}$ in the Bj limit,
\begin{align} \label{3ff12}
\quart M_{Bn}^2/V'-\halft m^2/V'\log(\halft M_{Bn}^2/V')+\arg\Gamma(1+im^2/2V') = n\cdot\pi
\end{align}
where $n$ is a large positive integer. Subtracting the lhs. from the arguments of the sin and cos functions in \eq{3ff11} gives a sign $(-1)^n$ to $\phi_{B0}$ and $\phi_{B1}$. Using also $-2E_B\,x = 2P_B^1\,x-2P_A^+(1-\xbj)x$ from \eq{2ff38} gives
\begin{align}\label{3ff13}
\phi_{B1}(|\tau_B|\to\infty) =& \frac{4V'(-1)^n}{\sqrt{\pi}\, m}\,\sqrt{e^{\pi m^2/V'}-1}\,\exp\big[-\theta(-\tau_B)\pi m^2/2V'\big] \nn\crt
&\times\sin\Big[\halft P_B^1x-\halft P_A^+(1-\xbj)x+\quart V'x^2-\frac{m^2}{2V'}\log\Big|1-\frac{V'x}{P_A^+(1-\xbj)}\Big|\Big] \nn\crt
\phi_{B0}(|\tau_B|\to\infty) =& -i\frac{4V'(-1)^n}{\sqrt{\pi}\, m}\,\sqrt{e^{\pi m^2/V'}-1}\,\exp\big[-\theta(-\tau_B)\pi m^2/2V'\big] \nn\crt
&\times\cos\Big[\halft P_B^1x-\halft P_A^+(1-\xbj)x+\quart V'x^2-\frac{m^2}{2V'}\log\Big|1-\frac{V'x}{P_A^+(1-\xbj)}\Big|\Big]
\end{align}

The asymptotically large phase $\halft P_B^1x$ must cancel the one in the factor $\sin\big(\halft q^1x\big)=\sin\big[\halft(P_B^1-P_A^1)x\big]$ of \eq{3ff9} to give a non-vanishing result. Using
\begin{align} \label{3ff14}
\sin\alpha\sin\beta &= \halft\big[\cos(\alpha-\beta)-\cos(\alpha+\beta)\big] \nn\crt
\sin\alpha\cos\beta &= \halft\big[\sin(\alpha+\beta)+\sin(\alpha-\beta)\big]
\end{align}
and defining the angle
\begin{align} \label{3ff15}
\varphi_B(x) \equiv \halft\big[P_A^+(1-\xbj)-P_A^1\big]x + \frac{m^2}{2V'}\log\Big|1-\frac{V'x}{P_A^+(1-\xbj)}\Big|-\quart V'x^2
\end{align}
the expression \eq{3ff9} for $G_{AB}$ gives \eq{3ff16} when terms with asymptotically large phases are neglected,

%%%%%%%%%%%
\subsection{Do the $x$-integral in \eq{3ff16} numerically for the parameters in \fig{f8}, and compare.} \label{e17}

Separate the integral in the form factor \eq{3ff16} into three parts,
\begin{align} \label{3ff24}
E_B G(q^2) &= (-1)^n\frac{16iV'}{\sqrt{\pi}m}\,\sqrt{e^{\pi m^2/V'}-1}\,(I_1+I_2+I_3) \crt
I_1 &= \int_0^\infty dx \big[i\sin\vphi_B\,\phi_{A0}(\tau_A)+\cos\vphi_B\,\phi_{A1}(\tau_A)\big]e^{-\theta(x-x_0)\pi m^2/2V'} \nn\crt
I_2 &= \int_0^{x_0} dx \,\cos\vphi_B\,\phi_{A1}(\tau_A)\,\frac{2m^2}{V'(x_0-x)(P_A^+-V'x)} \nn\crt
I_3 &= -\int_{x_0}^{\infty} dx \,\cos\vphi_B\,\phi_{A1}(\tau_A)\,\frac{2m^2}{V'(x-x_0)(P_A^+-V'x)}e^{-\pi m^2/2V'}
\end{align}

The $I_1$ integrand oscillates with constant amplitude at large $x$, where the approximation \eq{3ff11} for $\phi_{A}(\tau_A\to\infty)$ applies. $I_1$ is further divided into three parts. In $I_{1a}$ the range is $0 < x < x_1$, where $x_{1} > x_0$ and $\tau_A(x_1)>0$, ensuring that $\theta(-\tau_A) =0,\  \theta(-\tau_B) =1$ for $x>x_1$. $I_{1b}$ integrates over $x_1 \le x \le\infty$ with $\phi_{A0}$ and $\phi_{A1}$ replaced by the difference with their  large $x$ approximations, $\phi_{A}-\phi_{A}^{as}$. Finally $I_{1c}$ integrates $\phi_{A}^{as}$over $x_1 \le x \le\infty$:
\begin{align} \label{3ff25}
I_{1a} &= \int_0^{x_1} dx\big[i\sin\vphi_B\,\phi_{A0}(\tau_A)+\cos\vphi_B\,\phi_{A1}(\tau_A)\big]e^{-\theta(x-x_0)\pi m^2/2V'} \nn\crt
I_{1b} &= \int_{x_1}^\infty dx\big\{i\sin\vphi_B[\phi_{A0}(\tau_A)-\phi_{A0}^{as}(\tau_A)]+\cos\vphi_B [\phi_{A1}(\tau_A)-\phi_{A1}^{as}(\tau_A)]\big\}e^{-\pi m^2/2V'} \crt
I_{1c} &= \int_{x_1}^\infty dx\big\{i\sin\vphi_B\,\phi_{A0}^{as}(\tau_A)+\cos\vphi_B\,\phi_{A1}^{as}(\tau_A)\big\}e^{-\pi m^2/2V'} \nn
\end{align}
The oscillations in $I_{1b}$ at large $x$ are damped, allowing a numerical integration.
The phase in $\phi_A^{as}$ \eq{3ff11} is
\begin{align} \label{3ff26}
\vphi_{A} =\quart\tau_A-\halft m^2/V'\log(\halft |\tau_A|)+\arg\Gamma(1+im^2/2V')
\end{align}
The $I_{1c}$ integral reduces to
\begin{align} \label{3ff26b}
I_{1c} &=-\frac{4V'}{\sqrt{\pi}\,m}\sqrt{1-e^{-\pi m^2/V'}}\int_{x_1}^\infty dx\,\sin\vphi_C \nn\crt
\vphi_C &\equiv -(\vphi_A+\vphi_B) = \halft P_A^+\xbj x-\quart M_A^2/V'-\arg\Gamma(1+im^2/2V') +\frac{m^2}{2V'}\big[\log(\halft\tau_A)-\log(x/x_0-1)\big]
\end{align}
This integral is evaluated by rotating the contour in $u\equiv x-x_1$ by $\pi/2$ to ensure exponential convergence,
\begin{align} \label{3ff28}
\int_{x_1}^\infty dx\,\sin\vphi_C = \im\int_{x_1}^\infty dx\,e^{i\vphi_C(x)}=\im\int_{0}^{i\infty} du\,e^{i\vphi_C(u+x_1)}
\end{align}
%The $x_1$-independence of $I_1$ provides a numerical check of the calculation.
%Using Mathematica we checked (with $m=.5/\sqrt{V'}$) that the result for $I_1 = I_{1a}+I_{1b}+I_{1c}$ was independent of the choice of $x_1$ in the range $5 \le x_1 \le 50$ for $\xbj=.5$ in the rest frame {\tt [200412\_QED2\_DIS.nb]}, and for $10 \le x_1 \le 20$ with $\xbj=.1$ and $\xbj=.5$ in the frame with $\xi_a=1$ {\tt [200416\_QED2\_DIS.nb]}.

In $I_2$ the range $0<x<x_0$ is transformed into $0 < y < \infty$ through
\begin{align} \label{3ff29}
y = -\log(1-x/x_0) \hspace{2cm} x = x_0(1-e^{-y}) \hspace{2cm} \frac{dx}{x_0-x} = dy
\end{align}
The $y$-integration is further split into $0<y<y_2$ and $y_2<y<\infty$. The first path is finite and readily integrated numerically. The latter path is rotated by $\pi/2$, giving exponential convergence in $y_2 < y < y_2+i\infty$ when using $\cos\vphi_B = \re\exp(-i\vphi_B)$, due to the term $y m^2/2V'$ in $-\vphi_B$ \eq{3ff16}. $x \simeq x_0$ is constant on the complex path when $y_2$ is large, allowing the integral to be be evaluated analytically. The result for $I_2$ should be independent of $y_2$.
%$\phi_{a1}$ is an entire function. However, $\phi_{a1}[\tau_a(x)]$ can take very large values for short intervals of the complex contour in $y$, unless $y_2$ is sufficiently large and positive. In order to compensate for $x_0 = P_a^+(1-\xbj)$ growing with $\xi_a$ we chose $y_2 = 3\log P_a^+$ with $m=.5\,\sqrt{V'}$. We verified that $y_2 = 2\log P_a^+$ gave consistent results. The $0<y<y_2$ integral along the real axis was not problematic. With $m=.1\,\sqrt{2V'}$ good convergence for $.0001 < \xbj < .999$ required a larger $y_2$. It was convenient to choose $y_2 \ge 20$, so that to a good approximation $x=x_0$ is constant for $ y:\ y_2 \to y_2+ i\infty$. That part of the integration can then be done analytically.

In $I_3$ the integration contour is split into $x_0< x<2x_0$ and $2x_0< x< \infty$. In the first range the integration variable is changed to
\begin{align} \label{3ff30}
y = -\log(x/x_0-1) \hspace{2cm} x = x_0(1+e^{-y}) \hspace{2cm} \frac{dx}{x_0-x} = -dy
\end{align}
The $y$-range $0< y < \infty$ is further split into $0<y<y_3$ and $y_3<y<\infty$. The first is integrated numerically, and in the second the path is rotated by $\pi/2$, ranging over $y_3 < y < y_3+i\infty$. For large $y_3$ the value of $x \simeq x_0$ is constant on the complex contour, allowing an analytic integration. The integration over $2x_0< x< \infty$ is numerically stable, as the oscillations at large $x$ are damped.

%%%%%%%%%%%
\subsection{Derive the $q\bar qg$ potential \eq{qcd22}} \label{e18}

Using the commutators in \eq{qcd3} and \eq{qcd17} the operation of $\mE_a(\yv)$ \eq{qcd1} on the $\ket{q\bar qg}$ state \eq{qcd21} gives
\begin{align} \label{ex18.1}
\mE_a(\yv)\ket{q\bar qg} &= \big\{\bar\psi_{A'}(\xv_1)T_{A'A}^a A_b^i(\xv_g)T_{AB}^b\psi_B(\xv_2)\delta(\yv-\xv_1) \nn\crt
&+ \bar\psi_{A}(\xv_1)if_{abc}A_c^i(\xv_g)T_{AB}^b\psi_B(\xv_2)\delta(\yv-\xv_g) \nn\crt
&- \bar\psi_{A}(\xv_1)A_b^i(\xv_g)T_{AB}^b T_{BB'}^a\psi_{B'}(\xv_2)\delta(\yv-\xv_2) \big\}\ket{0}
\end{align}
When $\mE_a(\yv)$ and $\mE_a(\zv)$ in $\mH_V^{(0)}$ \eq{eII32} act on the same (quark or gluon) constituent we may use the previous results \eq{qcd5} and \eq{qcd18} showing that the coefficients of $\xv_1^2$ and $\xv_2^2$ are $C_F$ while that of $\xv_g^2$ is $N_c$, multiplied by the common factor $\big(\halft\kappa^2\intt d\xv + g\kappa\big)$. The new contributions are
\begin{align} \label{e18.2}
\xv_1\cdot\xv_2:& \hspace{1cm} -2\, \bar\psi_{A'}(\xv_1)T^a_{A'A}\,A_b^i(\xv_g)T^b_{AB}\,T^a_{BB'}\psi_{B'}(\xv_2)\ket{0} = \frac{1}{N_c}\ket{q\bar q g} \nn\crt
\xv_1\cdot\xv_g:& \hspace{1cm} 2\, \bar\psi_{A'}(\xv_1)T^a_{A'A}\,if_{abc}A_c^i(\xv_g)\,T^b_{AB}\psi_{B}(\xv_2)\ket{0} = -N_c\ket{qg\bar q} \nn\crt
\xv_2\cdot\xv_g:& \hspace{1cm} -2\bar\psi_{A}(\xv_1)\,if_{abc}A_c^i(\xv_g)\,T^b_{AB}\,T^a_{BB'}\psi_{B'}(\xv_2)\ket{0} = -N_c\ket{qg\bar q}
\end{align}
Altogether,
\begin{align} \label{e18.3}
\mH_V^{(0)}\ket{q\bar qg} &= \big(\halft\kappa^2\intt d\xv + g\kappa\big)\big[C_F(\xv_1^2+\xv_2^2)+N_c\xv_g^2 -N_c(\xv_1+\xv_2)\cdot\xv_g+\sfrac{1}{N_c}\xv_1\cdot\xv_2 \big]\ket{q\bar qg} \nn\crt
&= \big(\halft\kappa^2\intt d\xv + g\kappa\big)\big[d_{q\bar qg}(\xv_1,\xv_2,\xv_g)\big]^2\ket{q\bar qg} \nn\crt
d_{q\bar qg}(\xv_1,\xv_2,\xv_g) &\equiv \sqrt{\quart(N_c-\sfrac{2}{N_c})(\xv_1-\xv_2)^2+N_c(\xv_g-\halft\xv_1-\halft\xv_2)^2}
\end{align}
For the \order{\kappa^2} term to give the universal energy $E_\la$ \eq{qcd7} we need to choose the normalization of the homogeneous solution as
\begin{align} \label{e18.4}
\kappa_{q\bar qg} = \frac{\la^2}{g\sqrt{C_F}}\,\frac{1}{d_{q\bar qg}(\xv_1,\xv_g,\xv_2)}
\end{align}
The \order{g\kappa} contribution to $\mH_V$ gives the potential,
\begin{align} \label{e18.5}
V_{q\bar qg}^{(0)}(\xv_1,\xv_2,\xv_g) = g\kappa_{q\bar qg}\,\big[d_{q\bar qg}(\xv_1,\xv_2,\xv_g)\big]^2 = \frac{\la^2}{\sqrt{C_F}}\, d_{q\bar qg}(\xv_1,\xv_2,\xv_g)
\end{align}
When the self-energies are subtracted $\mH_V^{(1)}$ has contributions only from the three terms in \eq{e18.2},
\begin{align} \label{e18.6}
V_{q\bar qg}^{(1)}(\xv_1,\xv_2,\xv_g) = \halft\,\as\Big[\inv{N_c}\,\inv{|\xv_1-\xv_2|}-N_c\Big(\inv{|\xv_1-\xv_g|}+\inv{|\xv_2-\xv_g|}\Big)\Big]
\end{align}

%%%%%%%%%%%
\subsection{Derive the expression for $\mH_V^{(0)}\ket{8\otimes 8}$ in \eq{qcd36}} \label{e19}

Recall from \eq{qcd33},
\begin{align} \label{e19.1}
&\ket{8\otimes 8} = \bar\psi_A(\xv_1)T_{AB}^b\psi_B(\xv_2)\,\bar\psi_C(\xv_3)T_{CD}^b\psi_D(\xv_4)\ket{0} \nn\crt
&\bar\psi_A(\xv_1)\psi_B(\xv_2)\,\bar\psi_B(\xv_3)\psi_A(\xv_4)\ket{0} = 2\ket{8\otimes 8}+\sfrac{1}{N_c}\ket{1\otimes 1}
\end{align}
In the following I leave out the common factor $\big(\halft\kappa^2\intt d\xv + g\kappa\big)$ in $\mH_V^{(0)}$ \eq{qcd1},
\begin{align} \label{e19.2}
\mH_V^{(0)} &= \big(\halft\kappa^2\intt d\xv + g\kappa\big)\int d\yv\, d\zv\,\yv\cdot\zv\,\mE_a(\yv)\mE_a(\zv)
\end{align} 
and make use of the commutation relations \eq{qcd3},
\begin{align} \label{e19.3}
\comb{\mE_a(\xv)}{\bar\psi_A(\xv_1)} = \bar\psi_{A'}(\xv_1)T_{A'A}^a \delta(\xv-\xv_1) \hspace{2cm}
\comb{\mE_a(\xv)}{\psi_A(\xv_2)} = -T_{AA'}^a\psi_{A'}(\xv_2) \delta(\xv-\xv_2)
\end{align}
and of the SU($N_c$) generator relations \eq{qcd4}.

The commutators of $\mE_a(\yv)$ and $\mE_a(\zv)$ with the same quark at $\xv_i\ (i=1,\ldots 4)$ in $\ket{8\otimes 8}$ gives $\yv\cdot \zv=\xv_i^2$, color factor $T^a T^a = C_F\,I$ and state $\ket{8\otimes 8}$. The commutators with $\bar\psi(\xv_1)$ and $\psi(\xv_2)$ gives
\begin{flalign} \label{e19.4}
\xv_1\cdot\xv_2:& \hspace{1cm} -2\bar\psi_{A'}(\xv_1)T_{A'A}^a T_{AB}^b T_{BB'}^a \psi_{B'}(\xv_2)\bar\psi_C(\xv_3)T_{CD}^b\psi_D(\xv_4)\ket{0} = \sfrac{1}{N_c}\ket{8\otimes 8}&
\end{flalign}
The commutators with $\bar\psi(\xv_1)$ and $\bar\psi(\xv_3)$ give
\begin{flalign} \label{e19.5}
\xv_1\cdot\xv_3:& \hspace{1cm} 2\bar\psi_{A'}(\xv_1)T_{A'A}^a T_{AB}^b \psi_{B}(\xv_2)\bar\psi_{C'}(\xv_3)T_{C'C}^a T_{CD}^b\psi_D(\xv_4)\ket{0}&
\end{flalign}
The color factors
\begin{align} \label{e19.6}
\halft(\delta_{A'C}\delta_{AC'}-\sfrac{1}{N_c}\delta_{A'A}\delta_{C'C})T_{AB}^b T_{CD}^b
= \quart\delta_{A'C}\delta_{AC'}(\delta_{AD}\delta_{BC}-\sfrac{1}{N_c}\delta_{AB}\delta_{CD})-\sfrac{1}{2N_c}\delta_{A'A}\delta_{C'C}T_{AB}^b T_{CD}^b
\end{align}
give for the coefficient of $\xv_1\cdot\xv_3$ in \eq{e19.5},
\begin{flalign} \label{e19.7}
\big[\halft\bar\psi_{B}(\xv_1) \psi_{B}(\xv_2)\bar\psi_{D}(\xv_3)\psi_D(\xv_4)\ket{0}
- \sfrac{1}{2N_c}\bar\psi_{D}(\xv_1) \psi_{B}(\xv_2)\bar\psi_{B}(\xv_3)\psi_D(\xv_4)\big]\ket{0}-\sfrac{1}{N_c}\ket{8\otimes 8}
= \sfrac{C_F}{N_c}\ket{1\otimes 1}-\sfrac{2}{N_c}\ket{8\otimes 8}&
\end{flalign}
The commutators with $\bar\psi(\xv_1)$ and $\bar\psi(\xv_4)$ give
\begin{flalign} \label{e19.8}
\xv_1\cdot\xv_4:& \hspace{1cm} -2\bar\psi_{A'}(\xv_1)T_{A'A}^a T_{AB}^b \psi_{B}(\xv_2)\bar\psi_{C}(\xv_3) T_{CD}^b T_{DD'}^a\psi_{D'}(\xv_4)\ket{0}&
\end{flalign}
Now the color factors
\begin{align} \label{e19.9}
\halft(\delta_{A'D'}\delta_{AD}-\sfrac{1}{N_c}\delta_{A'A}\delta_{D'D})T_{AB}^b T_{CD}^b
= \quart\delta_{A'D'}\delta_{AD}(\delta_{AD}\delta_{BC}-\sfrac{1}{N_c}\delta_{AB}\delta_{CD})-\sfrac{1}{2N_c}\delta_{A'A}\delta_{DD'}T_{AB}^b T_{CD}^b
\end{align}
give for the coefficient of $\xv_1\cdot\xv_4$ in \eq{e19.8},
\begin{flalign} \label{e19.10}
\big(-\sfrac{N_c}{2}+\sfrac{1}{2N_c}\big)\bar\psi_{A'}(\xv_1) \psi_{B}(\xv_2)\bar\psi_{B}(\xv_3)\psi_{A'}(\xv_4)\ket{0}+\sfrac{1}{N_c}\ket{8\otimes 8}
= -\sfrac{C_F}{N_c}\ket{1\otimes 1}-\sfrac{N_c^2-2}{N_c}\ket{8\otimes 8}&
\end{flalign}
The coefficients of $\xv_2\cdot\xv_3,\ \xv_2\cdot\xv_4$ and $\xv_3\cdot\xv_4$ are the same as those of $\xv_1\cdot\xv_4,\ \xv_1\cdot\xv_3$ and $\xv_1\cdot\xv_2$, respectively. Altogether,
\begin{align} \label{e19.11}
\mH_V^{(0)}\ket{8\otimes 8} &= \big(\halft\kappa^2\intt d\xv + g\kappa\big)\Big\{\big[\halft N_c(\xv_1-\xv_4)^2 +\halft N_c(\xv_2-\xv_3)^2-\sfrac{1}{2N_c}(\xv_1-\xv_2)^2 \nn\crt
&-\sfrac{1}{2N_c}(\xv_3-\xv_4)^2-\sfrac{2}{N_c}(\xv_1-\xv_2)\cdot(\xv_3-\xv_4)\big]\ket{8\otimes 8}+\sfrac{C_F}{N_c}(\xv_1-\xv_2)\cdot(\xv_3-\xv_4)\ket{1\otimes 1}\Big\}
\end{align}
When expressed in terms of the separations \eq{qcd35} this gives \eq{qcd36}.

%%%%%%%%%%%
\subsection{Verify that the expression \eq{qcd60} for $\Phi_{-+}(\xv)$ satisfies the bound state equation \eq{qcd47} given the radial equation \eq{qcd59}.} \label{e20}

The BSE as in \eq{qcd47} applied to $\Phi_{-+}(\xv)$ in the alternative forms of \eq{qcd60},
\begin{align} \label{e20.1}
&\rh_-\Phi(\xv)+\Phi(\xv)\lh_- = 0 \nn\crt
&\Phi_{-+}(\xv) = \rh_+\gf\,F_1(r)Y_{j\lambda}(\hat\xv) = F_1(r)Y_{j\lambda}(\hat\xv)\,\gf \lh_+
\end{align}
allows the use of 
\begin{align}\label{e20.2}
\rh_-\rh_+ &= \frac{4}{(M-V)^2}(-\rnab^2+m^2)-1 +\frac{4iV'}{r(M-V)^3}\,\alv\cdot\xv \,(i\alv\cdot\rnab+m\gz) \nn \crt
\lh_+\lh_- &= (-\lnab^2+m^2)\frac{4}{(M-V)^2}-1 + (i\alv\cdot\lnab-m\gz)\, \alv\cdot\xv\,\frac{4iV'}{r(M-V)^3} 
\end{align}
Moving the $\gf$ to the right in the BSE,
\begin{align} \label{e20.3}
\rh_-\rh_+\gf F_1(r)Y_{j\lambda}(\hat\xv)&+ F_1(r)Y_{j\lambda}(\hat\xv)\gf\lh_+\lh_- \nn\crt
&=\Big[\frac{8}{(M-V)^2}(-\nv^2+m^2)-2+\frac{4iV'x^j}{r(M-V)^3}\acomb{\alpha_j}{i\rder_k\alpha_k+m\gz}\Big]F_1(r)Y_{j\lambda}(\hat\xv)\gf
\end{align}
Using $\acom{\alpha_j}{\alpha_k}=2\delta_{jk}$, $\acom{\alpha_j}{\gz}=0$, $x^j\rder_j = r{\overset{\rar}{\partial}_r\strut}$ and $\nv^2 = (1/r^2)\partial_r(r^2\partial_r)-\Lv^2/r^2$ with $\Lv^2Y_{j\lambda}(\hat\xv)=j(j+1)Y_{j\lambda}(\hat\xv)$ gives the radial equation \eq{qcd59}.

%%%%%%%%%%%
\subsection{Derive the coupled equations \eq{qcd100} from the bound state equation \eq{qcd98}.} \label{e21}

I make use of commutator identities such as,
\beqa
\com{A}{BC}&=&\com{A}{B}C+B\com{A}{C} \label{eB1} \\[2mm]
\acom{A}{BC}&=&\com{A}{B}C+B\acom{A}{C} = \acom{A}{B}C-B\com{A}{C} \label{eB2} \\[2mm]
\acom{A}{\acom{B}{C}} &=& -\com{B}{\com{A}{C}} \hspace{1.15cm} {\rm when}\ \ \acomb{A}{B}=0 \label{eB3} \\[2mm]
\acom{A}{\com{B}{C}} &=& -\acom{B}{\com{A}{C}} \hspace{1cm} {\rm when}\ \ \acomb{A}{B}=0 \label{eB4} \\[2mm]
\com{A}{\acom{B}{C}} &=& -\com{B}{\acom{A}{C}} \hspace{1cm} {\rm when}\ \ \acomb{A}{B}=0 \label{eB5} \\[2mm]
\acom{A}{\com{A}{C}} &=& \com{A}{\acom{A}{C}}=\com{A^2}{C}  \label{eB6} \\[2mm]
\acom{A}{\acom{A}{C}} &=& 2A\acom{A}{C} \hspace{1.5cm} {\rm when}\ \ A^2=1 \\[2mm] \label{eB7}
\com{A}{\com{A}{C}} &=& 2A\com{A}{C} \hspace{1.65cm} {\rm when}\ \ A^2=1  \label{eB8}
\eeqa

Taking the commutator $i\nv\cdot\com{\alv}{\rm BSE}$ of the bound state equation \eq{qcd98} gives
\begin{align} \label{eB9}
\com{i\nv\cdot \alv}{(E-V)\Phip} &= \com{i\nv\cdot \alv}{\acomb{i\nv\cdot\alv}{\Phip}}
-\halft\com{i\nv\cdot \alv}{\comb{\Pv\cdot\alv}{\Phip}}+m\com{i\nv\cdot \alv}{\comb{\gz}{\Phip}}
\end{align}
The first term on the rhs. vanishes due to the commutator identity \eq{eB6}, when we recall that $\nv$ in the BSE always operates on $\Phip$.
The identity \eq{eB3} implies for the third term on the rhs. of \eq{eB9},
\begin{align} \label{eB10}
m\com{i\nv\cdot \alv}{\comb{\gz}{\Phip}} = -m\acom{\gz}{\acomb{i\nv\cdot \alv}{\Phip}}
\end{align}
Using the original BSE \eq{qcd98} on the rhs. of \eq{eB10} we get
\begin{align} \label{eB11} 
m\com{i\nv\cdot \alv}{\comb{\gz}{\Phip}} &= -m\acom{\gz}{\halft \com{\Pv\cdot\alv}{\Phip}}+m^2\acom{\gz}{\com{\gz}{\Phip}}-m\acom{\gz}{(E-V)\Phip} \nn\crt
&\hspace{-1cm}= \halft m \acom{\Pv\cdot\alv}{\com{\gz}{\Phip}}-m(E-V)\acom{\gz}{\Phip} \nn\crt
&\hspace{-1cm}= \halft\acom{\Pv\cdot\alv}{-\acomb{i\nv\cdot\alv}{\Phip}+\halft\comb{\Pv\cdot\alv}{\Phip}+(E-V)\Phip}-m(E-V)\acom{\gz}{\Phip}
\end{align}
where I used \eq{eB4}, \eq{eB6} and in the last step expressed $m\com{\gz}{\Phi_P}$ using the BSE \eq{qcd98}. The second term on the rhs. of \eq{eB11} vanishes according to \eq{eB6}. Inserting this result in \eq{eB9} we have
\begin{align} \label{eB12} \hspace{-.4cm}
\com{i\nv\cdot \alv}{(E-V)\Phip} &= \nn\crt
& \hspace{-3.6cm}-\halft\com{i\nv\cdot \alv}{\com{\Pv\cdot\alv}{\Phip}}-\halft\acom{\Pv\cdot\alv}{\acom{i\nv\cdot \alv}{\Phip}} + \halft(E-V)\com{\Pv\cdot\alv}{\Phip}-m(E-V)\acom{\gz}{\Phip}
\end{align}
The sum of the first two terms on the rhs. simplifies. With $\nv\cdot \alv=\alpha^i\partial_i$ and $\Pv\cdot \alv=P^j\alpha^j$,
\begin{align} \label{eB13}
\com{\alpha^i}{\com{\alpha^j}{\partial_i\Phip}} &= \alpha^i(\alpha^j\partial_i\Phip- \partial_i\Phip\alpha^j)-(\alpha^j\partial_i\Phip- \partial_i\Phip\alpha^j)\alpha^i
\nn\crt
\acom{\alpha^j}{\acom{\alpha^i}{\partial_i\Phip}} &= \alpha^j(\alpha^i\partial_i\Phip+ \partial_i\Phip\alpha^i)+(\alpha^i\partial_i\Phip+ \partial_i\Phip\alpha^i)\alpha^j
\end{align}
so that
\begin{align} \label{eB14}
\com{\alpha^i}{\com{\alpha^j}{\partial_i\Phip}}+\acom{\alpha^j}{\acom{\alpha^i}{\partial_i\Phip}} &= (\alpha^i\alpha^j+\alpha^j\alpha^i)\partial_i\Phip+ \partial_i\Phip(\alpha^j\alpha^i+\alpha^i\alpha^j) = 4\partial_j\Phip
\end{align}
Using this in \eq{eB12} and dividing by $E-V$ gives
\begin{align} \label{eB15}
\inv{E-V}\com{i\nv\cdot \alv}{(E-V)\Phip}-\halft\acom{\Pv\cdot\alv}{\Phip}+m\acom{\gz}{\Phip} = -\frac{2i}{E-V}\Pv\cdot\nv\Phip
\end{align}
For a linear potential $i\nv\cdot\alv\, V'|\xv|= iV'\alv\cdot\xv/r$, where $r=|\xv|$. Bringing this derivative to the rhs. in \eq{eB15},
\begin{align} \label{eB16}
\com{i\nv\cdot \alv}{\Phip}-\halft\acom{\Pv\cdot\alv}{\Phip}+m\acom{\gz}{\Phip} = \inv{E-V}\Big(-2i\Pv\cdot\nv\Phip+\frac{V'}{r}\com{i\alv\cdot\xv}{\Phip}\Big)
\end{align}
The lhs. is now the same as in the original BSE \eq{qcd98}, with commutators and anticommutators interchanged. Adding and subtracting the two equations and dividing by $E-V$ we get equations \eq{qcd100}.

%%%%%%%%%%%
\subsection{Derive the frame dependence  \eq{qcd101} of $\Phip_{V=0}(\xv)$ using the boost generator $\mK_0^z$.} \label{e21b}

The action of $\mK_0^z(t=0)$ \eq{add1} on the state \eq{qcd97} with $\Pv=(0,0,P)$ and $V=0$,
\begin{align} \label{e21b.1}
\ket{M,P}_0 = \int d\xv_1 d\xv_2\,\bar\psi(\xv_1)e^{iP(z_1+z_2/2}\Phip_{V=0}(\xv_1-\xv_2)\psi(\xv_2)\ket{0}
\end{align}
is determined by
\begin{align} \label{e21b.2}
\com{\mK_0^z}{\bar\psi(\xv_1)} &= \psi^\dag(\xv_1)\big[z_1(-i\alv\cdot\lnab_1-m\gz)+\halft i\alz\big]\gz
= \bar\psi(\xv_1)\big[z_1(i\alv\cdot\lnab_1-m\gz)-\halft i\alz\big] \nn\crt
\com{\mK_0^z}{\psi(\xv_2)} &= -\big[z_2(i\alv\cdot\rnab_2-m\gz)+\halft i\alz\big]\psi(\xv_2)
\end{align}
Making the derivatives act on the wave function through partial integration,
\begin{align} \label{e21b.3}
\mK^z_0\ket{M,P}_0 = \int d\xv_1 d\xv_2\,\bar\psi(\xv_1)\Big\{
&\big[z_1(-i\alv\cdot\rnab_1-m\gz)-\halft i\alz\big]e^{iP(z_1+z_2/2}\Phip_{V=0}(\xv_1-\xv_2) \nn\crt
&-e^{iP(z_1+z_2/2}\Phip_{V=0}(\xv_1-\xv_2)\big[z_2(-i\alv\cdot\lnab_2-m\gz)+\halft i\alz\big]\Big\}\psi(\xv_2)\ket{0}
\end{align}
Using $z_2(-i\alv\cdot\lnab_2)+\halft i\alz = (-i\alv\cdot\lnab_2)z_2-\halft i\alz$ and then expressing $z_{1,2}=\halft(z_1+z_2) \pm \halft(z_1-z_2)$, the BSE \eq{qcd99} satisfied by $\Phip_{V=0}$,
\begin{align} \label{e21b.4}
\big(i\alv\cdot\rnab_1 -\halft P\alz+m\gz\big)\Phip_{V=0}(\xv_1-\xv_2)  + \Phip_{V=0}(\xv_1-\xv_2) \big(-i\alv\cdot\lnab_2 + \halft P\alz-m\gz\big)=E\,\Phip_{V=0}(\xv_1-\xv_2)
\end{align} 
reduces the coefficient of $\halft(z_1+z_2)$ to $-E$, with $E=M\cosh\xi = \sqrt{P^2+M^2}$. We have then
\begin{align} \label{e21b.5}
&\mK^z_0\ket{M,P}_0 = \int d\xv_1 d\xv_2\,\bar\psi(\xv_1)e^{iP(z_1+z_2/2}\Big\{-\halft(z_1+z_2)E\,\Phip_{V=0}(\xv_1-\xv_2)-\halft i\comb{\alz}{\Phip_{V=0}(\xv_1-\xv_2)} \crt
&-\halft(z_1-z_2)\big[(i\alv\cdot\rnab_1-\halft P\alz+m\gz)\Phip_{V=0}(\xv_1-\xv_2) + \Phip_{V=0}(\xv_1-\xv_2)(i\alv\cdot\lnab_2-\halft P\alz+m\gz)\big]\Big\}\psi(\xv_2)\ket{0} \nn
\end{align}
where $\rnab_1$ and $\lnab_2$ only differentiate $\Phip_{V=0}(\xv_1-\xv_2)$. Subtracting the two BSE equations \eq{qcd100} gives
\begin{align} \label{e21b.6}
\big(i\alv\cdot\rnab_1 -\halft P\alz+m\gz\big)\Phip_{V=0}(\xv_1-\xv_2)  + \Phip_{V=0}(\xv_1-\xv_2) \big(i\alv\cdot\lnab_2 - \halft P\alz+m\gz\big)=-2i\frac{P}{E}\,\rder_{z_1}\Phip_{V=0}(\xv_1-\xv_2)
\end{align}
Thus
\begin{align} \label{e21b.7}
-id\xi\mK^z_0\ket{M,P}_0 &= \int d\xv_1 d\xv_2\,\bar\psi(\xv_1)e^{iP(z_1+z_2/2}\Big\{\halft id\xi\,E(z_1+z_2)\Phip_{V=0}(\xv_1-\xv_2)-\halft d\xi \comb{\alz}{\Phip_{V=0}(\xv_1-\xv_2)}  \nn \crt
&+d\xi(z_1-z_2)\frac{P}{E}\,\partial_{z_1}\Phip_{V=0}(\xv_1-\xv_2)\Big\}\psi(\xv_2)\ket{0}
\end{align}
Let us now assume that the frame dependence \eq{qcd101} holds, and show that it agrees with the change in the wave function \eq{e21b.7} caused by the infinitesimal boost. According to \eq{qcd101},
\begin{align} \label{e21b.8}
\Phi^{(P+dP)}_{V=0}(\xv) &= e^{-(\xi+d\xi)\alz/2}\Phi^{(P=0)}_{V=0}(\xv_{R})e^{(\xi+d\xi)\alz/2} \nn\crt 
\xv_R &= (x,y,z\cosh(\xi+d\xi)) = (x,y,z\cosh\xi)+ (0,0,d\xi z\sinh\xi)
\end{align}
The first term within the $\{\ \}$ of \eq{e21b.7} reflects the change in the plane wave phase of $\ket{M,P}_0$,
\begin{align} \label{e21b.9}
e^{i(P+d\xi E)(z_1+z_2)/2} = e^{i(P(z_1+z_2)/2}\big[1+\halft i\,d\xi E(z_1+z_2)\big]
\end{align}
The second term is due to the $\exp[\mp(\xi+d\xi)\alz/2]$ factors in $\Phi^{(P+dP)}_{V=0}(\xv)$,
\begin{align} \label{e21b.10}
\exp[-(\xi+d\xi)\alz/2]\Phi^{(P=0)}_{V=0}(\xv_{1R}-\xv_{2R}) = (1-\halft d\xi\alz)\exp(-\xi\alz/2) \Phi^{(P=0)}_{V=0}(\xv_{1R}-\xv_{2R})
\end{align}
and similarly for $\Phi^{(P=0)}_{V=0}(\xv_{1R}-\xv_{2R})\exp[(\xi+d\xi)\alz/2]$. The third term in \eq{e21b.7} relates to the Lorentz contraction, \ie, the change in $\xv_R$ \eq{e21b.8},
\begin{align} \label{e21b.11}
e^{-\xi\alz/2}d\xi(z_1-z_2)\sinh\xi \frac{\partial}{\partial z_{1R}}\Phi^{(P=0)}_{V=0}(\xv_{1R}-\xv_{2R})e^{\xi\alz/2} = d\xi(z_1-z_2)\frac{\sinh\xi}{\cosh\xi}\,\frac{\partial}{\partial z_{1}}\Phip_{V=0}(\xv_1-\xv_2)
\end{align}
This confirms that the frame dependence of the state $\ket{M,P}_0$ \eq{e21b.1} implied by \eq{e21b.8} agrees with the transformation of a boost.

%%%%%%%%%%%
\subsection{Show that $\Phip(\tau)$ given by \eq{qcd108} satisfies the BSE \eq{qcd99} at $\xtr=0$.} \label{e22}

Denoting by $B$ the lhs. of \eq{qcd99} at $\xtr=0$ when the wave function $\Phip(\tau)$ is given by \eq{qcd108},
\begin{align} \label{e22.1}
e^{\zeta\alz/2}\,B\,e^{-\zeta\alz/2} &= e^{\zeta\alz/2}\big[i\rnab\cdot\alv-\halft(E-V+P\alz)+m\gz\big]e^{-\zeta\alz/2}\Phi^{(0)}(\tau) \nn\crt
&+ \Phi^{(0)}(\tau)e^{\zeta\alz/2} \big[i\lnab\cdot\alv-\halft(E-V-P\alz)-m\gz\big] e^{-\zeta\alz/2}
\end{align} 
We need to show that $B=0$. For the $i\partial_z$ terms,
\begin{align} \label{e22.2}
e^{\zeta\alz/2}i\rder_z\alz e^{-\zeta\alz/2} &= i\rder_z\alz -\halft i(\rder_z\zeta) \nn\crt
e^{\zeta\alz/2}i\lder_z\alz e^{-\zeta\alz/2} &= i\lder_z\alz +\halft i(\rder_z\zeta)
\end{align}
The contributions $\propto \partial_z\zeta$ cancel in \eq{e22.1}.
Transforming $\partial_z=-2(E-V)\partial_\tau = -2\sqrt{V'\tau}\cosh\zeta\,\partial_\tau$ (which requires both a linear potential and $\xtr=0$),
\begin{align} \label{e22.3}
i\rder_z\alz&= -2\sqrt{V'\tau}e^{\zeta\alz}\alz\, i\rder_\tau + 2\sqrt{V'\tau}\sinh\zeta\, i\rder_\tau \nn\crt
i\lder_z\alz&= -2i\lder_\tau\alz\sqrt{V'\tau}e^{-\zeta\alz} - 2i\lder_\tau\sqrt{V'\tau}\sinh\zeta
\end{align}
The terms $\propto \sinh\zeta$ cancel in \eq{e22.1}. 
Expressing $E-V\pm P\alz = \sqrt{V'\tau}\exp(\pm\zeta\alz)$, commuting the $\exp(\pm\zeta\alz/2)$ factors using 
$i\ntr\cdot\atr\,\exp(-\zeta\alz/2) = \exp(\zeta\alz/2)i\ntr\cdot\atr$ (since $\ntr\zeta=0$) and similarly for the $m\gz$ terms we get,
\begin{align} \label{e22.4}
e^{\zeta\alz/2}\,B\,e^{-\zeta\alz/2} &= e^{\zeta\alz}\big[-2\sqrt{V'\tau}\,i\rder_\tau\alz+i\rnab_\perp\cdot\atr-\halft\sqrt{V'\tau}+m\gz\big]\Phi^{(0)}(\tau) \nn\crt
&+ \Phi^{(0)}(\tau) \big[-2\,i\lder_\tau\alz\sqrt{V'\tau}+i\lnab_\perp\cdot\atr-\halft\sqrt{V'\tau}-m\gz\big] e^{-\zeta\alz}
\end{align} 
The terms in [\ ] depend only on $\tau$, \ie, they are as in the $\zeta=0$ BSE of $\Phi^{(0)}(\tau)$. Expressing $\exp(\pm\zeta\alz) = \cosh\zeta \pm \alz\sinh\zeta$ the coefficent of $\cosh\zeta$ is the rest frame BSE, which $\Phi^{(0)}(\tau)$ satisfies by definition. Consequently the two terms in [\ ] give equal and opposite contributions to the $\zeta=0$ BSE. Using this leaves an anticommutator with $\alz$. Transforming back $\tau\to z$ at $\zeta=0$ allows to identify the $\rh_-$ operator \eq{qcd45}, 
\begin{align} \label{e22.5}
e^{\zeta\alz/2}\,B\,e^{-\zeta\alz/2} &= \sinh\zeta\acomb{\alz}{\big(-2\sqrt{V'\tau}\,i\rder_\tau\alz+i\rnab_\perp\cdot\atr-\halft\sqrt{V'\tau}+m\gz\big)\Phi^{(0)}(\tau)} \crt
&= \sinh\zeta\acomb{\alz}{\Big[i\rnab\cdot\alv -\halft\big(M-V)+m\gz\Big]\Phi^{(0)}(0,0,z)}= \halft (M-V)\sinh\zeta\acomb{\alz}{\rh_-\Phi^{(0)}(0,0,z)} \nn
\end{align}
The expressions in \eq{qcd61}, \eq{qcd72} and \eq{qcd81} show that $\acomb{\alz}{\rh_-\Phi^{(0)}(0,0,z)}=0$ for all wave functions. Hence $B=0$ and $\Phip(\tau)$ given by \eq{qcd108} solves the BSE for all $P$ at $\xtr=0$.

%%%%%%%%%%%
\subsection{Prove the orthogonality relation \eq{qcd110} for states with wave functions satisfying the BSE \eq{qcd98}.} \label{e23}

I follow the proof presented in \cite{Dietrich:2012un}.
From the expression \eq{qcd97} for the states in terms of their wave functions,
\begin{align} \label{e23.1}
\bra{M_B,\Pv_B}M_A,\Pv_A\rangle &= \int d\xv_{1B}d\xv_{2B}d\xv_{1A}d\xv_{2A}\bra{0}\psi^\dag(\xv_{2B})e^{-i\Pv_B\cdot(\xv_{1B}+\xv_{2B})/2}\Phi_B^{(\Pv_B)\dag}(\xv_{1B}-\xv_{2B})\gz\psi(\xv_{1B}) \nn\crt
&\times \bar\psi(\xv_{1A})e^{i\Pv_A\cdot(\xv_{1A}+\xv_{2A})/2}\Phi_A^{(\Pv_A)}(\xv_{1A}-\xv_{2A})\psi(\xv_{1A})\ket{0}
\end{align}
The field contractions set $\xv_{1A}=\xv_{1B} \equiv \xv_{1}$ and $\xv_{2A}=\xv_{2B} \equiv \xv_{2}$. Then $\int d\xv_{1}d\xv_{2} = \int d[(\xv_{1}+\xv_{2})/2]d(\xv_{1}-\xv_{2})$ and the integral over $\xv_{1}+\xv_{2}$ sets $\Pv_A=\Pv_B\equiv \Pv$,
\begin{align} \label{e23.2}
\bra{M_B,\Pv_B}M_A,\Pv_A\rangle &= \int d\xv_{1}d\xv_{2}\,e^{i(\Pv_A-\Pv_B)\cdot(\xv_{1}+\xv_{2})/2}\tr\Big[\Phi_B^{(\Pv_B)\dag}(\xv_{1}-\xv_{2})\Phi_A^{(\Pv_A)}(\xv_{1}-\xv_{2})\Big] \nn\crt
&= (2\pi)^3\delta^3(\Pv_A-\Pv_B)\int d\xv\,\tr\Big[\Phi_B^{(\Pv)\dag}(\xv)\Phi_A^{(\Pv)}(\xv)\Big]
\end{align}
The BSE \eq{qcd98} for $\Phi_A^{(\Pv)}$ and $\Phi_B^{(\Pv)\dag}$ are
\begin{align} \label{e23.3}
i\nv\cdot\acomb{\alv}{\Phi_A^{(\Pv)}(\xv)}-\halft \Pv\cdot \comb{\alv}{\Phi_A^{(\Pv)}(\xv)}+m\comb{\gz}{\Phi_A^{(\Pv)}(\xv)} &= \big[E_A-V(\xv)\big]\Phi_A^{(\Pv)}(\xv) \nn\crt
-i\nv\cdot\acomb{\alv}{\Phi_B^{(\Pv)\dag}(\xv)}+\halft \Pv\cdot \comb{\alv}{\Phi_B^{(\Pv)\dag}(\xv)}-m\comb{\gz}{\Phi_B^{(\Pv)\dag}(\xv)} &= \big[E_B-V(\xv)\big]\Phi_B^{(\Pv)\dag}(\xv)
\end{align}
Multiplying the first equation by $\Phi_B^{(P)\dag}(\xv)$ from the left, the second by $\Phip_A(\xv)$ from the right and taking the trace of their difference the terms $\propto \halft\Pv$, $m$ and $V$ cancel, giving
\begin{align} \label{e23.4}
2i\tr\Big[\alv\cdot\nv\acomb{\Phi_B^{(\Pv)\dag}(\xv)}{\Phi_A^{(\Pv)}(\xv)}\Big]= \big[E_A-E_B\big)\tr\Big[\Phi_B^{(\Pv)\dag}(\xv)\Phi_A^{(\Pv)}(\xv)\Big]
\end{align}
Integrating both sides $\int_{-\infty}^\infty d\xv$ the lhs. vanishes due to the substitution at (one component of) $\xv=\pm\infty$. The vanishing of the rhs. implies (for $M_A \neq M_B$) the orthogonality of the states according to \eq{e23.2}.

%%%%%%%%%%%
\subsection{Verify the expression \eq{qcd113} for the global norm of $\Phi_{-+}(\xv)$ in terms of $F_1(r)$.} \label{e24}

According to \eq{e23.2} the bound state norm is proportional to the trace on the lhs. of \eq{qcd113}. 
The expression \eq{qcd60} of $\Phi_{-+}(\xv)$ implies
\begin{align} \label{e24.1}
N &\equiv \int d\xv\,\tr\big\{\Phi_{-+}^\dag(\xv)\Phi_{-+}(\xv)\big\} \nn \crt
&= \int dr\,r^2d\Omega\,\tr\Big\{Y_{j\lm}^*(\Omega)F_1^*(r)\gf\Big[(-i\alv\cdot\lnab+m\gz)\frac{2}{M-V}+1\Big]\Big[\frac{2}{M-V}(i\alv\cdot\rnab+m\gz)+1\Big] \gf F_1(r) Y_{j\lm}(\Omega)\Big\} \nn\crt
&= 4\int dr\,r^2d\Omega\,Y_{j\lm}^*F_1^* \Big[\lder_j\frac{4}{(M-V)^2}\rder_j+\frac{4m^2}{(M-V)^2}+1\Big] F_1 Y_{j\lm} \nn\crt
&= 4\int dr\,r^2d\Omega\,Y_{j\lm}^*F_1^* \Big[-\frac{4}{(M-V)^2}\rnab^2-\frac{8V'}{(M-V)^3}\partial_r+\frac{4m^2}{(M-V)^2}+1\Big] F_1 Y_{j\lm} 
\end{align}
Expressing $\rnab^2$ in spherical coordinates and using the radial equation \eq{qcd59},
\begin{align} \label{e24.2}
\rnab^2 F_1(r) Y_{j\lm}(\Omega) = \Big[F''_1+\frac{2}{r}F'_1-\frac{j(j+1)}{r^2}F_1\Big]Y_{j\lm}
= \Big[-\frac{V'}{M-V}F'_1-\quart(M-V)^2F_1+m^2F_1\Big]Y_{j\lm}
\end{align}
Substituting this into \eq{e24.1} and using $\int d\Omega\, |Y_{j\lm}(\Omega)|^2 = 1$ gives \eq{qcd113}.

%%%%%%%%%%%
\subsection{Verify the expressions \eq{qcd156} for radial functions $H_1(r),H_2(r)$ and $H_3(r)$.} \label{e25}

The structure \eq{qcd145} of the $J^{PC}=0^{++}$ wave function,
\begin{align} \label{e25.1}
\Phi_\sigma(\xv) = H_1(r) + i\,\alv\cdot\hat\xv\,H_2(r) + i\,\gz\alv\cdot\hat\xv\,H_3(r)
\end{align}
follows from its parity and charge conjugation quantum numbers, as listed in \eq{qcd55}. The three Dirac structures involving the orbital angular momentum operator $\Lv$ do not contribute for $j=0$, since $\Lv$ acts on $Y_{00}$ in \eq{qcd48}, which has no angular dependence. The BSE \eq{qcd143} involves the (anti)commutators
\begin{align} \label{e25.2}
\halft\acom{\alv}{\Phi_\sigma(\xv)} &= \alv H_1(r) + i\hat\xv H_2(r) +\gz\hat\xv\times\alv\,\gf H_3(r) \nn\crt
\halft\com{\gz}{\Phi_\sigma(\xv)} &= i\gz\alv\cdot\hat\xv H_2(r) + i\alv\cdot\hat\xv H_3(r)
\end{align}
Inserting the expression \eq{e25.1} into the BSE \eq{qcd143} with $M=0$ gives, using $\nv f(r)= \hat\xv f'(r)$, $\partial_i x^j = \delta_{ij}$ and $\nv\cdot\hat\xv\times\alv\, f(r)=0$,
\begin{align} \label{e25.3}
i\,\alv\cdot\hat\xv H_1' -\frac{2}{r}H_2-H_2' +im\gz\alv\cdot\hat\xv H_2 + im\alv\cdot\hat\xv H_3 +\halft V'r\,\Phi_\sigma(\xv) =0
\end{align}
The coefficients of the Dirac structures $1,\ i\alv\cdot\hat\xv$ and $i\gz\alv\cdot\hat\xv$ impose
\begin{align} \label{e25.4}
-\frac{2}{r}H_2-H_2'+\halft V'r\,H_1 &= 0\nn\crt
H_1'+mH_3+ \halft V'r\,H_2 &= 0\nn\crt
mH_2+\halft V'r\,H_3 &= 0
\end{align}
The first and third equations allow to express $H_1$ and $H_3$, respectively, in terms of $H_2$. Substituting them into the second equation gives the differential equation
\begin{align} \label{e25.5}
H_2'' + \inv{r}H_2' +\Big[\inv{4}(V'r)^2-m^2-\frac{4}{r^2}\Big]H_2 = 0
\end{align}
It is straightforward to verify that the expressions for $H_i(r)$ given in \eq{qcd156} satisfy the above equations. Their properties $H_1(r\to 0) \sim r^{0}$, $H_2(r\to 0) \sim r^{2}$ and $H_3(r\to 0) \sim r^{1}$ ensure that the wave function is locally normalizable at $r=0$.

\bibliography{210412_refs} 
\end{document}